PIERRE AUGER OBSERVATORY

# The Pierre Auger Observatory: Contributions to the 34th International Cosmic Ray Conference (ICRC 2015)

## The Pierre Auger Collaboration

A. Aab[41], P. Abreu[65], M. Aglietta[52], E.J. Ahn[80], I. Al Samarai[28], I.F.M. Albuquerque[16], I. Allekotte[1], P. Allison[85], A. Almela[11,8], J. Alvarez Castillo[58], J. Alvarez-Muñiz[75], R. Alves Batista[40], M. Ambrosio[43], A. Aminaei[59], G.A. Anastasi[45], L. Anchordoqui[79], S. Andringa[65], C. Aramo[43], F. Arqueros[72], N. Arsene[68], H. Asorey[1,24], P. Assis[65], J. Aublin[30], G. Avila[10], N. Awal[83], A.M. Badescu[69], C. Baus[35], J.J. Beatty[85], K.H. Becker[34], J.A. Bellido[12], C. Berat[31], M.E. Bertaina[52], X. Bertou[1], P.L. Biermann[38], P. Billoir[30], S.G. Blaess[12], A. Blanco[65], M. Blanco[30], J. Blazek[26], C. Bleve[47], H. Blümer[35,36], M. Boháčová[26], D. Boncioli[51], C. Bonifazi[22], N. Borodai[63], J. Brack[78], I. Brancus[66], T. Bretz[39], A. Bridgeman[36], P. Brogueira[65], P. Buchholz[41], A. Bueno[74], S. Buitink[59], M. Buscemi[43], K.S. Caballero-Mora[56], B. Caccianiga[42], L. Caccianiga[30], M. Candusso[44], L. Caramete[67], R. Caruso[45], A. Castellina[52], G. Cataldi[47], L. Cazon[65], R. Cester[46], A.G. Chavez[57], A. Chiavassa[52], J.A. Chinellato[17], J. Chudoba[26], M. Cilmo[43], R.W. Clay[12], G. Cocciolo[47], R. Colalillo[43], A. Coleman[86], L. Collica[42], M.R. Coluccia[47], R. Conceição[65], F. Contreras[9], M.J. Cooper[12], A. Cordier[29], S. Coutu[86], C.E. Covault[76], J. Cronin[87], R. Dallier[33,32], B. Daniel[17], S. Dasso[5,3], K. Daumiller[36], B.R. Dawson[12], R.M. de Almeida[23], S.J. de Jong[59,61], G. De Mauro[59], J.R.T. de Mello Neto[22], I. De Mitri[47], J. de Oliveira[23], V. de Souza[15], L. del Peral[73], O. Deligny[28], N. Dhital[82], C. Di Giulio[44], A. Di Matteo[48], J.C. Diaz[82], M.L. Díaz Castro[17], F. Diogo[65], C. Dobrigkeit[17], W. Docters[60], J.C. D'Olivo[58], A. Dorofeev[78], Q. Dorosti Hasankiadeh[36], R.C. dos Anjos[15], M.T. Dova[4], J. Ebr[26], R. Engel[36], M. Erdmann[39], M. Erfani[41], C.O. Escobar[80,17], J. Espadanal[65], A. Etchegoyen[8,11], H. Falcke[59,62,61], K. Fang[87], G. Farrar[83], A.C. Fauth[17], N. Fazzini[80], A.P. Ferguson[76], B. Fick[82], J.M. Figueira[8], A. Filevich[8], A. Filipčič[70,71], O. Fratu[69], M.M. Freire[6], T. Fujii[87], B. García[7], D. García-Gámez[29], D. Garcia-Pinto[72], F. Gate[33], H. Gemmeke[37], A. Gherghel-Lascu[66], P.L. Ghia[30], U. Giaccari[22], M. Giammarchi[42], M. Giller[64], D. Głas[64], C. Glaser[39], H. Glass[80], G. Golup[1], M. Gómez Berisso[1], P.F. Gómez Vitale[10], N. González[8], B. Gookin[78], J. Gordon[85], A. Gorgi[52], P. Gorham[88], P. Gouffon[16], N. Griffith[85], A.F. Grillo[51], T.D. Grubb[12], F. Guarino[43], G.P. Guedes[18], M.R. Hampel[8], P. Hansen[4], D. Harari[1], T.A. Harrison[12], S. Hartmann[39], J.L. Harton[78], A. Haungs[36], T. Hebbeker[39], D. Heck[36], P. Heimann[41], A.E. Hérve[36], G.C. Hill[12], C. Hojvat[80], N. Hollon[87], E. Holt[36], P. Homola[34], J.R. Hörandel[59,61], P. Horvath[27], M. Hrabovský[27,26], D. Huber[35], T. Huege[36], A. Insolia[45], P.G. Isar[67], I. Jandt[34], S. Jansen[59,61], C. Jarne[4], J.A. Johnsen[77], M. Josebachuili[8], A. Kääpä[34], O. Kambeitz[35], K.H. Kampert[34], P. Kasper[80], I. Katkov[35], B. Keilhauer[36], E. Kemp[17], R.M. Kieckhafer[82], H.O. Klages[36], M. Kleifges[37], J. Kleinfeller[9], R. Krause[39], N. Krohm[34], D. Kuempel[39], G. Kukec Mezek[71], N. Kunka[37], A.W. Kuotb Awad[36], D. LaHurd[76], L. Latronico[52], R. Lauer[90], M. Lauscher[39], P. Lautridou[33], S. Le Coz[31], D. Lebrun[31], P. Lebrun[80], M.A. Leigui de Oliveira[21], A. Letessier-Selvon[30], I. Lhenry-Yvon[28], K. Link[35], L. Lopes[65], R. López[53], A. López Casado[75], K. Louedec[31], A. Lucero[8], M. Malacari[12], M. Mallamaci[42], J. Maller[33], D. Mandat[26], P. Mantsch[80], A.G. Mariazzi[4], V. Marin[33], I.C. Mariş[74], G. Marsella[47], D. Martello[47], H. Martinez[54], O. Martínez Bravo[53], D. Martraire[28], J.J. Masías Meza[3], H.J. Mathes[36], S. Mathys[34], J. Matthews[81], J.A.J. Matthews[90], G. Matthiae[44], D. Maurizio[13], E. Mayotte[77], P.O. Mazur[80], C. Medina[77], G. Medina-Tanco[58], R. Meissner[39], V.B.B. Mello[22], D. Melo[8], A. Menshikov[37], S. Messina[60], M.I. Micheletti[6], L. Middendorf[39], I.A. Minaya[72], L. Miramonti[42], B. Mitrica[66], L. Molina-Bueno[74], S. Mollerach[1], F. Montanet[31], C. Morello[52], M. Mostafá[86], C.A. Moura[21], G. Müller[39], M.A. Muller[17,20], S. Müller[36], S. Navas[74], P. Necesal[26],

L. Nellen[58], A. Nelles[59,61], J. Neuser[34], P.H. Nguyen[12], M. Niculescu-Oglinzanu[66], M. Niechciol[41], L. Niemietz[34], T. Niggemann[39], D. Nitz[82], D. Nosek[25], V. Novotny[25], L. Nožka[27], L.A. Núñez[24], L. Ochilo[41], F. Oikonomou[86], A. Olinto[87], N. Pacheco[73], D. Pakk Selmi-Dei[17], M. Palatka[26], J. Pallotta[2], P. Papenbreer[34], G. Parente[75], A. Parra[53], T. Paul[79,84], M. Pech[26], J. Pękala[63], R. Pelayo[55], I.M. Pepe[19], L. Perrone[47], E. Petermann[89], C. Peters[39], S. Petrera[48,49], Y. Petrov[78], J. Phuntsok[86], R. Piegaia[3], T. Pierog[36], P. Pieroni[3], M. Pimenta[65], V. Pirronello[45], M. Platino[8], M. Plum[39], A. Porcelli[36], C. Porowski[63], R.R. Prado[15], P. Privitera[87], M. Prouza[26], E.J. Quel[2], S. Querchfeld[34], S. Quinn[76], J. Rautenberg[34], O. Ravel[33], D. Ravignani[8], D. Reinert[39], B. Revenu[33], J. Ridky[26], M. Risse[41], P. Ristori[2], V. Rizi[48], W. Rodrigues de Carvalho[75], J. Rodriguez Rojo[9], M.D. Rodríguez-Frías[73], D. Rogozin[36], J. Rosado[72], M. Roth[36], E. Roulet[1], A.C. Rovero[5], S.J. Saffi[12], A. Saftoiu[66], F. Salamida[28,c], H. Salazar[53], A. Saleh[71], F. Salesa Greus[86], G. Salina[44], J.D. Sanabria Gomez[24], F. Sánchez[8], P. Sanchez-Lucas[74], E.M. Santos[16], E. Santos[17], F. Sarazin[77], B. Sarkar[34], R. Sarmento[65], C. Sarmiento-Cano[24], R. Sato[9], C. Scarso[9], M. Schauer[34], V. Scherini[47], H. Schieler[36], D. Schmidt[36], O. Scholten[60,b], H. Schoorlemmer[88], P. Schovánek[26], F.G. Schröder[36], A. Schulz[36], J. Schulz[59], J. Schumacher[39], S.J. Sciutto[4], A. Segreto[50], M. Settimo[30], A. Shadkam[81], R.C. Shellard[13], G. Sigl[40], O. Sima[68], A. Śmiałkowski[64], R. Šmída[36], G.R. Snow[89], P. Sommers[86], S. Sonntag[41], J. Sorokin[12], R. Squartini[9], Y.N. Srivastava[84], D. Stanca[66], S. Stanič[71], J. Stapleton[85], J. Stasielak[63], M. Stephan[39], A. Stutz[31], F. Suarez[8,11], M. Suarez Durán[24], T. Suomijärvi[28], A.D. Supanitsky[5], M.S. Sutherland[85], J. Swain[84], Z. Szadkowski[64], O.A. Taborda[1], A. Tapia[8], A. Tepe[41], V.M. Theodoro[17], O. Tibolla[56], C. Timmermans[59,61], C.J. Todero Peixoto[14], G. Toma[66], L. Tomankova[36], B. Tomé[65], A. Tonachini[46], G. Torralba Elipe[75], D. Torres Machado[22], P. Travnicek[26], M. Trini[71], R. Ulrich[36], M. Unger[83,36], M. Urban[39], J.F. Valdés Galicia[58], I. Valiño[75], L. Valore[43], G. van Aar[59], P. van Bodegom[12], A.M. van den Berg[60], S. van Velzen[59], A. van Vliet[40], E. Varela[53], B. Vargas Cárdenas[58], G. Varner[88], R. Vasquez[22], J.R. Vázquez[72], R.A. Vázquez[75], D. Veberič[36], V. Verzi[44], J. Vicha[26], M. Videla[8], L. Villaseñor[57], B. Vlcek[73], S. Vorobiov[71], H. Wahlberg[4], O. Wainberg[8,11], D. Walz[39], A.A. Watson[a], M. Weber[37], K. Weidenhaupt[39], A. Weindl[36], C. Welling[39], F. Werner[35], A. Widom[84], L. Wiencke[77], H. Wilczyński[63], T. Winchen[34], D. Wittkowski[34], B. Wundheiler[8], S. Wykes[59], L. Yang[71], T. Yapici[82], A. Yushkov[41], E. Zas[75], D. Zavrtanik[71,70], M. Zavrtanik[70,71], A. Zepeda[54], B. Zimmermann[37], M. Ziolkowski[41], F. Zuccarello[45]

—————•—————

[1] Centro Atómico Bariloche and Instituto Balseiro (CNEA-UNCuyo-CONICET), San Carlos de Bariloche, Argentina
[2] Centro de Investigaciones en Láseres y Aplicaciones, CITEDEF and CONICET, Villa Martelli, Argentina
[3] Departamento de Física, FCEyN, Universidad de Buenos Aires and CONICET, Buenos Aires, Argentina
[4] IFLP, Universidad Nacional de La Plata and CONICET, La Plata, Argentina
[5] Instituto de Astronomía y Física del Espacio (IAFE, CONICET-UBA), Buenos Aires, Argentina
[6] Instituto de Física de Rosario (IFIR) – CONICET/U.N.R. and Facultad de Ciencias Bioquímicas y Farmacéuticas U.N.R., Rosario, Argentina
[7] Instituto de Tecnologías en Detección y Astropartículas (CNEA, CONICET, UNSAM), and Universidad Tecnológica Nacional – Facultad Regional Mendoza (CONICET/CNEA), Mendoza, Argentina
[8] Instituto de Tecnologías en Detección y Astropartículas (CNEA, CONICET, UNSAM), Buenos Aires, Argentina
[9] Observatorio Pierre Auger, Malargüe, Argentina
[10] Observatorio Pierre Auger and Comisión Nacional de Energía Atómica, Malargüe, Argentina
[11] Universidad Tecnológica Nacional – Facultad Regional Buenos Aires, Buenos Aires, Argentina
[12] University of Adelaide, Adelaide, S.A., Australia
[13] Centro Brasileiro de Pesquisas Fisicas, Rio de Janeiro, RJ, Brazil

[14] Universidade de São Paulo, Escola de Engenharia de Lorena, Lorena, SP, Brazil
[15] Universidade de São Paulo, Instituto de Física de São Carlos, São Carlos, SP, Brazil
[16] Universidade de São Paulo, Instituto de Física, São Paulo, SP, Brazil
[17] Universidade Estadual de Campinas, IFGW, Campinas, SP, Brazil
[18] Universidade Estadual de Feira de Santana, Feira de Santana, Brazil
[19] Universidade Federal da Bahia, Salvador, BA, Brazil
[20] Universidade Federal de Pelotas, Pelotas, RS, Brazil
[21] Universidade Federal do ABC, Santo André, SP, Brazil
[22] Universidade Federal do Rio de Janeiro, Instituto de Física, Rio de Janeiro, RJ, Brazil
[23] Universidade Federal Fluminense, EEIMVR, Volta Redonda, RJ, Brazil
[24] Universidad Industrial de Santander, Bucaramanga, Colombia
[25] Charles University, Faculty of Mathematics and Physics, Institute of Particle and Nuclear Physics, Prague, Czech Republic
[26] Institute of Physics of the Academy of Sciences of the Czech Republic, Prague, Czech Republic
[27] Palacky University, RCPTM, Olomouc, Czech Republic
[28] Institut de Physique Nucléaire d'Orsay (IPNO), Université Paris 11, CNRS-IN2P3, Orsay, France
[29] Laboratoire de l'Accélérateur Linéaire (LAL), Université Paris 11, CNRS-IN2P3, Orsay, France
[30] Laboratoire de Physique Nucléaire et de Hautes Energies (LPNHE), Universités Paris 6 et Paris 7, CNRS-IN2P3, Paris, France
[31] Laboratoire de Physique Subatomique et de Cosmologie (LPSC), Université Grenoble-Alpes, CNRS/IN2P3, Grenoble, France
[32] Station de Radioastronomie de Nançay, Observatoire de Paris, CNRS/INSU, Nançay, France
[33] SUBATECH, École des Mines de Nantes, CNRS-IN2P3, Université de Nantes, Nantes, France
[34] Bergische Universität Wuppertal, Fachbereich C – Physik, Wuppertal, Germany
[35] Karlsruhe Institute of Technology – Campus South – Institut für Experimentelle Kernphysik (IEKP), Karlsruhe, Germany
[36] Karlsruhe Institute of Technology – Campus North – Institut für Kernphysik, Karlsruhe, Germany
[37] Karlsruhe Institute of Technology – Campus North – Institut für Prozessdatenverarbeitung und Elektronik, Karlsruhe, Germany
[38] Max-Planck-Institut für Radioastronomie, Bonn, Germany
[39] RWTH Aachen University, III. Physikalisches Institut A, Aachen, Germany
[40] Universität Hamburg, II. Institut für Theoretische Physik, Hamburg, Germany
[41] Universität Siegen, Fachbereich 7 Physik – Experimentelle Teilchenphysik, Siegen, Germany
[42] Università di Milano and Sezione INFN, Milan, Italy
[43] Università di Napoli "Federico II" and Sezione INFN, Napoli, Italy
[44] Università di Roma II "Tor Vergata" and Sezione INFN, Roma, Italy
[45] Università di Catania and Sezione INFN, Catania, Italy
[46] Università di Torino and Sezione INFN, Torino, Italy
[47] Dipartimento di Matematica e Fisica "E. De Giorgi" dell'Università del Salento and Sezione INFN, Lecce, Italy
[48] Dipartimento di Scienze Fisiche e Chimiche dell'Università dell'Aquila and Sezione INFN, L'Aquila, Italy
[49] Gran Sasso Science Institute (INFN), L'Aquila, Italy
[50] Istituto di Astrofisica Spaziale e Fisica Cosmica di Palermo (INAF), Palermo, Italy
[51] INFN, Laboratori Nazionali del Gran Sasso, Assergi (L'Aquila), Italy
[52] Osservatorio Astrofisico di Torino (INAF), Università di Torino and Sezione INFN, Torino, Italy
[53] Benemérita Universidad Autónoma de Puebla, Puebla, México
[54] Centro de Investigación y de Estudios Avanzados del IPN (CINVESTAV), México, D.F., México
[55] Unidad Profesional Interdisciplinaria en Ingeniería y Tecnologías Avanzadas del Instituto Politécnico Nacional (UPIITA-IPN), México, D.F., México
[56] Universidad Autónoma de Chiapas, Tuxtla Gutiérrez, Chiapas, México
[57] Universidad Michoacana de San Nicolás de Hidalgo, Morelia, Michoacán, México
[58] Universidad Nacional Autónoma de México, México, D.F., México

[59] IMAPP, Radboud University Nijmegen, Nijmegen, Netherlands
[60] KVI – Center for Advanced Radiation Technology, University of Groningen, Groningen, Netherlands
[61] Nikhef, Science Park, Amsterdam, Netherlands
[62] ASTRON, Dwingeloo, Netherlands
[63] Institute of Nuclear Physics PAN, Krakow, Poland
[64] University of Łódź, Łódź, Poland
[65] Laboratório de Instrumentação e Física Experimental de Partículas (LIP) and Instituto Superior Técnico, Universidade de Lisboa (UL), Portugal
[66] "Horia Hulubei" National Institute for Physics and Nuclear Engineering, Bucharest-Magurele, Romania
[67] Institute of Space Science, Bucharest-Magurele, Romania
[68] University of Bucharest, Physics Department, Bucharest, Romania
[69] University Politehnica of Bucharest, Bucharest, Romania
[70] Experimental Particle Physics Department, J. Stefan Institute, Ljubljana, Slovenia
[71] Laboratory for Astroparticle Physics, University of Nova Gorica, Nova Gorica, Slovenia
[72] Universidad Complutense de Madrid, Madrid, Spain
[73] Universidad de Alcalá, Alcalá de Henares, Madrid, Spain
[74] Universidad de Granada and C.A.F.P.E., Granada, Spain
[75] Universidad de Santiago de Compostela, Santiago de Compostela, Spain
[76] Case Western Reserve University, Cleveland, OH, USA
[77] Colorado School of Mines, Golden, CO, USA
[78] Colorado State University, Fort Collins, CO, USA
[79] Department of Physics and Astronomy, Lehman College, City University of New York, Bronx, NY, USA
[80] Fermilab, Batavia, IL, USA
[81] Louisiana State University, Baton Rouge, LA, USA
[82] Michigan Technological University, Houghton, MI, USA
[83] New York University, New York, NY, USA
[84] Northeastern University, Boston, MA, USA
[85] Ohio State University, Columbus, OH, USA
[86] Pennsylvania State University, University Park, PA, USA
[87] University of Chicago, Enrico Fermi Institute, Chicago, IL, USA
[88] University of Hawaii, Honolulu, HI, USA
[89] University of Nebraska, Lincoln, NE, USA
[90] University of New Mexico, Albuquerque, NM, USA
[a] School of Physics and Astronomy, University of Leeds, Leeds, United Kingdom
[b] Also at Vrije Universiteit Brussels, Brussels, Belgium
[c] Currently at INFN Milano Bicocca, Milano, Italy

# Acknowledgments

The successful installation, commissioning, and operation of the Pierre Auger Observatory would not have been possible without the strong commitment and effort from the technical and administrative staff in Malargüe. We are very grateful to the following agencies and organizations for financial support:

Comisión Nacional de Energía Atómica, Agencia Nacional de Promoción Científica y Tecnológica (ANPCyT), Consejo Nacional de Investigaciones Científicas y Técnicas (CONICET), Gobierno de la Provincia de Mendoza, Municipalidad de Malargüe, NDM Holdings and Valle Las Leñas, in gratitude for their continuing cooperation over land access, Argentina; the Australian Research Council; Conselho Nacional de Desenvolvimento Científico e Tecnológico (CNPq), Financiadora de Estudos e Projetos (FINEP), Fundação de Amparo à Pesquisa do Estado de Rio de Janeiro (FAPERJ), São Paulo Research Foundation (FAPESP) Grants No. 2010/07359-6 and No. 1999/05404-3, Ministério de Ciência e Tecnologia (MCT), Brazil; Grant No. MSMT-CR LG13007, No. 7AMB14AR005, and the Czech Science Foundation Grant No. 14-17501S, Czech Republic; Centre de Calcul IN2P3/CNRS, Centre National de la Recherche Scientifique (CNRS), Conseil Régional Ile-de-France, Département Physique Nucléaire et Corpusculaire (PNC-IN2P3/CNRS), Département Sciences de l'Univers (SDU-INSU/CNRS), Institut Lagrange de Paris (ILP) Grant No. LABEX ANR-10-LABX-63, within the Investissements d'Avenir Programme Grant No. ANR-11-IDEX-0004-02, France; Bundesministerium für Bildung und Forschung (BMBF), Deutsche Forschungsgemeinschaft (DFG), Finanzministerium Baden-Württemberg, Helmholtz Alliance for Astroparticle Physics (HAP), Helmholtz-Gemeinschaft Deutscher Forschungszentren (HGF), Ministerium für Wissenschaft und Forschung, Nordrhein Westfalen, Ministerium für Wissenschaft, Forschung und Kunst, Baden-Württemberg, Germany; Istituto Nazionale di Fisica Nucleare (INFN), Istituto Nazionale di Astrofisica (INAF), Ministero dell'Istruzione, dell'Universitá e della Ricerca (MIUR), Gran Sasso Center for Astroparticle Physics (CFA), CETEMPS Center of Excellence, Ministero degli Affari Esteri (MAE), Italy; Consejo Nacional de Ciencia y Tecnología (CONACYT), Mexico; Ministerie van Onderwijs, Cultuur en Wetenschap, Nederlandse Organisatie voor Wetenschappelijk Onderzoek (NWO), Stichting voor Fundamenteel Onderzoek der Materie (FOM), Netherlands; National Centre for Research and Development, Grants No. ERA-NET-ASPERA/01/11 and No. ERA-NET-ASPERA/02/11, National Science Centre, Grants No. 2013/08/M/ST9/00322, No. 2013/08/M/ST9/00728 and No. HARMONIA 5 - 2013/10/M/ST9/00062, Poland; Portuguese national funds and FEDER funds within Programa Operacional Factores de Competitividade through Fundação para a Ciência e a Tecnologia (COMPETE), Portugal; Romanian Authority for Scientific Research ANCS, CNDI-UEFISCDI partnership projects Grants No. 20/2012 and No. 194/2012, Grants No. 1/ASPERA2/2012 ERA-NET, No. PN-II-RU-PD-2011-3-0145-17 and No. PN-II-RU-PD-2011-3-0062, the Minister of National Education, Programme Space Technology and Advanced Research (STAR), Grant No. 83/2013, Romania; Slovenian Research Agency, Slovenia; Comunidad de Madrid, FEDER funds, Ministerio de Educación y Ciencia, Xunta de Galicia, European Community 7th Framework Program, Grant No. FP7-PEOPLE-2012-IEF-328826, Spain; Science and Technology Facilities Council, United Kingdom; Department of Energy, Contracts No. DE-AC02-07CH11359, No. DE-FR02-04ER41300, No. DE-FG02-99ER41107 and No. DE-SC0011689, National Science Foundation, Grant No. 0450696, The Grainger Foundation, USA; NAFOSTED, Vietnam; Marie Curie-IRSES/EPLANET, European Particle Physics Latin American Network, European Union 7th Framework Program, Grant No. PIRSES-2009-GA-246806 and PIOF-GA-2013-624803; and UNESCO.

# Contributions

# 1

# Energy Spectrum

PIERRE
AUGER
OBSERVATORY

# The flux of ultra-high energy cosmic rays after ten years of operation of the Pierre Auger Observatory

**Inés Valiño**[*a] **for the Pierre Auger Collaboration**[b]
[a] *Departamento de Física de Partículas e IGFAE, Universidade de Santiago de Compostela, Spain*
[b] *Observatorio Pierre Auger, Av. San Martín Norte 304, 5613 Malargüe, Argentina*
*E-mail:* auger_spokespersons@fnal.gov
*Full author list:* http://www.auger.org/archive/authors_2015_06.html

The flux of ultra-high energy cosmic rays has been measured with unprecedented precision at the Pierre Auger Observatory. We report an update of the all-sky flux of cosmic rays above $3\times10^{17}$ eV obtained by combining four independent data sets. These measurements are based on data from the surface detector arrays (divided into two sets according to the shower zenith angle), from a nested, denser, detector array, and hybrid events measured simultaneously with both the fluorescence detector and the surface detector array. The spectral features are presented in detail and the systematic uncertainties are addressed. The huge amount of data collected to date, with a total exposure exceeding 50,000 $\mathrm{km}^2$ sr yr, together with the wide range of sky observed (in declination from $-90^\circ$ to $45^\circ$) also allow us to measure the energy spectrum from different regions of the sky. We present the results of the search for a dependence of the measured flux on the declination of the incoming directions.

*Speaker.

## 1. Introduction

Over a century after their discovery, the measurement of the energy spectrum of ultra-high energy cosmic rays (UHECRs) remains as one of the main issues within the field, being fundamental to the unveiling of the origin of these particles and understanding their propagation. The Pierre Auger Observatory [1, 2] has collected high-quality data for more than 10 years, which has already led to a measurement of the flux of UHECRs above $3\times10^{17}$ eV with unprecedented statistics. Two relevant spectral features have been established beyond doubt: the hardening in the spectrum at about $5\times10^{18}$ eV (the *ankle*), and a strong suppression of the flux at the highest energies. The accurate measurement of the spectrum, combined with results from the study of the mass composition and of the distribution of the arrival directions of the primaries over the sky, presents a challenge for astrophysical modelling of origin and propagation of UHECRs [3].

The energy spectrum can also be exploited to study the distribution of cosmic-ray sources by searching for a flux variation with declination ($\delta$) of the incoming directions. This study is of particular interest to the discussion of the difference seen in the suppression region between the spectra measured by Auger and by the Telescope Array (TA) experiment [4], which, despite being still compatible within the quoted systematic uncertainties of both experiments, is not understood so far. We also expect to find a $\delta$-dependence of the measured flux compatible with the hint of a dipole anisotropy for cosmic rays with energies above $8\times10^{18}$ eV recently reported in [5, 6].

This paper deals with the energy spectrum of UHECRs obtained by combining the measurements of the surface detector array (SD) and the fluorescence detector (FD). The SD, spread over an area of 3000 km$^2$, is composed of a baseline array of 1600 water-Cherenkov detectors separated by 1500 m in a hexagonal grid, and a smaller nested array of 49 additional detectors spaced by 750 m covering an area of 24 km$^2$. The FD comprises 27 telescopes at 5 perimeter buildings viewing the atmosphere over the array. The hybrid technique developed exploits the large aperture of the SD, operating continuously, as well as the calorimetric measurement of the shower energy deposited in the atmosphere obtained with the FD which, by contrast, has duty cycle limited to clear moonless nights (13%). This allows energy-spectrum measurements weakly reliant upon shower simulations.

## 2. Measurements of the cosmic-ray energy

The FD allows the measurement of the electromagnetic energy released by the shower in the atmosphere as a function of the atmospheric depth, $\mathrm{d}E/\mathrm{d}X$. The total primary energy is then derived by integrating this longitudinal profile over the $X$-range and adding an estimate of the so-called "invisible energy" carried into the ground by high-energy muons and neutrinos. The shower-energy estimated with the FD, $E_{\mathrm{FD}}$, has a total systematic uncertainty of 14% [7]. The hybrid measurement is based on the selection and reconstruction of showers observed by the FD in coincidence with at least one SD station, which enables an accurate determination of the shower geometry and consequently of the energy of the primary particle. To ensure good energy reconstruction, only events that satisfy strict quality criteria are accepted [8].

The SD samples the shower particles that reach the ground. The intensities of the signals registered in the stations of the SD are used to quantify the shower size and the impact point of the shower axis on the ground. The reconstruction technique used depends upon the zenith angle ($\theta$) of the incoming direction which defines the amount of atmosphere traversed by the shower, and therefore the level of attenuation of the shower components. We distinguish between cosmic-ray showers with $\theta<60°$, defined as *vertical* events, and those with $60°<\theta<80°$, defined as *inclined*.

| | SD-1500 m | | SD-750 m | Hybrid |
|---|---|---|---|---|
| | vertical | inclined | | |
| Data-taking period | 01/2004–12/2014 | 01/2004–12/2013 | 08/2008–12/2014 | 11/2005–12/2013 |
| Exposure [km$^2$ sr yr] | 42500±1300 | 10900±300 | 150±5 | 1500±20 at $10^{19}$ eV |
| Zenith angle [deg] | 0-60 | 60-80 | 0-55 | 0-60 |
| Threshold energy | $3\times10^{18}$ eV | $4\times10^{18}$ eV | $3\times10^{17}$ eV | $10^{18}$ eV |
| Number of events | 102901 | 15614 | 61130 | 9346 |
| Number of hybrid events | 1731 | 255 | 469 | |
| Energy scale (A) | (0.1871 ± 0.004) EeV | (5.71±0.09) EeV | (12.87± 0.63) PeV | |
| Energy scale (B) | 1.023 ±0.006 | 1.01±0.02 | 1.013±0.013 | |
| Energy resolution [%] | 15.3±0.4 | 19±1 | 13±1 | |

**Table 1:** Summary of the experimental parameters describing the different data sets used to measure the energy spectrum at the Pierre Auger Observatory.

For vertical events, the energy estimator is the observed signal $S(r_{\rm opt})$ at an optimal distance $r_{\rm opt}$ from the shower axis [2]. The energy estimators are $S(1000)$ and $S(450)$ for the 1500 m and 750 m arrays respectively. For a given energy, the value of $S(r_{\rm opt})$ decreases with $\theta$, due to the attenuation of the shower particles in the atmosphere and geometrical effects. The *Constant Intensity Cut* method is used to correct the energy estimator $S(1000)$ ($S(450)$) for the $\theta$-dependence and estimate the signal $S_{38}$ ($S_{35}$) that the shower would have produced at the median zenith angle of 38° (35°). Inclined events are reconstructed using a different procedure [9] since muons dominate the SD signals, developing asymmetric footprints at ground due to the geomagnetic field. The energy estimator, $N_{19}$, is defined as the normalisation of the muon content of a particular event relative to a reference 2D muon distribution at ground, derived from simulated proton showers with an energy of $10^{19}$ eV for a given arrival direction. $N_{19}$ is thus independent of the zenith angle. To ensure a good reconstruction, only events well-contained in the SD array are selected. This fiducial trigger requires that the detector with the highest signal is enclosed in a hexagon of 6 active stations.

The absolute calibration of the SD is inferred from a high-quality subset of hybrid events used to calibrate the SD energy estimators with the calorimetric energies measured with the FD (full details in [9, 10, 11]). Only events with energies in the range of full efficiency of the SD (see values in Table 1) are used in the calibration. Here we update the SD energy scale using hybrid data up to 31 Dec 2013, increasing the data samples by about 20% with respect to those used previously. The correlations between the different SD energy estimators and $E_{\rm FD}$ are well described by a simple power-law function $E_{\rm FD} = A\,(\hat{S})^B$ with $\hat{S} = S_{38}$, $S_{35}$ or $N_{19}$. We fit this function to the selected data using a tailored maximum-likelihood method [12] (see Fig. 1). The best-fit parameters are given in Table 1. Although statistical uncertainties of the calibration constants $A$ and $B$ affect the SD energy scale, their contribution is small (at the few % level), decreasing as the number of events increases. The main contribution to the systematic uncertainty in the SD energy scale comes from the uncertainties on $E_{\rm FD}$ that are correlated between different showers. This means that the SD shares the uncertainty of the FD energy scale of 14%.

The resolution in the SD energy is computed from the distribution of the ratio $A(\hat{S})^B/E_{\mathrm{FD}}$ for the hybrid events used for the calibration, assuming a fixed FD energy resolution of 7.6%. The resulting average resolutions are reported in Table 1.

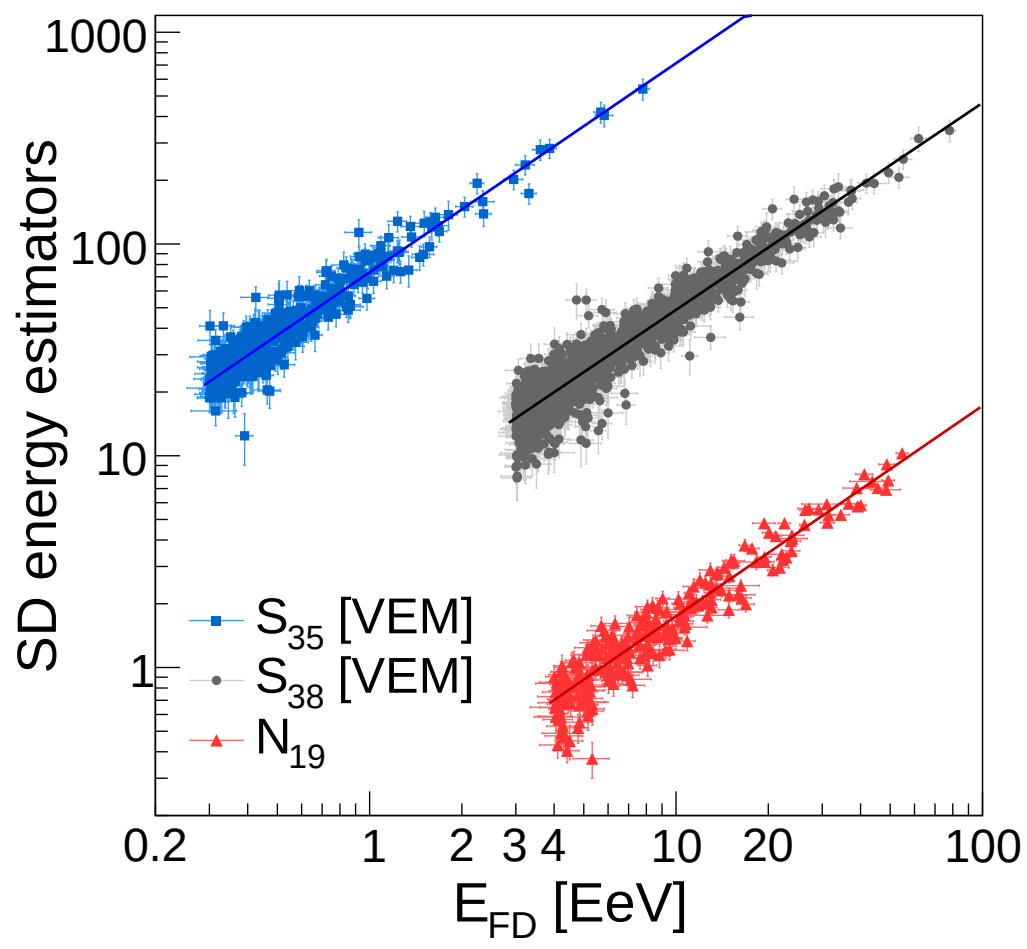

**Figure 1:** Correlation between the energy estimators (see text) and the energy FD energy. $S_{38}$ and $S_{35}$ are given in units of *Vertical Equivalent Muon* or VEM, corresponding to the signal produced by a vertical muon traversing the detector through its center. Since $N_{19}$ is a scaling factor it is dimensionless.

## 3. Energy spectrum

The final step in measuring the energy spectrum is a precise determination of the exposure for the observations. Above the energy for full detector efficiency, the calculation of the SD exposure is based solely on the determination of the geometrical aperture of the array for the corresponding zenith-angle interval and of the observation time. The choice of a fiducial trigger based on active hexagons allows one to exploit the regularity of the array, and to compute the aperture simply as the sum of the areas of all active hexagons. The calculation of the hybrid exposure is more complex. It relies on a detailed time-dependent Monte Carlo (MC) simulation which exactly reproduces the data taking conditions and includes the response of the Hybrid detector [8]. The result is an exposure growing with shower energy above the threshold energy of $10^{18}$ eV.

A correction must be applied to the measured flux to account for the effect of the finite resolution in the energy determination, responsible for bin-to-bin event migration. For a steeply-falling spectrum, upward movements of reconstructed energies into a given bin are not compensated by movements from the opposite direction. The net effect is that the measured spectrum is shifted towards higher energies with respect to the true one. For the hybrid measurement, this is corrected by calculating the exposure as a function of the reconstructed energy instead of the input energy in the MC. For the SD measurements, a forward-folding approach is applied. MC simulations are used to generate a bin-to-bin migration matrix that accounts for all the resolution effects and physical fluctuations in shower development. The matrix is then used to find a flux parameterisation that fits the measured data when forward-folded, using a binned-maximum likelihood approach assuming Poisson statistics. The forward-folded spectrum is finally divided by the input flux to obtain the correction factor which is in turn applied to the measured binned spectrum to obtain the true spectrum. This correction is slightly energy dependent but is below 15% over all of the $E$-range.

Here we present an update of the measurements of the energy spectrum derived from vertical SD data sets recorded by both the 750 m and 1500 m arrays up to 31 Dec 2014, and hybrid data up to 31 Dec 2013. Moreover, we report the spectrum derived from inclined events recorded by the 1500 m array up to 31 Dec 2013, recently published in [13]. Values of the corresponding exposures are given in Table 1, together with other experimental parameters describing the data. Note that the exposure for the vertical SD-750 m data set is double the value reported previously in [14].

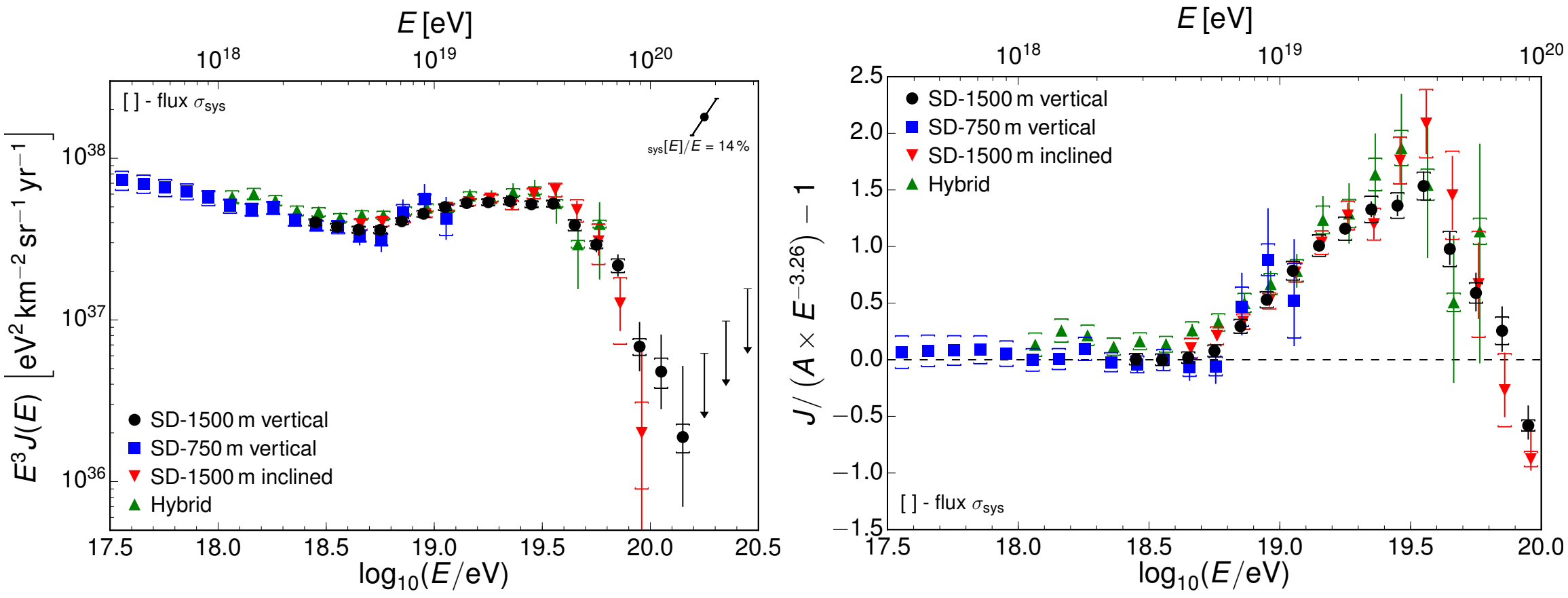

**Figure 2:** Left: energy spectra derived from SD and hybrid data recorded at the Pierre Auger Observatory. The error bars represent statistical uncertainties. The upper limits correspond to the 84% C.L. Right: fractional difference between the Auger spectra and a reference spectrum with an index of 3.26.

However, the number of hybrid events does not reflect the increase of exposure accumulated in 2013 due to the adoption of more stringent selection criteria from [8].

The four independent measurements of the energy spectrum of cosmic rays are shown in Fig. 2. The differential fluxes are also displayed as fractional differences with respect to a reference spectrum with an index of 3.26 [1]. The comparison shows that all spectra are in agreement within uncertainties. The four independent measurements of the energy spectrum of cosmic rays are then combined using a method that takes into account the systematic uncertainties of the individual measurements. The systematic uncertainties of the SD-1500 vertical and inclined fluxes are 5.8% and 5%, respectively. The one of the SD-750 m flux decreases from 14% at $10^{17.5}$ eV to less than 7% above $10^{18.5}$ eV. Similarly, the hybrid flux's uncertainty decreases from 10% at $10^{18}$ eV to less than 6% above $10^{19}$ eV. In this procedure, the flux normalisations are used as additional constraints to derive the flux scaling factors needed to match them: $(5.7 \pm 0.2)\%$ for the vertical spectrum, $(-0.1 \pm 0.8)\%$ for the inclined spectrum, $(1.8 \pm 4.3)\%$ for the SD-750 m spectrum and $(-5.8 \pm 2.4)\%$ for the hybrid spectrum.

The characteristic features of the combined energy spectrum, shown in Fig 3, have been quantified by fitting a model that describes a spectrum by a power-law below the ankle $J(E) = J_0\,(E/E_{\rm ankle})^{-\gamma_1}$ and a power-law with a smooth suppression at the highest energies:

$$J(E) = J_0 \left(\frac{E}{E_{\rm ankle}}\right)^{-\gamma_2} \left[1 + \left(\frac{E_{\rm ankle}}{E_{\rm s}}\right)^{\Delta\gamma}\right] \left[1 + \left(\frac{E}{E_{\rm s}}\right)^{\Delta\gamma}\right]^{-1}. \tag{3.1}$$

Here, $\gamma_1$ and $\gamma_2$ are the spectral indices below and above the ankle energy $E_{\rm ankle}$, respectively, $E_{\rm s}$ is the energy at which the differential flux falls to one-half of the value of the power-law extrapolation from the intermediate region, $\Delta\gamma$ gives the increment of the spectral index beyond the suppression region, and $J_0$ is the normalisation of the flux, taken as the value of the flux at $E = E_{\rm ankle}$. The result of the best fit is shown in Fig. 3 and the corresponding parameters are presented in Table 2, quoting both statistical and systematic uncertainties.

[1] Reference spectrum: $J_{\rm ref} = 2.51 \times 10^{42}\,(E/{\rm eV})^{-3.26}\,{\rm eV}^{-1}\,{\rm km}^{-2}\,{\rm sr}^{-1}\,{\rm yr}^{-1}$, fitted to the SD-1500 m vertical differential flux in the energy bin corresponding to $\log_{10}(E/{\rm eV}) = 18.55$ (bin width of 0.1), which contains 29371 events.

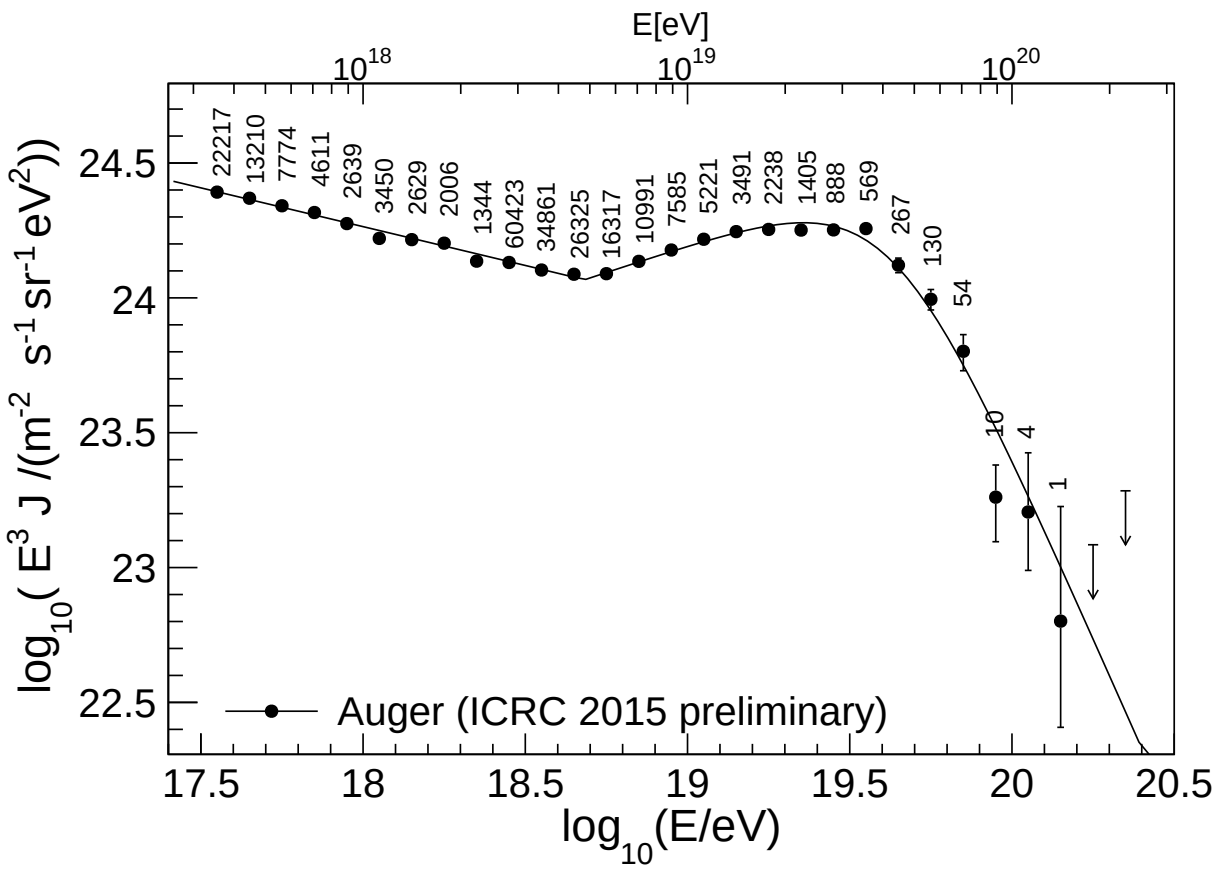

**Figure 3:** The combined energy spectrum of cosmic-rays as measured by the Auger Observatory, fitted with a flux model (see text). Only statistical uncertainties are shown. The systematic uncertainty on the energy scale is 14%. The number of events is given above the points, which are positioned at the mean value of $\log_{10}(E/\mathrm{eV})$. The upper limits correspond to the 84% C.L.

| $J_0$ [eV$^{-1}$km$^{-2}$sr$^{-1}$yr$^{-1}$] | $E_{\rm ankle}$ [EeV] | $E_{\rm s}$ [EeV] | $\gamma_1$ | $\gamma_2$ | $\Delta\gamma$ |
|---|---|---|---|---|---|
| $(3.30\pm0.15\pm0.20)\times10^{-19}$ | $4.82\pm0.07\pm0.8$ | $42.09\pm1.7\pm7.61$ | $3.29\pm0.02\pm0.05$ | $2.60\pm0.02\pm0.1$ | $3.14\pm0.2\pm0.4$ |

**Table 2:** Best-fit parameters, with statistical and systematic uncertainties, for the combined energy spectrum measured at the Pierre Auger Observatory.

The combined spectrum shows a flattening above the ankle, $E_{\rm ankle} = 4.8\times10^{18}$ eV, up to the onset of the flux suppression. This suppression is clearly established with a significance of more than $20\,\sigma$ (the null hypothesis that the power law above the ankle continues beyond the suppression point can be rejected with such confidence). The spectral index in the region of the suppression is less certain due the low number of events and large systematic uncertainties.

A spectral observable in the GZK [15, 16] region that can be used to discriminate between different UHECR source-composition models is the energy $E_{1/2}$ at which the integral spectrum drops by a factor of two below what would be expected with no cutoff. The corresponding value derived from the Auger data, computed as the integral of the parameterisation given by eq. (3.1) with the parameters reported in Table 2, is $E_{1/2} = (2.47\pm0.01^{+0.82}_{-0.34}(\mathrm{sys}))\times10^{19}$ eV. This result, for instance, differs at the level of $3.4\,\sigma$ from the value of $\approx 5.3\times10^{19}$ eV predicted in [17] under the assumption that the sources of UHECRs are uniformly distributed over the universe and that they accelerate protons only. Note that, in reality, sources are discrete and in the GZK region the shape of the spectrum will be dominated by the distribution of sources around us (see [18] for example).

## 4. Declination-dependence of the energy spectrum

Given the location of the Auger Observatory at a latitude $-35.2°$, events arriving with $\theta<60°$ cover a wide range of declinations from $-90°$ to $+25°$, corresponding to a sky fraction of 71%, and therefore become an excellent data set to search for declination dependence of the measured energy spectrum. Although the inclusion of inclined events would extend the declination range to $+45°$, only vertical showers with energy above $3\times10^{18}$ eV are considered for this first analysis.

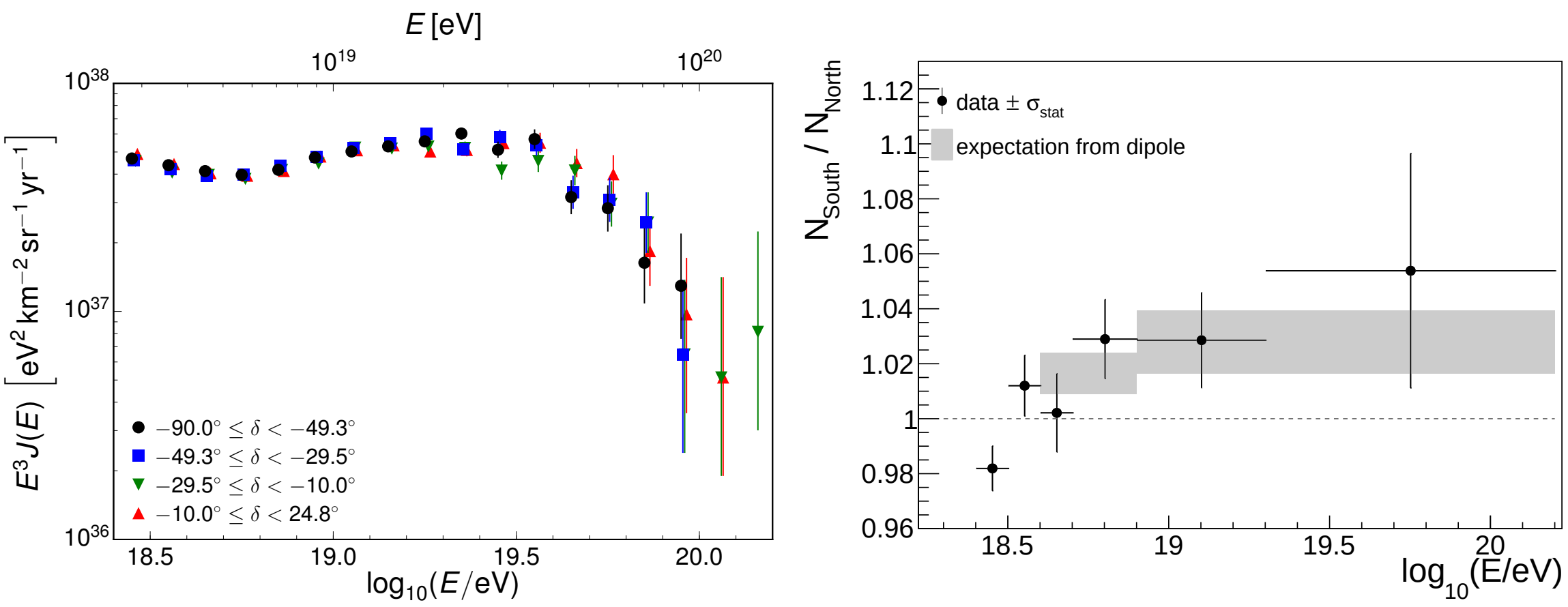

**Figure 4:** Left: The SD vertical energy spectrum in different declination bins. Right: the ratio of the fluxes of cosmic rays arriving from southern ($\delta < -29.47°$) and northern ($\delta > -29.47°$) directions derived from vertical events compared to the expectation from the dipolar modulation of the flux measured with Auger data with $\theta < 80°$ in the energy ranges $4 < E < 8$ EeV and $E > 8$ EeV [5, 6]. The shaded boxes correspond to the propagation of the statistical uncertainties in the amplitude of the reconstructed North-South dipole component.

To search for a variation with declination of the cosmic-ray flux, it is important to carefully account for spurious effects that can modulate the flux that arise from experimental, atmospheric and geomagnetic effects. We take into account these effects following the same procedure adopted to study the large angular scale distribution of arrival directions of cosmic rays recorded at the Auger Observatory [19, 5]. Firstly, the observed part of the sky is divided into four $\delta$-bands each with approximately the same exposure. The sub-spectra are shown in the left panel of Fig. 4. Given the small relative differences found between them and the all-sky spectrum (<5% at energies below $E_s$ and <13% above), there is no significant indication of a dependence on $\delta$. As a consequence, the difference seen at the highest energies (suppression region) between the spectra measured by the Auger and TA Observatories [4] can not be explained by a $\delta$-dependence of the measured flux, unless the flux measured by TA is substantially larger above declination $+25°$ than below.

Recent studies of the distribution of arrival directions of both vertical and inclined events above $4\times10^{18}$ eV recorded up to 31 Dec 2013 at the Auger Observatory [5, 6] have reinforced the hint of a dipole anisotropy. After performing two Rayleigh analyses in the right ascension and azimuth angles in two different energy bins, the observed amplitude in right ascension above $8\times10^{18}$ eV suggests a large-scale anisotropy with a significance exceeding $4\,\sigma$ (lower statistical significance for events with energies between 4 and 8 EeV). The reconstructed dipole points to $(\alpha,\delta) = (95°\pm13°, -39°\pm13°)$ in the higher energy range, and $(\alpha,\delta) = (15°\pm115°, -81°\pm17°)$ in the lower. Here we investigate if this hint is also observable in the measured flux. For this purpose, the observed sky is divided only into two bands of declination. Then the ratio of the corresponding sub-spectra is computed and compared to the expectation from this dipole anisotropy as shown in the right panel of Fig. 4, demonstrating good agreement between both results.

## 5. Summary

The energy spectrum above $3\times10^{17}$ eV has been measured with unprecedented precision and statistics using the data collected by the Auger Observatory for more than 10 years. The results can be described by a power-law spectrum with spectral index 2.6 above $4.8\times10^{18}$ eV and clearly show a steepening of the cosmic-ray flux above an energy around $4.2\times10^{19}$ eV. The dominant systematic uncertainty of the spectrum stems from the overall uncertainty in the energy scale of 14%.

Differences between the recent Auger and TA spectra have motivated the search for a declination dependence of the flux of cosmic rays. No significant variation in the flux measured with the SD in four declination bands were found that could account for the discrepancy between spectra measured from different hemispheres. The differences found between the measurements in two separate declination bands are compatible with the variations expected from a dipolar modulation of the flux.

# 2

# Arrival Directions and Anisotropies

PIERRE
AUGER
OBSERVATORY

# Arrival directions of the highest-energy cosmic rays detected with the Pierre Auger Observatory

**Julien Aublin**[*a] **for the Pierre Auger Collaboration**[b]
[a]*Université Pierre et Marie Curie (UPMC), Laboratoire de Physique Nucléaire et de Hautes Energies (LPNHE), CNRS-IN2P3, Paris, France*
[b]*Observatorio Pierre Auger, Av. San Martín Norte 304, 5613 Malargüe, Argentina*
*E-mail:* auger_spokespersons@fnal.gov
*Full author list:* http://www.auger.org/archive/authors_2015_06.html

We present the results of a search for small to intermediate scale anisotropies in the distribution of arrival directions of ultra-high energy cosmic rays recorded at the Pierre Auger Observatory. The data set, gathered in ten years of operation, includes arrival directions with zenith angles up to 80°, and is about three times larger than that used in earlier studies. We update the test based on correlations with active galactic nuclei (AGNs) from the Véron-Cetty and Véron catalog, which does not yield a significant indication of anisotropy with the present data set. We perform a blind search for localized excess fluxes and for self-clustering of arrival directions at angular scales up to 30° and for different energy thresholds between 40 EeV and 80 EeV. We also examine the correlation of arrival directions with relatively nearby galaxies in the 2MRS catalog, AGNs detected by Swift-BAT,and a sample of radio galaxies with jets and with the Centaurus A galaxy. None of the searches shows a statistically significant evidence of anisotropy. The two largest departures from isotropy that were found have a post-trial probability $\approx 1.4\%$. One is for cosmic rays with energy above 58 EeV that arrive within 15° of the direction toward Centaurus A. The other is for arrival directions within 18° of Swift-BAT AGNs closer than 130 Mpc and brighter than $10^{44}$ erg/s, with the same energy threshold.

*Speaker.

## 1. Introduction

The determination of the origin of the very high energy cosmic rays is a difficult task, mostly because of the very small value of their flux at Earth, together with the fact that they experience magnetic deflections during propagation. Nevertheless, the distribution of arrival directions might contain crucial information about the cosmic ray sources, provided that their distribution is not uniform and that the deflections are small enough. Such conditions could be fulfilled by low-Z cosmic rays with energies above $\sim 40$ EeV: the suppression of the flux that is observed [1] in the energy spectrum is compatible with a GZK [2] mechanism, thus potentially limiting the distance from which a source can contribute. Independently of the origin of this flux suppression, the recent upper-limits on the primary cosmic ray photon flux [3] severely constrain top-down models, thus favoring an astrophysical origin for the sources of cosmic rays.

The present paper describes the latest analysis of the distribution of arrival directions of cosmic rays with energies above 40 EeV detected at the Pierre Auger Observatory in 10 years of operation, with a total exposure of about 66,000 km$^2$ sr yr. After a brief description of the data set, we first present an update of the correlation analysis performed with the AGNs from the Véron-Cetty and Véron catalog [4]. The result of this test does not confirm the initial evidence of anisotropy that was observed [5] for cosmic rays with energies above 57 EeV. Consequently, we searched for the presence of a possible anisotropy in the distribution of arrival directions for events with energies above 40 EeV, applying different types of tests.

In section 2 we first analyze the distribution of arrival directions in the data set without using any external information such as a catalog of candidate sources. This search of intrinsic anisotropy is performed with the auto-correlation method, together with a search for excess of events in circular windows over the whole exposed sky.

In section 3, we study the cross-correlation between the cosmic ray arrival directions and the position of candidate sources extracted from catalogs. We describe the cross-correlation analysis performed with the 2MRS catalog of galaxies detected in IR, the 70 months Swift-BAT catalog of AGNs detected in X-rays, and with a catalog of radio galaxies. We describe in addition the analysis of the distribution of events around the direction of the Cen A galaxy. The details of all these analyses together with the list of arrival directions and energies of the events can be consulted in [6].

**Data set** The data set used in the present analysis consists of 602 events with energy above 40 EeV measured by the Surface Detector (SD) of the Pierre Auger Observatory between the 1$^{st}$ January 2004 and the 31$^{st}$ March 2014. The Pierre Auger Observatory is a 3000 km$^2$ array of water-Cherenkov detectors with 1.5 km spacing, overlooked by 24 fluorescence telescopes located on its periphery. A recent and detailed description of the detector can be found in [7].

The total data set is the combination of 454 events with zenith angle $\theta < 60^\circ$ (vertical events) and 148 events with $60^\circ < \theta^\circ < 80^\circ$ (inclined events). The extension of the zenith angle range, compared to previous anisotropy searches that were limited to vertical events, has two important advantages: a 30% increase of statistics and a higher fraction of the sky covered, ranging from $-90^\circ$ to $+45^\circ$ in declination.

The properties of the signal measured at ground being zenith angle dependent, the selection and reconstruction procedures are different for vertical and inclined events. The vertical events are selected if the water-Cherenkov detector that measured the highest signal is surrounded by at least four other operational detectors in the closest range. In addition, the reconstructed shower core position at ground must lie within a triangle of contiguous operational detectors. This event selection ensures an accurate event reconstruction and increases the number of vertical events by 14% with respect to the previous selection used in our analyses. The selection of inclined events requires the presence of 5 operational detectors around the one with the highest signal.

For both data sets, the detection and selection efficiency is 100% for the energies considered here, the exposure is therefore determined only by the geometry of the array and amounts to 51,753 and 14,699 $km^2$ sr yr for the vertical and inclined samples respectively.

The angular resolution, defined as the 68% containment radius around the true arrival direction, is better than 0.9° above 10 EeV [8], where a high number of detectors participate in an event. The ground estimator for the energy determination is different for vertical and inclined events: the vertical reconstruction uses the fitted signal at 1000 m from the shower axis whereas the inclined reconstruction estimates the muon content relative to a simulated proton shower with energy $10^{19}$ eV. In both cases, the final energy estimation uses the cross-calibration with the Fluorescence Detector (FD) that provide a quasi-calorimetric measurement. The statistical uncertainty in the energy determination is better than 12% above 10 EeV [9] and the systematic uncertainty in the absolute energy scale is 14%. As a consequence of the recent update of the absolute energy scale [10], the energy threshold of 55 EeV used in our previous publication [11] now corresponds to approximately 53 EeV.

**Note on the anisotropy test with the VCV catalog** The Véron-Cetty and Véron catalog of active galactic nuclei [4] has been previously used to search for correlation with potential cosmic ray extragalactic sources. The number of CR events with $E > E_{\rm th}$ that arrive within an angular distance $\Psi$ of an AGN with redshift $z < z_{\rm max}$ is measured and compared to isotropic expectations. After an initial scan over the parameters with vertical events collected between the $1^{st}$ January 2004 and the $26^{th}$ May 2006, the most significant excess was found for $E_{\rm th} = 57$ EeV, $\Psi = 3.1°$ and a maximum redshift corresponding to a distance of 75 Mpc. This set of parameters has been used on subsequent independent data, leading to a correlation fraction of 61% with a $1.7 \times 10^{-3}$ probability of happening by chance [5]. The analysis has been performed with increased statistics, leading to a much lower $(38^{+7}_{-6})\%$ correlation fraction [11]. An update of this analysis with the present vertical data set described previously yields a correlation fraction of $(28.1^{+3.8}_{-3.6})$ %, which is only 2 standard deviations above the isotropic expectation of 21%. We conclude that the present level of the correlation fraction does not provide a significant indication of anisotropy.

## 2. Intrinsic anisotropy tests

**Search for a localized excess flux over the exposed sky** We searched for an excess in the arrival directions of cosmic rays by counting the number of events that fall inside circular windows of varying radius $\Psi$ from 1° to 30° in 1° steps. The centers of those windows cover the whole exposed sky and are located on a regular $1° \times 1°$ grid. An energy threshold $E_{\rm th}$ is applied to the events, and is varied between 40 EeV and 80 EeV with 1 EeV steps. The number $n_{obs}$ of observed events is compared to that expected from an isotropic flux $n_{exp}$: the isotropic value $n_{exp}$ is estimated by numerically integrating the geometric exposure function in the corresponding angular window. For each window we compute the binomial probability $p$ of observing by chance in an isotropic flux a greater or equal number of events than that found in the data. A scan using the combined vertical and inclined data set was performed on the parameters $\Psi$ and $E_{th}$, leading to a minimum probability of $p = 5.9 \times 10^{-6}$ for an excess of $n_{obs}/n_{exp} = 14/3.23$ with $E_{th} = 54$ EeV and $\Psi = 12^\circ$.

The Li-Ma significance [12] of event excesses with $E \geq 54$ EeV in windows of 12° radius is shown in Figure 1 (left). The highest significance region ($4.3\sigma$) is found to be close to the Super-Galactic Plane and to the CenA radiogalaxy. To quantify the significance of this excess, we simulated 10,000 isotropic data sets with the same number of events and applied the full parameter scan described above. In 69% of isotropic simulations an excess with $p$ smaller than $5.9 \times 10^{-6}$ can be found, hence indicating that the data are compatible with isotropic expectations.

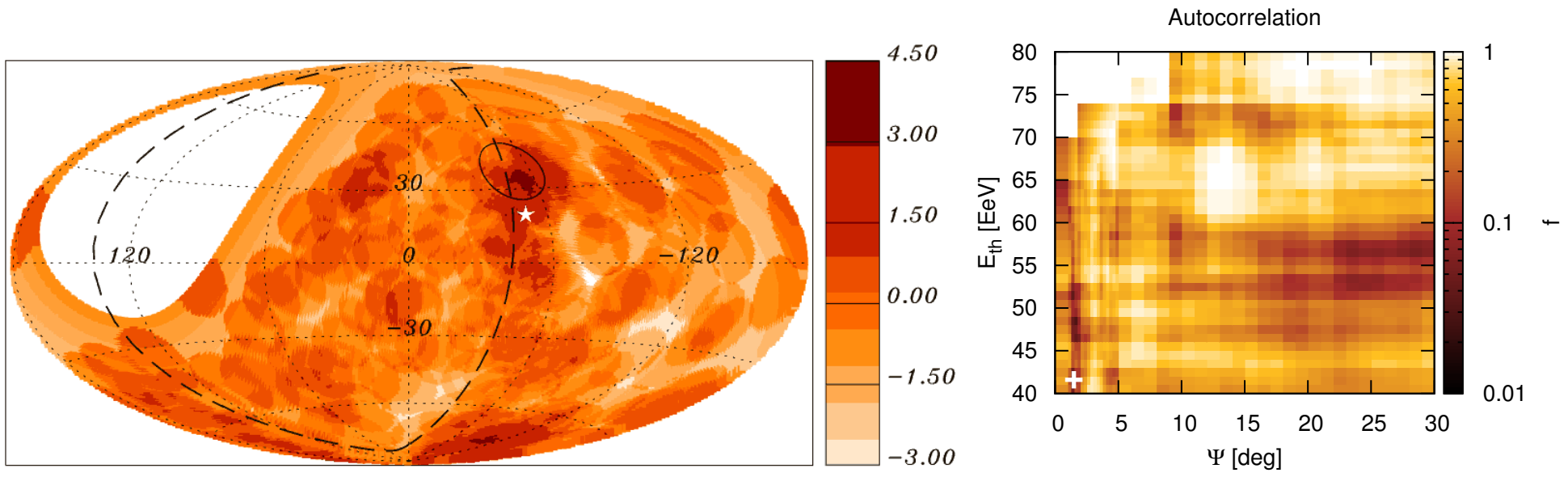

**Figure 1:** *Left*: map in galactic coordinates of the Li-Ma significances of excesses in 12°-radius windows for the events with $E \geq 54$ EeV. Also indicated are the Super-Galactic Plane (dashed line) and Centaurus A (white star). *Right*: Fraction $f$ obtained in the autocorrelation of events versus $\psi$ and $E_{th}$, white cross indicating the minimum.

**The autocorrelation of events** The angular auto-correlation analysis is a simple method to test for self clustering in the arrival directions distribution. The principle is to count the number of pairs of events $N_p(\psi, E_{th})$, above a given energy threshold $E_{th}$ that are within a certain angular distance $\psi$. Using the full data set, we performed a scan in energy threshold from 40 EeV up to 80 EeV and in angle from 1° to 30 ° in which we compare the number of pairs $N_p(\psi, E_{th})$ measured in data to the isotropic expectation. The result of the scan is shown in Figure 1 (right), where the fraction $f(\psi, E_{th})$ of isotropic simulations that have a higher or equal number of pairs than the data is represented. The minimum of this fraction is found at $f_{min} = 0.027$, for $\psi = 1.5^\circ$ and $E_{th} = 42$ EeV, where 30 pairs are expected on average and 41 are observed. To quantify the significance of this minimum, we computed the penalized fraction $P$ of isotropic data sets that lead to a lower or equal value of $f_{min}$ under a similar scan. The resulting value $P \simeq 70$ % indicates that the auto-correlation function is compatible with isotropic expectations.

## 3. Search for cross-correlations with astrophysical sources

We investigate in this section the possible correlation between the arrival directions of cosmic rays with the position of nearby extra-galactic sources (for additional searches of correlation with the Galactic and Super-galactic planes see [6]). We used three complementary catalogs that have an almost uniform coverage and that are complete above a given luminosity: namely the 2MRS catalog of galaxies [13], the Swift-BAT [14] X-ray catalog of AGNs, and a catalog of radio galaxies with jets compiled in [15].

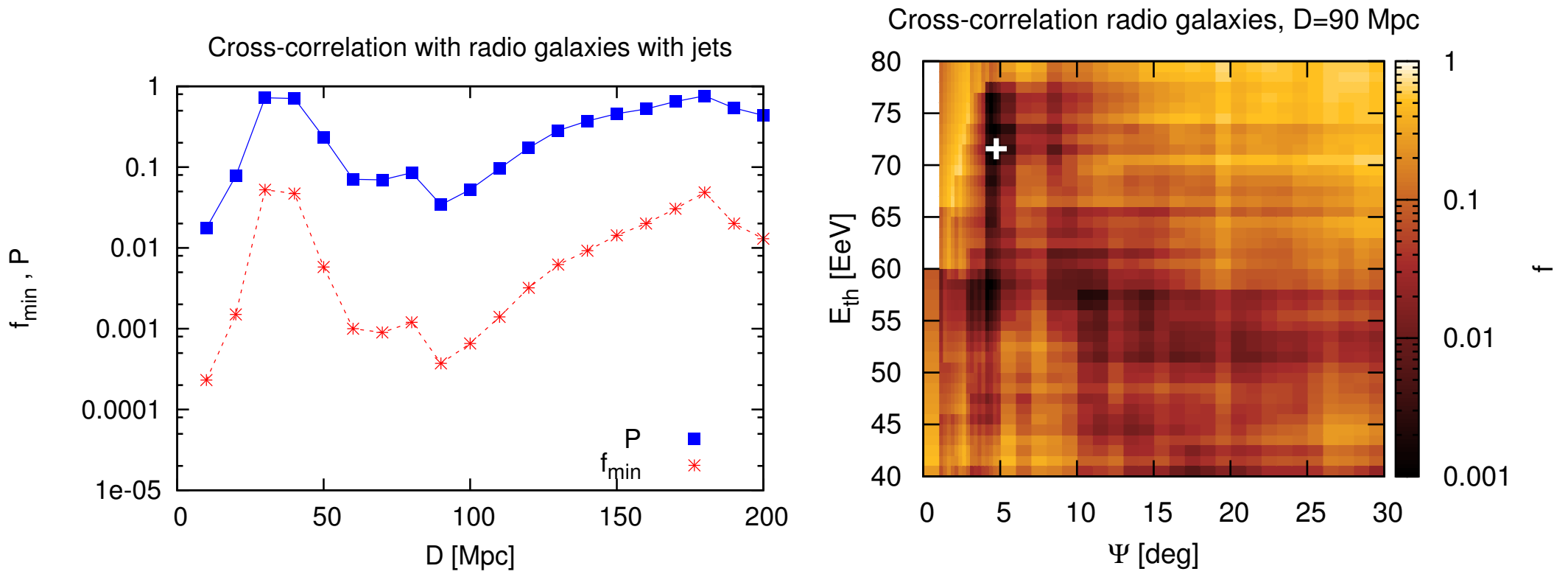

**Figure 2:** *Left*: Cross-correlation of events with the AGNs in the catalog of radio galaxies with jets. Values of $f_{\rm min}$ and $P$ are shown as a function of the maximum distance $D$ to the AGNs considered. *Right*: Results of the scan in $\psi$ and $E_{\rm th}$ for the value $D = 90$ Mpc corresponding to the (second) minimum in the left plot.

The 2MRS catalog maps the local distribution of galaxies detected in the infrared domain, which could be associated with the location of gamma-ray bursts and newborn pulsars. The catalog is 97.6% complete above magnitudes brighter than $K_s = 11.75$ and contains more than 37,000 galaxies within 200 Mpc and about 16,000 galaxies within 100 Mpc.

The Swift-BAT catalog contains 1210 sources detected in X-rays after 70 months of operation, among which 705 are identified as AGNs and have a measured redshift. The catalog is complete for 90% of the sky above an X-ray flux of $13.4 \times 10^{-12}$ erg cm$^{-2}$ s$^{-1}$ in the 14-195 keV range. This cut in flux selects 296 AGN-like galaxies within 200 Mpc and 160 within 100 Mpc.

The radio galaxy catalog that we use is a compilation from the 1.4 GHz NRAO VLA Sky Survey [16] and the 843 MHz Sydney University Molonglo Sky Survey [17] produced by Van Velzen et al. [15]. The catalog is quasi-complete for fluxes above 213 mJy at 1.4 GHz and 289 mJy at 843 MHz, and contains 205 radio galaxies with jets within 200 Mpc, and 56 within 100 Mpc. The overlap between the Swift-BAT and the radio galaxies catalog is only 5%, the majority of the Swift-BAT galaxies being of spiral type whereas the radio galaxies are mostly elliptical.

As a first approach, we use these catalogs in a cross-correlation analysis where all the objects with the above flux limits are selected. This selection corresponds to the assumption that all sources contribute to the cosmic ray flux in the proportion of their apparent luminosity. In a second approach, we select only the brightest sources from the catalogs by applying a cut in intrinsic luminosity which is motivated by the expectation that the maximum energy $E_{\rm max}$ achieved by the cosmic rays could be related to the intrinsic electromagnetic bolometric luminosity $\mathscr{L}$ of the object ($E^2_{\rm max} \propto \mathscr{L}$). In that case, we test the assumption that only the brightest sources can produce a cosmic ray flux above an energy threshold $E_{\rm th}$.

**Cross-correlation with flux-limited catalogs** The cross-correlation analysis is similar to the auto-correlation method, and consists in counting the number of pairs of a given angular separation between cosmic ray events and objects in a sources catalog. To find an excess, we compare the number of pairs with the expectation of an isotropic simulation. We scan over the energy threshold $E_{th}$ of the events from 40 EeV up to 80 EeV and in angular scale $\Psi$ between 1° and 30°. For the sources in the catalogs, we impose a maximum distance cut $D$, that can vary from 10 Mpc up to 200 Mpc in 10 Mpc steps.

| Objects | $E_{th}$ [EeV] | $\Psi$ [°] | $D$ [Mpc] | $\mathscr{L}_{min}$ [erg/s] | $f_{min}$ | $\mathscr{P}$ |
|---|---|---|---|---|---|---|
| 2MRS Galaxies | 52 | 9 | 90 | - | $1.5\times10^{-3}$ | 24% |
| Swift AGNs | 58 | 1 | 80 | - | $6\times10^{-5}$ | 6% |
| Radio galaxies | 72 | 4.75 | 90 | - | $2\times10^{-4}$ | 8% |
| Swift AGNs | 58 | 18 | 130 | $10^{44}$ | $2\times10^{-6}$ | 1.3% |
| Radio galaxies | 72 | 4.75 | 90 | $10^{39.33}$ | $5.1\times10^{-5}$ | 11% |
| Centaurus A | 58 | 15 | - | - | $2\times10^{-4}$ | 1.4% |

**Table 1:** Summary of the parameters of the minima found in the cross-correlation analyses.

For each value of $D$, we compute the fraction $f(\psi, E_{th})$ of isotropic simulations having an equal or higher number of pairs than the data, and search for its minimum $f_{min}$. We calculate the associated post-trial probability $P$ as the fraction of isotropic realizations that lead to a lower or equal value of $f_{min}$ under a similar scan on $E_{th}$ and $\Psi$.

The evolution of $f_{min}$ and $P$ as a function of the maximum distance $D$ is shown in Figure 2 (left) for the radio galaxies catalog. The absolute minimum is $f_{min} = 2 \times 10^{-4}$ with a penalized value of $P = 1.4\%$ obtained for a distance $D = 10$ Mpc. The only object in that distance range is Cen A , the closest radio galaxy located at 4.2 Mpc. As we discuss the correlation with Cen A in a separate section, we describe here only the second minimum of $f_{min} = 4 \times 10^{-4}$ and $P = 3.4\%$ that occur for $D = 90$ Mpc. The result of the scan in energy and angular radius for $D = 90$ Mpc is shown in Figure 2 (right). The minimum (indicated by a white cross) corresponds to $E_{th} = 72$ EeV and $\Psi = 4.75°$, where 13 pairs are observed in data and 3.2 are expected in average from isotropy. The penalized probability $\mathscr{P}$ of getting a lower or equal value of $P = 1.4\%$ (that is the absolute minimum of the scan) when repeating the same scan in distance $D$ with isotropic samples is $\mathscr{P} = 8$ %.

The same analysis is applied for the 2MRS and Swift-BAT catalogs, and the results are summarized in table 1. The penalized probabilities $\mathscr{P}$ are of the order of a few percent, indicating that no significant excess is observed in this cross-correlation test.

**Cross-correlation with bright AGNs** We describe here the results of an additional scan on the minimum luminosity $\mathscr{L}_{min}$ of the sources in the catalogs. The cross-correlation analysis was applied on the Swift-BAT AGNs where we used the luminosity $\mathscr{L}_X$ measured in the X-ray band, and for the radio galaxies with the radio luminosity $\mathscr{L}_R$ computed at 1.1 GHz. For Swift-BAT we scan from $\mathscr{L}_X = 10^{42}$ erg/s up to $10^{44}$ erg/s, while for the radio galaxies we scan from $\mathscr{L}_R = 10^{39}$ erg/s up to $10^{41}$ erg/s, considering three logarithmic steps per decade, for a total of 7 luminosity values in each case. These values are designed to cover most of the luminosity range of the AGNs that are present in the catalogs.

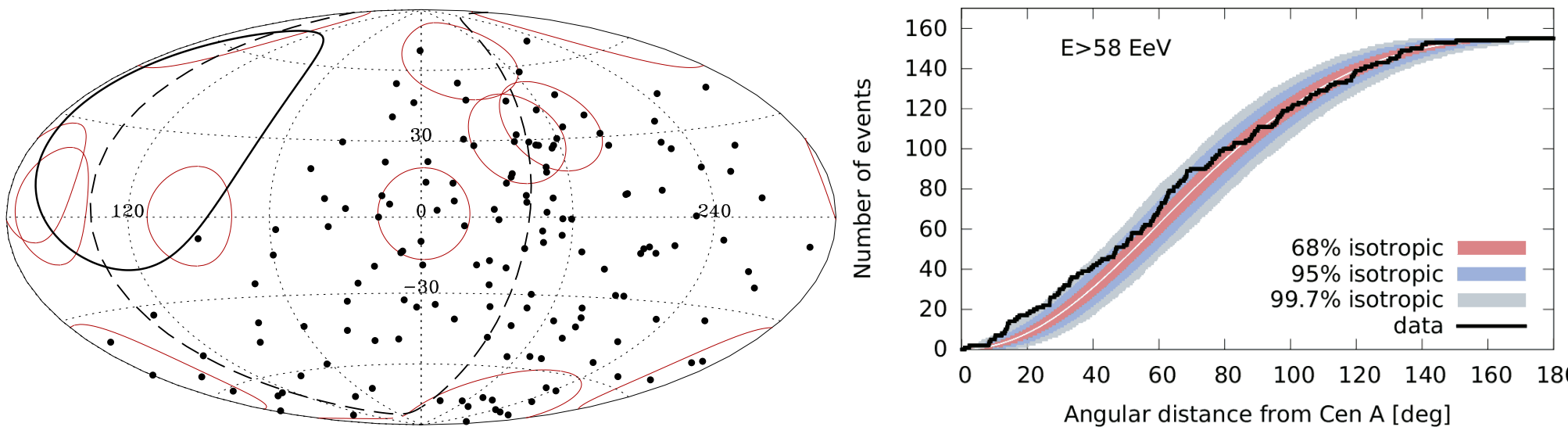

**Figure 3:** *Left*: Map in galactic coordinates showing the events (black dots) with $E \geq 58$ EeV together with the Swift AGNs brighter than $10^{44}$ erg/s and closer than 130 Mpc, indicated with circles of 18° radius. *Right*: cumulative number of events around the Cen A radio galaxy, for the threshold $E_{\rm th} = 58$ EeV, exploring the whole angular range.

The analysis procedure is similar to what was described in the previous section: we search for the minimum of the fraction $f(\psi, E_{\rm th})$ of isotropic simulations having an equal or higher number of pairs than the data, for each value of $D$ and $\mathscr{L}_{\rm min}$. The resulting parameters that correspond to the minimum are included in table 1.

For Swift-BAT bright AGNs, the minimum $f_{\rm min} = 2 \times 10^{-6}$ is obtained for $D = 130$ Mpc and $\mathscr{L} > 10^{44}$ erg/s, with a threshold energy of $E_{\rm th} = 58$ EeV and an angular radius $\Psi = 18°$. For those parameters, 62 pairs are observed between 155 cosmic rays and 10 AGNs (with $\mathscr{L}_{\rm X} > \mathscr{L}_{\rm min}$) while 32.8 are expected from isotropy. A sky map is shown in Figure 3 (left) representing these events and AGNs in galactic coordinates. The penalized probability to find in isotropic simulations $f_{\rm min}$ values lower or equal than $2 \times 10^{-6}$ under the same scan on $(\Psi, E_{\rm th}, \mathscr{L}_{\rm min}, D)$ is $\mathscr{P} \simeq 1.3\%$.

For the radio galaxies, two equivalent minima are found when scanning in luminosity, both for a maximum distance $D = 90$ Mpc. The first one is obtained for a luminosity $\mathscr{L}_{min} = 10^{39.33}$ erg/s, with ($E_{\rm th} = 72$ EeV, $\Psi = 4.75°$), and corresponds to a fraction $f_{\rm min} = 5.1 \times 10^{-5}$. For those parameters, 13 pairs are observed while 2.4 are expected from isotropy. The second minimum has $\mathscr{L}_{\rm min} = 10^{40}$ erg/s, for ($E_{\rm th} = 58$ EeV, $\Psi = 12°$), with a fraction $f_{\rm min} = 5.6 \times 10^{-5}$. The penalized probability corresponding to the first minimum is found to be $\mathscr{P} \simeq 11\%$.

**The Cen A region** At a distance of only 4 Mpc, Cen A is the closest radio-loud active galaxy, being a natural candidate source for the acceleration of high energy cosmic rays. We also note that the Centaurus cluster, which contains of a large number of galaxies, lies at a distance of 50 Mpc and is approximately in the same direction as Cen A. We thus searched for a correlation between the cosmic ray arrival directions and the location of Cen A, counting the number of events within an angular radius $\Psi$ between 1° and 30° for an energy threshold $E_{\rm th}$ ranging from 40 EeV up to 80 EeV. The significance of a potential excess is evaluated by computing the fraction $f$ of isotropic simulations that give a higher or equal number of events than the data.

The minimum of this fraction is found to be $f_{\rm min} = 2 \times 10^{-4}$ for $E_{\rm th} = 58$ EeV and $\Psi = 15°$, where 14 events are observed while 4.5 are expected. The right panel of Figure 3 shows the number of events with $E > 58$ EeV as a function of the angular distance from Cen A for the whole angular range, indicating also the 68, 95 and 99.7% intervals obtained with isotropic simulations. The penalized probability to find a smaller $f_{\rm min}$ value in isotropic simulations under the same scan is $\mathscr{P} \simeq 1.4\%$.

## 4. Discussion

The distribution of arrival directions of cosmic rays detected by the Pierre Auger Observatory in 10 years of operation has been studied by several complementary methods. An update of the fraction of events with $E > 53$ EeV correlating with AGNs from the VCV catalog has been performed, leading to a value of $(28.1^{+3.8}_{-3.6})$ %, which is only 2 standard deviations above the isotropic expectation of 21%.

We then searched for intrinsic anisotropies in the distribution of arrival directions of cosmic rays with energies above 40 EeV by computing the angular auto-correlation function and by looking at potential excesses in circular windows all across the exposed sky. Both tests give results that are compatible with isotropic expectations.

The cross-correlation with nearby (within 200 Mpc) sources from three complementary astrophysical catalogs has been performed, together with a specific analysis of the arrival directions around the Cen A radio galaxy. The results are summarized in table 1. The penalized probabilities $\mathscr{P}$ accounting for the scan on parameters are of the order of a few percent, and can reach the 1% level when selecting only the brightest AGNs of the Swift-BAT catalog or with the Cen A radio galaxy. We note that all minima, despite being not statistically significant, occur for a value of the maximum distance $D \simeq 80 - 90$ Mpc.

# Indications of anisotropy at large angular scales in the arrival directions of cosmic rays detected at the Pierre Auger Observatory

**Imen Al Samarai*[a] for the Pierre Auger Collaboration[b]**
[a] *LPNHE, CNRS/IN2P3 & Université Pierre et Marie Curie, Paris, France*
[b] *Observatorio Pierre Auger, Av. San Martín Norte 304, 5613 Malargüe, Argentina*
*E-mail:* auger_spokespersons@fnal.gov
*Full author list:* http://www.auger.org/archive/authors_2015_06.html

The large-scale distribution of arrival directions of high-energy cosmic rays carries major clues to understanding their origin. The Pierre Auger Collaboration have implemented different analyses to search for dipolar and quadrupolar anisotropies in different energy ranges spanning four orders of magnitude. A common phase $\approx 270^\circ$ of the first harmonic modulation in right-ascension was found in adjacent energy intervals below 1 EeV, and another common phase $\approx 100^\circ$ above 4 EeV. A constancy of phase measurements in ordered energy intervals originating from a genuine anisotropy is expected to appear with a smaller number of events than those needed to achieve significant amplitudes. This led us to design a prescribed test aimed at establishing whether this consistency in phases is real at 99% CL. The test required a total independent exposure of 21,000 $\mathrm{km}^2$ sr yr. We report on the status of this prescription. We also report the results of the search for a dipole anisotropy for cosmic rays with energies above 4 EeV using events with zenith angles between $60^\circ$ and $80^\circ$. Compared to previous analyses of events with zenith angles smaller than $60^\circ$, this extension increases the size of the data set by 30%, and enlarges the fraction of exposed sky from 71% to 85%. The largest departure from isotropy is found in the energy range above 8 EeV, with an amplitude for the first harmonic in right ascension $r_1 = (4.4 \pm 1.0) \times 10^{-2}$, that has a chance probability $P(\geq r_1) = 6.4 \times 10^{-5}$, reinforcing the hint previously reported with vertical events alone.

*Speaker.

## 1. Introduction

Establishing the energy at which the flux of extragalactic Cosmic Rays (CRs) starts to dominate the flux of Galactic ones would provide an important step forward in understanding the demands that must be placed upon their accelerators in the Galaxy, and the origin of ultra-high energy CRs. To this aim, the large-scale distribution of arrival directions of CRs as a function of their energy is a relevant tool to study generic signatures of anisotropies in the framework of Galactic scenarios around 1 EeV, and extragalactic ones at the highest energies.

The Pierre Auger Observatory [1] provides the largest number of events ever collected to scrutinise such anisotropies in this energy range. The Observatory combines two techniques to detect the extensive air showers resulting from the interaction of CRs with the atmosphere. The longitudinal development of the air showers is measured by fluorescence detectors, while the lateral distribution of the secondary particles at ground level is measured by the Surface Detector Array (SD). Two arrays of SDs are operating: an arrangement of water Cherenkov detectors separated by 1500 m distributed over an area of 3000 km$^2$, reaching full detection efficiency at $3\times10^{18}$ eV, and an array with detectors separated by 750 m distributed over an area of 23.5 km$^2$, reaching full detection efficiency at $3\times10^{17}$ eV.

The different analysis methods used in determining the amplitude and phase of the modulation in Right Ascension (RA) in different energy bins are described in section 2. Using data from both the 750 m array and the 1500 m array up to the end of 2014, we present in section 3 the updated status of a prescribed test on the phases of the first harmonic modulation in RA previously reported in [2]. Studies of the flux modulation in declination and in RA from the analysis of events with zenith angles smaller than 60° have been reported in [3, 4]. By including events with zenith angles between 60° and 80°, an increase of 30% in the number of events is obtained [5]. In section 4, we present the results of the reconstructed dipolar modulations in RA and in declination above 4 EeV for this data set. Finally, the complete picture of large-scale anisotropy studies over four decades in energy using the Pierre Auger Observatory data is presented in the last section.

## 2. Harmonic analysis in right ascension

Harmonic analysis of the RA distribution is a classical tool for picking up and for characterising any modulation in this coordinate system [6]. The low amplitudes of the first harmonic modulation expected from the limits set in many historic studies, motivate a careful consideration of possible spurious modulations that arise from experimental or atmospheric effects. In particular, due to the steepness of the energy spectrum, even small changes in the energy estimator as a function of time or the local angular coordinates would distort significantly the counting rate of events above a given energy.

Atmospheric conditions are known to affect the observed shower size for a given primary energy. For a larger (smaller) pressure, an air shower will be at a more (less) advanced stage of development when it arrives at the ground since the column density traversed would be larger (smaller). Also the air density affects the Molière radius and hence the lateral profile of the showers. These atmospheric effects are accounted for by correcting the energy estimator of the events according to the weather conditions at the time each event was recorded [7]. Such corrections guarantee that the observed event rate with time is controlled by the instantaneous exposure only.

**Rayleigh analysis.** For a directional exposure $\omega(\alpha)$, where $\alpha$ is the RA, the flux $\Phi(\alpha)$ can be decomposed in terms of a harmonic expansion from the observed distribution of arrival directions $dN/d\alpha$ as:

$$\Phi(\alpha) = \frac{1}{\omega(\alpha)}\frac{dN}{d\alpha} = a_0^\alpha + \sum_{n>0} a_n^\alpha \cos n\alpha + \sum_{n>0} b_n^\alpha \sin n\alpha. \tag{2.1}$$

The first harmonic coefficients ($n = 1$) in RA are enough to reconstruct the equatorial dipole component under the hypothesis that contributions from higher-order multipoles are negligible, while the second harmonic coefficients ($n = 2$) are sensitive to the quadrupole component (and to eventual higher-order multipoles). The Fourier coefficients of the flux can be estimated as:

$$a_n^\alpha = \frac{2}{\tilde{N}}\sum_{i=1}^{N} w_i \cos(n\alpha_i)\ ,\ b_n^\alpha = \frac{2}{\tilde{N}}\sum_{i=1}^{N} w_i \sin(n\alpha_i), \tag{2.2}$$

where the sums run over the number of events $N$ in the energy range considered, and the normalization factor is $\tilde{N} = \sum_{i=1}^{N} w_i$. The factors $w_i$ account for the variations in the operating size of the array as a function of time that lead to small modulations in the exposure. Their determination is optimised within the energy range where the study is performed and is detailed in the following sections. The amplitude can then be expressed as $r_n = \sqrt{(a_n^\alpha)^2 + (b_n^\alpha)^2}$, and the phase as $\phi_n = 1/n\ \arctan(b_n^\alpha/a_n^\alpha)$. In case of an underlying isotropy, the amplitude follows a Rayleigh distribution while the phase follows a uniform distribution. The probability P($\geq r_n$) that an amplitude equal to or larger than $r_n$ arises from an isotropic distribution can be approximated by the cumulative distribution function of the Rayleigh distribution $P(\geq r_n) = \exp(-\tilde{N} r_n^2/4)$.

**East-West analysis.** Alternatively to modelling the event rate and controlling the variations of the exposure with time, the modulation of the flux in RA can be revealed using the East-West (E-W) method, though it is less sensitive than the Rayleigh analysis by a factor $\simeq 2.5$ [8]. The counting rate of the events observed in either the Eastern or the Western half field of view of the array is subjected to variations during a sidereal day that can be either due to experimental effects and/or to real variations in the primary CR flux from different parts of the sky. Systematic effects of experimental origin are independent of the incoming direction so that they can be removed by subtracting the counting rates of events coming from each of the sectors. On the other hand, in the presence of a genuine equatorial dipole, the difference in the E-W counting rate would show modulations that are expected to be related to those of the genuine dipole. More specifically, the amplitude $r_{EW}$, and phase $\phi_{EW}$ can be calculated from the arrival times of $N$ events using the standard first harmonic analysis, slightly modified to account for the subtraction of the Western sector to the Eastern one [8]. The amplitude of the first harmonic, $r_1$, and the phase, $\phi_1$, of the RA modulation determined with the Rayleigh formalism are related to $r_{EW}$ and $\phi_{EW}$ through the relations $r_1 = r_{EW}\ \pi\langle\cos\delta\rangle/2\langle\sin\theta\rangle$, and $\phi_1 = \phi_{EW} + \pi/2$ [8], where $\langle\cos\delta\rangle$ is the mean value of the cosine of the declinations of events, and $\theta$ is the zenith angle.

**Reconstruction of the equatorial dipole component.** In the case of an underlying pure dipole, the relationship between $r_1$ and the projection of the dipole on the Earth equatorial plane $d_\perp$, which is the quantity of interest to compare the results of different experiments, depends on the latitude of the observatory and on the range of zenith angles considered. To first order, the relationship reads $d_\perp \simeq r_1/\langle\cos\delta\rangle$, with $\langle\cos\delta\rangle \simeq 0.78$ for events with zenith angles below 60° at the Auger Observatory [9].

## 3. Phase of the first harmonic from $\simeq 10$ PeV to the highest energies

Harmonic analyses in RA have been previously reported in [2] and [9]. Although no significant departure from isotropy was revealed in the amplitude, notable features concerning the evolution of the phases with energy from $10^{16}$ eV up to the highest energies were pointed out. In this section, we report on the status of these features with independent data.

In [9], a Rayleigh analysis of the data collected at the 1500 m array above 1 EeV was presented in detail. In this analysis, the weight factors $w_i$ make use of the number of active cells $n_{\rm cell}(t)$ (number of active detectors surrounded by six active neighbours) constantly monitored at the Observatory. The total number of active cells, $N_{\rm cell}$, as a function of the sidereal time $\alpha_0$ (measured by the right ascension of the zenith at the center of the array) and its relative variations, $\Delta N_{\rm cell}$, are obtained as $N_{\rm cell}(\alpha_0) = \sum_j n_{\rm cell}(\alpha_0 + j.T_{sid})$ and $\Delta N_{\rm cell}(\alpha_0) = \frac{N_{\rm cell}(\alpha_0)}{\langle N_{\rm cell}\rangle}$, where $T_{sid}$ is the duration of the sidereal day, and $\langle N_{\rm cell}\rangle = T_{sid}^{-1}\int_0^{T_{sid}} d\alpha_0 N_{\rm cell}(\alpha_0)$. The weighting factor for each event is thus derived as $w_i = \Delta N_{\rm cell}^{-1}(\alpha_0^i)$. Below 1 EeV, additional spurious modulations of the event rate arise from the variation of the detection efficiency with time (through the impact of weather effects). For this reason, amplitudes and phases are derived from the E-W method. In [2], amplitudes and phases as obtained from the data collected at the 750 m array were also derived using the E-W method.

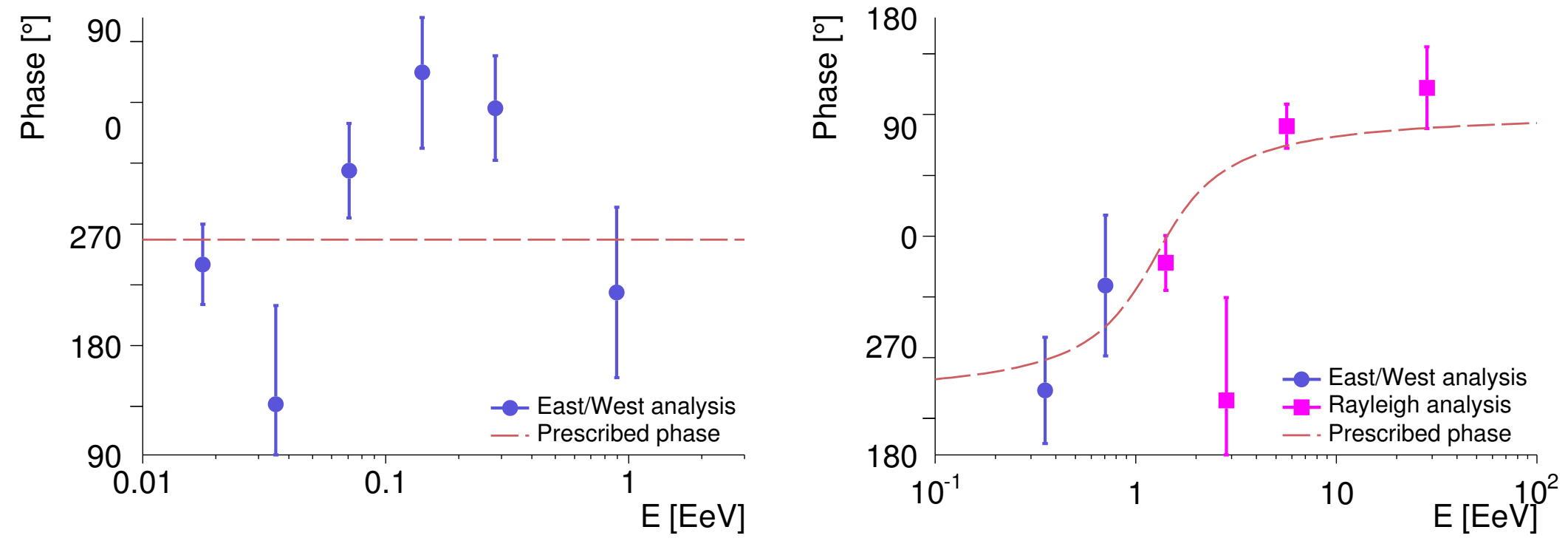

**Figure 1:** Left: Status of the prescription on phases for the 750 m array data using the E-W method, the prescribed test on phases ($\Phi$= 263°) is shown with a dashed line. Right: Status of the prescription on phases for the 1500 m array data, the prescribed test on phases is shown with a dashed line.

Applying these particular analysis methods to data prior to 25 June 2011, a constant phase was observed around $\approx 270°$ for energies below 1 EeV, while a change of phase to $\approx 100°$ was observed at higher energies. This may be indicative of a real anisotropy, since a constancy of phases in ordered energy bins is expected to appear with a smaller number of events than necessary for the amplitudes to stand significantly above the background noise [9, 10]. Since the phases were not defined *a priori*, a prescribed test was set to establish with data posterior to 25 June 2011 whether this effect is real at 99% CL. Once an additional exposure of 21,000 km$^2$ sr yr is reached for the 1500 m array, a positive anisotropy signal will be claimed with a global threshold of 1% upon the realization of one or two predefined conditions. The conditions require an alignment of phases detected by a likelihood ratio test with a chance probability less than 0.5% around a predefined phase value, assuming a signal amplitude of 0.5% for the 750 m array data and an amplitude comparable to the mean noise for the 1500 m array data over the whole energy range. From the date when the prescription started up to the end of 2014, the 750 m array has reached an exposure of 124 km$^2$ sr yr, while the 1500 m array has accumulated an exposure of 19,100 km$^2$ sr yr. The phases measured in each energy bin are shown in figure 1 for the 750 m array (left panel) and for the 1500 m array (right panel). At almost the end of the prescribed test, the phases do not appear aligned around the predefined values. The current p-values obtained from the likelihood test that the hypothesis of isotropy can account for the observations with no need of the alternative hypothesis are 40% for the 750 m array and 7% for the 1500 m array. Below 1 EeV, the phase alignment observed with data prior to June 2011 was suggestive of an overall signal amplitude at the percent level or so around $10^{17}$ eV. For lower amplitudes, such as 0.5% for instance, the current sensitivity of the analysis method to pick up a genuine alignment with a threshold of 0.5% is only $\simeq$ 50%. Many more events and an improved sensitivity in the analysis method are thus required to reveal the large-scale structure of arrival directions in this energy range. On the other hand, the test making use of the 1500 m array data is challenged by the observed phase between $2\times10^{18}$ eV and $4\times10^{18}$ eV which value stands significantly far from the prescribed phase. Since the corresponding amplitude ($2.5\times10^{-3}$) is lower than the mean noise ($6.7\times10^{-3}$) expected from statistical fluctuations, the measurement of the phase in this energy bin is quite uncertain.

The final status of the prescribed test will be given once an additional exposure of 1,900 km$^2$ sr yr is reached. This is expected from data recorded up to mid-2015.

## 4. Dipole search above 4 EeV

The Pierre Auger Collaboration have reported studies of the flux modulation in RA [2, 9] as well as in both declination and right ascension [3, 4] from the analysis of events with zenith angles smaller than 60°. Another study recently reported in [5] has been performed by including for the first time inclined events with zenith angles between 60° and 80°, enabling the extension of the covered sky from 71% to 85%. Large-scale angular modulations of the flux are studied by performing two Rayleigh analyses, one on the right ascension in the same manner as presented in section 2, and the other on the azimuth distribution, by replacing the expression of the Fourier coefficients given as a function of $\alpha$ in eqn. 2.2 by the azimuth angles $\varphi$, the latter being sensitive to modulations in declination. The analysis has been performed in the two same energy bins as previous reports in this energy range. In the lower energy bin, between 4 and 8 EeV, harmonic coefficients are consistent with zero within their uncertainties, and there is no evidence for departures from isotropy in the right ascension distribution. In the highest energy bin where all events with energies greater than 8 EeV are gathered, the first harmonic has an amplitude $r_1 = (4.4\pm1.0)\times10^{-2}$, that has a chance probability $P(\geq r_1) = 6.4\times10^{-5}$. The phase $\phi_1$ points to $95^\circ\pm13^\circ$. The amplitude of the second harmonic is less significant, with a 2% probability of arising by chance. The azimuthal distribution - after having accounted for the modulation induced by the tilt of the array and the geomagnetic effect - is expected to be uniform for energies above full efficiency for an isotropic distribution of CRs. The Fourier coefficients retrieved from the distribution in azimuth thus give information on the dipolar component along the Earth's rotation axis, $d_z$. The largest departure from isotropy appears for the $b_1^{\phi}$ Fourier coefficient in both energy bins. The negative values found indicate a dipolar component $d_z$ pointing to the south, although with low statistical significance ($4 < E < 8$ EeV: 2.4%, $E > 8$ EeV: 1.5% probability).

The reconstruction of the dipole components from the Rayleigh analysis has been done for the case where only a dipole contribution to large-scale anisotropies is relevant, and for the case where a possible quadrupole contribution is present. In the first case, the equatorial component $d_\perp$ is retrieved in the same manner as given in section 2, while the dipole component along the Earth rotation axis is retrieved through $d_z = b_1^{\phi}/(\cos\ \ell_{\rm obs}\langle\sin\theta\rangle)$, where $\ell_{\rm obs}$ denotes the latitude of the Observatory. The total dipole amplitude for the higher energy bin is $0.073 \pm 0.015$ pointing to $(\alpha,\delta) = (95^\circ \pm 13^\circ, -39^\circ \pm 13^\circ)$. In the second case, a combination of a dipole plus a quadrupole was considered. It was found that the dipole is consistent with results from the first case with larger uncertainties, and the quadrupole components are not significant. The exposure-weighted average of the differential flux smoothed in angular windows of 45° radius in equatorial coordinates is shown in figure 2 [1] for the two energy bins considered. The maximum flux difference in the lower energy bin is just 8%, while for the highest energy bin, it reaches 21%.

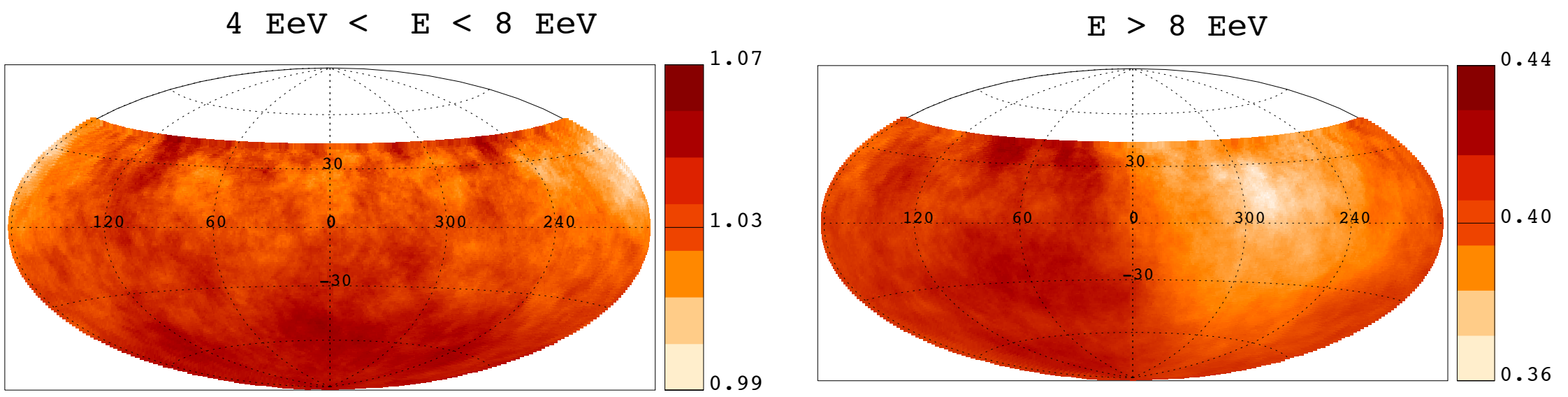

**Figure 2:** Sky map in equatorial coordinates of flux, in km$^{-2}$ yr$^{-1}$ sr$^{-1}$ units, smoothed in angular windows of 45° radius, for observed events with energies 4< $E$ <8 EeV (left) and $E$ >8 EeV (right).

## 5. Conclusion

Different approaches have been explored by the Pierre Auger Collaboration to reveal large-scale anisotropies imprinted on the CR arrival directions. These analyses take advantage of the large number of events provided by the two arrays, even below full detection efficiency. Using the cumulative data sets, a summary of these analyses is given in table 1 and figure 3.

Upper limits on amplitudes are reported in the right panel of figure 3. In the two energy intervals where the p-values for the amplitudes are $1.5 \times 10^{-4}$ and $6.4 \times 10^{-5}$ (between 1 and 2 EeV, and for the integral bin above 8 EeV (mean energy of 14.5 EeV) respectively, amplitudes are also shown. The observed amplitude above 8 EeV suggests that a large-scale anisotropy is imprinted on the CR arrival directions of extragalactic CRs towards $\simeq 95^\circ$ in right ascension. It is interesting to note that this phase is roughly in the opposite direction to the one suggested in the summary phase plot shown in the left panel for CRs with energies below 1 EeV and which is in the general direction of the Galactic Centre. An interesting possibility to explain the low amplitudes over the wide energy range would thus be that a progressive cross-over might be taking place between a component of Galactic origin and another one of extragalactic origin. The global dipole anisotropy is then the sum of two vectors with opposite directions, providing then a natural mechanism to reduce significantly the amplitude of the vector describing the arrival directions of the whole population of CRs. Continued scrutiny of the large-scale distribution of arrival directions of CRs with increased sensitivity will provide further insights to reveal the origin of CRs in this energy range.

[1] A rectification of the analogous figure published in [5] is shown here. Figure 3 in [5] had the flux incorrectly normalised to the exposure limited to events with zenith angles smaller than 60° (37,142 km$^2$ sr yr) while here it is correctly normalised to the total exposure including events with zenith angles between 60° and 80° (48,029 km$^2$ sr yr).

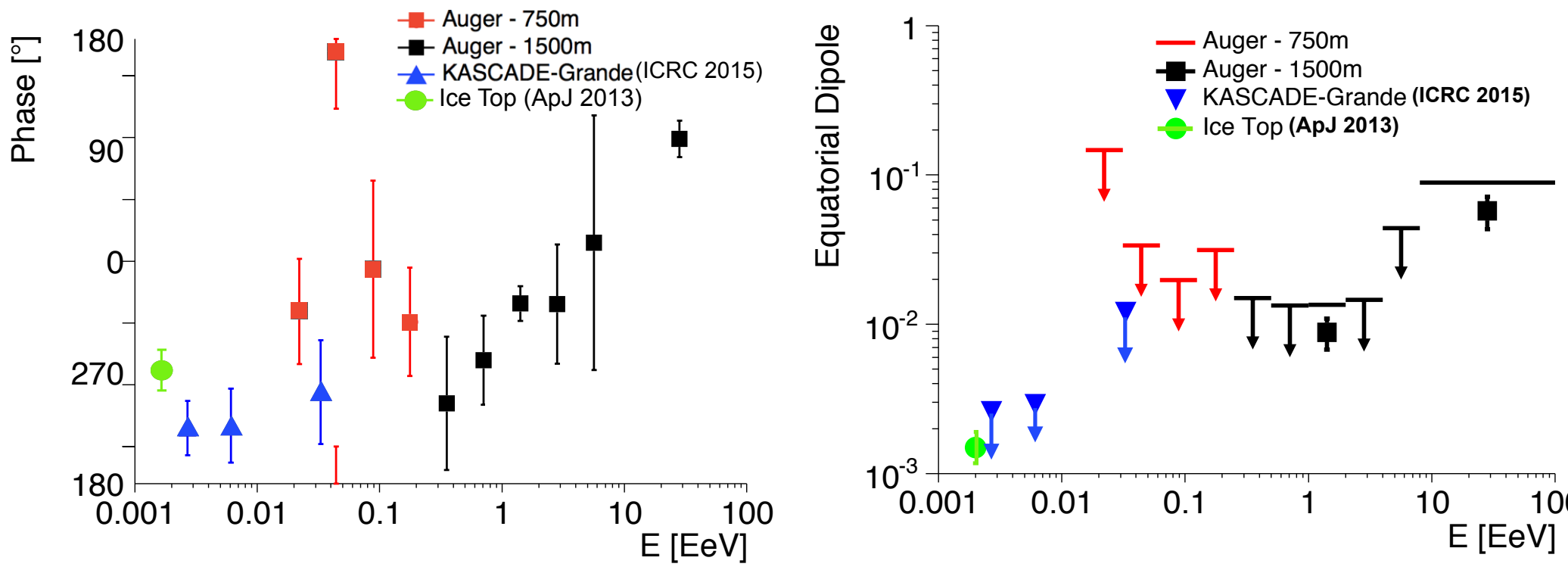

**Figure 3:** Left: Measured phases of the first harmonic modulation in RA. Right: Upper limits of the dipole equatorial component. Amplitudes are also reported in the two energy bins when the corresponding p-value expected from isotropy is below $10^{-3}$.

| | $\Delta E$ [EeV] | $N$ | $d_\perp \pm \Delta_{d_\perp}$ [%] | $\phi \pm \Delta_\phi [°]$ | $P(>d_\perp)$ [%] | $d_\perp^{\mathrm{UL}}$ [%] |
|---|---|---|---|---|---|---|
| 750 m [E-W] | $0.015 - 0.03$ | 32,244 | $6.4 \pm 3.8$ | $319 \pm 42$ | 25 | 14.5 |
| 750 m [E-W] | $0.03 - 0.06$ | 393,846 | $1.4 \pm 0.9$ | $169 \pm 46$ | 30 | 3.3 |
| 750 m [E-W] | $0.06 - 0.12$ | 581,313 | $0.5 \pm 0.6$ | $353 \pm 71$ | 73 | 2.0 |
| 750 m [E-W] | $0.12 - 0.25$ | 268,728 | $1.4 \pm 0.8$ | $310 \pm 43$ | 27 | 3.1 |
| 750 m [E-W] | $0.25 - 0.5$ | 68,782 | $2.8 \pm 1.5$ | $325 \pm 39$ | 20 | 6.0 |
| 750 m [E-W] | $0.5 - 1$ | 14,324 | $7.2 \pm 3.3$ | $233 \pm 31$ | 10 | 14.5 |
| 1500 m [E-W] | $0.25 - 0.5$ | 918,247 | $0.58 \pm 0.45$ | $245 \pm 54$ | 45 | 1.5 |
| 1500 m [E-W] | $0.5 - 1$ | 1,464,390 | $0.65 \pm 0.33$ | $279 \pm 36$ | 15 | 1.3 |
| 1500 m [R] | $1 - 2$ | 738,683 | $0.90 \pm 0.2$ | $326 \pm 14$ | $1.5 \times 10^{-2}$ | - |
| 1500 m [R] | $2 - 4$ | 196,992 | $0.60 \pm 0.38$ | $325 \pm 48$ | 45 | 1.45 |
| 1500 m [R,*] | $4 - 8$ | 50,417 | $0.40 \pm 0.80$ | $15 \pm 103$ | 88 | 4.0 |
| 1500 m [R,*] | $> 8$ | 19,797 | $5.7 \pm 1.3$ | $95 \pm 13$ | $6.4 \times 10^{-3}$ | - |

**Table 1:** Summary of the harmonic analysis in different energy intervals in terms of the equatorial component of the dipole. In the left column, [E-W] and [R] stand for the selected methods used to obtain the results, East-West or Rayleigh respectively. Data used are from 01/01/04 to 31/12/14, except for the two last bins indicated with [R,*], where events with zenith angles larger than 60° are included and where the ending date is 31/12/13, as reported in [5].

The measured dipolar anisotropy at the few % level above 8 EeV could result from the diffusive propagation of extragalactic cosmic rays in the extragalactic turbulent magnetic field. This could happen if the amplitude of the field is large and/or if the cosmic rays have a component with large electric charge [13]. A large-scale anisotropy is also expected in the case that magnetic deflections are small if the cosmic ray sources are distributed similarly to the matter in the universe, due to the fact that in our local neighbourhood, matter is distributed inhomogeneously.

# Search for energy dependent patterns in the arrival direction of cosmic rays at the Pierre Auger Observatory

**Tobias Winchen**[*a] **for the Pierre Auger Collaboration**[b]
[a]*Bergische Universität Wuppertal, Department of Physics, 42119 Wuppertal, Germany*
[b]*Observatorio Pierre Auger, Av. San Martín Norte 304, 5613 Malargüe, Argentina*
*E-mail:* auger_spokespersons@fnal.gov
*Full author list:* http://www.auger.org/archive/authors_2015_06.html

Energy-dependent patterns in the arrival directions of cosmic rays are expected from deflections in galactic and extragalactic magnetic fields. We report on searches for such patterns in the data of the surface detector of the Pierre Auger Observatory at energies above $E = 5\,\mathrm{EeV}$ in regions within approximately $15^\circ$ around events with energy $E > 60\,\mathrm{EeV}$. No significant patterns are found with this analysis which can be used to constrain parameters in propagation scenarios.

[*]Speaker.

## 1. Introduction

The sources of ultra-high energy cosmic rays (UHECR) have not been identified so far, presumably because of the deflection of the charged cosmic rays in the galactic and extragalactic magnetic fields. However, from such deflections the distribution of UHECRs arrival directions may show energy-dependent patterns. In particular, a circular 'blurring' of the sources is expected from deflection in turbulent magnetic fields, while energy-dependent linear structures are expected from deflection in coherent magnetic fields. To search for such patterns in the data collected with the Pierre Auger Observatory [1], we investigate the local Regions Of Interest (ROI) around cosmic rays with $E \geq 60\,\mathrm{EeV}$. We analyzed the cosmic rays with energies above $E = 5$ EeV arriving within an angular separation of 0.25 rad using two independent methods [2]. In one method we study energy-energy correlations between pairs of cosmic rays depending on their angular separation from the center of the region. This method is sensitive to the circular patterns expected from particle deflection in turbulent magnetic fields [3, 4]. In the second method we decompose the directional energy distribution of the cosmic rays along its principal axes. This method is sensitive to clusters of cosmic rays as well as to linear patterns expected from deflections in coherent magnetic fields [5, 6].

## 2. Methods

### 2.1 Energy-Energy Correlations

The Energy-energy correlation $\Omega_{ij}$ is calculated for every pair of UHECRs $ij$ within a ROI using

$$\Omega_{ij} = \frac{(E_i - \langle E(\alpha_i)\rangle)(E_j - \langle E(\alpha_j)\rangle)}{E_i E_j} \tag{2.1}$$

where $E_i$ is the energy of the UHECR $i$ with the angular separation $\alpha_i$ to the center of the ROI and $\langle E_i(\alpha_i)\rangle$ is the average energy of all UHECRs at the angular separation $\alpha_i$ from the center of the ROI. A pair of cosmic rays $ij$ can contribute positively or negatively to the distribution of $\Omega_{ij}$. If one particle has an energy above the corresponding average energy and the other below the corresponding average energy, the contribution is negative. A pair with both energies being below average as well as a pair with both being above average contribute positively. If there is a pattern in the energy distribution there are more positive correlations than expected from isotropically distributed cosmic rays.

### 2.2 Principal Axes

The system of principal axes of the energy distribution $\vec{n}_{k=1,2,3}$ is calculated by successive maximization of the quantity

$$T_k = \max_{\vec{n}_k} \left( \frac{\sum_i |\omega_i^{-1}\, \vec{p}_i \cdot \vec{n}_k|}{\sum_i |\omega_i^{-1}\, \vec{p}_i|} \right) \tag{2.2}$$

with respect to the axes $\vec{n}_k$ starting with $k = 1$. Here $\vec{p}_i$ is the momentum and $\omega_i$ the exposure of the detector [7] in the direction of particle $i$. The resulting values of $T_{k=1,2,3}$ quantify the strength of the collimation of the particle momenta along each of the three axes $\vec{n}_{k=1,2,3}$ of the principal system.

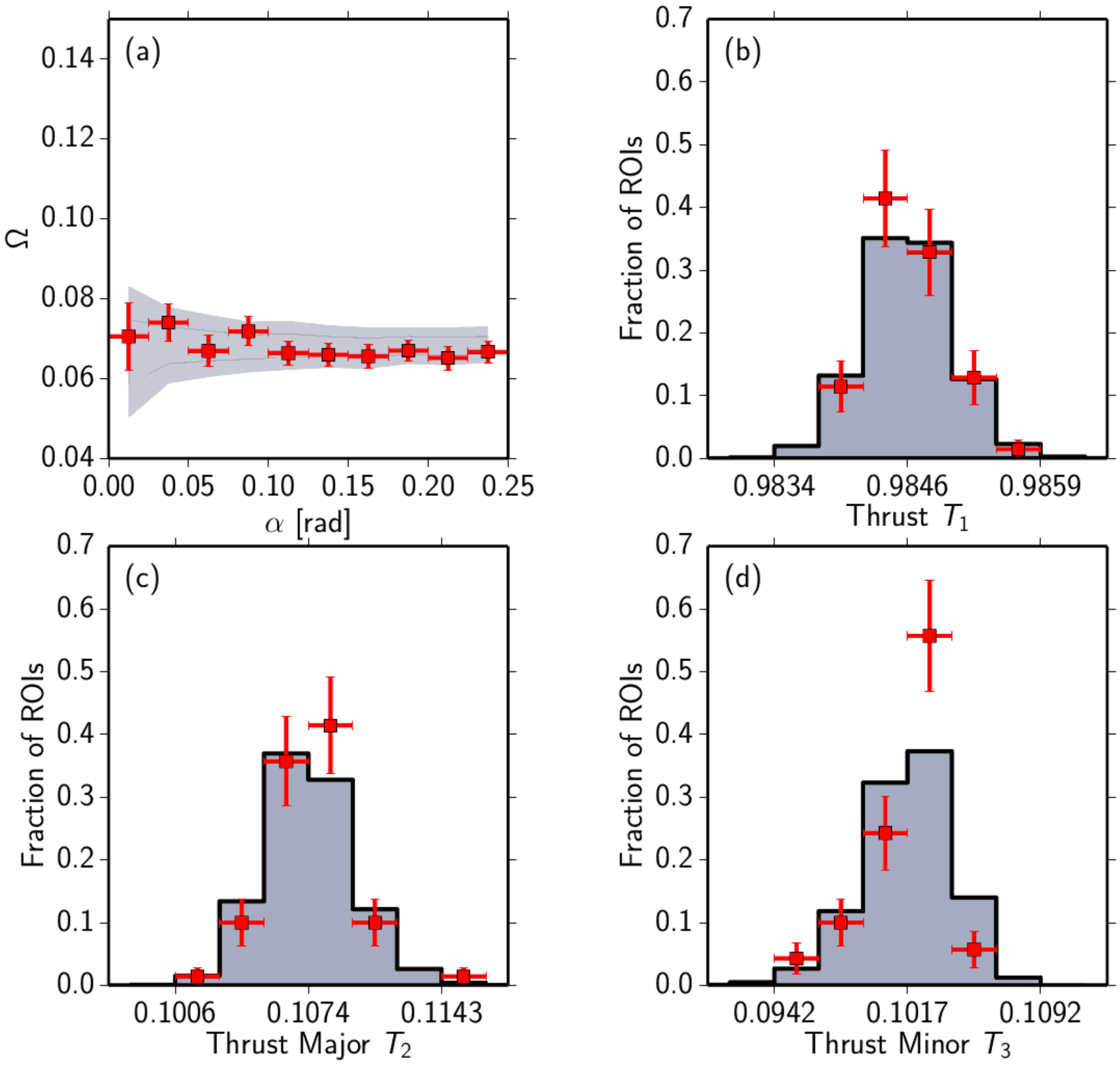

**Figure 1:** Measurement of the **(a)** energy-energy correlation Ω and **(b-d)** observables $T_{1,2,3}$ with the Pierre Auger Observatory (red squares and error bars). The measurements are compared to distributions without patterns in the arrival directions of UHECRs (filled distributions).

The first principal axis $\vec{n}_1$ is the radial unit vector $\vec{e}_r$ pointing to the local barycenter of the energy distribution. The $T_1$ value is thus a measure for the energy-weighted strength of clustering of the events. It is $T_1 = 1$ for no dispersion of the particles in the region, whereas for an isotropic distribution of UHECRs in a region the expectation value is determined by the size of the region [6].

The second and third principal axes $\vec{n}_2$ and $\vec{n}_3$ can be written as linear combination of the unit vectors $\vec{e}_\phi$ and $\vec{e}_\theta$. The $T_2$ value becomes maximal if $\vec{n}_2$ is aligned with a linear distribution of UHECR arrival directions and can be thus used as a generalized multiplet analysis. It thus points along threadlike structures in the energy distribution of UHECRs. As the $\vec{n}_2$ axis is chosen perpendicular to $\vec{n}_1$ and $\vec{n}_2$ it has no additional physical meaning. However, the $T_3$ value contains information as it denotes the collimation strength perpendicular to the $\vec{n}_2$ axis.

## 3. Results

We measured the EEC and $T_{1,2,3}$ distributions using 30,664 events recorded with the surface detector of the Pierre Auger Observatory above 5 EeV. For the selected events we further required that the zenith angle of the events is smaller than 60° and that the detector stations surrounding the station with the highest signal are active [8] to obtain a sample with minimum potential biases. Of the selected events, 70 have an energy $E \geq 60$ EeV and are at least 0.25 rad inside the field of view of the Pierre Auger Observatory. These events mark the ROIs used in this analysis.

In Figure 1 the distributions of the EEC and the $T_{1,2,3}$ observables are shown together with the distributions expected from isotropic arrival directions of UHECRs. The measured distributions of all four observables reveal no local patterns in the arrival directions of UHECRs. The goodness-of-fit of the measurements compared to expected distributions without structure in the arrival directions of UHECRs using a $\chi^2$ test, yields $p$-values which are all above $p = 0.2$ except for the $T_3$ distribution with $p(T_3) = 0.01$. However, this low $p$-value results from a lack of signal-like regions in the data which are expected to broaden the distribution, and thus does not indicate significant patterns.

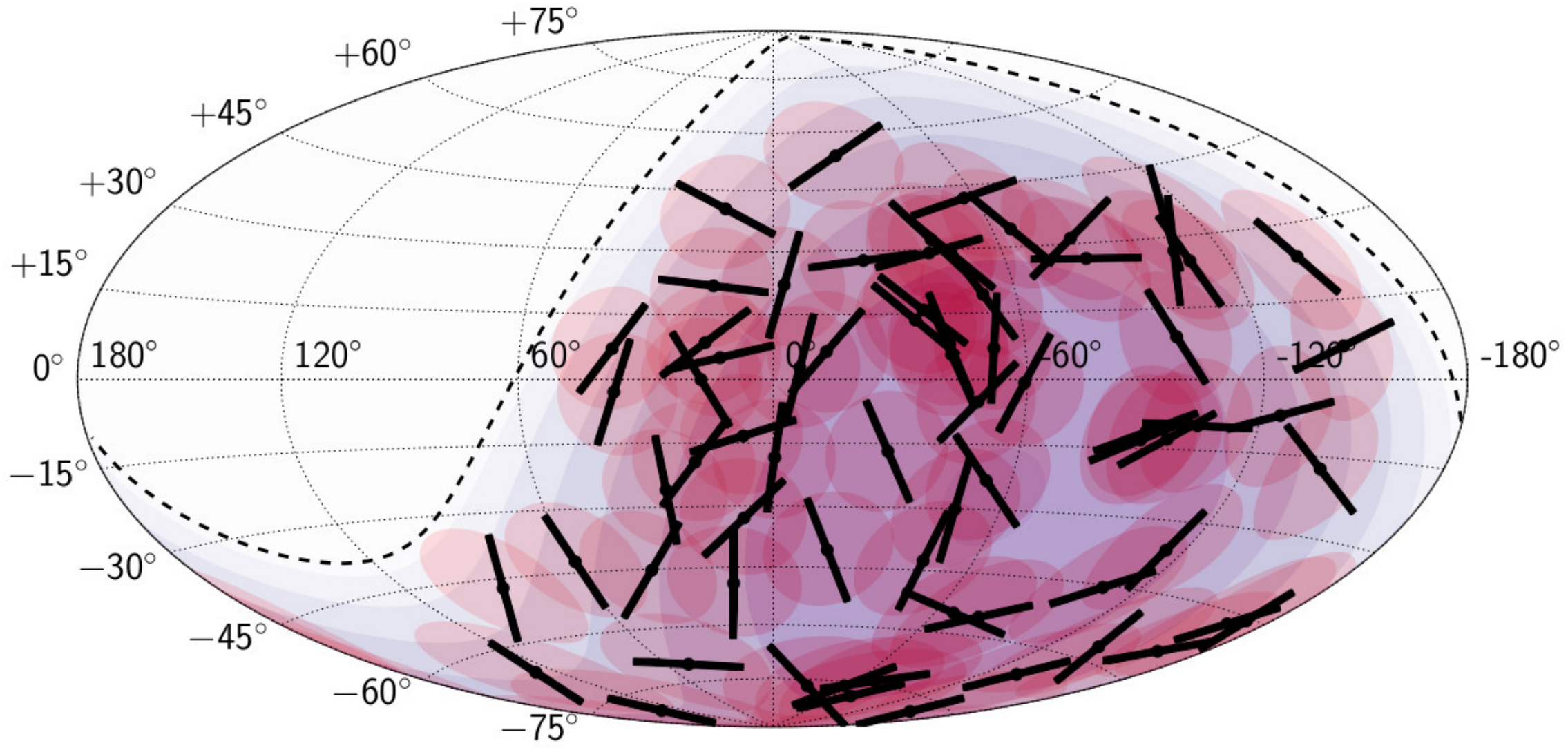

**Figure 2:** Map of principal axes of the directional energy distribution using a Hammer projection and galactic coordinates. The red shaded areas represent the regions of interest. Black lines denote the second principal axes $\vec{n}_2$ and black dots mark the positions of axes $\vec{n}_1$. The blue shading indicates the exposure of the Pierre Auger Observatory; the dashed line marks the extent of its field of view.

In addition to the scalar distributions, the direction of the second principal axes $\vec{n}_2$ can be displayed as the map shown in figure 2. If the axes are non-trivial, this map displays the directions of deflection of UEHCR in coherent cosmic magnetic fields. In simulation studies it was shown that the distribution of principal axes can contain information, even though the $T_{1,2,3}$ scalar values are compatible with isotropy [6]. We tested the reproducibility of the axes in subsets of the data by splitting the dataset into 12 independent subsets by chance and analyzing the variance of the directions in the subsets. We found no reproducibility of the axes in this analysis.

## 4. Discussion and Conclusions

The non observation of significant patterns can be used to exclude all scenarios that predict otherwise. To illustrate the required procedure we simulated a simple model for extragalactic propagation of protons based on parameterizations as implemented in version 1.2 of the PARSEC software [9]. In particular we assume here, that the UHECRs originate from isotropically distributed point sources with equal luminosity and that the root-mean-square of the deflections of UHECRs with energies $E$ from a source in distance $D$ can be parametrized as $\delta_{\mathrm{RMS}} = C_{\mathrm{E}} \frac{\sqrt{D}}{E}$. The calculations further account for energy losses of the UHECRs from interaction with extragalactic-photon backgrounds, effects from the expansion of the universe, and the deflection in the galactic magnetic field using the model proposed by Jansson and Farrar [10, 11].

We scanned the density of point sources and the strength of the deflection in the extragalactic magnetic field $C_{\mathrm{E}}$ in this scenario and derived a combined limit on both parameters using the $CL_S$ method [12]. Within this simplified scenario, we found that the deflection in the extragalactic magnetic field has to be larger than $C_{\mathrm{E}} = 10 - 120^{\circ}\,\mathrm{Mpc}^{-1/2}\,\mathrm{EeV}$ for source densities smaller than $10^{-3}\,\mathrm{Mpc}^{-3}$. For protons with an energy $E = 10\,\mathrm{EeV}$ from a source at $16\,\mathrm{Mpc}$ this translates to a required strength of the deflection in extragalactic space of more than $4^{\circ}$ if the source density is smaller than $10^{-3}\,\mathrm{Mpc}^{-3}$ and more than $25^{\circ}$ if the source density is smaller than $10^{-4}\,\mathrm{Mpc}^{-3}$.

# 3

# Composition

PIERRE
AUGER
OBSERVATORY

# Measurements of $X_{max}$ above $10^{17}$ eV with the fluorescence detector of the Pierre Auger Observatory

**Alessio Porcelli**[*,a] **for the Pierre Auger Collaboration**[b]
[a]*Karlsruhe Institute of Technology, Karlsruhe, Germany,*
*currently at the Department of Nuclear Physics, University of Geneva, Geneva, Switzerland*
[b]*Observatorio Pierre Auger, Av. San Martín Norte 304, 5613 Malargüe, Argentina*
*E-mail:* auger_spokespersons@fnal.gov
*Full author list:* http://www.auger.org/archive/authors_2015_06.html

For the first time the Pierre Auger Collaboration presents $\langle X_{max} \rangle$ and $\sigma(X_{max})$ measurements covering nearly three decades of energy. In this analysis we include new $X_{max}$ data obtained with the High Elevation Auger Telescopes (HEAT) enhancement. The HEAT telescopes cover a field of view ranging from 30° to 60° of elevation and are located next to one of the standard fluorescence detector sites (Coihueco). The combination of the HEAT and Coihueco telescopes covers a field of view from ∼2° up to ∼60° of elevation. Thus, the combination can sample the longitudinal profile of nearby lower energy showers, allowing us to extend the energy range down to $10^{17}$ eV.

*Speaker.

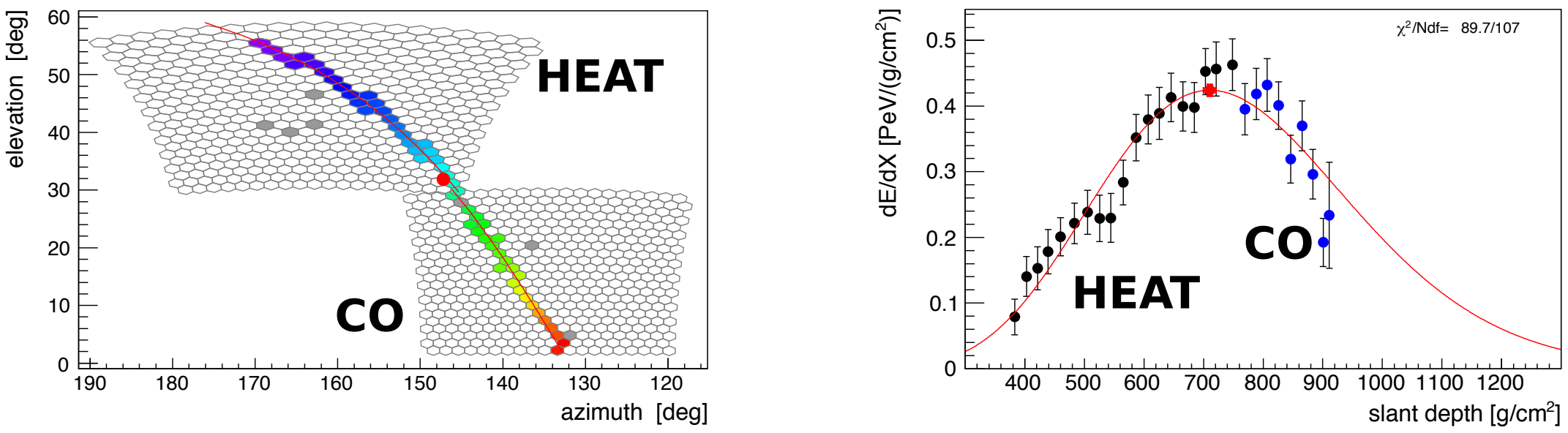

**Figure 1:** Example of a HeCo event with an energy of $(3.7 \pm 0.1) \times 10^{17}$ eV. Left: the camera view, the timing of the pixel pulses is color-coded (early = blue, late = red). Right: the measured longitudinal profile (black dots — HEAT, blue dots — Coihueco) with the Gaisser-Hillas fit (red line). The red point in both panels indicates the $X_{\max}$ position.

## 1. Introduction

The knowledge of the composition of cosmic rays in the energy range of 0.1 to 1 EeV is a key ingredient to identify a possible transition from galactic to extra-galactic sources and for understanding the nature of the energy spectrum features (ankle at $\approx 4$ EeV and cut-off at $\approx 40$ EeV).

The depth at which the number of secondary air-shower particles reaches its maximum, $X_{\max}$, is one of the most robust observables for studying the mass composition [1]. Experimentally, the longitudinal profile of the shower development can be measured using fluorescence light emitted by molecules of atmospheric nitrogen excited by Extensive Air Shower (EAS) particles. At the Pierre Auger Observatory, which is continuously taking data since 01.2004, such measurements are performed using the fluorescence detector (FD) consisting of 24 telescopes placed at 4 locations and, since 06.2010, using the High Elevation Auger Telescopes (HEAT). With these telescopes the Field of View (FoV) of the Coihueco (CO) site is expanded from $2 \div 30°$ up to $2 \div 60°$ of elevation, which allows one to observe nearby low energy showers ($E < 10^{17.8}$ eV). In Figure 1 an example of a low energy event in the enlarged FoV is shown: the track on the camera (left) and the longitudinal profile with the Gaisser-Hillas fit (right).

The determination of the primary composition with FD data is performed using the characteristics of measured $X_{\max}$ distributions of EAS. The first two moments of the $X_{\max}$ distribution ($\langle X_{\max} \rangle$ and $\sigma(X_{\max})$) are related to the first two moments of the distribution of the logarithm of masses of primary particles ($\ln A$ and $\sigma(\ln A)$) [2]:

$$\langle X_{\max} \rangle = \langle X_{\max} \rangle_p + f_E \langle \ln A \rangle \quad (1.1)$$

$$\sigma^2(X_{\max}) = \langle \sigma_{sh}^2 \rangle + f_E^2 \sigma^2(\ln A). \quad (1.2)$$

$\langle X_{\max} \rangle_p$ and $\langle \sigma_{sh}^2 \rangle$ are the mean $X_{\max}$ for protons and the composition-averaged shower-to-shower fluctuations[1], and $f_E$ is a parameter depending on details of hadronic interactions, properly parametrized from the interaction models for energies $\geq 10^{17}$ eV.

[1] $\langle \sigma_{sh}^2 \rangle = \sum_i f_i \sigma_i^2(X_{\max})$ where $f_i$ is the relative fraction of mass $A_i$

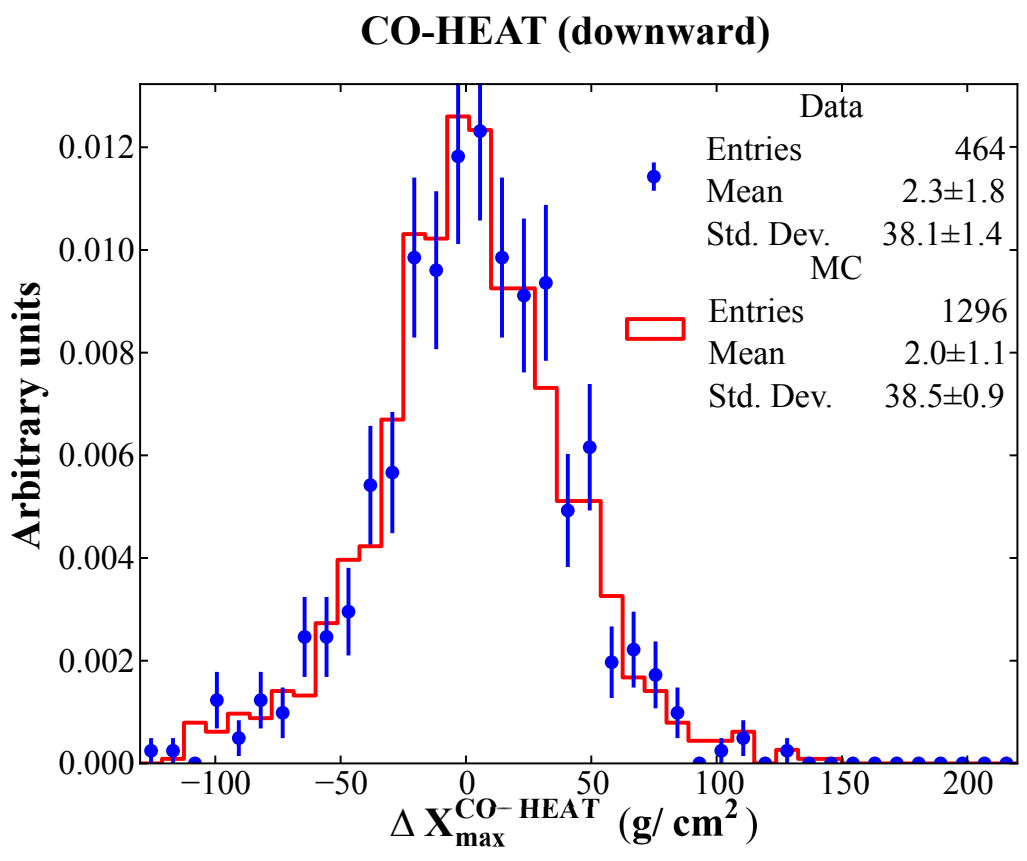

**Figure 2:** $X_{\rm max}$ difference between CO and HEAT (in downward mode) in MC simulation (red histogram) and data (blue dots).

In this paper, nearly two years of calibrated HEAT data, from 01.06.2010 to 15.08.2012, are used to extend the previous measurement of the $X_{\rm max}$ moments [1] from $10^{17.8}$ eV down to $10^{17}$ eV.

## 2. Data analysis

The analysis presented in this paper is based on two datasets. The data collected by the standard FD telescopes during the period from 01.12.2004 to 31.12.2012 (published in [1]), and the data collected with HEAT and Coihueco telescopes (HeCo) during the period from 01.06.2010 to 15.08.2012.

HEAT can be operated in upward and downward modes. The downward mode is when the telescopes are oriented such that their elevation angle extends up to 30° (same as the standard Auger telescopes). The upward mode is when they cover an elevation angle ranging from 30° to 60° (this is the HEAT standard operation mode). The HEAT downward mode is used for systematic cross checks, because it allows one to observe the same showers in coincidence with telescopes from the Coihueco site. In Figure 2 the $X_{\rm max}$ difference between CO and HEAT in downward mode is shown. Data (blue dots) and simulations (red lines) are in agreement, which implies a good knowledge of the detector.

The standard FD dataset contains events with energies above $10^{17.8}$ eV and the HeCo one contains events with energies above $10^{17.0}$ eV. HeCo runs out of statistics at energies beyond $E < 10^{18.3}$ eV. In order to combine two statistically independent datasets, we have removed from the standard FD dataset all Coihueco events with energies below $E < 10^{18.3}$ eV and recorded during the HeCo period (from 01.06.2010 to 15.08.2012). This cut has reduced the standard FD data by only 1377 events (out of 19759 events).

### 2.1 Data selection

The analysis is based on hybrid events, i.e. on events with geometries reconstructed using information on arrival times of both light in the cameras of FD telescopes and of the shower front at ground as measured by the ground station closest to the shower axis. We selected data recorded during stable running conditions and good atmospheric conditions [1]. In addition to these selection criteria a set of fiducial FoV cuts are applied to reduce to the minimum the detector effects in the sampled $X_{\text{max}}$ distributions (as explained in Section 2.2).

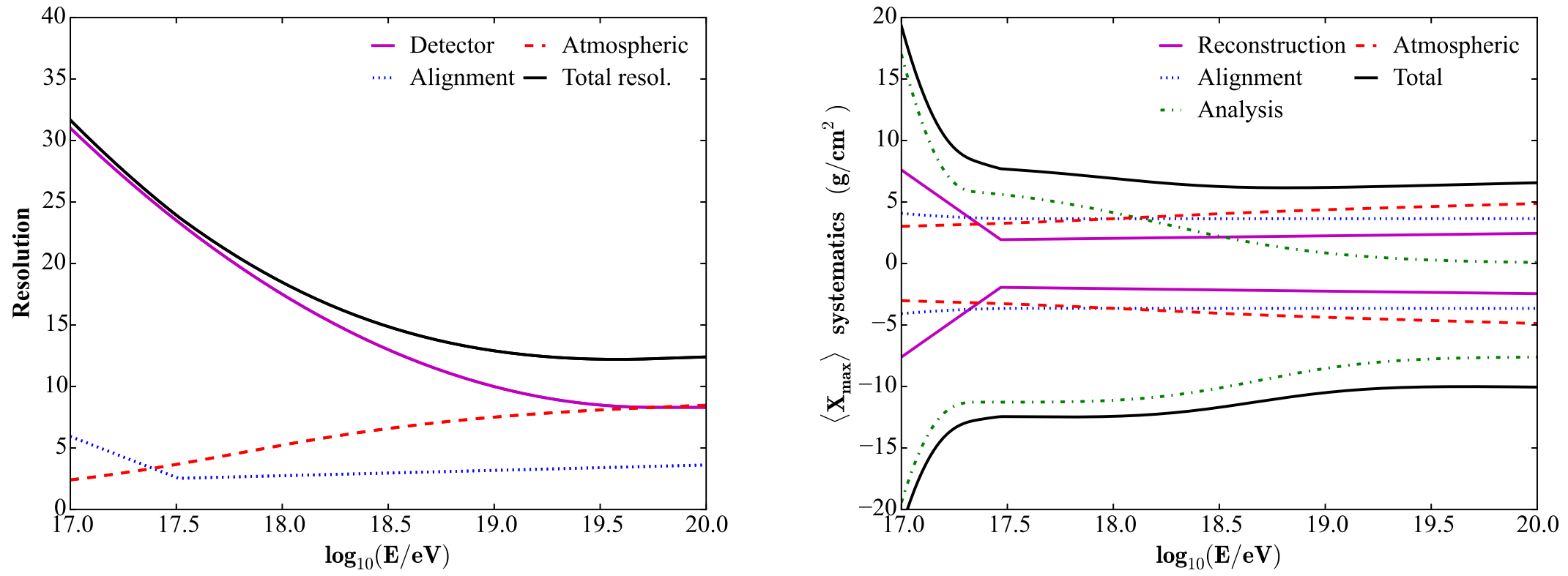

**Figure 3:** $X_{\text{max}}$ resolution (left) and systematic uncertainties in the $X_{\text{max}}$ scale (right) as functions of energy.

For the HeCo dataset, specific corrections to $\langle X_{\text{max}} \rangle$ and $\sigma(X_{\text{max}})$ are applied to reduce residual biases, especially for the lower energy bins ($E < 10^{17.5}$ eV).

### 2.2 FoV selection criteria

A shower is reconstructed accurately only if its $X_{\text{max}}$ is within the FoV. Shallow or deep events are more likely to have their $X_{\text{max}}$ values outside the FoV and have larger chances to be excluded from the analysis. In general, at the lower energies where the showers are closer to the telescopes, the limited FoV biases the sample towards lighter composition (i.e. towards deeper $X_{\text{max}}$ values).

For data satisfying the selection criteria explained in Section 2.1, a fiducial FoV is derived. This fiducial range is characterized by the lower $X_{\text{low}}$ and upper $X_{\text{up}}$ boundaries. These parameters define the slant depth range where $X_{\text{max}}$ of each event would be reconstructed with a resolution better than 40 g cm$^{-2}$. To have higher quality events, the $X_{\text{max}}$ value must fall inside these boundaries. Furthermore, if the values of $X_{\text{low}}$ and $X_{\text{up}}$ are not within certain limits (i.e. $X_{\text{low}}$ and $X_{\text{up}}$ should enclose the bulk of the $X_{\text{max}}$ distribution), the event is also excluded.

The processes to calculate $X_{\text{low}}$ and $X_{\text{up}}$ parameters, and the limits on them, are explained in detail in [1].

### 2.3 Estimating the $X_{\text{max}}$ moments

After the application of all selection criteria, the moments of the $X_{\text{max}}$ distribution are estimated as described in [1]. Reconstruction and residual acceptance biases are estimated through simulations and corrected for. The observed width of the distribution is corrected by subtracting the detector resolution (Figure 3, left) in quadrature to obtain $\sigma(X_{\text{max}})$.

The systematic uncertainty in the $X_{\text{max}}$ scale is displayed in Figure 3 (right). At low energies it is dominated by uncertainties of the analysis procedure, at high energies atmospheric uncertainties give as well a significant contribution.

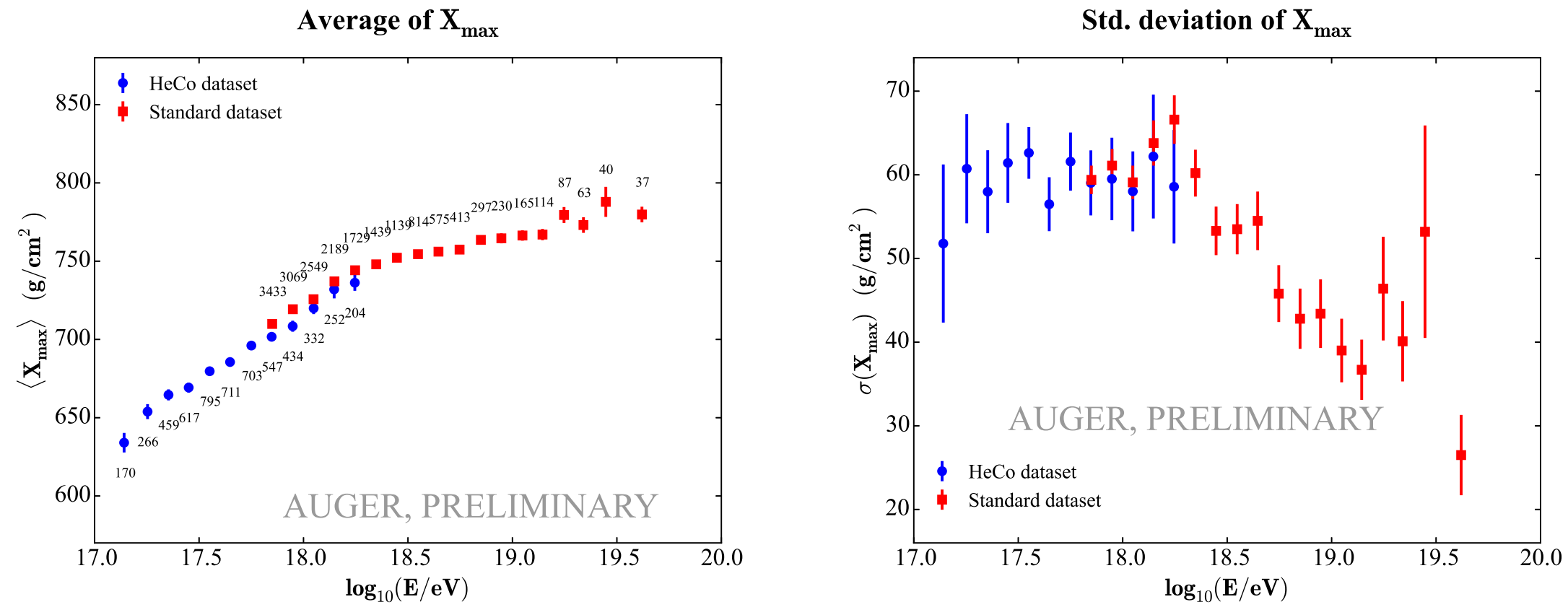

**Figure 4:** The mean (left) and standard deviation (right) of measured $X_{\mathrm{max}}$ distributions of the two independent datasets: HeCo (blue circles) and the standard FD (red squares).

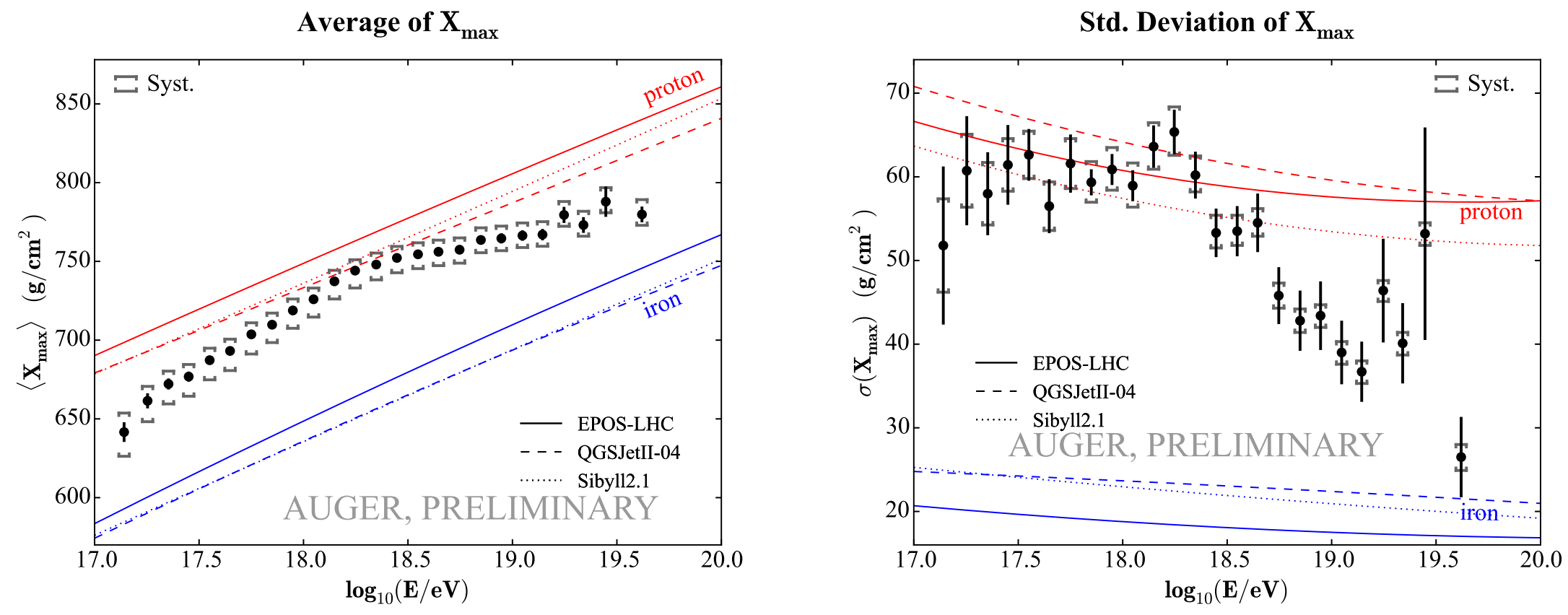

**Figure 5:** The mean (left) and the standard deviation (right) of the measured $X_{\mathrm{max}}$ distributions (combining HeCo and standard datasets) as a function of energy compared to air-shower simulations for proton and iron primaries.

## 2.4 Results and Interpretation

In Figure 4 the $X_{\mathrm{max}}$ moments estimated using HeCo and the standard FD datasets are compared. While $\langle X_{\mathrm{max}} \rangle$ differs by $\sim 7$ g cm$^{-2}$ between datasets (within the uncorrelated systematics of the two analyses), the second moments $\sigma(X_{\mathrm{max}})$ are found to be in a good agreement. For the combination of the datasets the HeCO $\langle X_{\mathrm{max}} \rangle$ is shifted by $+7$ g cm$^{-2}$ and the resulting $\langle X_{\mathrm{max}} \rangle$ and $\sigma(X_{\mathrm{max}})$ are shown in Figure 5.

Between $10^{17.0}$ and $10^{18.3}$ eV $\langle X_{\mathrm{max}} \rangle$ increases by around 85 g cm$^{-2}$ per decade of energy (Figure 5, left). This value, being larger than the one expected for a constant mass composition ($\sim 60$ g cm$^{-2}$/decade), indicates that the mean primary mass is getting lighter. Around $\approx 10^{18.3}$ eV the observed rate of change of $\langle X_{\mathrm{max}} \rangle$ becomes significantly smaller ($\sim 26$ g cm$^{-2}$/decade) indicating that the composition is becoming heavier. The fluctuations of $X_{\mathrm{max}}$ (Figure 5, right) start to decrease at around the same energy $\approx 10^{18.3}$ eV.

The mean value of $\ln A$ and its variance $\sigma^2(\ln A)$, determined from Equations (1.1) and (1.2),

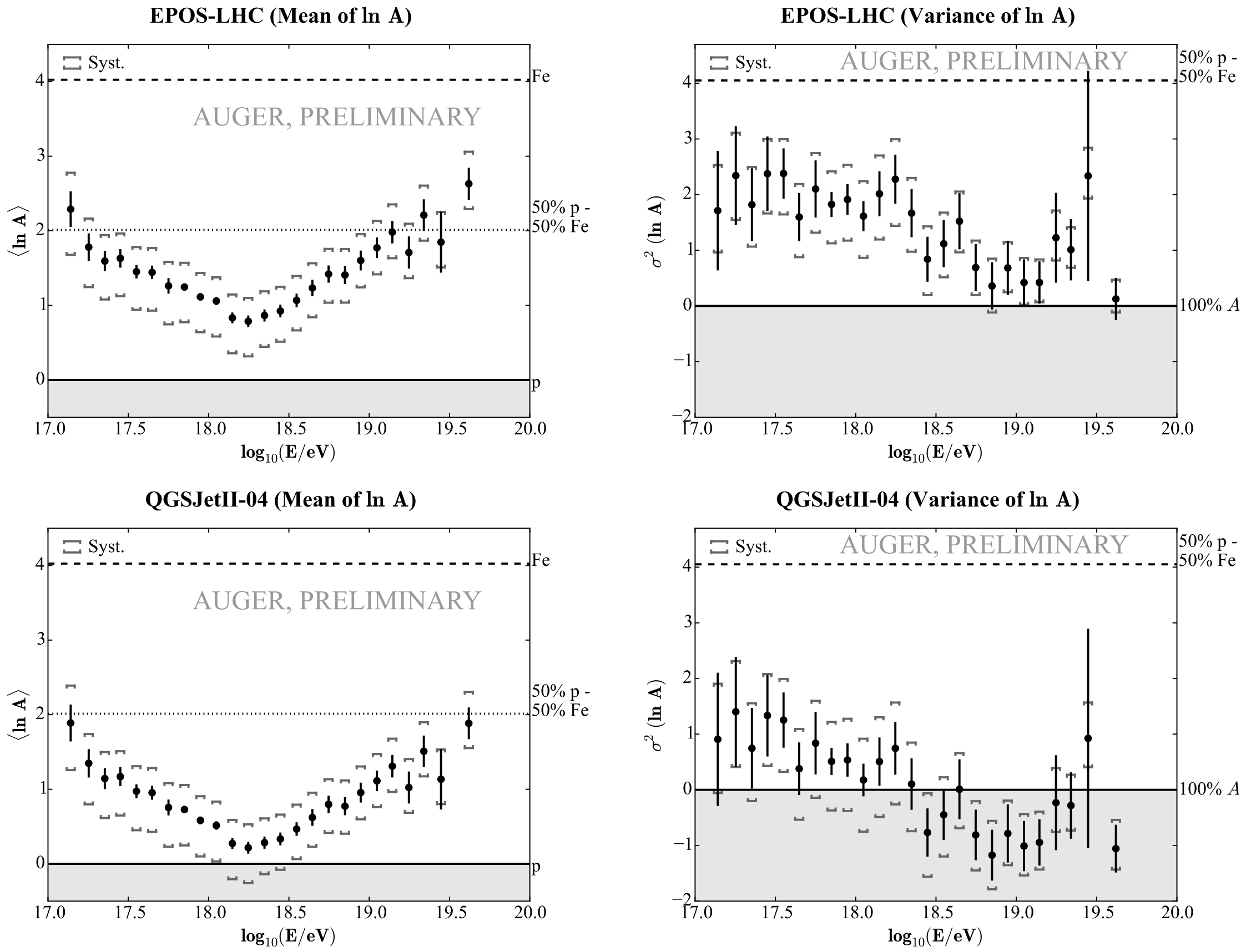

(a) $\langle \ln A \rangle$ (b) $\sigma^2(\ln A)$

**Figure 6:** The mean (left) and the variance (right) of ln$A$ estimated from data with EPOS-LHC (up) and QGSJetII-04 (down).

are shown in Figure 6. For the parameters $\langle X_{\max} \rangle_p$, $f_E$ and $\langle \sigma^2_{sh} \rangle$, EPOS-LHC and QGSJetII-04 hadronic interaction models are used.

For both models the similar trends with energy for $\langle \ln A \rangle$ and $\sigma^2(\ln A)$ are observed. The primary mass is decreasing reaching the minimal values at around $10^{18.3}$ eV and starts to increase for the higher energies. The spread of the masses is almost constant till $\approx 10^{18.3}$ eV and then starts to decrease, together with the behavior of $\langle \ln A \rangle$ that might be an indication that the relative fraction of protons becomes smaller for the energies above $\approx 10^{18.3}$ eV (see [3]).

# Composition at the "ankle" measured by the Pierre Auger Observatory: pure or mixed?

**Alexey Yushkov**[*,a] **for the Pierre Auger Collaboration**[b]
[a] *University of Siegen, Department of Physics, 57072, Siegen, Germany*
[b] *Observatorio Pierre Auger, Av. San Martín Norte 304, 5613 Malargüe, Argentina*
*E-mail:* auger_spokespersons@fnal.gov
*Full author list:* http://www.auger.org/archive/authors_2015_06.html

We report for the first time on the measurement of the correlation between the depth of shower maximum and the signal in water-Cherenkov stations for events reconstructed by both the fluorescence and the surface detectors of the Pierre Auger Observatory. Such a correlated measurement is a unique feature of a hybrid air-shower observatory and allows us to determine the purity of the cosmic-ray composition. The observed correlation in the energy range around the "ankle" $\lg(E/\mathrm{eV}) = 18.5 - 19.0$ differs significantly from the expectations for pure beams, indicating that the primary composition in this range is mixed, unless the hadronic interactions at these energies behave very differently than in conventional, LHC-tuned event generators.

*Speaker.

## 1. Introduction

The determination of the mass composition of ultra-high-energy cosmic rays (UHECRs) is one of the most challenging problems for extensive air-shower (EAS) experiments; for a recent review, see [1]. Parameters of hadronic interactions at these energies are only loosely constrained by accelerator data and thus the evolution of the behavior of EAS properties with energy can in general be interpreted in terms of both changes of the primary mass and/or of the characteristics of the particle interactions. One way to try to resolve this ambiguity is to find shower parameters or their combinations that rely on more general aspects of EAS physics and are thus relatively insensitive to the uncertainties in the properties of the hadronic interactions.

In particular, in [2] it was proposed to use the correlation between the depth of the shower maximum $X_{\max}$ and the number of muons $N_\mu$ of the EAS for the determination of the degree of purity of the beam, i.e., whether it is composed of several or just one nuclear species. In the present work we adapt this idea to the conditions of the Pierre Auger Observatory [3]. In place of $N_\mu$ we use the total signal in water-Cherenkov detectors at 1000 meters from the core, $S(1000)$, a substantial fraction of which is due to muons: from 40% to 90% for zenith angles from 20° to 60° [4]. We show that the correlation $(X_{\max}, S(1000))$ for pure primary beams for all current interaction models turns out to be close to zero or positive, while for well-mixed compositions with a large spread of masses it becomes negative (see [2]). Thus the correlation coefficient can be used to determine the dispersion, $\sigma(\ln A)$, of primary masses, given by $\sigma(\ln A) = \sqrt{\langle \ln^2 A\rangle - \langle \ln A\rangle^2}$ where $\langle \ln A\rangle = \sum_i f_i \ln A_i$ and $\langle \ln^2 A\rangle = \sum_i f_i \ln^2 A_i$ with $f_i$ being the relative fraction of mass $A_i$.

An estimation of the degree of purity of the primary beam in the energy range $\lg(E/\mathrm{eV}) = 18.5 - 19.0$ is of particular interest as a test of the 'dip' scenario [5]. In this scenario the break in the energy spectrum at around $\lg(E/\mathrm{eV}) = 18.7$ results from electron-positron pair-production by extragalactic protons interacting with the cosmic microwave background. The 'dip' is well pronounced only if the fraction of heavier nuclei at the acceleration site is $\lesssim 15\%$ (see [6] and references therein).

## 2. Data and simulations

The analysis is based on the same hybrid events as used in [7] recorded by both fluorescence (FD) and surface detectors (SD) of the Pierre Auger Observatory during the period from 01.01.2004 to 31.12.2012. The procedure of data selection, described in [7], guarantees that only high quality events are included in the analysis and that the mass composition of the selected sample is unbiased. The use of the signal in ground stations requires an additional application of the fiducial trigger cut [8] (the station with the highest signal should have at least 5 working neighbour stations), and exclusion of events with stations having saturated signal traces. The final data set for energies $\lg(E/\mathrm{eV}) = 18.5 - 19.0$ and zenith angles $\theta = 0° - 65°$ contains 1376 events.

Monte-Carlo (MC) simulations are performed with CORSIKA [9] for high-energy interaction models QGSJetII-04 [10], Epos-LHC[11] and Sibyll 2.1 [12]. FLUKA [13] is used to treat low energy interactions. For CORSIKA events, full detector simulation and reconstruction procedures with the Auger $\overline{\mathrm{Off}}\underline{\mathrm{line}}$ software [14] are performed, and the event selection follows that applied to data. After selection, the proton samples for all models contain $\simeq 10^4$ showers; for heavier nuclei the samples vary from $5\cdot 10^3$ to $10^4$ showers.

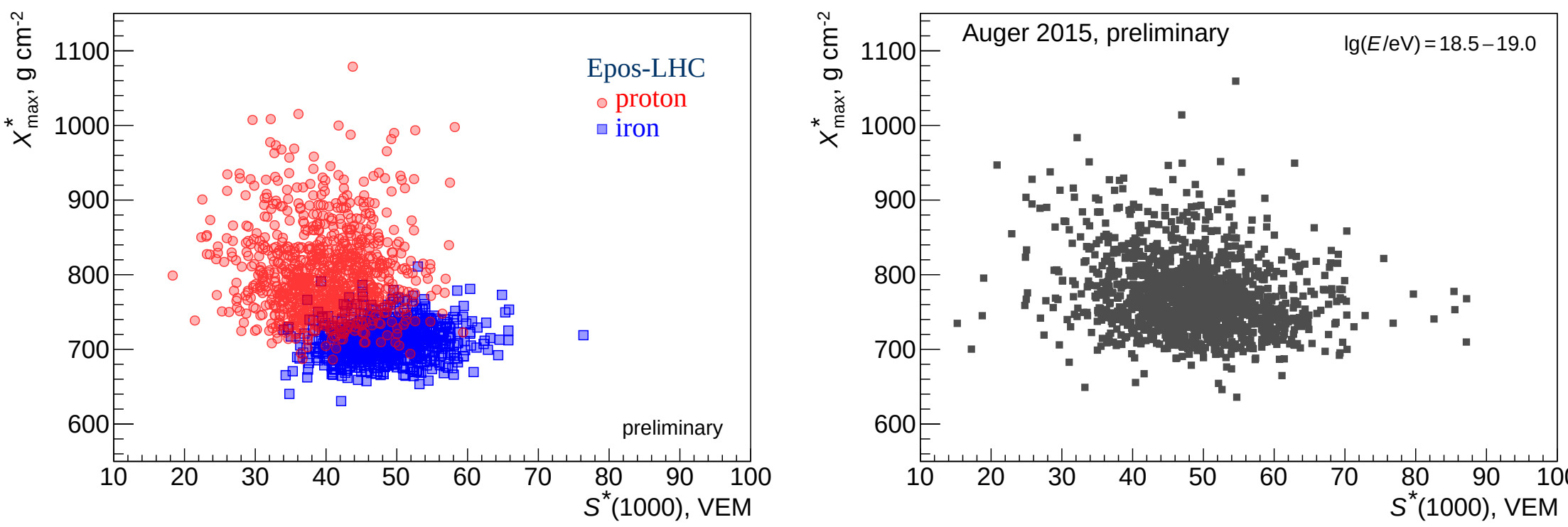

**Figure 1:** Scatter plot of $X^*_{\rm max}$ vs $S^*(1000)$ for protons and iron of Epos-LHC from full detector simulations (left) and for data (right) for $\lg(E/{\rm eV}) = 18.5-19.0$.

Since $S(1000)$ and $X_{\rm max}$ of an air shower depend on the energy and, in the case of $S(1000)$, also on the zenith angle, we scale $S(1000)$ and $X_{\rm max}$ to a reference energy and zenith angle. In this way, a decorrelation between the observables from combining different energies and zenith angles in the data set is avoided. We scale $S(1000)$ to 38° and 10 EeV using the calibration curves from [15] and $X_{\rm max}$ to 10 EeV using an elongation rate of 58 g cm$^{-2}$/decade. These scaled quantities will be marked with an asterisk: $X^*_{\rm max}$, $S^*(1000)$. Fig. 1 (right panel) shows the correlation between $X^*_{\rm max}$ and $S^*(1000)$ observed in data. Also shown, for illustration purposes, are the simulations for proton and iron primaries with Epos-LHC.

## 3. Method and results

As a measure of the correlation between $X^*_{\rm max}$ and $S^*(1000)$ we take the ranking coefficient $r_{\rm G}$ introduced by Gideon and Hollister in [16]. All events are ranked in both $X^*_{\rm max}$ and $S^*(1000)$, and the measured values of these observables are replaced by ranks for calculating the correlation. Further, the values of ranks are not used directly to calculate $r_{\rm G}$, but rather the general statistical dependence between $X^*_{\rm max}$ and $S^*(1000)$ is estimated counting numbers of events with ranks deviating from the expectations for perfect correlation and anti-correlation. With respect to the classical Pearson and Spearman coefficients, $r_{\rm G}$ provides a more robust estimate of the correlation [16, 17]. In particular, $r_{\rm G}$ is less sensitive to the removal of the most influential events or to outliers. We also note that the difference between correlation coefficients found in data and MC simulations for pure beams gets larger using Pearson and Spearman coefficients (or a number of other correlation coefficients considered in [17]) compared to using $r_{\rm G}$ so that the choice of $r_{\rm G}$ can be also viewed as conservative. The statistical uncertainty of $r_{\rm G}$ is determined using dedicated simulations and for the sample of size $N$ it is $\sigma_{\rm stat}(r_{\rm G}) \approx 0.9/\sqrt{N}$.

In Table 1 we present the $r_{\rm G}$ values for data and for the simulations for pure beams. Compared to data, where the correlation is significantly negative $r_{\rm G}(X^*_{\rm max}, S^*(1000)) = -0.125 \pm 0.024\,({\rm stat})$, the smallest difference is found for Epos-LHC protons and it is around $5\sigma_{\rm stat}$. Pre-LHC versions of Epos and QGSJetII produce values of correlations similar to Epos-LHC and QGSJetII-04. The differences between data and simulations are larger for pure beams other than protons. Using Pearson and Spearman coefficients one gets the same or slightly more positive values for pure beams as with $r_G$, and more negative correlation for data: $r(\text{Pearson}) = -0.210 \pm 0.038\,(\text{stat})$; $r(\text{Spearman}) = -0.199 \pm 0.027\,(\text{stat})$. This result shows that the composition in the considered energy range is not pure but mixed.

**Table 1:** $r_G(X^*_{max}, S^*(1000))$ for data and for MC simulations of pure beams (preliminary). Statistical uncertainties on the MC values are $\sigma_{stat} \approx 0.01$.

| data | $-0.125 \pm 0.024\,(\text{stat})$ | | |
|---|---|---|---|
| | Epos-LHC | QGSJetII-04 | Sibyll 2.1 |
| $p$ | 0.00 | 0.08 | 0.07 |
| He | 0.08 | 0.15 | 0.15 |
| O | 0.09 | 0.15 | 0.14 |
| Fe | 0.08 | 0.12 | 0.12 |

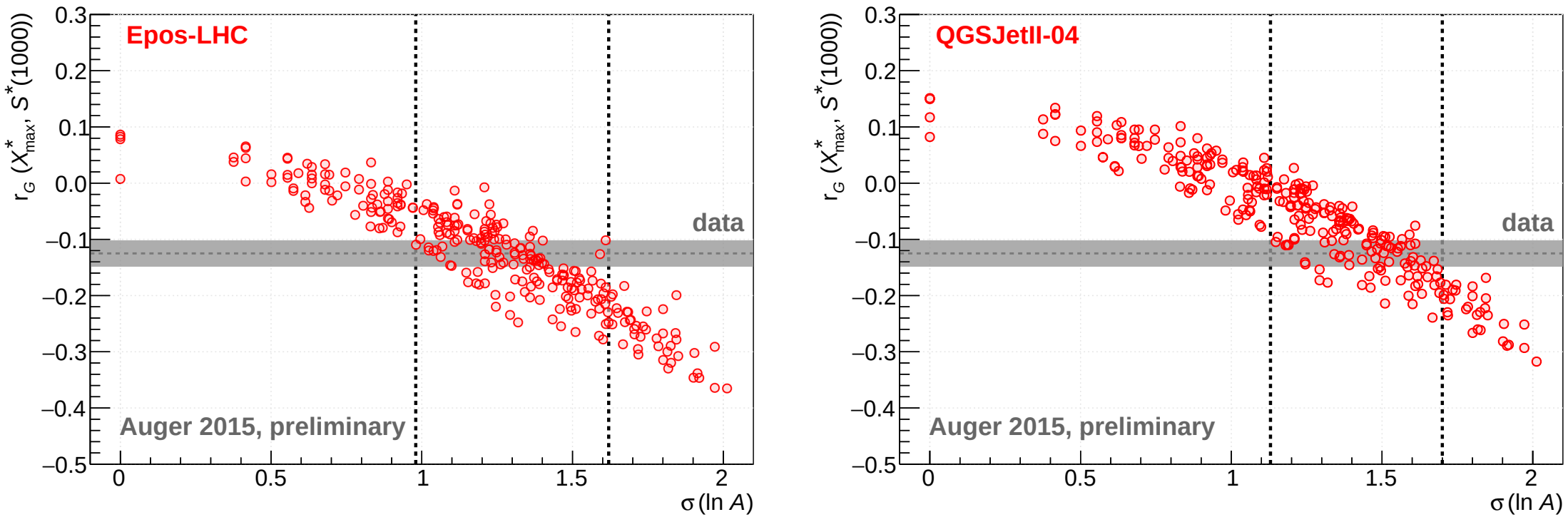

**Figure 2:** Dependence of the correlation coefficients $r_G$ on $\sigma(\ln A)$ for Epos-LHC (left), QGSJetII-04 (right). Each MC point corresponds to a mixture with different fractions of protons, helium, oxygen and iron, the relative fractions change with 0.1 steps (4 points for pure beams are grouped at $\sigma(\ln A) = 0$). The shaded band shows the $1\sigma_{stat}$ interval for data. Vertical dotted lines indicate the range of $\sigma(\ln A)$ in simulations compatible with the observed correlation in data.

Fig. 2 presents the dependence of the correlation $r_G(X^*_{max}, S^*(1000))$ on the dispersion of primary masses $\sigma(\ln A)$. Each MC point in this plot represents a mixture containing different fractions of protons, helium, oxygen and iron. The relative fractions $f_i$ of each species change with 0.1 steps between different mixtures. There are four points corresponding to beams of pure $p$, He, O, and Fe, grouped on the left side at $\sigma(\ln A) = 0$; of these, the proton beam has the smallest $r_G$ (cf. Table 1). The maximum possible value of $\sigma(\ln A) \simeq 2.01$ corresponds to the $0.5\,p - 0.5\,\text{Fe}$ mix.

The value of the correlation in data, indicated with the shaded band, is compatible with the MC samples with dispersions of primary masses $\sigma(\ln A) \gtrsim 1$. The conclusions on $\sigma(\ln A)$ are similar for all models considered (for Sibyll 2.1 one gets almost identical results to QGSJetII-04, cf. Table 1) and thus is weakly sensitive to the uncertainties in the description of the high-energy hadronic interactions.

The robustness of the presented approach makes it suitable for testing the self-consistency of the hadronic interaction models. For example, using the fractions of primary nuclei obtained from the fits of Auger $X_{\rm max}$ distributions [18] for QGSJetII-04 and Sibyll 2.1, which in the $\lg(E/{\rm eV}) = 18.5-19.0$ energy range are close to $0.5\,p-0.5\,{\rm He}$ ($\sigma(\ln A)\approx 0.7$), one gets $r_{\rm G}\approx 0.07-0.09$. The incompatibility of this value with the results of the present correlation analysis may be an indication of deficiencies in these two interaction models. The composition found in [18] from $X_{\rm max}$ fits with Epos-LHC is close to $\approx 0.35\,p-0.30\,{\rm He}-0.35\,{\rm O}$ mix ($\sigma(\ln A)\approx 1.17$), and the corresponding correlation $r_{\rm G} = -0.084$ is within $2\sigma$ from the $r_{\rm G}$ value in data.

## 4. Uncertainties

A number of standard tests were performed for estimation of the robustness of the obtained results. These checks include analysis of data recorded in various time periods and by different FD telescopes, separation of data in several angular ranges, and study of $r_{\rm G}$ in smaller energy bins. Results were consistent in all cases.

The ranking correlation coefficients are invariant with respect to any transformations not affecting ranks of the events. Thus $r_{\rm G}$ is insensitive to the systematic effects on $X^*_{\rm max}$ or $S^*(1000)$ that might lead to shift or multiplication of these observables by a constant value. In particular we have checked that the recent changes in Auger energy and $X_{\rm max}$ scales [19, 7] do not change the observed correlation. The same insensitivity of $r_{\rm G}$ was observed with respect to the application of various FD selection cuts which have been used in our publications from 2010 [20] until 2014 [7]. Finally, we introduced arbitrary energy and zenith angle dependent biases in $X^*_{\rm max}$ (up to 10 g cm$^{-2}$) and $S^*(1000)$ (up to 10%) and this changed $r_{\rm G}$ by $\simeq 0.01$. We take that value as a conservative estimate of the systematic error on $r_{\rm G}$.

We checked whether moderate changes of hadronic interaction parameters could make the value of $r_{\rm G}$ predicted for a pure proton composition consistent with observations. Using the approach described in [21] we performed simulations with Epos-LHC modifying the cross-section, multiplicity, elasticity and pion charge ratio in proton–air interactions by a factor $f_{19} = 1.5$, i.e. increasing them by a factor linearly growing with $\lg E$ from 1.0 at $10^{15}$ eV to 1.5 at $10^{19}$ eV with respect to the nominal values [21]. CONEX [22] with 3D option was used for approximate estimation of the signal in Auger stations at 1000 meters from the core. It turned out that $r_{\rm G}$ is practically insensitive to the modifications of these interaction parameters decreasing only by $\Delta r_{\rm G}\lesssim 0.03$. The change in $r_{\rm G}$ due the increase of cross-section, still being small compared to the difference between data and pure proton expectations, is found to be zenith angle dependent and it would also lead to zenith angle dependent (and thus contradictory) conclusions on $\sigma(\ln A)$. Such a scenario is constrained additionally by other Auger findings (e.g. regarding the proton-air cross section derived for $\lg(E/{\rm eV}) = 17.8-18.5$ [23, 24]), making it implausible as an explanation of our observations.

A possible under-production of muons by the current interaction models [25, 26, 27] could lead to changes in the ordering of events in the ($X^*_{\rm max}$, $S^*(1000)$) plane due to the presence of events with largely varying muon fractions of $S^*(1000)$. We performed a number of studies using CORSIKA showers and showers fully reconstructed with Offline with the numbers of muons increased by the factors ranging from 1.1 to 1.9 and we have found that for the muon scaling by a factor $\approx 1.3$, as suggested by data for Epos-LHC [25, 26], the $r_G$ value decreseas by $\lesssim 0.03$.

## 5. Summary

The observed correlation $r_G(X^*_{max}, S^*(1000))$ between depth of shower maximum and total signal at 1000 meters from the core differs significantly from the correlations for any pure beam for simulations with Epos-LHC, QGSJetII-04 and Sibyll 2.1. The result is invariant with respect to additive and multiplicative scale transformations of the two variables and to any other transformations of $X_{max}$ and $S(1000)$ which leave ranks of events unchanged, and hence is robust against many possible experimental systematic uncertainties. Several modifications of hadronic interactions were studied. The conclusions remain robust also with regard to hadronic uncertainties, unless hadronic interactions at these energies behave very differently than in conventional, LHC-tuned event generators. The results are compatible with a mixed primary composition around the 'ankle' with the dispersion of masses $1.0 \lesssim \sigma(\ln A) \lesssim 1.7$ and question the 'dip' scenario.

# Search for Ultra-relativistic Magnetic Monopoles with the Pierre Auger Observatory

**Toshihiro Fujii**[*a] **for the Pierre Auger Collaboration**[b]
[a] *Kavli Institute for Cosmological Physics, University of Chicago, Chicago, 60637 Illinois, USA*
[b] *Observatorio Pierre Auger, Av. San Martín Norte 304, 5613 Malargüe, Argentina*
*E-mail:* auger_spokespersons@fnal.gov
*Full author list:* http://www.auger.org/archive/authors_2015_06.html

Ultra-relativistic magnetic monopoles, possibly a relic of phase transitions in the early universe, would deposit an amount of energy comparable to UHECRs in their passage through the atmosphere, producing highly distinctive air shower profiles. We have performed a search for ultra-relativistic magnetic monopoles in the sample of air showers with profiles measured by the fluorescence detector of the Pierre Auger Observatory. No candidate was found to satisfy our selection criteria and we establish upper limits on the flux of ultra-relativistic magnetic monopoles - the first from an UHECR detector - improving over previous results by up to an order of magnitude.

*Speaker.

## 1. Introduction

In 1931, Dirac showed that the existence of magnetic monopoles would explain the quantization of the electric charge from the relation between the unit of electric charge $e$ and the unit of magnetic charge $g_M = Ne/2\alpha$ [1], where $\alpha$ is the fine-structure constant and $N$ is an integer. In addition, it would bring Maxwell's equations to a complete symmetry. Within the framework of Grand Unified Theories (GUT), supermassive magnetic monopoles ($M \approx 10^{26}$ eV/c$^2$) may be produced in the early Universe as intrinsically stable topological defects at the phase transition corresponding to the spontaneous symmetry breaking of the unified fundamental interactions [2]. Lower mass monopoles may come from later phase transitions at lower energy scales. These particles may be present in the Universe today as a relic of these transitions. Experimental searches for magnetic monopoles are based on their velocity-dependent interactions with matter, with a wide range of velocities allowed for GUT monopoles. Supermassive GUT magnetic monopoles should be gravitationally bound to the Galaxy (or to the Sun and Earth) with non-relativistic virial velocities [2]. Lighter magnetic monopoles can reach relativistic velocities through acceleration in coherent domains of the galactic and intergalactic magnetic fields, as well as in the vicinity of astrophysical objects (e.g. neutron stars) [3]. Kinetic energies of the order of $10^{25}$ eV have been predicted [4], which result in ultra-relativistic velocities for intermediate mass monopoles (IMMs, $M \sim 10^{11} - 10^{20}$ eV/c$^2$).

There is a long history of experimental searches for magnetic monopoles with a variety of experiments such as MACRO [5], Amanda [6], Baikal [7], SLIM [8], RICE [9], ANITA [10] and IceCube [11]. The strongest upper limit on the flux of non-relativistic magnetic monopoles ($\beta = v/c < 0.5$) comes from the MACRO experiment at $\approx 1.5 \cdot 10^{-16}$ (cm$^2$ sr s)$^{-1}$ (90% C.L.) [5]. At relativistic velocities ($\beta \approx 0.9$), the IceCube Observatory has placed the best limit at $10^{-19}$ (cm$^2$ sr s)$^{-1}$ [11]. The best limit on the flux of ultra-relativistic IMM (Lorentz factor $\gamma \geq 10^{10}$) is reported by the ANITA-II experiment at $10^{-19}$ (cm$^2$ sr s)$^{-1}$ [10].

The Pierre Auger Observatory is the largest detector to observe ultra-high energy cosmic rays currently in operation [12] located in the southern hemisphere in central Argentina, just north east of the town of Malargüe (69°W, 35°S, 1400 m a.s.l.), and covers a ground area of 3,000 km$^2$. The Pierre Auger Observatory consists of a surface detector array (SD) of 1660 individual water-Cherenkov surface detectors [13] overlooked by a fluorescence detector (FD) of 24 individual fluorescence telescopes grouped in units of 6 at four locations [14]. Since the FD steadily observes UHECRs using a huge target volume in the atmosphere with high precision measurements, it would be a suitable detector to search for signals generated from ultra-relativistic IMMs. Therefore, we search for ultra-relativistic IMMs with data collected by the FD between 01.12.2004 and 31.12.2012.

## 2. Air Shower Simulations for the IMM Search

Electromagnetic interactions of magnetic monopoles have been extensively investigated [15]. The electromagnetic energy loss of a magnetic monopole in air is shown in Figure 1(a) as a function of its Lorentz factor. Collisional energy loss of a magnetic monopole is the dominant contribution for $\gamma \leq 10^4$. At higher Lorentz factors, pair production and photonuclear interactions become the main cause of energy loss. Bremsstrahlung is highly suppressed by the large monopole mass. An ultra-relativistic IMM would deposit a large amount of energy in its passage through the Earth's atmosphere, comparable to that of an UHECR. For example, an IMM with $\gamma = 10^{11}$ loses $\approx 400$ PeV/(g/cm$^2$) (cf. Figure 1(a)), which sums up to $\approx 10^{20.5}$ eV when integrated over an atmospheric depth of $\approx 1000$ g/cm$^2$. This energy will be dissipated by the IMM through production of secondary showers along its path.

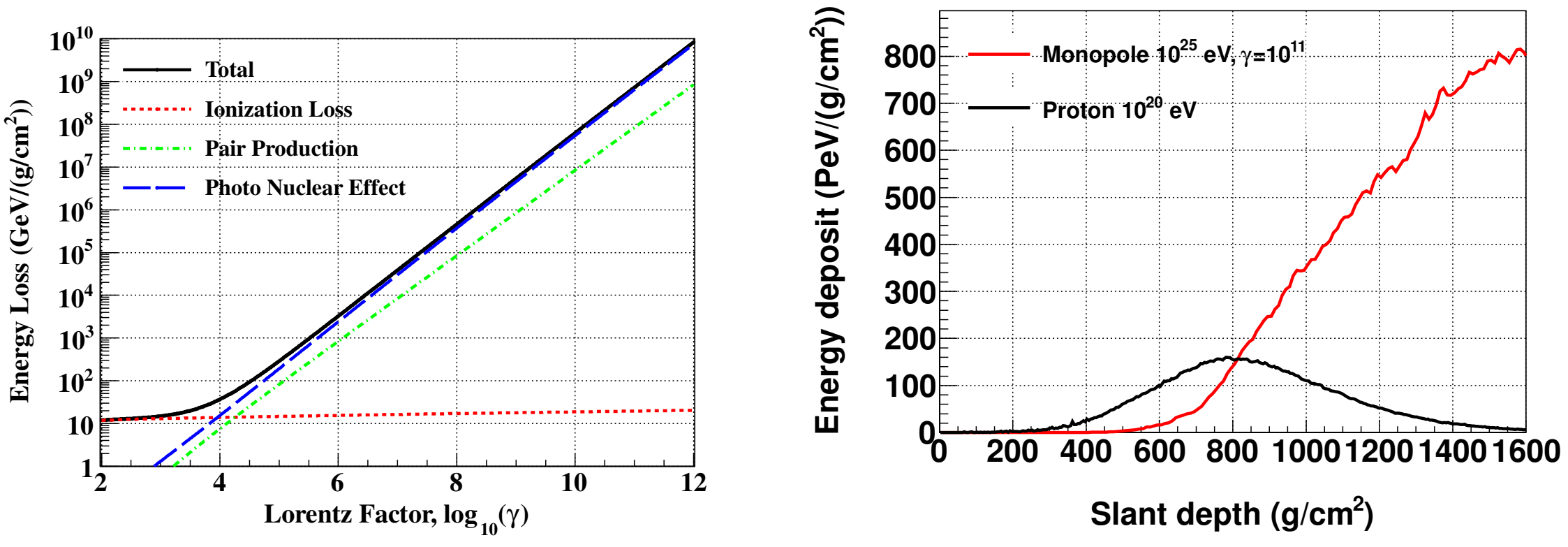

(a) Energy loss (b) IMM profile

**Figure 1:** (a) Energy loss of a magnetic monopole in air as a function of its Lorentz factor $\gamma$. (b) Longitudinal profile of the energy deposited by an ultra-relativistic IMM of $E_{\mathrm{mon}} = 10^{25}$ eV, $\gamma = 10^{11}$ and zenith angle of 70° (red solid line). The profile of an UHECR proton shower of energy $10^{20}$ eV is shown as a black solid line.

The longitudinal profile of the energy deposited by an ultra-relativistic IMM of $E_{\mathrm{mon}} = 10^{25}$ eV, $\gamma = 10^{11}$ and zenith angle of 70° is shown in Figure 1(b). When compared with a standard UHECR proton shower of energy $10^{20}$ eV (black solid line in Figure 1(b)), the IMM shower presents a much larger energy deposit and deeper development, due to the superposition of many showers uniformly produced by the IMM along its path in the atmosphere. This distinctive feature will be used in our analysis, which is based on the shower development measured by the FD and SD events.

Monte Carlo samples of ultra-relativistic IMM were simulated for Lorentz factors in the range $\gamma = 10^8 - 10^{12}$ at a fixed monopole energy $E_{\mathrm{mon}}$ of $10^{25}$ eV. While we used a fixed $E_{\mathrm{mon}}$ in the simulations, the results can be readily applied to a much larger range of monopole energies, since in the ultra-relativistic regime of this search the monopole energy loss does not depend on $E_{\mathrm{mon}}$ but rather on $\gamma$. To estimate the background from UHECRs, we simulated proton showers with energy $E_{\mathrm{p}}$ between $10^{18}$ eV and $10^{21}$ eV by three different models, QGSJetII-04, Sibyll 2.1 and Epos-LHC. We used three different models to account for uncertainties in the hadronic interactions. Events were simulated according to an $E_{\mathrm{p}}^{-1}$ energy spectrum, to ensure sufficient Monte Carlo statistics at the highest energy, and then appropriately weighted to reproduce the energy spectrum measured by the Pierre Auger Observatory [16]. For both the IMM and UHECR simulations, we used the CORSIKA package [17] to generate an isotropic distribution of showers above the horizon, and the Auger $\overline{\text{Off}}\underline{\text{line}}$ software [18] to produce the corresponding FD and SD events.

| Event selection criteria | # events | (%) |
|---|---|---|
| Reconstructed events | 376084 | — |
| Zenith angle $< 60°$ | 360159 | 95.8 |
| Distance from nearest SD$<$1500 m | 359467 | 99.8 |
| Number of FD pixels $> 5$ | 321293 | 89.4 |
| Slant depth interval $> 200$ g/cm$^2$ | 205165 | 63.9 |
| Gaps in profile $< 20\%$ | 199625 | 97.3 |
| profile fit $\chi^2$/ndf $< 2.5$ | 197293 | 98.8 |
| $dE/dX\|_{X\mathrm{up}} > 3.0$ PeV/(g/cm$^2$) | 6812 | 3.5 |
| $X_{\mathrm{max}} > X_{\mathrm{up}}$ | 352 | 5.2 |
| $X_{\mathrm{up}} > 1080$ g/cm$^2$ or $dE/dX\|_{X\mathrm{up}} > 150$ PeV/(g/cm$^2$) | 0 | 0.0 |

**Table 1:** Event selection criteria and unblinded result. The number of events passing each selection criteria is reported, together with the corresponding fraction.

## 3. Event Selection

Before shower selection, only time periods with a good status of FD telescopes and a high quality calibration of the gains of PMTs are selected using information in Auger databases. Additional cuts are applied to assure good atmospheric conditions due to aerosols and clouds. A further set of selection criteria was applied to ensure good quality showers as summarized in Table 1. Those criteria for IMM selection were established from Monte Carlo simulations described in Section 2.

The important parameters for the IMM search are the slant depth at the upper field-of-view boundary, $X_{\mathrm{up}}$, and the energy deposited at $X_{\mathrm{up}}$, $dE/dX|_{X\mathrm{up}}$. The requirement $dE/dX|_{X\mathrm{up}} >3.0$ PeV/(g/cm$^2$) is equivalent to an energy threshold of $\approx 10^{18.5}$ eV, where the SD is fully efficient. When $X_{\mathrm{max}} > X_{\mathrm{up}}$ is required, the number of proton backgrounds is drastically reduced and become constrained in a much smaller region, as shown in Figure 2(a). On the other hand, the reconstructed $X_{\mathrm{max}}$ will always be outside the FD field of view for ultra-relativistic IMM showers, independently of the shower's $X_{\mathrm{up}}$. This is apparent in Figure 2(b), where the correlation of $dE/dX|_{X\mathrm{up}}$ with $X_{\mathrm{up}}$ is shown for ultra-relativistic IMM simulated events. The background from UHECRs is almost eliminated by excluding an appropriate region of the ($X_{\mathrm{up}}$, $dE/dX|_{X\mathrm{up}}$) plane. We optimized the selection to achieve less than 0.1 background events expected in the data set of this search. The final requirement, $X_{\mathrm{up}} > 1080$ g/cm$^2$ or $dE/dX|_{X\mathrm{up}} > 150$ PeV/(g/cm$^2$), is shown in Figure 2 as a dashed box, and results in an expected background of 0.07 events in the search period data set.

## 4. Data Analysis and Results

The search for ultra-relativistic IMM was performed following a blind procedure. The selection criteria described in Section 3 were optimized using Monte Carlo simulations and a small fraction (10%) of the data. Then, the selection was applied to the full sample of data collected between 01.12.2004 and 31.12.2012. The number of events passing each of the selection criteria is reported in Table 1. The correlation of $dE/dX|_{X\mathrm{up}}$ with $X_{\mathrm{up}}$ for events passing the shower quality criteria and $X_{\mathrm{max}} > X_{\mathrm{up}}$ is shown in Figure 3(a).

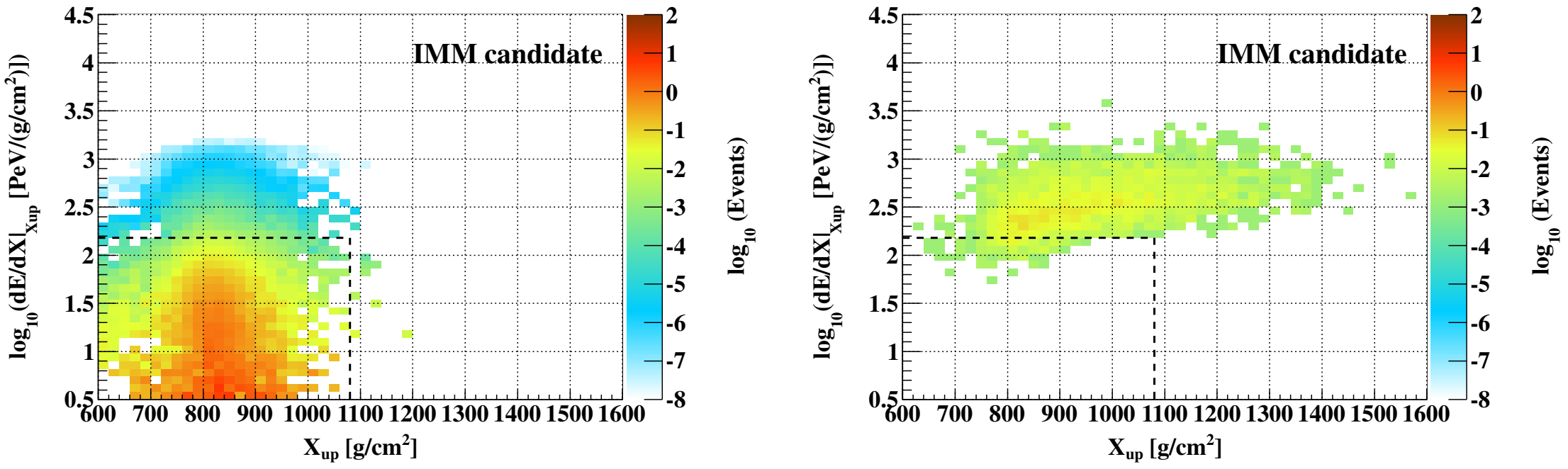

(a) Proton background

(b) Monopole signal, $\gamma = 10^{11}$

**Figure 2:** (a) Correlation of $dE/dX|_{X\mathrm{up}}$ with $X_{\mathrm{up}}$ for simulated UHECR proton showers passing the quality selection criteria. The color-coded scale indicates the number of events expected in the search period data set. Only events outside the dashed box are kept in the final selection for ultra-relativistic IMM. (b) Correlation of $dE/dX|_{X\mathrm{up}}$ with $X_{\mathrm{up}}$ for simulated ultra-relativistic IMM of energy $10^{25}$ eV with Lorentz factor $\gamma = 10^{11}$. The color-coded scale indicates the number of events expected in the search period data set assuming a flux of $10^{-20}$ (cm$^2$ sr s)$^{-1}$. Only events outside the dashed box are kept in the final selection for ultra-relativistic IMM.

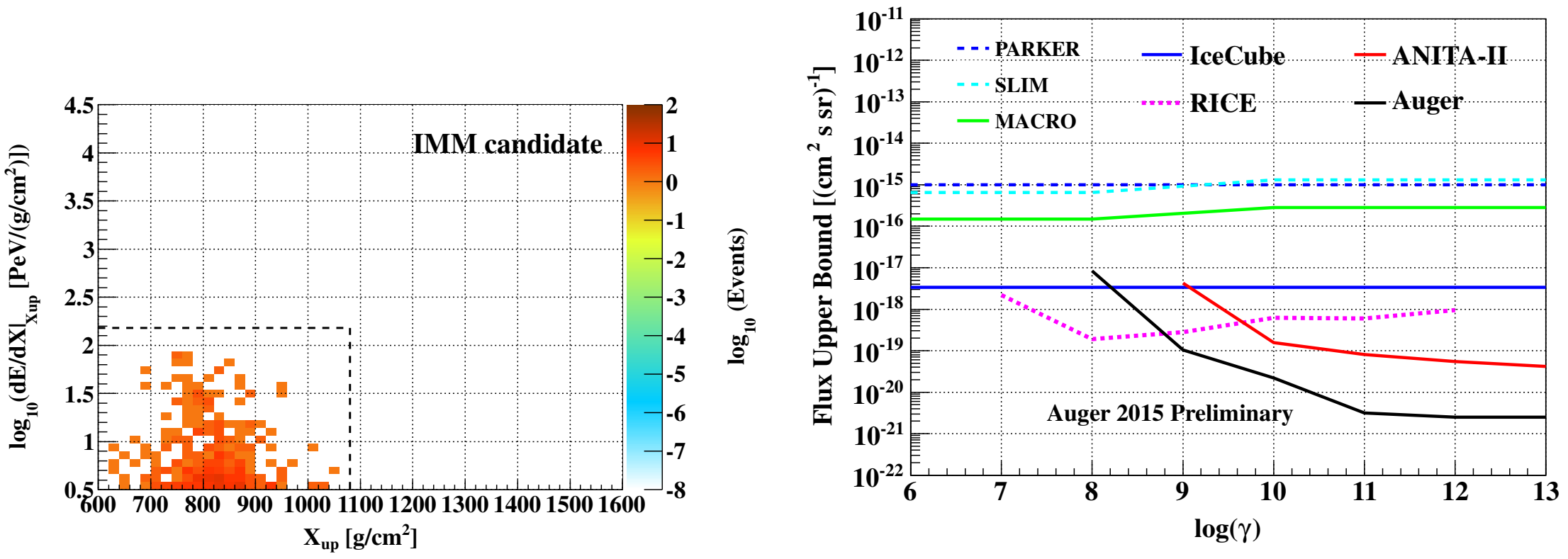

(a) Unblinded result

(b) 90% C.L. upper limit

**Figure 3:** (a) Correlation of $dE/dX|_{X\mathrm{up}}$ with $X_{\mathrm{up}}$ for the data sample passing the shower quality selection criteria and $X_{\mathrm{max}} > X_{\mathrm{up}}$. The color-coded scale indicates the number of events. No event is found outside the dashed box in the final selection for ultra-relativistic IMM. (b) 90% C.L. upper limits on the flux of ultra-relativistic IMM: this work (black solid line); Parker bound (blue dashed line) [19]; SLIM (sky blue dashed line) [8], MACRO (green solid line) [5], IceCube (blue solid line) [11], RICE (pink dotted line) [9] and ANITA-II (red line) [10]. The MACRO and SLIM limits above $\gamma = 10^9$ were weakened by a factor of two to account for the IMM attenuation through Earth.

Given the null result of the search, a 90% C.L. upper limit on the flux of ultra-relativistic IMM was derived. The flux $\Phi$ of ultra-relativistic IMM of Lorentz factor $\gamma$ is given by $\Phi(\gamma) = k/\mathscr{E}(\gamma)$ with a value $k = 2.44$ (the Feldman-Cousins upper limit [20] for zero candidates and zero background events because the background events are conservatively assumed to be 0 instead of 0.07 expected events described in Section 3). The exposure, $\mathscr{E}(\gamma)$, determined by *Time Dependent Detector Simulation* [21], which makes use of slow control information and atmospheric measurements recorded during the data taking with considering the efficiency in IMM detection under the event selection criteria. The systematic uncertainty of the limit originating from the uncertainty of the exposure and $dE/dX|_{X\mathrm{up}}$ cut is 21%. The propagation of uncertainties from the event statistics and systematics would lead to limit that is worse by a factor of 1.05 [22]. Here we follow a conservative approach and account for the systematics without propagation. We increase the limit by a factor $f = 1 + n \times 0.21$, where $n = 1.28$ corresponds to 90% C.L. value. The corresponding upper limits for different Lorentz factors are shown in Figure 3(b), together with results from previous experiments. Our result represents the best limit on ultra-relativistic IMM for $\gamma \geq 10^9$, with a factor of ten improvement with respect to previous experiments for $\gamma \geq 10^{10}$.

## 5. Conclusions

We have reported an analysis procedure and an event selection to search for signals produced by ultra-relativistic IMMs with FD of the Pierre Auger Observatory. Using this analysis, there is no candidate for IMMs signal in the data collected by the FD from 2004 to 2012. Therefore, we have evaluated a flux limit of $\Phi_{90\%\mathrm{C.L.}} \sim 10^{-20}$ (cm$^2$ s sr)$^{-1}$ for $\log_{10}(\gamma) \geq 10$ which is a order of magnitude stronger than all previous published experimental limits.

# Update of the neutrino and photon limits from the Pierre Auger Observatory

**Carla Bleve**[*,a] **for the Pierre Auger Collaboration**[b]
[a]*INFN and Department of Mathematics and Physics, University of Salento, Lecce, Italy*
[b]*Observatorio Pierre Auger, Av. San Martín Norte 304, 5613 Malargüe, Argentina*
*E-mail:* auger_spokespersons@fnal.gov
*Full author list:* http://www.auger.org/archive/authors_2015_06.html

Ultra-high energy neutrinos and photons, with energies above 1 EeV and 10 EeV respectively, can be detected with the Surface Detector array (SD) of the Pierre Auger Observatory. Downward-going neutrinos of all flavours interacting in the atmosphere at zenith angles $\theta > 60^\circ$, upward-going tau neutrinos ("Earth-skimming"), as well as photons in the zenith range $30^\circ - 60^\circ$ can be identified through the broad time-structure of the signals expected to be induced in the SD stations. An additional signature for photon-induced air showers is the steeper lateral distribution of secondary particles at ground with respect to the nucleonic showers.
Stringent limits are set to the diffuse flux of ultra-high energy (UHE) neutrinos and photons, using data collected between 2004 and mid-2013, under the hypothesis of an $E^{-2}$ spectrum for signal primaries.

[*]Speaker.

## 1. Introduction

Searches for ultra-high energy (UHE) photons and neutrinos are amongst the methods used to unravel the mystery of the origin of cosmic rays of the highest energy. Protons and nuclei interacting with the universal low energy photon background (CMB) are expected to produce a flux of UHE photons that can propagate for a few tens of Mpc without being absorbed and neutrinos (from the decay of charged pions, muons and neutrons) that can travel to the observer with no interaction or deflection. The expected cosmogenic fluxes depend on the composition and maximum energy of CRs at the sources and the emissivity, distribution and cosmological evolution of the acceleration sites. Thus, observing UHE photons or neutrinos, can pose constraints on the UHECR origin and properties of the sources.
The Surface Detector array (SD) of the Pierre Auger Observatory covers ~3,000 km$^2$ with a triangular grid of water-Cherenkov Detectors (WCD) [1] providing a very large exposure for the search of UHE neutrinos and photons. We report here upper limits to the diffuse fluxes obtained with the analysis of data collected between 2004 and mid-2013.

## 2. Search method

SD observables for the searches of neutrino and photons are identified on the base of the expected physical differences between signal and background, quantified using simulations of air showers initiated by signal particles. The background of showers induced by nuclei is not simulated, instead a fraction of the data set is used as a training sample to define the selection for candidate signal events. The search sample consists of the remaining events.

### 2.1 UHE neutrinos

The search for neutrinos exploits the extremely small cross-section of the signal particles. At large zenith angles ($\theta$ >60°) the thickness of the atmosphere traversed is large enough to absorb almost completely the electromagnetic component of showers initiated by nucleons or even photons, leaving their signal dominated by muons. Showers initiated by neutrinos very deep in the atmosphere, on the other hand, have a considerable amount of the electromagnetic component remaining ("young" showers). Two types of neutrino-induced showers are sought:
**(1) Earth-Skimming (ES)** showers ( $90° < \theta < 95°$, induced by $\nu_\tau$ travelling upward with respect to the vertical at the ground) can skim the crust of the Earth and interact close to the surface, producing a $\tau$ lepton which can decay in flight in the atmosphere close to the SD. At $10^{18}$ eV the mean decay length of the $\tau$ lepton is ~50 km.
**(2) Downward-Going (DG)** showers ($60° < \theta < 90°$) initiated by neutrinos of all flavours interacting in the atmosphere close to the SD through neutral current or charged current interactions, as well as showers produced by $\nu_\tau$ interacting in the mountains surrounding the observatory.
The $\nu$-nucleon interactions for DG neutrinos in the atmosphere are simulated with HERWIG [2]; the $\tau$ propagation in Earth and in the atmosphere with TAUOLA [3] (with $\nu$-nucleon cross-section from [4]); the subsequent air shower is simulated with AIRES [5]. To identify neutrinos we search for very inclined "young" showers. Signatures of inclined showers are: large ratio length/width (L/W) of the major/minor axis of the ellipse encompassing the footprint of the shower (Fig. 1 (i)) and the distribution of apparent speeds of the trigger time between stations being required to have an average value close to the speed of light and a small RMS. Large values of the Area-over-Peak ratio (AoP, Fig. 1 (ii)) in the time traces indicate a large contribution of the electromagnetic component. For all the channels the observable used to identify neutrinos is generally based on the AoP of stations: the full selection strategy is described in [6]. The region for neutrino candidates is defined using a training data sample ($\sim$20% of the whole data set). From the distribution of the data in the training set, the range of the separation variable in which 1 event is expected in 50 yr on the full SD array is defined. Any event in this range is considered as a neutrino candidate. Fig. 2 illustrates the exponential fit of the tail of the data distribution for the ES channel, using the average AoP of early stations as the discriminating variable [6]. The results on the search sample are also shown.

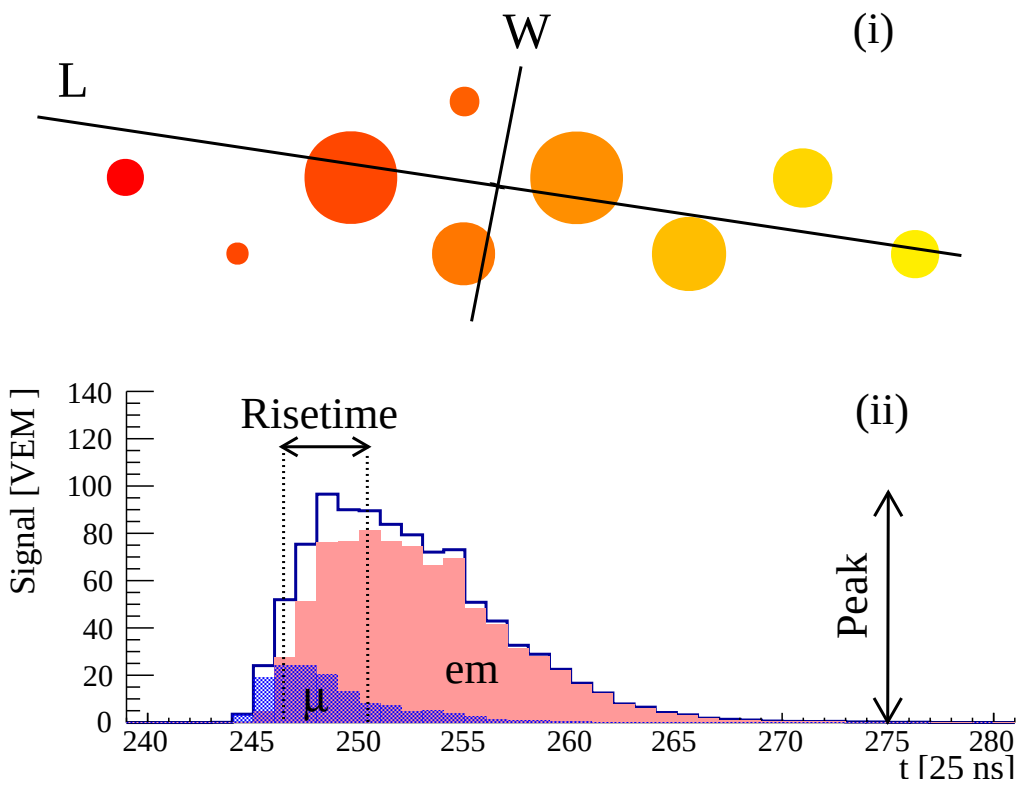

**Figure 1:** The footprint of an inclined event (i) on the SD detector: colours from light to dark represent trigger times from early to late, circle areas are proportional to the logarithm of the total signal in individual stations. L, the direction of arrival projected on the detector plane, is the major axis of the ellipse encompassing the footprint, W the minor axis. An example of a time trace in a SD station digitised with a FADC in 25 ns bins is shown in (ii): a larger fraction of electromagnetic (em) signal produces a larger Area over Peak (AoP) ratio and a larger risetime.

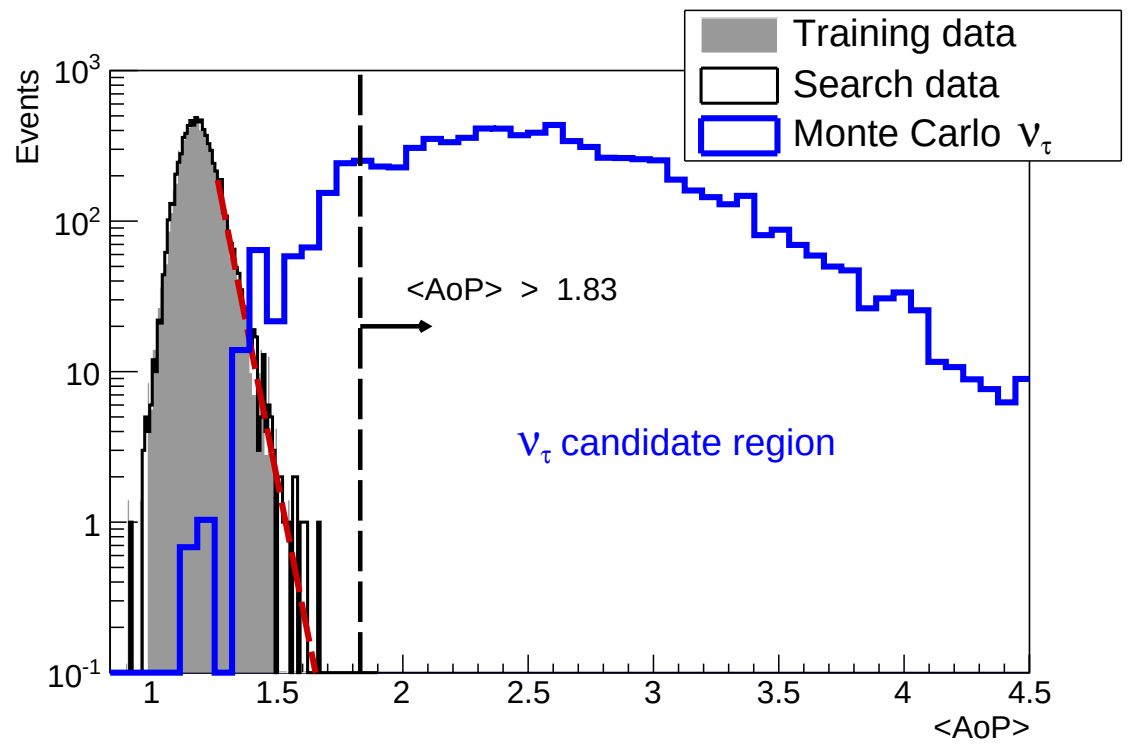

**Figure 2:** Distributions of ⟨AoP⟩ (the variable used to identify neutrinos in the ES selection for data after 01/06/2010) after applying the inclined shower selection. The distributions for data in the training period (full grey histogram) and search period (black histogram) are normalised to the same number of events. The blue histogram shows the simulated ES $\nu_\tau$ events. The vertical line represents the cut on ⟨AoP⟩ above which an event is regarded as a neutrino candidate. An exponential fit to the tail of the distribution of training data is also shown as a red dashed line - see text for details.

### 2.2 Photons ($E_\gamma$ >10 EeV)

Showers induced by photons are characterised by a lower content of muons and larger average depth of maximum longitudinal development ($X_{\mathrm{max}}$) than showers initiated by nuclei with the same energy. This is due to the radiation length being more than two orders of magnitude smaller than the mean free path for photo-nuclear interaction, causing a reduced transfer of energy to the hadron/muon channel, and to the development of the shower being delayed by the typically small multiplicity of electromagnetic interactions. The Landau-Pomeranchuk-Migdal (LPM) effect becomes important beyond 10 EeV. At $E_\gamma$ >50 EeV - for the site of the Pierre Auger Observatory - photons can also convert in the geomagnetic field producing a pre-shower [7], with the probability of this occurring increasing with energy and depending on the arrival direction with respect to the field. The resulting air-shower is, in this case, a superposition of cascades initiated by lower energy electrons and photons with smaller $\langle X_{\rm max} \rangle$ and larger ratio of muonic to electromagnetic content, at ground level, than showers induced by non-converted photons of the same energy [8].
The effects of pre-shower and LPM are both included in the simulations of photon showers used for this analysis, generated with an energy spectrum $E^{-1}$. Air showers are simulated with CORSIKA [9] (including PRESHOWER [10]), with QGSjetII.03 [11] as hadronic generator.
Photons are searched for in the $30^\circ < \theta < 60^\circ$ zenith range[1], the reconstruction of the energy-related observable $S(1000)$, the signal in VEM (vertical equivalent muons) at 1000 m from the shower axis, is the same as for the SD spectrum analysis [12], where the function used to fit the lateral distribution of the signal (LDF) has a shape parameterised to describe the bulk of Auger data. The conversion of ($S(1000)$,$\theta$) to energy is specific for the photon search. An iterative procedure similar to the one described in [13] is used, assuming the universality of $S(1000)/E^{\alpha}$ as a function of $DX = X_{\rm obs} - X_{\rm max}$ ($DX$ is the difference in atmospheric depth between the observation level and the shower $X_{\rm max}$, used as an indicator of the shower age, i.e. stage of development of the cascade). To avoid analysing showers not fully developed, we require $X_{\rm max}$ to be no more than 50 g cm$^{-2}$ below the ground level ($DX > -50$ g cm$^{-2}$). The universal profile is described with a Gaisser-Hillas function whose parameters, together with $\alpha$, are calibrated with photon simulations. The reduced muon content of photon showers with respect to data ($\sim 15\%$ compared to protons at 10 EeV, for QGSjetII.03) produces a steeper lateral distribution of the signal observed in WCDs at ground level. At large distances from the axis, photon showers produce typically smaller signals than expected from the data LDF (Fig. 3 (i)). We thus define an event observable measuring the departure from the average data LDF as the logarithm of the average deviation of the station signal from the event LDF: $L_{LDF} = log_{10}(\frac{1}{N}\sum_i S_i/LDF(r_i))$ where $i$ runs over the stations with radial distance from the shower axis $r_i > 1000$ m, $S_i$ is the total signal of the i-th station, $LDF(r_i)$ is the signal value at distance $r_i$ according to the LDF fit. $L_{LDF}$ is expected to be negative for photons.
The spread in the arrival time of secondary particles in individual stations can be measured defining the risetime as the difference between the 50% and 10% time quantiles of the FADC time trace (Fig. 1 (ii) on page 63). The risetime is not only increased by a larger contribution of the electromagnetic component, it also increases when the difference in depth between $X_{\rm max}$ and the observation level becomes smaller, for geometrical reasons. Being sensitive to both the deeper $X_{\rm max}$ and limited muon content of photon showers, the risetime is a suitable variable for the search of photons. The raw risetime is corrected for azimuthal asymmetry effects (i.e. difference between earlier and later triggering stations in an inclined event) obtaining $t_{1/2}$. The correction is based on the average effect observed in data. A "Data Benchmark" is produced to describe the average risetime of data. Sampling fluctuations, $\sigma_{t_{1/2}}$, are also estimated from the data, using the difference of the measurements of station doublets (a station in the regular SD grid plus a second station off-grid deployed close to it) or station pairs (two stations in the same event with similar distance from the axis and total signal). For each station, $\delta_i = (t^i_{1/2} - t^{Bench}_{1/2})/\sigma^i_{t_{1/2}}$ measures the deviation of the corrected risetime of the station from the data benchmark in units of expected standard deviation (Fig. 3 (ii)). For photon searches an event observable $\Delta = (\sum_i \delta_i)/N$ is defined, where the sum runs over stations with $S > 6$ VEM, and radial distance in the range 600-2000 m. A minimum of 4 selected stations is required. Δ is expected to average ∼0 for data and to be significantly positive for air showers initiated by photons.

[1] The $\langle X_{\rm max} \rangle$ of photons at 10 EeV is already $\sim$ 100 g cm$^{-2}$ larger than the atmospheric vertical depth at the site. Selecting a minimum zenith of 30° (observation depth $\simeq$1020 g cm$^{-2}$) guarantees that most UHE photon-induced showers are fully developed at the depth of observation.

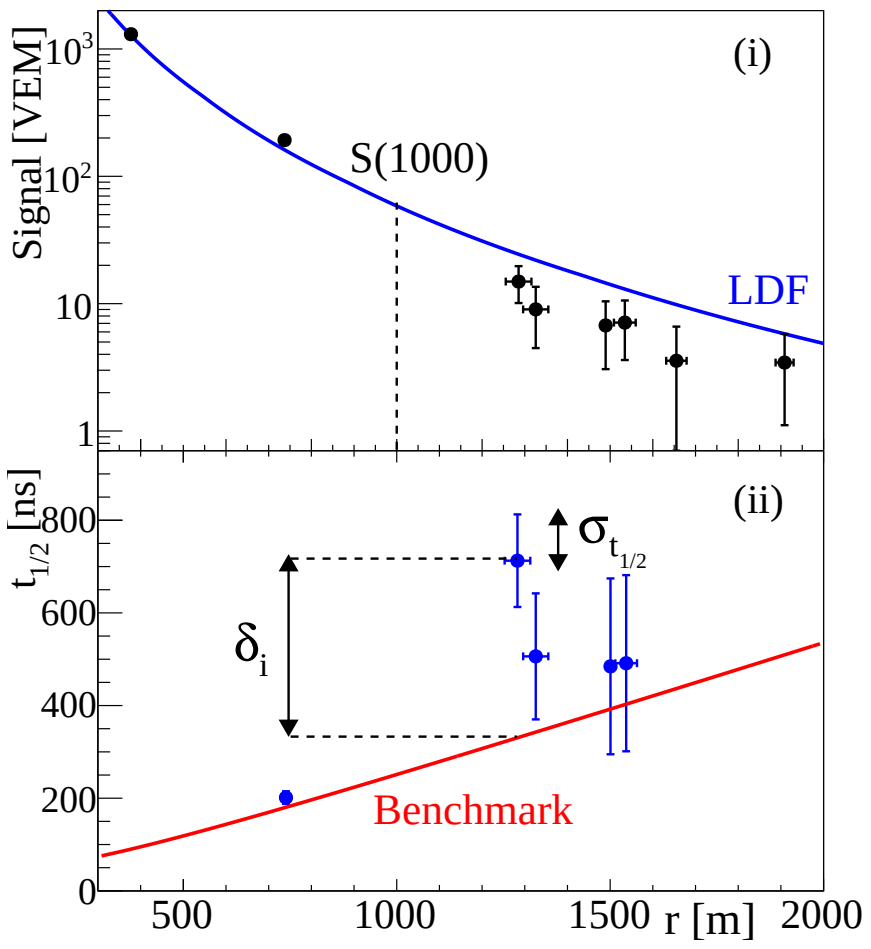

**Figure 3:** A shower induced by a photon of ∼20 EeV simulated with a zenith angle of 45°: (i) the lateral distribution of signals in the WCDs is steeper than the LDF obtained from data (solid line) and (ii) the risetime of selected stations is larger than the average data benchmark (solid line) - see text for details.

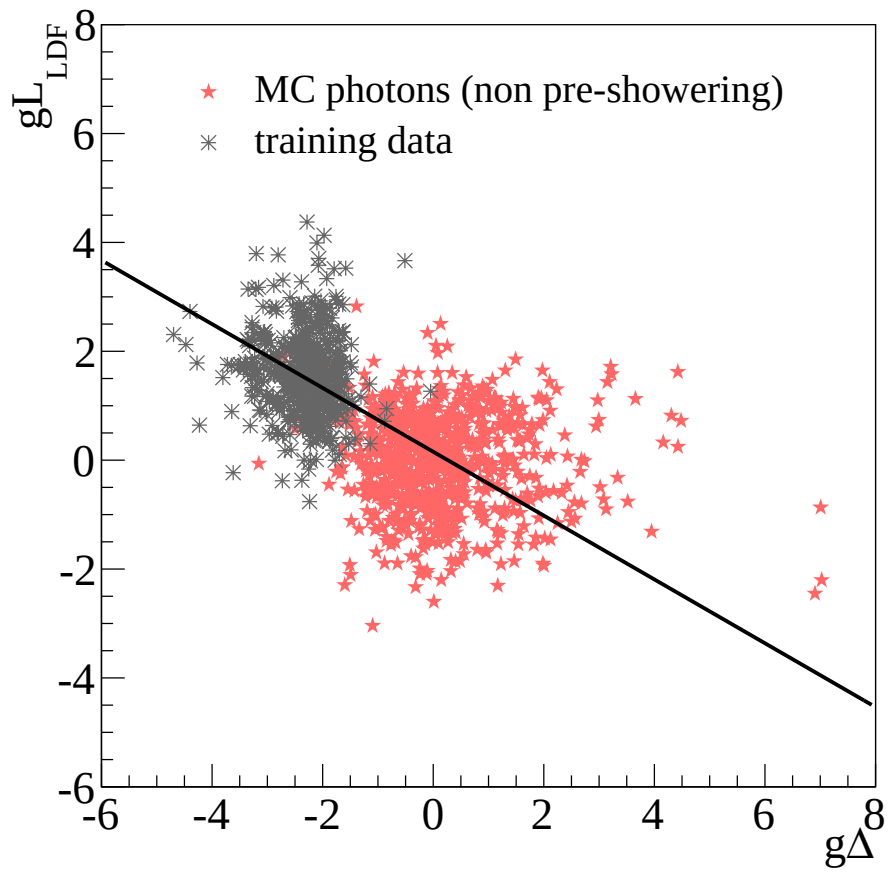

**Figure 4:** Principal component analysis for the photon search. The training data sample and a set of simulated photons, not undergoing pre-showering, are used to find the principal component (black line), i.e. the linear combination of gΔ and g$L_{LDF}$ that maximises the separation of the two sets. The principal component is used as separation variable for the photon search.

The observables $L_{LDF}$ and Δ are redefined to obtain a distribution with mean zero and standard deviation 1 for photon showers. Taking $x = L_{LDF}$ or $x = \Delta$, we define: $gx = (x - \bar{x}_\gamma(E_\gamma, \theta))/\sigma_\gamma(E_\gamma, \theta)$. The resulting distribution is shown in Fig. 4 for the training data set (grey) consisting of 2% of data, and photon simulations (red). Only events with reconstructed photon energy > 10 EeV, and photons not converting in the geomagnetic field are considered[2]. A Principal Component Analysis (PCA) on the training data set and unweighted photon simulations defines the first component, used for the selection of photon candidate events (Fig. 4). Its distribution is shown in Fig. 5 for the search data set (98% of data collected between 01/01/2004 and 15/05/2013) and photon simulations weighted to an $E^{-2}$ spectrum. The photon candidate cut is defined, with an "a priori" choice, to be the value of the median of the weighted distribution of non-pre-showering photon simulations. Any event with principal component larger than the median cut value is considered as a photon candidate.

## 3. Results

After application of the selection criteria to the data, no event collected between 01/01/2004 and 20/06/2013 is selected with the neutrino cuts, while 4 events survive the photon cuts in the pe-riod between 01/01/2004 and 15/05/2013. The corresponding exposure is determined by application of the same criteria to simulated showers induced by signal primaries. Assuming a differential flux $dN(E) = k \cdot E^{-2}$ for both neutrinos and photons, upper limits to their flux are derived.

[2]The distribution of the observables is different for photons pre-showering or not, as a consequence of the differences in $\langle X_{max} \rangle$ and ratio of muonic to electromagnetic signal at ground. Non-preshowering photons initiate cascades from the interaction of a primary photon with the atmosphere and they are the larger subset in the energy range considered (relative contributions are visible in Fig. 5).

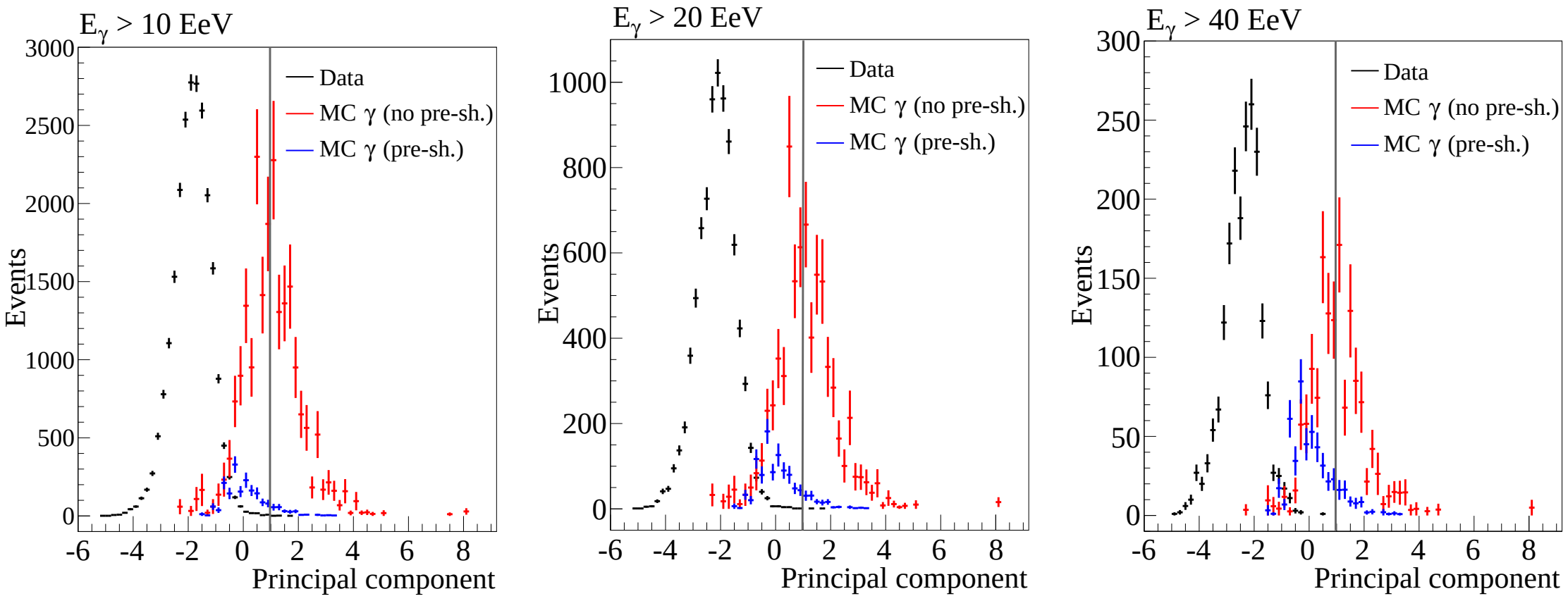

**Figure 5:** Photon search: distributions of the principal component for events in the search sample (black) photons non pre-showering (red) and pre-showering (blue) for three different energy thresholds. Photon simulations are weighted to an $E^{-2}$ spectrum and the median of the non pre-showering photon distribution is defined as the photon candidate cut (vertical line). The photon distribution is normalised to the number of data. Only 4, 2, 0 events survive the selection.

### 3.1 Limits to the flux of neutrinos

An upper limit on the value of $k$ is obtained at a confidence level CL:

$$k_\nu^{\mathrm{CL}} = \frac{N_\nu^{\mathrm{CL}}}{\int_{E_\nu} E_\nu^{-2} \mathscr{E}_{\mathrm{tot}}(E_\nu) dE_\nu} \tag{3.1}$$

where $N_\nu^{\mathrm{CL}}$ is the upper limit at confidence level CL to the number of neutrino events, assuming conservatively that no background event is expected. $N_\nu^{\mathrm{CL}}$ is determined from the number of candidates using a semi-Bayesian extension [14] of the Feldman-Cousins approach [15] to include the uncertainties in the calculation of the exposure $\mathscr{E}_{\mathrm{tot}}(E_\nu)$ combined for all the search channels. The relative contribution of the selection channels to the total expected event rate is ES:DG$\sim$0.84:0.16, under the hypothesis $\nu_e : \nu_\mu : \nu_\tau = 1 : 1 : 1$. Details of the neutrino exposure calculation and values as a function of $E_\nu$ are reported in [6]. The single-flavour limit to the normalisation factor of the diffuse flux of neutrinos is:

$$k_\nu^{90\%} < 6.4 \times 10^{-9} \text{ GeV cm}^{-2} \text{ s}^{-1} \text{ sr}^{-1} \tag{3.2}$$

for $E_\nu = 0.1 - 25$ EeV, defined as the energy range containing symmetrically 90% of the expected events for an $E_\nu^{-2}$ flux. An extensive study of systematic uncertainties is included [6].
The SD of the Auger Observatory is the first air shower array to set a limit below the Waxman-Bahcall bound [19]. As shown in Fig. 6, cosmogenic $\nu$ models that assume a pure primary proton composition at the sources for strong evolution (FRII-type) of the sources [17] and constrained by the GeV observations of Fermi-LAT [16] are disfavoured. The current Auger limit is approaching the fluxes predicted under a range of assumptions for the composition of the primary flux, source evolution and model for the transition from galactic to extragalactic cosmic-rays [18]. A 10-fold increase in the exposure will be needed to reach the most optimistic predictions in case of a pure iron composition at sources, out of the range of the current configuration of the observatory.

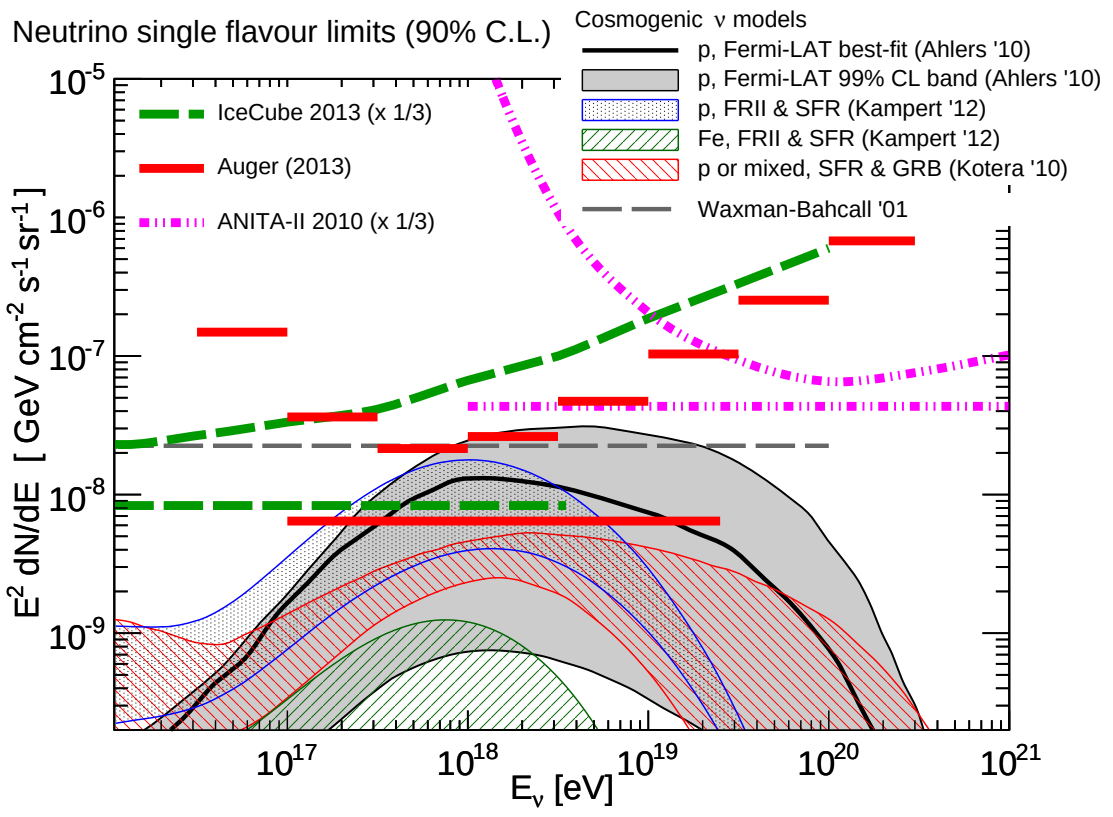

**Figure 6:** Upper limits to the diffuse flux of UHE neutrinos at 90% C.L. in integrated (horizontal lines) and differential form. Limits described in this work (red lines) are compared with cosmogenic neutrino models [16, 17, 18], the Waxman-Bahcall bound [19], and limits from IceCube [20] and ANITA [21]. All neutrino limits and fluxes are converted to single-flavour.

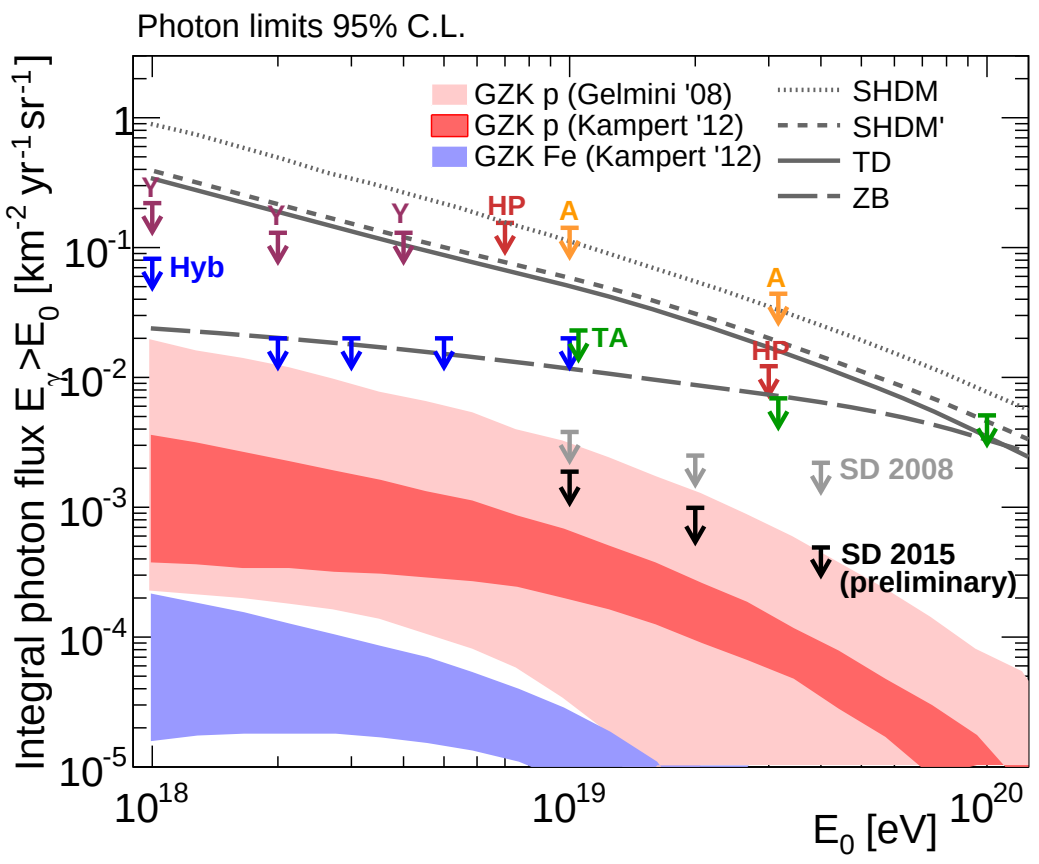

**Figure 7:** Upper limits at 95% C.L. to the diffuse flux of UHE photons derived in this work (black) shown together with previous results from the Pierre Auger Observatory with hybrid (Hyb) and SD data [22], Telescope Array (TA) [23], Yakutsk (Y) [24], Haverah Park (HP) [25], AGASA (A) [26] and predictions from several top-down [27, 28] and cosmogenic photon models [27, 17].

### 3.2 Limits to the integrated photon flux

The upper limits on the integral flux of photons, for $E_\gamma > E_0$, are defined as:

$$F_\gamma^{\mathrm{CL}}(E_\gamma > E_0) = \frac{N_\gamma^{\mathrm{CL}}}{\langle \mathscr{E} \rangle} \tag{3.3}$$

where $E_\gamma$ is assigned according to the photon energy reconstruction; $N_\gamma^{\mathrm{CL}}$ is the Feldman-Cousins upper limit to the number of photon events computed at a confidence level CL in the hypothesis of no background event expected; $\langle \mathscr{E} \rangle$ is the spectrum-weighted average exposure in the energy range $E_\gamma > E_0$. In the period of data taking considered, the value of $\langle \mathscr{E} \rangle$ is 5200, 6800, 6300 km$^2$ sr yr, for $E_\gamma >$10, 20, 40 EeV respectively. The limits to the integral flux are:

$$F_\gamma^{95\%}(E_\gamma > 10,\ 20,\ 40\ \mathrm{EeV}) < 1.9,\ 1.0,\ 0.49 \times 10^{-3}\ \mathrm{km}^{-2}\ \mathrm{yr}^{-1}\ \mathrm{sr}^{-1}. \tag{3.4}$$

The limits to the diffuse flux of photons obtained with the Auger Observatory are the most stringent currently available above 1 EeV (Fig. 7). Top-down models of photon production from the decay of heavy primordial particles [27, 28] are strongly disfavoured. Preliminary limits derived in this work for $E_\gamma > 10$ EeV start constraining the most optimistic predictions of cosmogenic photon fluxes in the assumption of a pure proton composition at the sources [27]. Cosmogenic models using a primary spectral index of -2 and maximum energy of $10^{21}$ eV at the sources [17] predict an integrated photon flux above 10 EeV $\sim$4 times lower than the current limits in the case of proton primaries, $\sim$2 orders of magnitude lower if iron nuclei are injected at the sources.

# Azimuthal asymmetry in the risetime of the Surface Detector signals of the Pierre Auger Observatory

**I.A. Minaya**[*a] **for the Pierre Auger Collaboration**[b]
[a]*Department of Atomic, Molecular and Nuclear Physics, Complutense University of Madrid, Madrid, Spain*
[b]*Observatorio Pierre Auger, Av. San Martín Norte 304, 5613 Malargüe, Argentina*
*E-mail:* auger_spokespersons@fnal.gov
*Full author list:* http://www.auger.org/archive/authors_2015_06.html

The azimuthal asymmetry in the risetime of signals in Auger surface detector stations is a source of information on shower development. The azimuthal asymmetry is due to a combination of the longitudinal evolution of the shower and geometrical effects related to the angles of incidence of the particles into the detectors. The magnitude of the effect depends upon the zenith angle and state of development of the shower and thus provides a novel observable sensitive to the mass composition of cosmic rays above $3 \times 10^{18}$ eV. By comparing measurements with predictions from shower simulations, we find for both of our adopted models of hadronic physics (QGSJetII-04 and Epos-LHC) an indication that the mean cosmic-ray mass increases with energy, as has been inferred from other studies. However the absolute values derived for the mass are dependent on the shower model and on the range of distance from the shower core selected. Thus the method has uncovered further deficiencies in our understanding of shower modelling that must be resolved before the mass composition can be inferred from $(\sec\theta)_{\max}$.

*Speaker.

## 1. Introduction

The measurement of the mass composition of ultra-high energy (UHE) cosmic rays ($E \gtrsim 10^{18}$ eV) is one of the greatest challenges in this field, not only due to the large fluctuations resulting from the stochastic nature of the extensive air-shower (EAS) development but also because it is necessary to make assumptions about the hadronic physics in regions of phase space not covered by measurements at accelerators. The Pierre Auger Observatory [1] uses two complementary techniques for the detection of EAS from UHE cosmic rays: a Fluorescence Detector (FD) that directly registers their longitudinal development and a Surface Detector Array (SD) that samples the shower tail by means of water-Cherenkov detectors. Although the FD provides a well-established technique for mass composition via the direct measurement of the depth of shower maximum, $X_{\rm max}$ [2, 3], its duty cicle is $< 13\%$. Therefore different techniques using the SD, which operates $\sim 100\%$ of time, are being developed [4, 5, 6]. This paper is focused on an SD observable that is related to the azimuthal asymmetry found in the risetime of the signals with respect to the direction of the incoming air-shower. This asymmetry in its turn is related to the stage of the shower development and, as will be shown, has the potential to give alternative insights into the matching of data to mass and hadronic models.

## 2. Concept of azimuthal asymmetry in the risetime

The Auger water-Cherenkov detectors measure the spread in arrival times of the signals produced by the different components of an EAS. The risetime $t_{1/2}$, defined as the time of increase from 10% to 50% of the total integrated signal, is a useful parameter in composition studies. While in vertical showers the SD signals show a perfect circular symmetry around the polar angle $\zeta$ in the shower plane, in inclined showers, particles reaching late detectors ($\pi/2 < \zeta < -\pi/2$, see Fig. 1) traverse longer atmospheric paths than those arriving to early detectors ($-\pi/2 < \zeta < \pi/2$). Consequently for inclined showers, both the magnitude and risetime of the signals depend on $\zeta$. The method described here uses the above-mentioned azimuthal asymmetry.

The observed azimuthal asymmetry is due to two effects. On the one hand, a contribution comes from the quenching of the electromagnetic signal. Since the particles that reach late detectors traverse longer atmospheric paths, we expect a bigger attenuation of electrons and photons as compared to early detectors. On the other hand, there are also contributions to the asymmetry from geometrical effects. In this case, not only is the electromagnetic component important, but muons also play a role. Although track length and solid angle effects are compensated by the near-cubical design of the detectors, the angular distributions of muons impinging on the detectors are different, as late detectors record more muons emitted closer to the shower axis. Geometrical effects predominate at low zenith angles, while for showers with $\theta > 30^\circ$ attenuation effects are the main contribution.

The azimuthal asymmetry of the $t_{1/2}$ depends on the distance to the core $r$ in the plane of the shower front, since the larger is $r$ the larger is the difference in atmospheric paths between early and late stations. In addition this asymmetry depends on the zenith angle of the shower, $\theta$. No asymmetry is expected for vertical showers: it grows with $\theta$ and finally this trend reaches a point where the EM component is quenched due to the longer atmospheric path travelled and the shower is mainly composed of penetrating muons. As will be demonstrated, for a given primary energy $E$, the time asymmetry dependence upon $\sec\theta$ shows a maximum which is sensitive to the average primary mass.

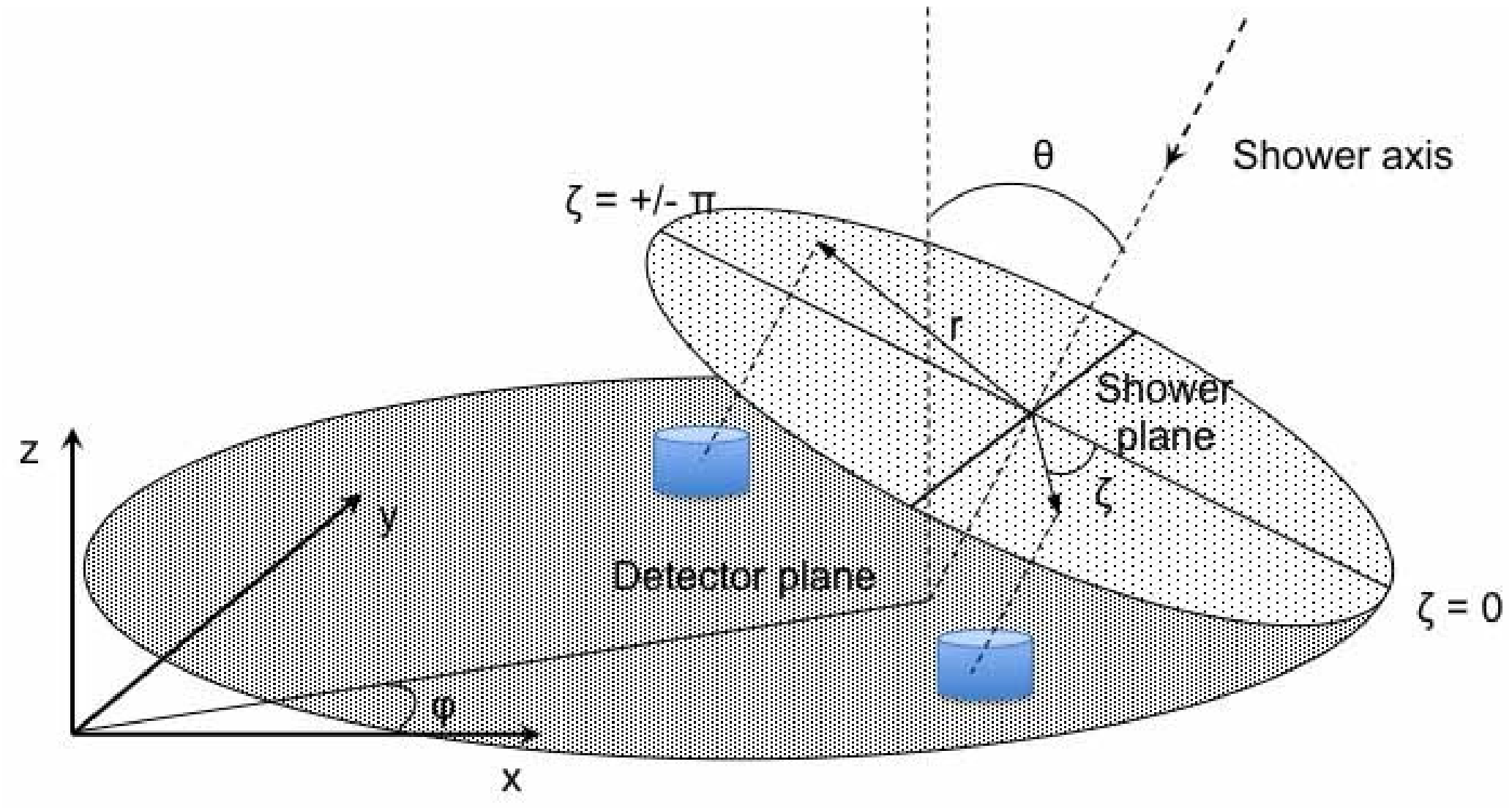

**Figure 1:** Schematic view of the shower geometry with the notation used in this paper. The incoming direction of the primary particle defines two regions: early ($-\pi/2 < \zeta < \pi/2$) and late ($\pi/2 < \zeta < -\pi/2$). Note the different amount of atmosphere traversed by the particles reaching the detectors in each region.

## 3. Azimuthal asymmetry analysis using Auger data

For the present analysis we have used data collected with the SD of the Pierre Auger Observatory from 01.01.04 to 29.10.14 with $E > 3 \times 10^{18}$ eV and $\theta < 62°$ to assure 100% detection efficiency, and with $\theta > 30°$ where the contribution of the attenuation to the azimuthal asymmetry dominates over other sources. These events are required to satisfy the standard trigger levels described in [7]. Additionally, quality cuts at station level have been applied: the recorded signal must be larger than 10 VEM (1 VEM is the signal produced by a vertical and central through-going muon) to assure 100% probability for single-detector triggering, and stations are required to be at a core distance $500 < r < 2000$ m to assure an accurate determination of $t_{1/2}$. After application of these cuts a total of 191534 signals from 54584 events remain.

The first step in the analysis is to measure the azimuthal asymmetry of the $t_{1/2}$ for fixed $E$ and $\theta$ values. This measurement cannot be done on a shower-by-shower basis because it is not possible to sample the whole range of the polar angle, from early to late regions, in a single event. Thus, a statistical approach is applied to characterize the azimuthal asymmetry of $t_{1/2}$ using all the stations from the events at a given $E$ and $\theta$. To this end we use the variable $t_{1/2}/r$ since in this way the risetime information of all stations within given $E$ and $\theta$ bins can be used, independently of the core distance, to compute an average $\langle t_{1/2}/r\rangle$ value as a function of $\zeta$. Since the asymmetry depends on $r$, the analysis has been carried out independently for two $r$-intervals, i.e., 500 - 1000 m and 1000 - 2000 m each containing a similar number of signals and events. As an example, we show in the left panel of Fig. 2 $\langle t_{1/2}/r \rangle$ vs $\zeta$ for $19.2 < \log(E/eV) < 19.5$ and eight $\theta$ values. For each $\theta$ band the data are fitted to the function $\langle t_{1/2}/r \rangle = a + b\cos\zeta + c\cos^2\zeta$. The asymmetry with respect to $\zeta$ is evident and the ratio $b/(a+c)$, the so-called asymmetry factor, will be used to quantify the asymmetry. In right panel of Fig. 2 $\langle t_{1/2}/r \rangle$ vs $\zeta$ is displayed for both core distance intervals for showers with $\log(E/eV) = 19.0 - 19.2$ and $\theta = 51°$. As can be observed, the amplitude of the asymmetry for 500 - 1000 m is smaller. This is due to the fact that, close to the core, there is a higher electromagnetic contribution to the risetime and a smaller difference in the paths travelled by the particles.

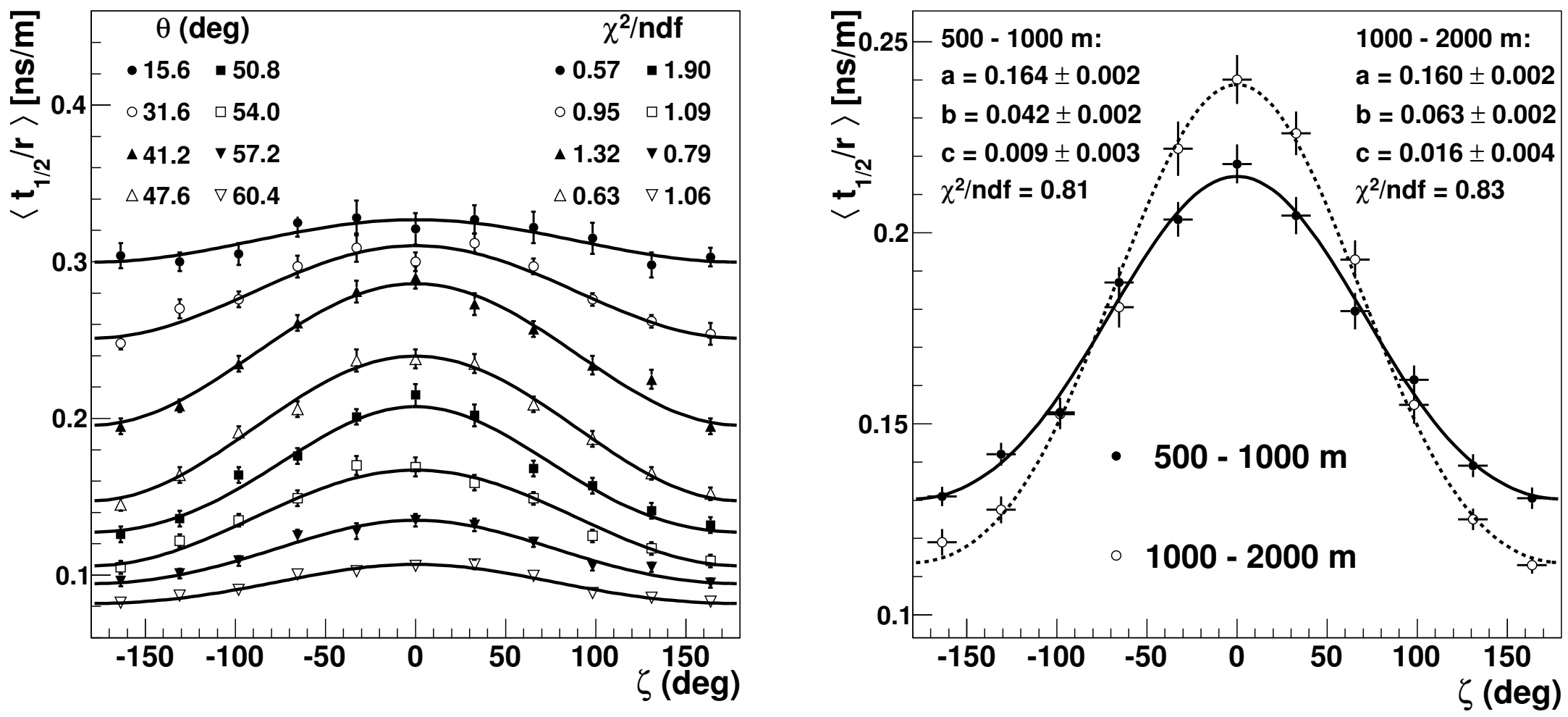

**Figure 2:** Asymmetry plots (dependence of $\langle t_{1/2}/r \rangle$ vs the polar angle in the shower plane $\zeta$) for different cases. Left panel: for $\log(E/eV) = 19.2 - 19.5$ in the 500 - 1000 m interval at different $\theta$ bins. Right panel: for $\log(E/eV) = 19.0 - 19.2$ in the two chosen core distance intervals with $\theta = 51°$; results of the fitted parameters (see text) are shown for each $r$-interval.

Next, this asymmetry factor is represented as a function of atmospheric depth, measured by $\sec\theta$. For each energy interval, $b/(a+c)$ vs $\ln(\sec\theta)$ is fitted to a Gaussian function to obtain its maximum. An example is shown in Fig. 3 for the same energy bin in each $r$-interval. The value of $\sec\theta$ for which the asymmetry parameter maximizes, $(\sec\theta)_{\max}$, will be used as the observable for mass composition. Once the value of $(\sec\theta)_{\max}$ for each energy bin has been obtained in each core distance interval, we can study the evolution of the dependence of $(\sec\theta)_{\max}$ on the primary energy. In Fig. 4 the result for both $r$-intervals is shown. As expected, for a given energy, $(\sec\theta)_{\max}$ is larger in the interval 500 - 1000 m than in the 1000 - 2000 m one since closer to the core the asymmetry is smaller, and thus, the zenith angle at which the muon component starts to dominate (and the asymmetry starts to decrease) is higher. Note that, in principle, this dependence on the radial interval should not limit the capability of the asymmetry method for mass analysis provided that Monte Carlo simulations follow the same behavior. This will be discussed in section 4.

A first estimation of the systematic uncertainty of $(\sec\theta)_{\max}$ has been carried out for both core distance intervals. For that, we have taken into account the effect of the uncertainty in the reconstruction of the shower core, in the risetime and also the effect of the energy scale uncertainty, as well as the use of different parameterizations in the dependency of the risetime with $r$ and different algorithms to integrate the signal. Additionally, we have developed different cross-checks on the stability of the results, considering potential bias due to the selection cuts in the signal intensity and in the zenith interval, studying the dependence of the lateral width of the shower (in particular of the electromagnetic component) on pressure and temperature, and checking the possible effect of ageing of the SD detectors on the results. Considering all these sources we obtain a systematic uncertainty of $\pm 0.013$ and $\pm 0.017$ in units of $(\sec\theta)_{\max}$ for the 500 - 1000 m and 1000 - 2000 m core distance ranges respectively.

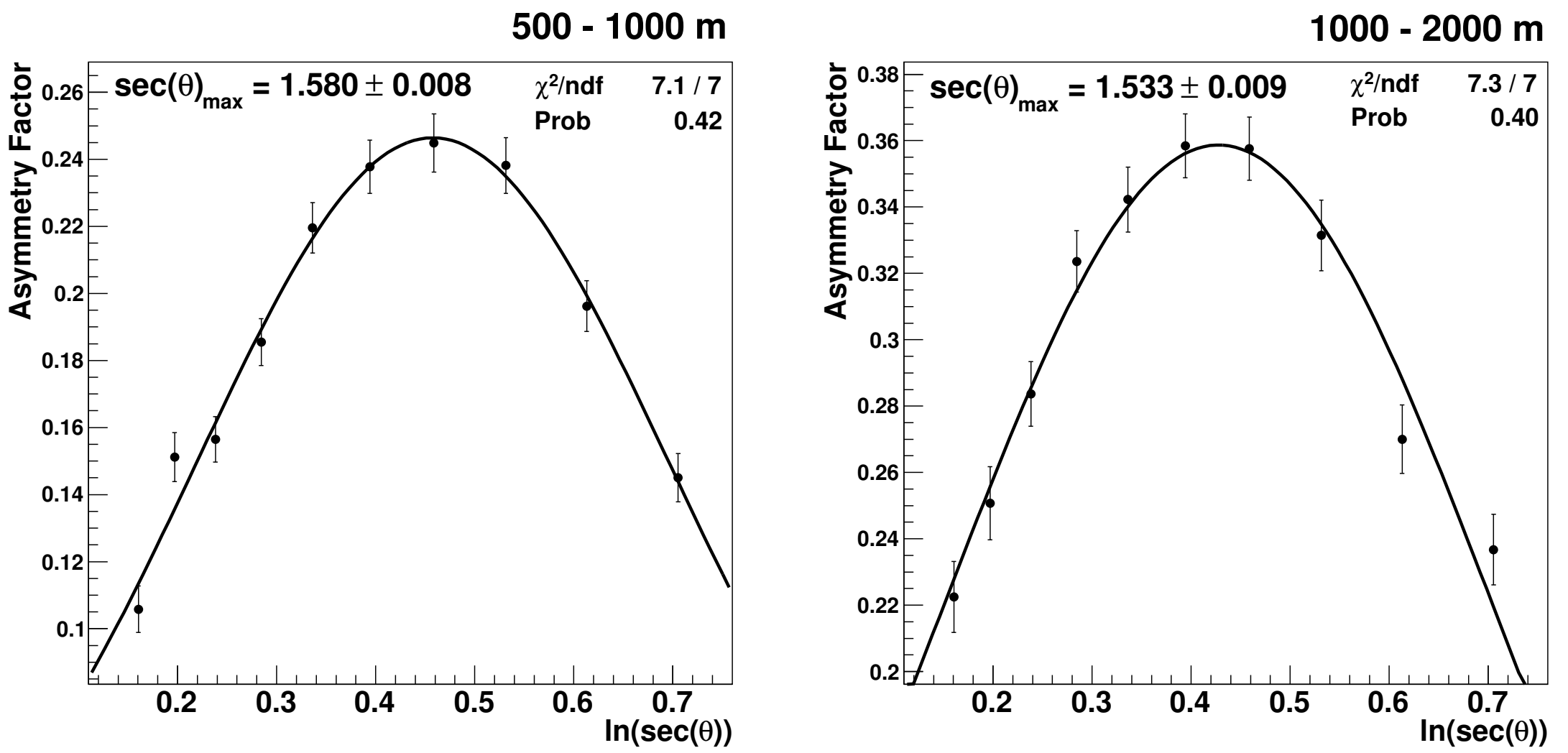

**Figure 3:** Longitudinal development of the asymmetry factor (b/(a+c), see text) for $\log(E/eV) = 19.0 - 19.2$ in both core distance intervals (500 - 1000 m in the left panel, 1000 - 2000 m in the right panel).

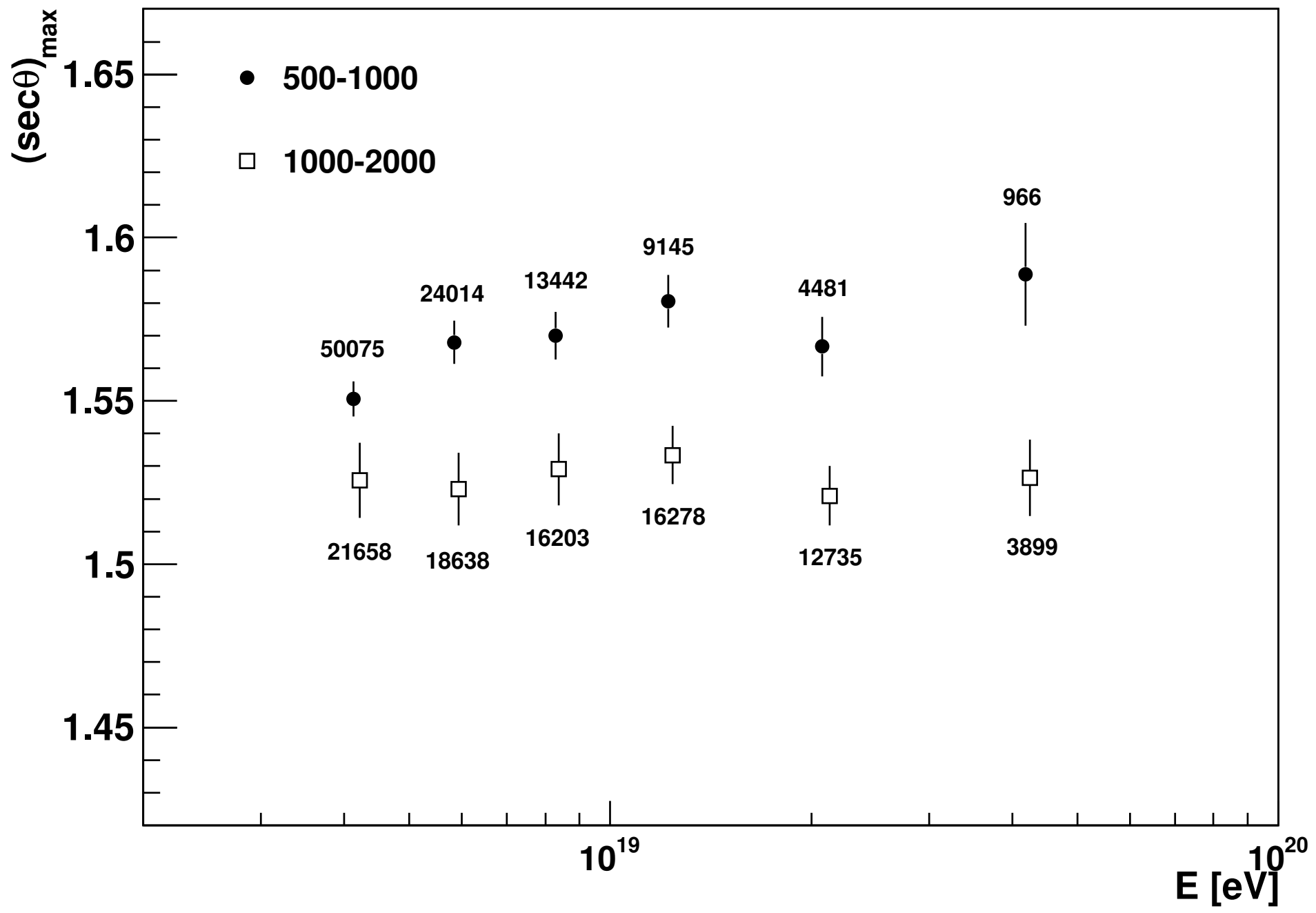

**Figure 4:** Energy dependence of $(\sec\theta)_{\max}$ for both intervals of core distance 500 - 1000 m and 1000 - 2000 m. Number of stations available for the analysis are indicated.

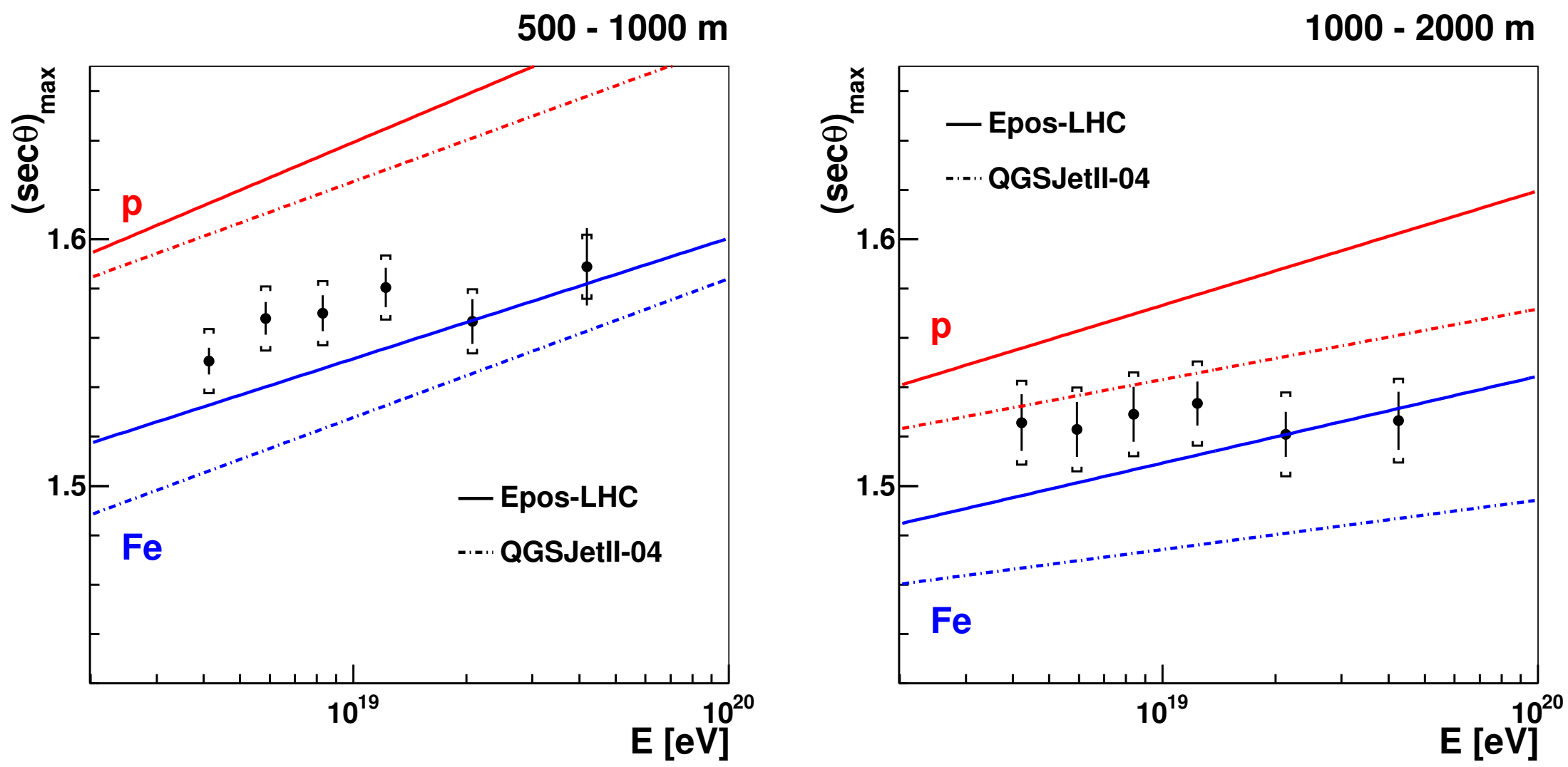

**Figure 5:** Comparison between $(\sec\theta)_{\max}$ for data (black points) and MC predictions using Epos-LHC (solid line) and QGSJetII-04 (dashed line) hadronic models for proton (red) and iron (blue) primaries in both $r$-intervals. Brackets represent the systematic uncertainty.

## 4. Discussion

The mass analysis has to rely on comparison with theoretical predictions using updated models of hadronic interactions extrapolated to these energies. To this end, a sample of Monte Carlo (MC) events generated with the CORSIKA [8] code have been produced using the Epos-LHC [9] and QGSJetII-04 [10] hadronic models for proton and iron primaries. A total of 77000 events (38500 of each primary) have been produced for each interaction model. The $\log(E/\mathrm{eV})$ values ranged from 18.00 to 20.25 in bins of 0.25 with eleven discrete $\theta$ values between 18° and 63°.

In Fig. 5 the $(\sec\theta)_{\max}$ measurements as a function of energy are shown and compared with predictions made using the Epos-LHC and QGSJetII-04 models for proton and iron primaries. The dependence on energy is small and it is difficult to draw strong conclusions as rather contrasting predictions follow from the two models. However in both cases there is an indication that the mean mass increases slowly with energy in line with other Auger studies [2, 5].

A further illustration of the dependence of the conclusions on the choice of model is shown in Fig. 6 where the variations of $\langle \ln A \rangle$ with energy, deduced from Fig. 5, are summarised and compared with the values of $\langle \ln A \rangle$ found from the measurements of the depth of shower maximum ($X_{\max}$) made with the fluorescence detectors [2] and using the muon production depth (MPD) technique [5].

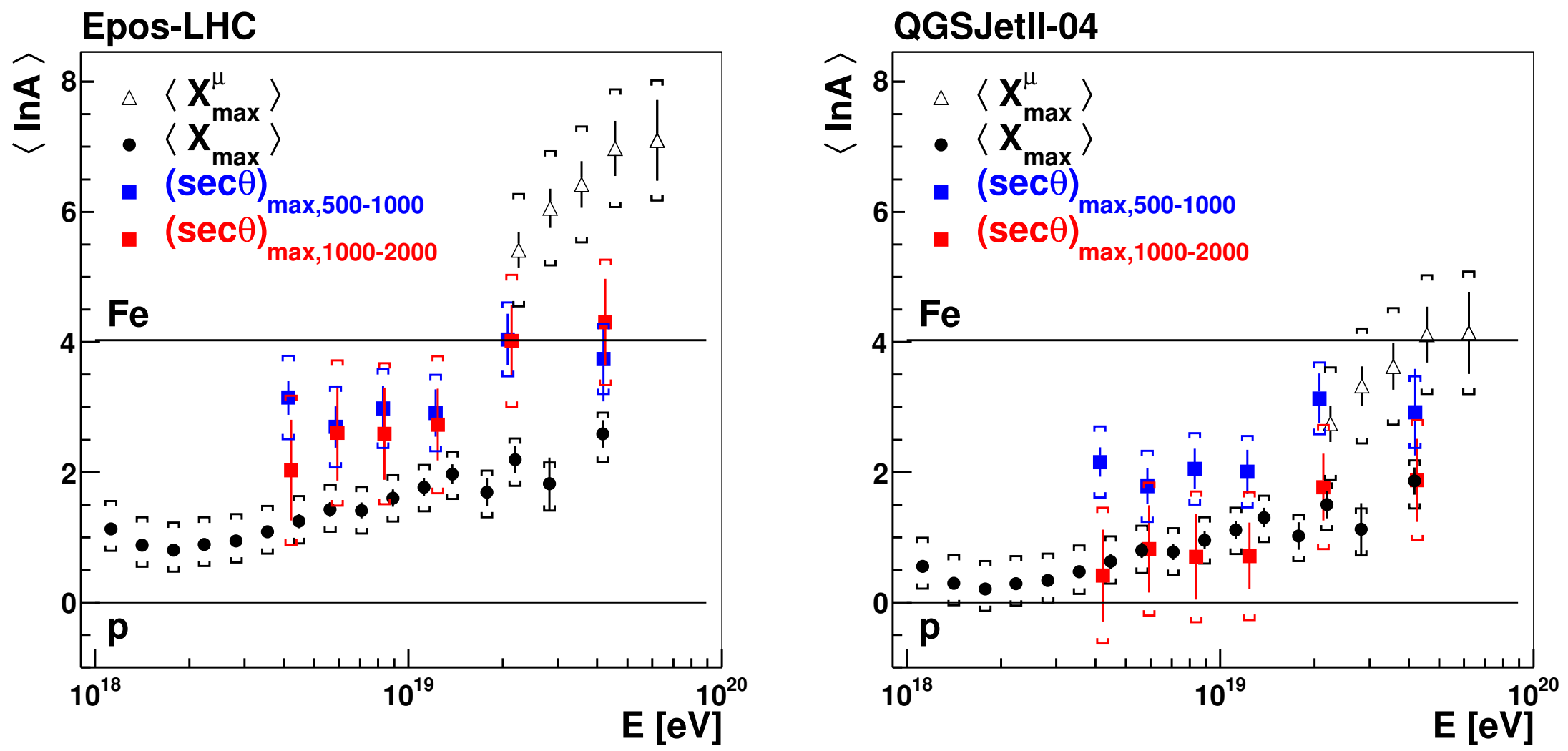

**Figure 6:** $\langle \ln A \rangle$ vs $E$ as predicted by Epos-LHC and QGSJetII-04. Results from the time asymmetry method in both $r$-intervals are compared with those from the elongation curve [2] and the MPD method [5].

Several inferences can be drawn from these plots. Within the uncertainties, the values of $\langle \ln A \rangle$ derived from the Epos-LHC model are consistent for the two distance ranges. However for QGSJetII-04 model this consistency is not so evident. We note that the MPD studies are of a parameter that is solely dependent on the muon content of the shower while the $X_{\max}$ observations are dominated by the electromagnetic component. As explained above, the complex nature of the zenith angle behavior of the time-asymmetry is a reflection of the contributions of both components. Thus, while the Epos-LHC is favoured over QGSJetII-04, neither model provides an accurate description of muons in air-showers. A similar conclusion has been drawn from the study of inclined showers made at the Pierre Auger Observatory [11] from which it was deduced that showers contain more muons than predicted by Epos-LHC model. Indeed no model provides a completely satisfactory description of the wide range of data using the Auger Observatory.

## 5. Conclusions

The azimuthal dependence of the $t_{1/2}$ values obtained from the FADC traces registered by the SD detector of the Pierre Auger Observatory has been used to obtain a mass sensitive parameter, $(\sec\theta)_{\max}$. Recent hadronic interaction models have been used to compute the $\langle \ln A \rangle$ predictions of the method described here and compare the results with those obtained from $X_{\max}$ and the MPD method. The results show a hint of a slow increase of the mean mass with energy in agreement with other studies. However, model-dependent discrepancies between data and Monte Carlo results on both core distance intervals have been found, and therefore, further deficiencies in our understand-ing of shower modelling have to be resolved before the mass composition can be inferred from $(\sec\theta)_{\max}$.

# 4

# Hadronic Interactions

PIERRE
AUGER
OBSERVATORY

# Measurement of the muon content in air showers at the Pierre Auger Observatory

**Laura Collica**[*a] **for the Pierre Auger Collaboration**[b]
[a] *INFN Torino, Italy*
[b] *Observatorio Pierre Auger, Av. San Martín Norte 304, 5613 Malargüe, Argentina*
*E-mail:* auger_spokespersons@fnal.gov
*Full author list:* http://www.auger.org/archive/authors_2015_06.html

The muon content of extensive air showers is an observable sensitive to the primary composition and to the hadronic interaction properties. We present here different methods which allow us to estimate the muon number at the ground level and the muon production depth by exploiting the measurement of the longitudinal, lateral and temporal distribution of particles in air showers recorded at the Pierre Auger Observatory. The results, obtained at about $10^{19}$ eV ($E_{CM} \sim 140$ TeV center-of-mass energy for proton primaries), are compared to the predictions of LHC-tuned hadronic interaction models with different primary masses and suggest a deficit in the muon content at the ground predicted by simulations. The Pierre Auger Observatory uses water-Cherenkov detectors to measure particle densities at the ground and therefore has a good sensitivity to the muon content of air showers. Moreover, due to its hybrid design, the combination of muon measurements with other independent mass composition analyses such as $X_{max}$ provides additional constraints on hadronic interaction models.

*Speaker.

## 1. Introduction

The spectrum and arrival directions of Ultra High Energy Cosmic Rays (UHECRs) have been recently measured with unprecedented precision [1, 2]. However, the origin of these particles is still not well understood and remains one of the greatest priorities of the field. Establishing their composition is a crucial step to discriminate between different acceleration and propagation scenarios. Information about the composition of the primary cosmic rays has been obtained using the Fluorescence Detector (FD) of the Pierre Auger Observatory. The FD allows the measurement of the depth at which the electromagnetic component of the air shower reaches its maximum number of particles, $X_{\max}$ [3]. However, the interpretation of these measurements is hampered by uncertainties of the hadronic interaction models, which extrapolate interaction details from measurements in the accelerators domain to much higher energies and to different kinematic regions which are difficult to separate from the effect of the primary composition. Moreover, the FD data suffer from small statistics due to the low FD duty cycle ($\sim$15%).

The design of the Pierre Auger Observatory includes also a Surface Detector (SD) consisting of an array of water-Cherenkov detectors [4]. Different methods, which are also sensitive to primary composition, have been developed exploiting the 100% duty cycle of the SD. Among them, the study of the muon content at the ground [5] and the study of the atmospheric depth at which the muon production rate reaches a maximum in air showers [6].

Muon measurements are sensitive to the details of the hadronic component of the air shower and provide a handle to study the mass composition independently of $X_{\max}$. The hybrid nature of the Auger observatory provides redundancy which allows for the combination of different measurements sensitive to the primary mass to place constraints on hadronic interaction models.

## 2. Measurement of muon number

Recently we developed techniques to reconstruct inclined showers [7, 8] that can be used to extract the muon content of air showers. They rely on the fact that the electromagnetic component of inclined showers is largely absorbed in the atmosphere before reaching the ground. Once the shower direction is obtained using the arrival times of the SD signals, it has been shown that the number of muons per unit area at the ground level, $\rho_\mu(\theta,\phi;x,y)$, has a shape which is practically independent of energy, composition or hadronic interaction model [9, 10]. As a result it can be expressed:

$$\rho_\mu = N_{19}\,\rho_{\mu,19}(\theta,\phi;x-x_c,y-y_c), \tag{2.1}$$

where $\rho_{\mu,19}$ are the reference functions for the number densities of muons, expressed in terms of position in a plane $(x,y)$ relative to the shower core $(x_\mathrm{c},y_\mathrm{c})$ and $N_{19}$ is a scale factor. The reference distributions for each arrival direction are conventionally obtained from proton simulations at $10^{19}$ eV using the QGSJETII-03 model for hadronic interactions.

The fitted value of $N_{19}$ gives the number of muons per unit area relative to the reference density. The total number of muons can be estimated as $N_\mu^{\mathrm{est}} = N_{19}\,N_{\mu,19}$ where $N_{\mu,19}$ is the surface integral of the $\rho_{\mu,19}$. For example, at the mean zenith angle of the data set, $\langle\theta_{\mathrm{data}}\rangle = 67^\circ$, $N_{19} = 1$ would correspond to about $1.5\times10^7$ muons at the ground with energies above 0.3 GeV.

The procedure has been extensively tested with simulations comparing $N_{19}$ to the true ratio $R_\mu^{\mathrm{MC}} = N_\mu/N_{\mu,19}$, computed as the total number of muons in the simulated shower relative to the total number of muons in the reference model. The average difference between $N_{19}$ and $R_\mu^{MC}$ for proton or iron simulated using QGSJET01, QGSJETII-04 and EPOS-LHC is always below 5%. To get an unbiased estimator, $N_{19}$ is corrected for the average bias of all the simulations. In the following the corrected estimator is called $R_\mu$. By combining the uncertainty of the reference model with that of the simulated response of the SD stations to muons, we conservatively estimate the systematic uncertainty of $R_\mu$ to be 11%.

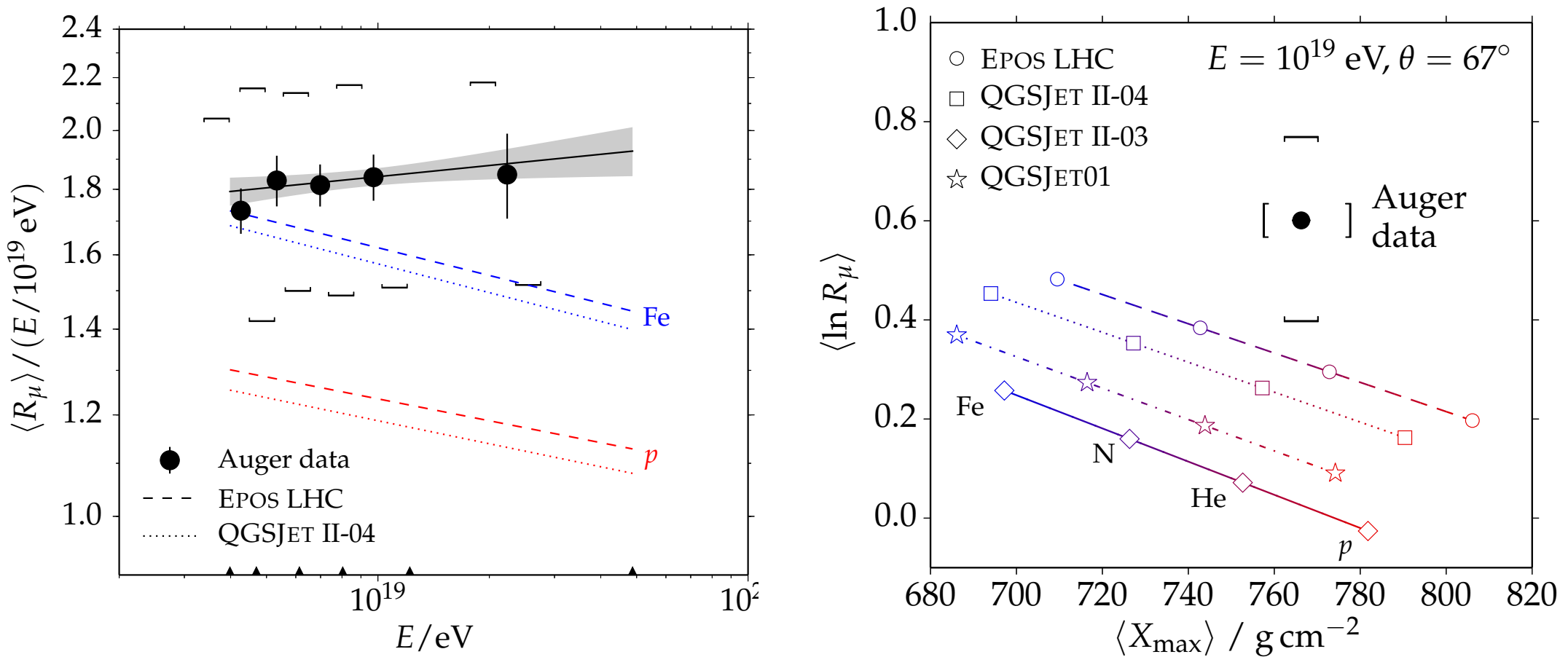

Figure 1: Left) $\langle R_\mu \rangle$ vs. primary energy, compared to air shower simulations. Right) $\langle \ln R_\mu \rangle$ vs. $\langle X_{max} \rangle$. Representative primary masses are indicated by open symbols. [5]

The method is applied to hybrid events with zenith angles $62° < \theta < 80°$ for which a simultaneous measurement of muon number with SD and of shower energy with FD is possible. Strict selection criteria are applied to get a high quality sample: the events must be well contained, i.e. the station closest to the fitted core and its six adjacent stations need all to be active, and only events with energy above $4 \times 10^{18}$ eV are taken to ensure a 100% SD trigger probability. Quality cuts are applied for the FD to ensure an accurate reconstruction of the arrival direction and of the longitudinal profile, minimizing composition bias [11]. Out of 29722 hybrid events recorded from 1 January 2004 to 1 January 2013, 174 are accepted after quality cuts.

The relative number of muons $R_\mu$ is correlated to the shower energy by a power law

$$R_\mu = a(E/10^{19} \text{ eV})^b \tag{2.2}$$

with parameters

$$\begin{aligned} a &= \langle \ln R_\mu \rangle (10^{19} \text{ eV}) = (1.841 \pm 0.029 \pm 0.324 \text{ (sys.)}) \\ b &= \mathrm{d}\langle \ln R_\mu \rangle / \mathrm{d} \ln E = (1.029 \pm 0.024 \pm 0.03 \text{ (sys.)}) \end{aligned} \tag{2.3}$$

The $a$ parameter represents the average muon content at $10^{19}$ eV while $b$ is the logarithmic gain of muons with increasing energy.

The systematic uncertainties are estimated from the dispersion of the different models and compositions studied with simulated events, from variations of the quality cuts on the FD and of fitting methods applied. They are dominated by the uncertainties on the energy scale (~14%) [11].

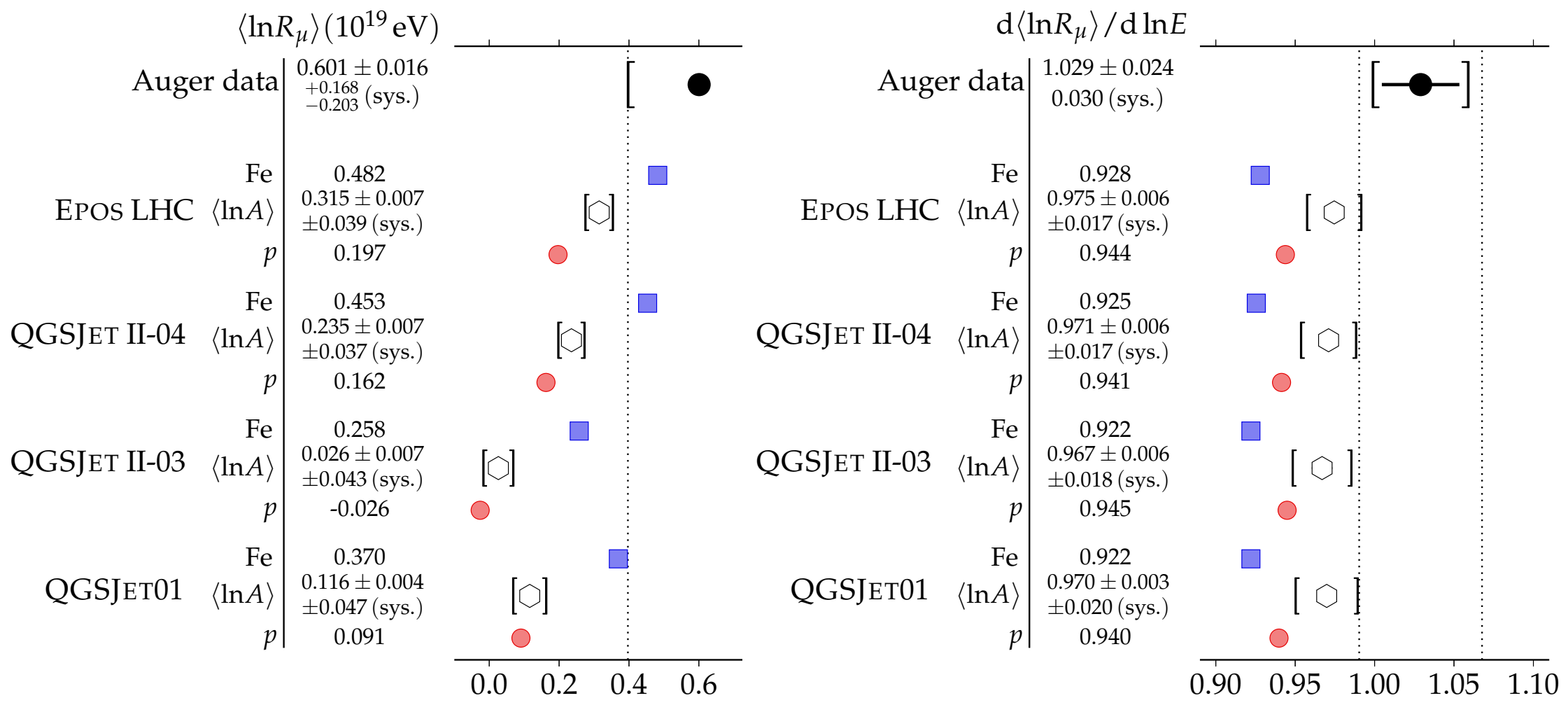

Figure 2: Comparison of $\langle \ln R_\mu \rangle$ (left) and $d\langle \ln R_\mu \rangle / d\ln E$ (right) between $4\times10^{18}$ eV and $5\times10^{19}$ eV with predictions for air shower simulation models for a pure proton, a pure iron and a mixed composition compatible with the FD measurements (labeled as $\langle \ln A \rangle$). [5]

The average values of $R_\mu$, divided by the energy, are plotted for five energy bins in Fig. 1(left). Data points are compared to simulations for proton and iron showers, performed with QGSJETII-04 and EPOS-LHC hadronic models at $\langle \theta_{\text{data}} \rangle = 67°$. The predictions for protons and iron nuclei are well separated, illustrating the power of $\langle R_\mu \rangle$ as a composition estimator. The measured muon number is higher than that expected in pure iron showers, a result not in agreement with studies based on the depth of shower maximum [12] which points to an average logarithmic mass $\langle \ln A \rangle$ being between proton and iron in this energy range.

The tension between the $X_{\text{max}}$ and $R_\mu$ measurements with respect to the expectations from different models is shown in Fig. 1 (right) which displays Auger data at $10^{19}$ eV compared to the predictions for different hadronic models and primary masses. The expectations for $\langle \ln R_\mu \rangle$ and $d\langle \ln R_\mu \rangle / d\ln E$ are compared to our measurement in Fig. 2. For QGSJETII-03, QGSJETII-04 and EPOS-LHC, $\langle \ln R_\mu \rangle$ and $d\langle \ln R_\mu \rangle / d\ln E$ are calculated for three different compositions: pure proton, pure iron and a composition with an average value of the logarithmic mass, $\ln A$, as obtained from the measurement of $X_{\text{max}}$ [3]. The QGSJET01 model was not considered in that reference so that an estimate of $\ln A$ was made using the data [12] and some simple assumptions [5]. These values of $\ln A$ in turn correspond to a mean value of $\ln R_\mu$. Assuming the generalized Heitler model of hadronic air [5], it is possible to convert the estimated $\langle \ln A \rangle$ into a prediction of the logarithmic muon content which can in turn be related to $\langle R_\mu \rangle$.

When we consider the values of $\ln A$ deduced from $X_{\text{max}}$, the measured values of $R_\mu$ indicate that the mean number of muons in the simulations have a deficit of 30% to 80 $^{+17}_{-20}$ (sys.) % at $10^{19}$ eV depending on the model. The measurement of the logarithmic gain is slightly higher than the prediction but the discrepancy is smaller than for $R_\mu$ for all the models. Assuming that the logarithmic gain of real showers is well reproduced by simulations, which is supported by the fact that the four models agree on this parameter, the measured value disfavors a pure composition hypothesis. Deviations from a constant proton (iron) composition are observed at the level of 2.2 (2.6) $\sigma$.

Different independent methods [13, 14] for vertical showers with $\theta < 60^\circ$ have been used to derive the fraction of the signal due to muons at 1000 m from the shower core with the SD array. In [13] the different features of the temporal distribution of the EM and muonic signals measured with the SD array are exploited to obtain information about the muon number. In [14] hybrid events are exploited: for each of them, a set of simulated proton and iron showers matching their longitudinal profile is produced using different hadronic interaction models. They show that the total and muonic signals are not well reproduced by the shower simulations using the most recent hadronic interaction models. Within uncertainties they are compatible with the results described above for the study of inclined showers.

## 3. Measurement of muon production depth

The time structure of the muon component reaching the ground can be exploited to obtain the distribution of muon production distances along the shower axis. Following a phenomenological model for muon time distributions in Extensive Air Showers (EASs) developed in [15] [16], the muon production height $z$ of muons recorded at the ground at distance $r$ from the core and arriving at time $t$ can be written as

$$z \simeq \frac{1}{2}\left(\frac{r^2}{c(t-\langle t_\varepsilon\rangle)} - c(t-\langle t_\varepsilon\rangle)\right) + \Delta - \langle z_\pi\rangle \tag{3.1}$$

where $t_g \simeq t - \langle t_\varepsilon\rangle$ is the *geometric delay*, due to deviation of muon trajectories with respect to the shower axis, $t_\varepsilon$ is the *kinematic delay*, due to the subluminal muon velocities, $\Delta = r\tan\theta\cos\xi$ is the distance from the ground impact point to the shower plane and $\langle z_\pi\rangle$ takes into account the decay length of the parent pion.

Since SD stations do not allow to measure the energy carried by each single muon, the kinematic delay cannot be directly measured and needs to be parameterized. A modeling of the muon energy distributions was then exploited to this aim [15].

By means of Eq. (3.1), a mapping between $z$ and $t$ at the ground is thus provided. The muon production depth (MPD) $X^\mu$, i.e. the total amount of traversed matter in $g/cm^2$, is obtained integrating the atmospheric density over the range of production distances. The MPD distribution is derived adding all MPDs recorded in each of the SD stations of the event. A fit to the MPD distribution with a Gaisser-Hillas function allows us to derive the muonic shower maximum $X^\mu_{max}$, i.e. the point in the shower development where the maximum number of muons is produced.

The reconstruction of the MPD distribution requires the removal of the EM contribution to the total signal. Only a residual EM contamination is contributing to the total signal at zenith angle around 60°. Since the EM signals are smaller and broader than muonic ones, a cut for $S_{threshold} = 15\%$ of the maximum (peak) of the recorded signal guarantees an efficient reduction of the background and muon fractions above 85%, regardless of the energy and mass of the primary particle.

Besides, the FADC sampling frequency (40 MHz) gives rise to an uncertainty in the $z$ reconstruction that decreases with $r^2$ and increases with $X^\mu$. To keep the distortion of the reconstructed MPD small, a cut in core distance, $r_{cut}$, is thus mandatory. An optimal value $r_{cut} = 1700$ m was derived from Monte Carlo simulations.

Another issue to be taken into account is that the light propagation inside the detector and the electronic response smears the muon arrival times. To compensate for this detector effect, a time offset $t_{\text{shift}}$ is subtracted to each time bin. The offset value depends on $S_{\text{threshold}}$: a value of 73 ns is derived from simulations.

The mean bias $[X^{\mu}_{\text{max}}(\text{rec.}) - X^{\mu}_{\text{max}}(\text{true})]$ stays within 10 g/cm$^2$, for all energies, masses and hadronic models used in simulations. The resolution, given by the standard deviation of the same distribution, ranges from 100 (80) g/cm$^2$ to about 50 g/cm$^2$ for increasing energy and proton (iron) showers. The improvement of the resolution with the energy is a direct consequence of the increase in the number of sampled muons.

The SD data of Pierre Auger Observatory between 1 January 2004 and 31 December 2012 have been used in this analysis. The selected events must satisfy the T5 trigger condition, which requires that the detector with the highest signal has all six closest neighbours operating. We considered events with zenith angle in the range [55°, 65°] and energy greater than 20 EeV, the latter allowing the reconstruction of the MPD distribution. Furthermore, the relative uncertainty $\delta X^{\mu}_{\text{max}}/X^{\mu}_{\text{max}}$ must be small enough to guarantee the accuracy in the estimation of $X^{\mu}_{\text{max}}$. Monte Carlo studies show that the chosen cuts introduce a negligible composition bias, smaller than 2 g/cm$^2$. 481 events out of 500 meet the required quality cuts.

The total systematic uncertainty on $X^{\mu}_{\text{max}}$ amounts to ≈17 g/cm$^2$, which corresponds to about 25% of the proton-iron separation. The most relevant contributions come from reconstruction, differences in the hadronic interaction models, unknown primary mass and seasonal effects. A systematic underestimation by ≈4.5 g/cm$^2$ in the $X^{\mu}_{\text{max}}$ determination has been found due to random accidental signals and has been corrected for.

Finally, we discovered that our simulations introduced an underestimation of the muon delay with respect to the arrival time of the shower front [17]. The effect is due to the resampling procedure, which is needed to be applied to showers simulated with a thinning method [18]. The latter is mandatory in the EASs simulations, since full simulations require large amounts of CPU time and disk space. A procedure to undo the thinning is thus necessary to have a fair representation of the signals collected by the water-Cherenkov detectors. The strategy consists in the estimation of the local distribution of particles at the position of the detectors by averages over extended areas around this position. We found that the value chosen for the sampling area was not optimal and introduced an underestimation of the muon delay. While in [6] the bias due to the resampling procedure was assumed to be negligible, recent simulation studies have shown that this effect is significant, making the total bias on $X^{\mu}_{\rm max}$ reconstruction $24\,{\rm g/cm^2}$. This bias is corrected for in simulations in order to obtain an unbiased analysis. However this correction has to be accounted for in the data (i.e. all measured $X^{\mu}_{\rm max}$ values must be reduced by an amount equal to the estimated bias).

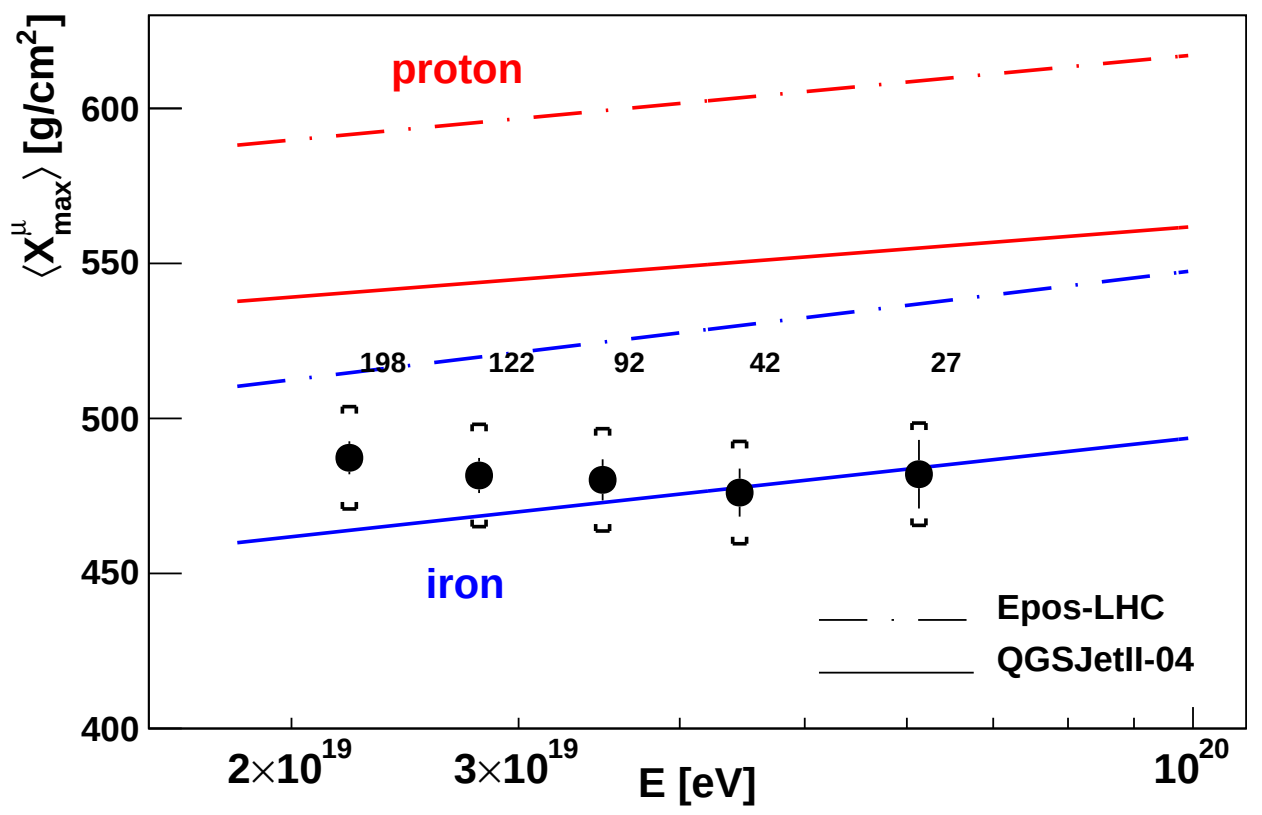

Figure 3: Evolution of $X^{\mu}_{\text{max}}$ with energy. The number of events is indicated in each energy bin. Brackets represent the systematic uncertainty [17].

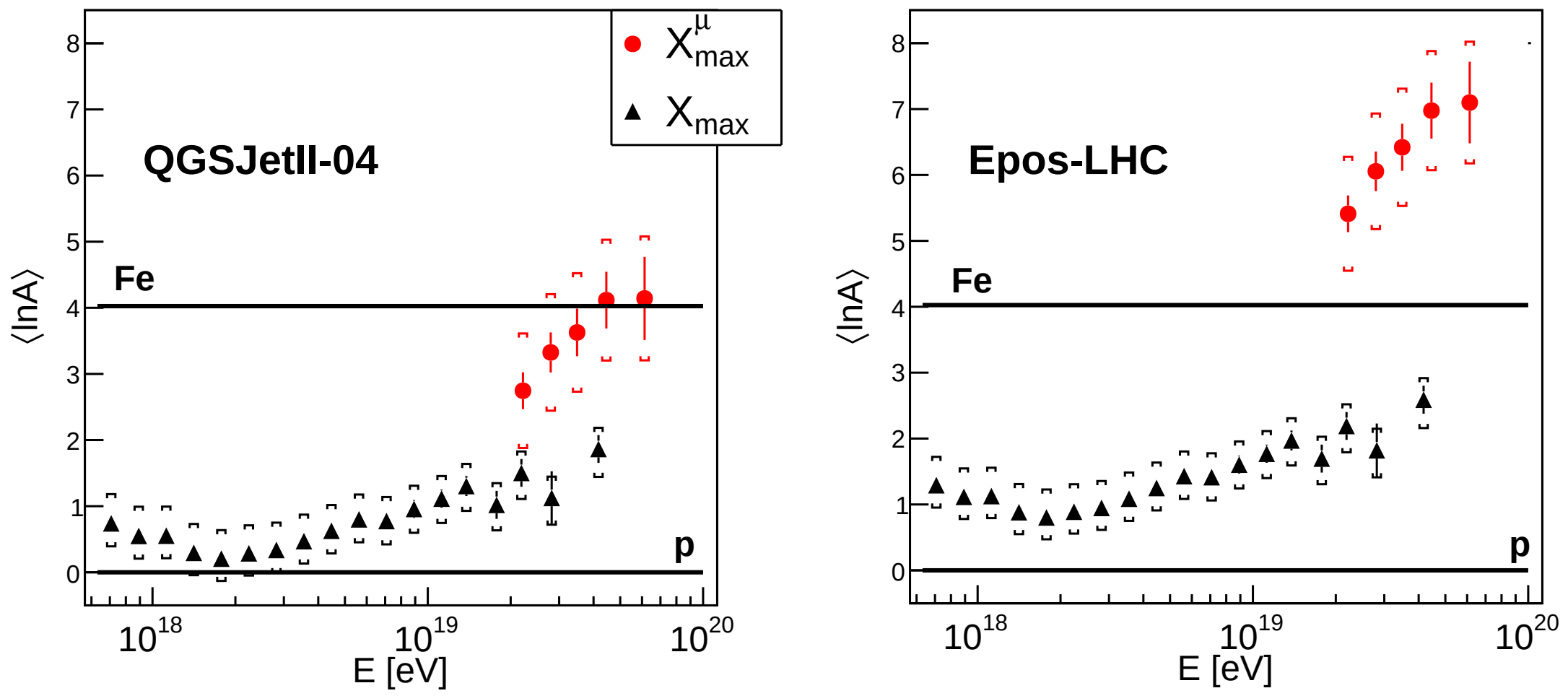

Figure 4: $\langle \ln A \rangle$ from $\langle X_{\rm max} \rangle$ (triangles) [3] and $\langle X^{\mu}_{\rm max} \rangle$ (circles) [17] vs. energy. QGSJETII-04 (left) and EPOS-LHC (right) are used as reference models. Brackets correspond to the systematic uncertainties.

The evolution of the measured $X^{\mu}_{\rm max}$ with energy is shown in Fig. 3. Data are grouped in five energy bins of width 0.1 in $\log_{10}(E/{\rm eV})$, except for the last bin, which contains all events with energy above $\log_{10}(E/{\rm eV}) = 19.7$ ($E = 50\,{\rm EeV}$). The interpretation of data in terms of mass composition requires a comparison with air shower simulations. As shown in Fig. 3, the two models predict a similar evolution of $X^{\mu}_{\rm max}$ for proton and iron but a considerable difference in its absolute value. While Auger data are bracketed by QGSJETII-04, they fall below the EPOS-LHC prediction for iron, thus demonstrating the power of the MPD analysis to constrain hadronic interaction models [19].

$X_{\rm max}$ and $X^{\mu}_{\rm max}$ are both correlated with the primary mass [20, 21]. Both observables can thus be converted into $\langle \ln A \rangle$ using the same interaction model.

The results of this conversion for two different hadronic models are shown in Fig. 4. With QGSJETII-04, we obtain compatible values for $\ln A$ within $1.5\,\sigma$, while in the case of EPOS-LHC the results from $\langle X^{\mu}_{\rm max} \rangle$ indicate primaries heavier than iron and the measurements are incompatible at a level of at least $6\,\sigma$. It should be noted however that QGSJETII-04 has problems to describe in a consistent way the first two moments of the $\ln A$ distribution obtained from the $X_{\rm max}$ values measured with the FD [3].

## 4. Conclusions

Two different analyses of the muon component have been described. By measuring the total muon content of UHE inclined showers, we observe a muon deficit in the simulations. Our data strongly disfavor light composition at $10^{19}$ eV.

The arrival time of muons at the ground was used to reconstruct the muon production depth distribution on an event-by-event basis. This analysis has established a new approach to study the longitudinal development of the hadronic component of EASs. The current level of systematic uncertainties does not allow to draw conclusions on composition. However, the described measurements, in correlation with the information from the EM shower profile, set valuable constraints on the most recent LHC-tuned interaction models, QGSJETII-04 and EPOS-LHC. In particular, we found that none of the interaction models provides a consistent description of both the electromagnetic and muonic shower profiles as measured by the Pierre Auger Observatory.

# Extension of the measurement of the proton-air cross section with the Pierre Auger Observatory

**Ralf Ulrich**[*a] **for the Pierre Auger Collaboration**[b]
[a] *Karlsruhe Institute of Technology, Germany*
[b] *Observatorio Pierre Auger, Av. San Martín Norte 304, 5613 Malargüe, Argentina*
*E-mail:* auger_spokespersons@fnal.gov
*Full author list:* http://www.auger.org/archive/authors_2015_06.html

With hybrid data of the Pierre Auger Observatory it is possible to measure the cross section of proton-air collisions at energies far beyond the reach of the LHC. Since the first measurement by the Pierre Auger Collaboration the event statistics has increased significantly. The proton-air cross section is now estimated in the two energy intervals in $\lg(E/\mathrm{eV})$ from 17.8 to 18 and from 18 to 18.5. These energies are chosen so that they maximise the available event statistics and at the same time lie in the region most compatible with a significant primary proton fraction. Of these data, only the 20% of most proton-like events are considered for the measurement. Furthermore, with a new generation of hadronic interaction models which have been tuned to LHC data, the model-dependent uncertainties of the measurement are re-visited.

*Speaker.

## 1. Introduction

The cross section of cosmic ray protons with air is derived from data collected with the Pierre Auger Observatory [1] at energies far beyond the reach of human-made particle accelerators. The evolution of the hadronic cross section with energy is a fundamental property of nature and can still not be calculated from first principles of the theory of quantum chromodynamics. The measurement of general particle-collision properties at ultra-high energies is a tool for the search of new physics.

In this paper the extension of the measurement of the proton-air cross section of the Pierre Auger Collaboration at 57 TeV [2] is presented. The technique has been kept identical. Significantly more data has become available and, due to improved external input, several sources of systematic uncertainty are re-visited. The 20% most deeply penetrating showers are used for the analysis in order to reduce the impact of primary cosmic ray nuclei heavier than protons. Since the measurement is sensitive only to the cross section of interactions that produce secondary particles, the quasi-elastic excitation of the target nuclei is not included in the measured cross section. The observable cross section is, thus, defined as $\sigma_{\rm prod} = \sigma_{\rm inel} - \sigma_{\rm q-el}$. Hybrid events are selected with a uniform acceptance over the full phase-space of the measurement.

The number of hybrid events available for the measurement has increased in total by a factor of about four. The range in primary energies is increased and the result is presented in the two regions $10^{17.8} - 10^{18}$ eV and $10^{18} - 10^{18.5}$ eV. The reason to chose this limited energy region is driven by the fact that in this interval the data of the Pierre Auger Observatory is compatible with a very high content of primary protons [3]. Besides the fact of more events and thus a smaller statistical uncertainty, also the systematic uncertainties are re-visited. Most importantly the impact of the first models tuned to LHC data (EPOS-LHC [4] and QGSJetII.4 [5]) on the analysis is demonstrated in respect to the same models before they were tuned to LHC data (EPOS-1.99 [6] and QGSJetII.3 [7]). The SIBYLL 2.1 [8] model is kept as a constant reference. A new release of SIBYLL, which will also be tuned with LHC data, will eventually complete this study and will be done as soon as the model becomes available for air shower simulations.

## 2. Experiment, data and simulations

The Pierre Auger Observatory currently has the largest aperture for the detection of ultra-high energy cosmic ray particles in the energy range from just above the second knee up to the highest accessible energies. The observatory is located near the town of Malargüe in Argentina and consists of two major components: Firstly, the surface detector, build by 1660 autonomous water-Cherenkov stations, is spread over a surface area of 3000 km$^2$ on a triangular grid with 1.5 km spacing. Secondly, five fluorescence telescope sites are overlooking the surface detector with a total of 27 Schmidt-optics telescopes. Light is focused by spherical mirrors of 12 m$^2$ area and 3.7 m radius on cameras build of 440 PMTs. The 24 telescopes with a field of view ranging from 1.5 to 30 degree in elevation are used for this analysis.

Hybrid data contains information from the fluorescence telescopes as well as from at least one surface detector station. This includes events below the trigger threshold of the surface detector by initiating the readout based on the trigger from the fluorescence telescopes. The reconstruction of hybrid events uses the timing from the surface detector to precisely determine the geometry of the shower. The light collected by the telescopes is corrected for all known effects to yield the energy deposit profile along the shower axis in the atmosphere [9]. The slant depth along the shower axis where the maximum energy deposit is observed, $X_{\rm max}$, is the main observable of this study.

The dataset used for the measurement comprises events collected from Dec 2004 to Dec 2012. Simulations of air showers are performed with the CONEX simulation program [10]. In these simulations the initial part of the air shower cascade is simulated in full 3D Monte Carlo mode, and as soon as secondary particle energies drop below $0.001E_0$, where $E_0$ is the primary cosmic ray energy, the simulation is completed with cascade equations. Hadronic interactions can be simulated with different event generators. For this purpose EPOS-LHC and QGSJetII.4, which both have been tuned to LHC data up to $\sqrt{s} = 8\,$TeV, as well as SIBYLL 2.1, which has not yet been tuned to LHC data, are used. The impact of the extrapolation of the description of hadronic interactions in air showers on the analysis is estimated based on these different models, and ad-hoc modifications of them. Each model comprises a set of different but self-consistent phenomenological and theoretical assumptions to describe hadronic interactions. EPOS is build from a parton-based Gribov-Regge theory with energy-conservation considered during the multiple-pomeron exchange. Many features of hadronic interactions like diffraction and remnant fragmentation are added in a phenomenological approach. QGSJetII is driven by the theory of multi-Pomeron amplitues calculated to all orders. This yields a theoretically clean description of diffraction up to high masses. SIBYLL is a minijet model combined with a Glauber calculation, where diffraction is added in phenomenological way. It is a big advantage to use a set of models which are internally so different in nature. It allows us to derive an estimate of the systematic uncertainties related to details of the description of hadronic interactions on the proton-air cross section measurement.

The limited acceptance caused by the field-of-view of the telescope detectors as well as by the absorption of light in the atmosphere is taken into account as described in Ref. [11]. In this way a measurement of the complete $X_{\rm max}$-distribution is obtained free of acceptance effects. This is achieved by accepting showers only when the range in slant depth along the shower axis, where the shower can be detected with an expected $X_{\rm max}$-resolution of better than $40\,{\rm g/cm}^2$, does fully comprise the $X_{\rm max}$-interval required for the measurement. Furthermore, the resolution of the detector is taken into account by folding the Monte Carlo simulations with the parameterisation given in Ref. [11].

## 3. Analysis strategy and event selection

The analysis consists of two subsequent parts. The first step is the dedicated measurement of the observable $\Lambda_\eta$, which is a measure of the attenuation length of air showers in the atmosphere. The advantage of this well defined and dedicated measurement of $\Lambda_\eta$ is that this observable can be eventually used also in a different context, and it can be re-evaluated at any later time with new air shower simulation models. The second analysis part is related to the conversion of $\Lambda_\eta$ into the proton-air cross section $\sigma_{\rm p-air}$. This depends on the simulation of air showers and hadronic interactions therein.

The total number of high-quality hybrid events in the data sample used for the measurement is 39360. Event quality cuts are applied as described in Ref. [11]. The depth of the maximum energy deposit, $X_{\rm max}$, can be reconstructed with a resolution of $25.0 \pm 1.1$ g/cm$^2$ at $10^{17.8}$ eV and $18.6 \pm 1.1$ g/cm$^2$ at $10^{18.5}$ eV. This includes uncertainties e.g. from the atmospheric density profile.

**Table 1:** Measurement of $\Lambda_\eta$ and $\sigma_{\rm p-air}$ in the two energy regions. Statistical uncertainties are quoted in the same line, while systematic uncertainties are listed explicitly.

| | $10^{17.8} - 10^{18}$ eV | $10^{18} - 10^{18.5}$ eV |
|---|---|---|
| Number of high-quality hybrid events | 18090 | 21270 |
| **Determination of the 20% tail range** | | |
| Range of 99.8% central $X_{\rm max}$-values (g/cm$^2$) | 556.6 – 1009.7 | 573.3 – 1030.1 |
| Fiducial selection of 99.8% central range, events | 1818 | 2807 |
| Start of 20% tail range, $X_{\eta,\rm start}$ (g/cm$^2$) | 762.2 | 782.4 |
| Fiducial selection of 20% tail range, number of events | 4847 | 6906 |
| **$\Lambda_\eta$ determination** | | |
| Number of events in tail range | 1196 | 1384 |
| Power-law slope of energy distribution | $-0.65 \pm 0.31$ | $1.85 \pm 0.28$ |
| Average energy (eV) | $10^{17.90}$ | $10^{18.22}$ |
| Corresponding $\sqrt{s_{\rm pp}}$ (TeV) | 38.7 | 55.5 |
| Energy scale uncertainty on $\sqrt{s_{\rm pp}}$ (TeV) | 2.5 | 3.6 |
| $\Lambda_\eta$ (g/cm$^2$) | 60.7±2.1 | 57.4±1.8 |
| $\Lambda_\eta$, systematic uncertainties (g/cm$^2$) | 1.6 | 1.6 |
| **$\sigma_{\rm p-air}$ determination** | | |
| EPOS-LHC (mb) | 466.1 | 494.1 |
| QGSJetII.04 (mb) | 458.7 | 487.9 |
| SIBYLL 2.1 (mb) | 447.8 | 475.3 |
| Central value, all models (mb) | 457.5±17.8 | 485.8±15.8 |
| **$\sigma_{\rm p-air}$ uncertainties** | | |
| $\Lambda_\eta$, systematic uncertainties (mb) | 13.5 | 14.1 |
| Hadronic interaction models (mb) | 10 | 10 |
| Energy scale uncertainty, $\Delta E/E = 14\%$ (mb) | 2.1 | 1.3 |
| Conversion of $\Lambda_\eta$ to $\sigma_{\rm p-air}$ (mb) | 7 | 7 |
| Photons (mb) | 4.7 | 4.2 |
| Helium, 25% (mb) | -17.2 | -15.8 |
| Total systematic uncertainty on $\sigma_{\rm p-air}$ (mb) | +19/-25 | +19/-25 |

The available data sample is divided into two energy intervals, one ranging from $10^{17.8}$ to $10^{18}$ eV and the other from $10^{18}$ to $10^{18.5}$ eV with 18090 and 21270 events, respectively. All steps and results of the analysis are summarized in Tab. 1.

## 4. Measurement of $\Lambda_\eta$

In both energy ranges selected for the measurement the $X_{\rm max}$-range is determined independently in a two step procedure. First, the $X_{\rm max}$-interval containing the 99.8% most central events is identified, and only showers with an unbiased $X_{\rm max}$ measurement within this range are considered. This provides the best possible estimate of the shape of the whole $X_{\rm max}$-distribution, but with a significant cost in terms of available event statistics. This distribution is used to determine the $X_{\rm max}$-intervals containing the 20% most deeply penetrating showers.

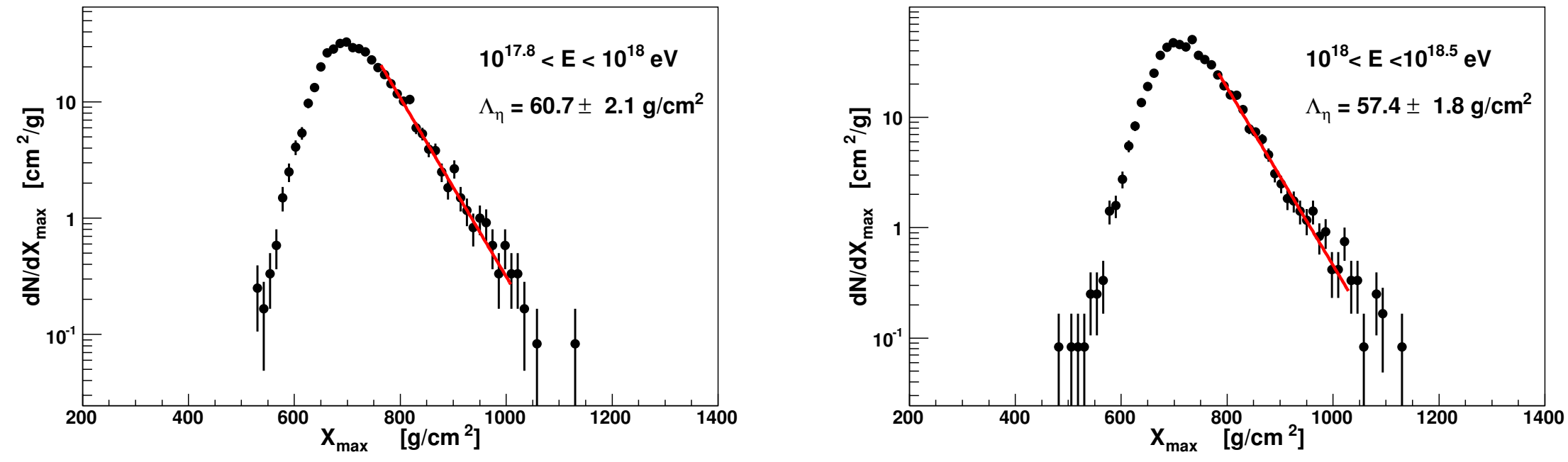

**Figure 1:** The $X_{\rm max}$-distributions in the two energy intervals. The result of the unbinned log-likelihood fit to derive $\Lambda_\eta$ is shown in the range of the tail fit.

Given this $X_{\rm max}$-interval, the event selection is updated by only requiring an unbiased $X_{\rm max}$-measurement in the tail region of the distribution. This step increases the available statistics for the measurement of $\Lambda_\eta$ by a factor of about three. At this stage the $X_{\rm max}$-distributions exist containing the unbiased tail from $X_{\eta,\rm start} = 762.2\,{\rm g/cm^2}$ to $X_{\eta,\rm end} = 1009.7\,{\rm g/cm^2}$ for the $10^{17.8} - 10^{18}$ eV range and $X_{\eta,\rm start} = 782.4\,{\rm g/cm^2}$ to $X_{\eta,\rm end} = 1030.1\,{\rm g/cm^2}$ for the $10^{18} - 10^{18.5}$ eV range. The upper end of the fit-range, chosen to exclude 0.1% of all available showers, also reduces the sensitivity to any possible primary photon contribution.

Due to the nature of the analysis, where the exponential tail of a distribution is measured, it is crucial to consider the Poissonian fluctuations of the data. This is achieved by numerically optimizing the following unbinned log-likelihood function for the $\Lambda_\eta$ parameter

$$\log L = \sum_{i=1}^{N_{\rm evts}} \log p(X_{{\rm max},i};\Lambda_\eta) \qquad \text{with} \tag{4.1}$$

$$p(X_{\rm max};\Lambda_\eta) = \left[\Lambda_\eta \left({\rm e}^{-X_{\eta,\rm start}/\Lambda_\eta} - {\rm e}^{-X_{\eta,\rm end}/\Lambda_\eta}\right)\right]^{-1} {\rm e}^{-X_{\rm max}/\Lambda_\eta} \ . \tag{4.2}$$

The statistical uncertainty of the result is determined using the values of $\Lambda_\eta$ where the likelihood exceeds $\log L_{\rm min}$+0.5. For simulated showers the default choice of $X_{\eta,\rm end} = \infty$ is used, which analytically yields the optimal result $\Lambda_\eta^{\rm opt,MC} = \sum_{i=0}^{N_{\rm evts}} (X_{{\rm max},i} - X_{\eta,\rm start})/N_{\rm evts}$, and the uncertainty can be derived from error propagation. The fit-range as well as the result is shown in Fig. 1.

The stability of the measurement of $\Lambda_\eta$ from data is tested by subdividing the data sample according to the zenith angle and to the distance of showers. The event selection cuts are changed within their experimental uncertainties. The observed variation of $\Lambda_\eta$ are consistent with statistical fluctuations. The standard deviation of these various observed deviations is considered as a systematic uncertainty for the measurement of $\Lambda_\eta$.

## 5. Determination of $\sigma_{\rm p-air}$

The value of $\sigma_{\rm p-air}$ is derived from the comparison of $\Lambda_\eta^{\rm MC}$, as calculated from full Monte

Carlo simulations of air showers, with the measured value of $\Lambda_\eta$. Air showers are generated with the same energy distribution as found in data as well as smeared with the $X_{\mathrm{max}}$ and energy resolution of the Pierre Auger Observatory. By default, none of the hadronic interaction models used for the air shower simulations is able to reproduce the measurement. For the purpose of the proton-air cross section measurement we exploit the fact that only the extrapolation of $\sigma_{\mathrm{p-air}}$ can significantly affect $\Lambda_\eta^{\mathrm{MC}}$. It is an important feature of this analysis that a change of $\sigma_{\mathrm{p-air}}$ is not just considered for the first interaction of the cosmic ray primary particle in the atmosphere, but also for all subsequent hadronic interactions at lower energies in a self-consistent way. For this purpose the energy-dependent scaling factor [12]

$$F(E, f_{19}) = 1 + (f_{19} - 1)\ \frac{\log(E/E_{\mathrm{thr}})}{\log(10^{19}\,\mathrm{eV}/E_{\mathrm{thr}})} \tag{5.1}$$

is used, where $f_{19}$ is the value of the scaling at $10^{19}\,$eV and $E_{\mathrm{thr}}$ is the threshold above which $F(E, f_{19}) \neq 1$. EPOS-LHC and QGSJetII.04 are both tuned up to cross sections measured by the TOTEM experiment at $\sqrt{s} = 8\,$TeV [13], while SIBYLL 2.1 is tuned to Tevatron at $\sqrt{s} = 1.96\,$TeV. The latter corresponds to primary cosmic ray protons of $E_{\mathrm{cr}} \approx 10^{16.5}\,$eV and the former to $E_{\mathrm{cr}} \approx 10^{15}\,$eV. Thus, for EPOS-LHC and QGSJetII.04 $E_{\mathrm{thr}} = 10^{16.5}\,$eV and for SIBYLL 2.1 $E_{\mathrm{thr}} = 10^{15}\,$eV is used. The results obtained with these simulations are shown in Fig. 2 and are used to convert the measured $\Lambda_\eta$ into values of $\sigma_{\mathrm{p-air}}$. The results for EPOS-LHC, QGSJetII.04 and SIBYLL 2.1 are summarized in Tab. 1. The central value for all three models is 457.5 mb at $10^{17.9}$ eV and 485.8 mb at $10^{18.2}\,$eV. The model uncertainty is estimated to be 10 mb for both energies. As long as the LHC-tuned SIBYLL model is not available, SIBYLL 2.1 is used as the third model in order to estimate the model dependence of the analysis approach. The modelling of interactions in air shower cascades is still a process with significant inherent uncertainties. Many phenomenological assumptions and parameters are part of any interaction model. The physics of diffraction, fragmentation, inelastic intermediate states, nuclear effects, QCD saturation, etc. are all described at different levels using different, but self-consistent, approaches in these models. It is not known whether by using these three different interaction models the true range of uncertainties of the modelling of hadronic interactions in air showers is covered. Thus, additional studies are done to investigate this. Characteristic features of hadronic interactions that are known to be particularly important for the air shower development have been studied independently. The secondary multiplicity, the inelasticity, the pion charge-ratio, as well as a separate scaling of the $\sigma_{\pi-\mathrm{air}}$ have been modified inside the models to determine the impact on $\Lambda_\eta^{\mathrm{MC}}$. Of those parameters, only the elasticity is found to have a potential relevance. The current level of systematic uncertainties of $\sigma_{\mathrm{p-air}}^{\mathrm{prod}}$ of 10 mb corresponds to altering the elasticity by about 10 %. Furthermore, the extreme assumption that the cross section of pions with air is modified in a different way as for protons with air by a factor two, is also checked. The effect on $\sigma_{\mathrm{p-air}}^{\mathrm{prod}}$ analysis is $< 1\,\%$ and is, thus, negligible.

For the conversion of $\Lambda_\eta$ to $\sigma_{\mathrm{p-air}}$ also other parameters are important. Scaling the simulated showers by the energy scale uncertainty of the Pierre Auger Observatory of 14% leads to a different conversion as well as changing the $X_{\mathrm{max}}$- and energy-resolution. Varying these within their precision yields an overal effect on $\sigma_{\mathrm{p-air}}^{\mathrm{prod}}$ of less than 7 mb for both energy regions.

While the composition of primary cosmic rays in the energy range under investigation is compatible with being dominated by protons, a contamination with Helium cannot be excluded. In earlier studies it was shown that primary particles heavier than Helium have only negligible impact on the analysis. The consequence of helium on the result is studied with simulations by producing samples of mixed proton-helium composition and testing the response of the analysis. There are indications that the helium content in the used data is not larger than on the order of 25% [3], which is also the number used in the past for this purpose. The impact of 25% helium on the cross section result is thus considered as systematic uncertainty towards smaller values of $\sigma_{\mathrm{p-air}}$. The contamination with primary photons is excluded to be larger than 0.5% in the energy range under investigation [14] and the impact on the cross section is added as systematic uncertainty towards larger values of $\sigma_{\mathrm{p-air}}$.

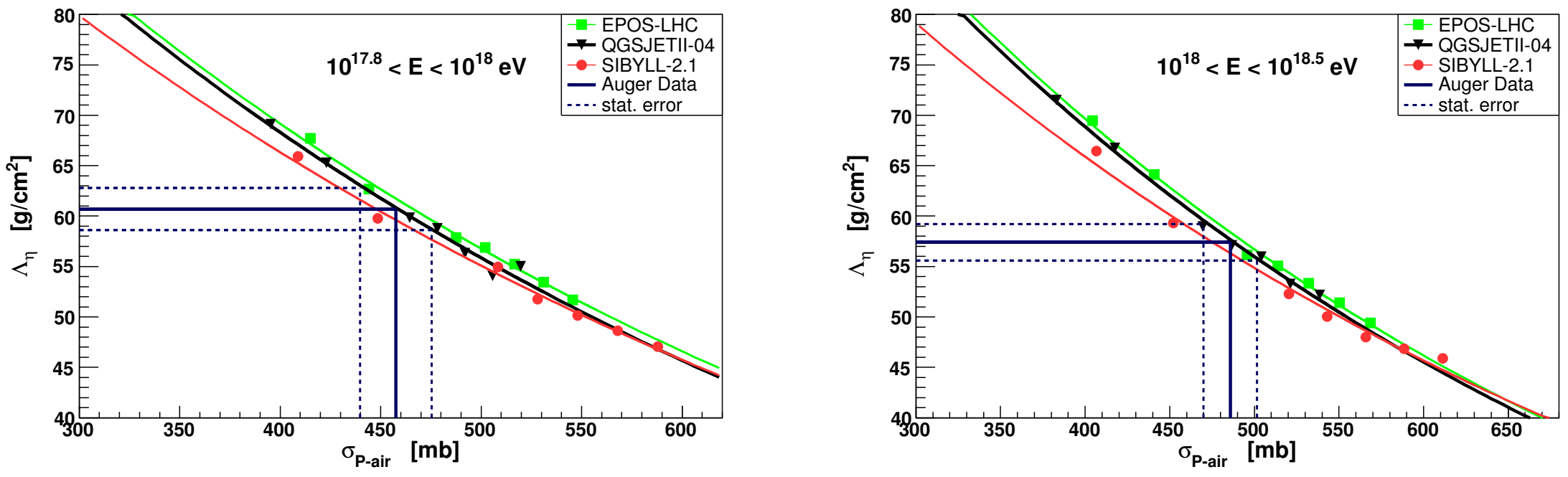

**Figure 2:** Conversion of $\Lambda_\eta$ to $\sigma_{\mathrm{p-air}}$. The simulations includes all detector resolution effects, while the data is corrected for acceptance effects. The solid and dashed lines show the $\Lambda_\eta$ measurement and its projection to $\sigma_{\mathrm{p-air}}$ as derived using the average of all models.

## 6. Results and summary

An updated measurement of the proton-air cross section with hybrid data of the Pierre Auger Observatory is presented. The result is shown in Fig. 3 and compared to previous measurements and model predictions. With respect to the previous measurement, the number of events is increased by about a factor of four. The measured value of $\Lambda_\eta = 57.4 \pm 1.8\,\mathrm{g/cm^2}$ in the energy range $10^{18} - 10^{18.5}$ eV is within 0.5 standard-deviations from the previous measurement. The statistical uncertainty of the measurement is consistent with a scaling by $1/\sqrt{N}$.

New hadronic interaction models, EPOS-LHC and QGSJetII.04, which are tuned to LHC data, are used for the conversion of $\Lambda_\eta$ to $\sigma^{\mathrm{prod}}_{\mathrm{p-air}}$. It is interesting to note, that the difference between these two models has changed by almost a factor of two with respect to the models prior to tuning to LHC data (EPOS-1.99 and QGSJetII.03). However, currently we keep also the SIBYLL 2.1 model as part of the analysis in order to get a more diverse estimation of the underlying modeling uncertainties. Since SIBYLL has not changed with respect to the previous analysis and both EPOS-LHC as well as QGSJetII.04 consitently predict larger values of $\sigma_{\mathrm{p-air}}$, the use of SIBYLL 2.1 leads to a slightly smaller central value of the final measurement and, even more relevant, a larger model-dependence. This will be revisited as soon as the next version of SIBYLL, also tuned to the LHC data, will be released for air shower simulations. It is a very interesting question, whether the trend observed with EPOS and QGSJetII continues and the overal model-dependence is further reduced.

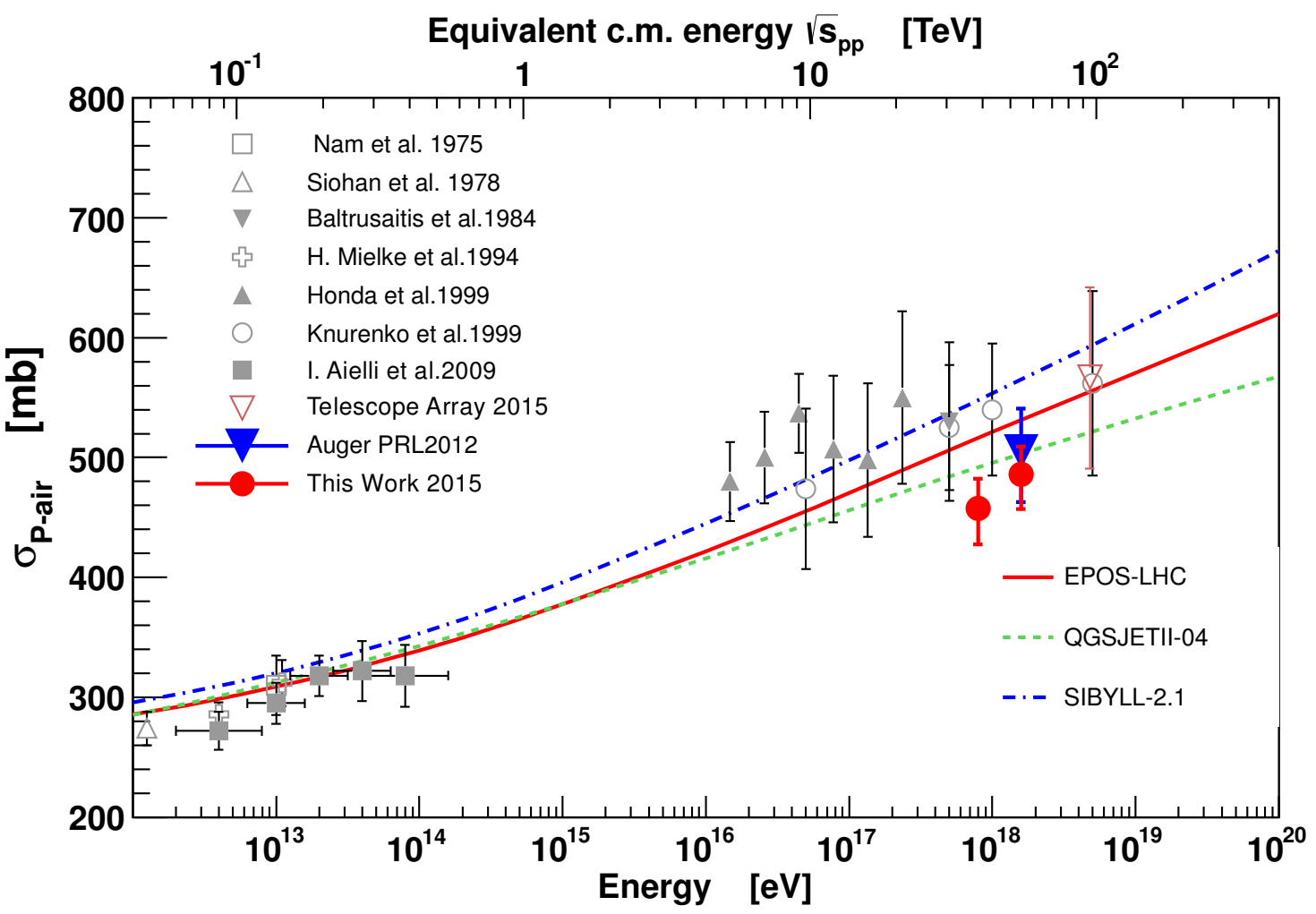

**Figure 3:** The $\sigma_{\mathrm{p-air}}$-measurement compared to previous data and model predictions. For references see [2] and [15].

For the present measurement the data is split in two energy intervals. The data is consistent with a rising cross section with energy, however, the statistical precision is not yet sufficient to make a statement on the functional form.

# Measurement of the average electromagnetic longitudinal shower profile at the Pierre Auger Observatory

**Francisco Diogo**[*a] **for the Pierre Auger Collaboration**[b]
[a] *LIP, Av. Elias Garcia, 14 1°, Lisboa, Portugal*
[b] *Observatorio Pierre Auger, Av. San Martín Norte 304, 5613 Malargüe, Argentina*
*E-mail:* auger_spokespersons@fnal.gov
*Full author list:* http://www.auger.org/archive/authors_2015_06.html

In addition to the standard $X_{\text{max}}$ and energy, the longitudinal profiles of extensive air showers contain some more interesting information. For energies above $10^{17.8}$ eV, we present the average profiles as a function of atmospheric depth measured for the first time at the Pierre Auger Observatory. The profile shapes for different energy ranges are all well reproduced by a Gaisser-Hillas function within the range studied. A detailed analysis of the systematic uncertainties is performed using data and a full detector simulation, and the results are compared with predictions of hadronic interaction models for different primaries.

*Speaker.

## 1. Introduction

Ultra High Energy Cosmic Rays (UHECRs) are the most energetic particles in the universe. The study of the cascades resulting from their interaction with the atmosphere nuclei gives us the only glimpse into hadronic interaction properties at these energies, more than an order of magnitude above those attained in man-made colliders.

While the atmospheric depth of shower maximum has long been used as a composition variable, the shape of the profile has remained largely untested.

In this paper we will present the first measurement of the average longitudinal profile in atmospheric depth (it has been previously been measured in age by HiRes/MIA and HiRes-II collaborations) for energies above $10^{17.8}$ eV. With the unprecedented statistics and experimental resolution of the Pierre Auger Observatory [1] at the highest energies, we can study the shower development in data, check the consistency of the reconstruction method, and also make a comparison with expectations from hadronic interaction models.

## 2. Event reconstruction

The Pierre Auger Observatory is a hybrid detector, consisting of a 3000 km$^2$ Surface Detector array (SD) overlooked by the Fluorescence Detector (FD). The FD consists of four sites with 6 telescopes each and a fifth site with 3 telescopes (HEAT). The field of view of each telescope spans 30° in azimuth and ranges from 1.5° to 30° in elevation (except for HEAT, where the elevation can be switched also to span 30° to 60°). In this analysis we have not used information from HEAT. In addition to both detectors, there are also atmosphere monitoring tools which measure aerosol content, clouds and the temperature and density height profile.

The first step in the profile reconstruction is the determination of the shower geometry. Firstly, the SDP plane spanned by the pointing directions of pixels in the shower image is calculated. The shower axis within this plane is obtained using the timing information of each pixel, as well as the timing of the closest SD station (hybrid reconstruction). The opening angle in the perpendicular direction to the SDP ($\zeta$) used for light integration is calculated by maximizing the signal to background ratio on an event-by-event basis. The light at the emission point is calculated taking into account the atmospheric characteristics measured autonomously.

Only the fluorescence component is proportional to the energy deposit in the shower volume considered, while Cherenkov light is dependent on the total number of particles above the emission energy threshold. Therefore, the complete profile has to be known for the calculation of the Cherenkov component. To extrapolate the directly observed profile to the earlier stages of development we use the Gaisser-Hillas (GH) function [2]

$$f_{\mathrm{GH}}(X) = \left(\frac{\mathrm{d}E}{\mathrm{d}X}\right)_{\max} \left(\frac{X - X_0}{X_{\max} - X_0}\right)^{\frac{X_{\max}-X_0}{\lambda}} \exp\left(\frac{X_{\max}-X}{\lambda}\right) \tag{2.1}$$

which has four parameters: the maximum energy deposit,$(\mathrm{d}E/\mathrm{d}X)_{\max}$, the depth at which this maximum is reached, $X_{\max}$, and shape parameters $X_0$ and $\lambda$. Given the shower geometry w.r.t. the telescope, both the fluorescence and Cherenkov components can be simultaneously determined [3].

## 3. Average Longitudinal Shower Profile

The maximum energy deposit of the longitudinal profile, $(\mathrm{d}E/\mathrm{d}X)_{\mathrm{max}}$, is proportional to the energy of the primary particle and varies three orders of magnitude in the energy range studied in this work. $X_{\mathrm{max}}$ is characteristic of the primary mass, but also varies greatly within each primary group , mainly due to the stochastic nature of shower development and $X_1$, the depth at which the first interaction occurs.

To separate the information on these two parameters ($(\mathrm{d}E/\mathrm{d}X)_{\mathrm{max}}$ and $X_{\mathrm{max}}$) from the profile shape, which is the focus of this work, we first divide each measured shower profile by its fitted $(\mathrm{d}E/\mathrm{d}X)_{\mathrm{max}}$, rescaling all showers to have maxima at 1. Then, we shift the $X$ axis, translating the atmospheric depth by $X_{\mathrm{max}}$, i.e., $X' \equiv X - X_{\mathrm{max}}$, thus centering all profiles at zero. In previous analysis, done by the Hires/MIA [4] and HiRes-II Collaborations [5], the $X$ axis was scaled using shower age ($s = 3X/(X + 2X_{\mathrm{max}})$), and the resulting profiles were found to be compatible with a gaussian having RMS equal to $\sigma_{\mathrm{age}}$. This width, however, is convolved with (and dominated by) the $X_{\mathrm{max}}$ value[1], hence in this work we choose to translate the profiles in atmospheric depth as it keeps the measured event-by-event shape unchanged.

We can write the Gaisser-Hillas function with these normalized variables, as a function of parameters $R$ and $L$ [6]:

$$\begin{aligned}\left(\frac{dE}{dX}\right)' &= \exp\left(-\frac{1}{2}\left[\frac{X'}{L}\right]^2\right)\prod_{n=3}^{\infty}\exp\left(-\frac{R^{n-2}}{n}\left[-\frac{X'}{L}\right]^n\right)\\ &= \left(1+R\frac{X'}{L}\right)^{R^{-2}}\exp\left(-\frac{X'}{RL}\right)\end{aligned} \tag{3.1}$$

where $(\mathrm{d}E/\mathrm{d}X)' = (\mathrm{d}E/\mathrm{d}X)/(\mathrm{d}E/\mathrm{d}X)_{\mathrm{max}}$, $R = \sqrt{\lambda/|X_0'|}$ and $L = \sqrt{|X_0'|\lambda}$ (note that $X_0' \equiv X_0 - X_{\mathrm{max}}$). We choose these parameters because they are much less correlated than $X_0$ and $\lambda$ and have a more clear meaning: looking at equation (3.1) we see the GH function is a Gaussian with standard deviation $L$, multiplied by a term that distorts it, with the asymmetry governed by $R$ (i.e., if $R = 0$, the function is a Gaussian). An equivalent parametrization as a function of the Full Width at Half Maximum, $f_{\mathrm{FWHM}}$, and asymmetry, $f$, has been reached independently [8].

The motivation for measuring the average profile shape is two-fold. First, it gives us a tool to control the quality of our reconstruction, and also to cross-check whether the assumption that showers are well described by a Gaisser-Hillas function is valid. Secondly, the profile shape carries information about the high energy hadronic interactions at the top of the atmosphere [7]. While the differences in $R$ and $L$ between proton and iron, or between different high energy hadronic models, are much smaller than those in $X_{\mathrm{max}}$, these variables give an independent measurement of the properties of the primary and its interacton in the atmosphere.

[1] We can write $s = 1 + \frac{2X'}{3X_{\mathrm{max}} - X'}$. If we say the RMS in age corresponds $X' = \pm L$ and invert the equation, we get an approximation for $L$ as a function of $X_{\mathrm{max}}$ and $\sigma_{\mathrm{age}}$: $L_\sigma = \frac{3X_{\mathrm{max}}}{\sigma_{\mathrm{age}}}(\sqrt{(1+\sigma_{\mathrm{age}}^2)} - 1)$. This agrees with the true L within 0.5% at all energies. Re-writing this expression for $\sigma_{\mathrm{age}}$ and making the derivatives w.r.t. $L$ and $X_{\mathrm{max}}$ we find that $\left(\frac{\partial\sigma_{\mathrm{age}}}{\partial L}\Delta_L\right) / \left(\frac{\partial\sigma_{\mathrm{age}}}{\partial X_{\mathrm{max}}}\Delta_{X_{\mathrm{max}}}\right) = \frac{X_{\mathrm{max}}\Delta_L}{L\Delta_{X_{\mathrm{max}}}} \approx 1/6$, so the majority of the separation in $\sigma_{\mathrm{age}}$ comes from $X_{\mathrm{max}}$

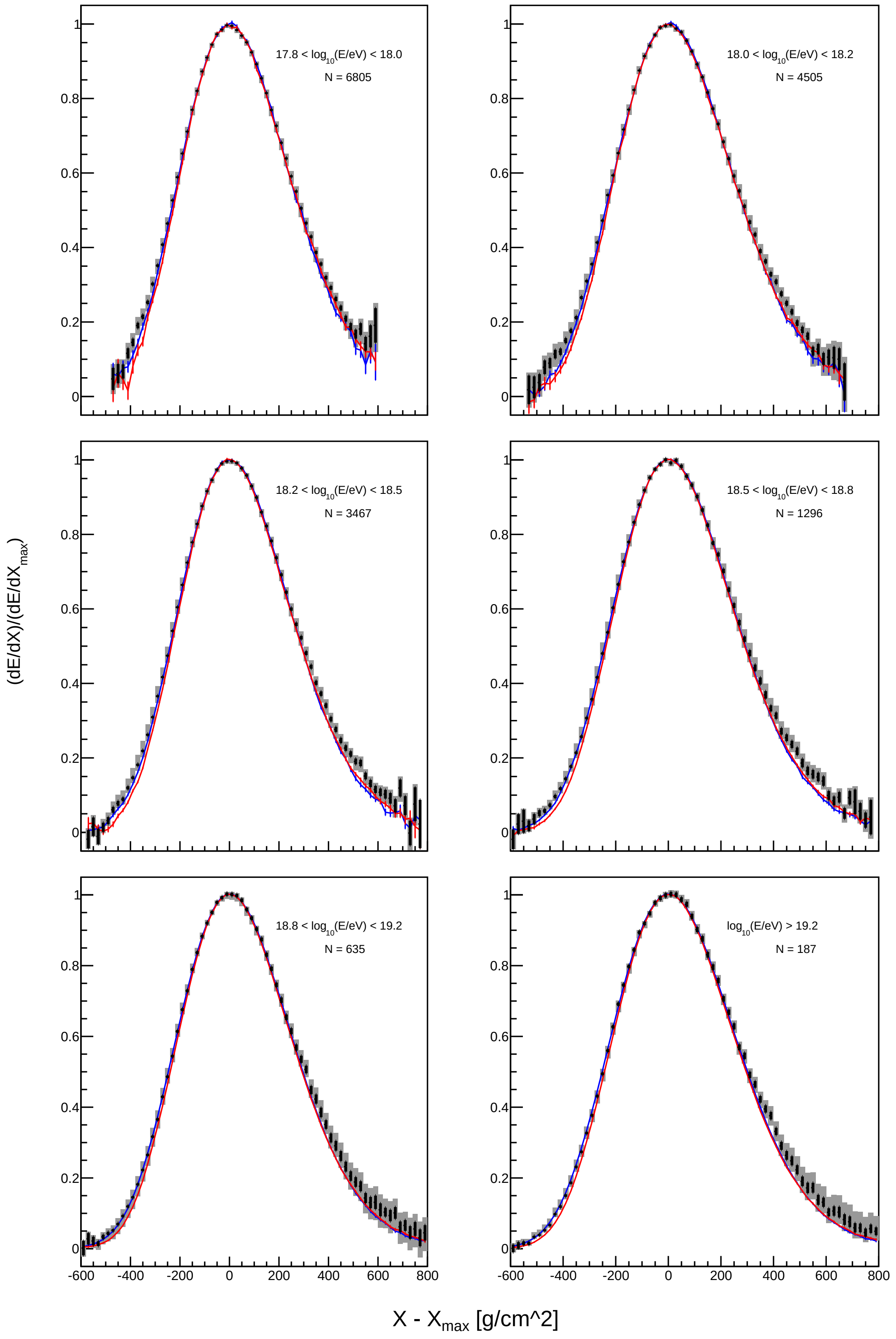

**Figure 1:** Average profiles for all the energy bins used in this work. The statistical error is shown as a black line and an estimation of the systematic uncertainty bin-by-bin as a gray area. Reconstructed profiles for MC are show in blue (proton) and re (iron). The high energy model used is QGSJETII03.

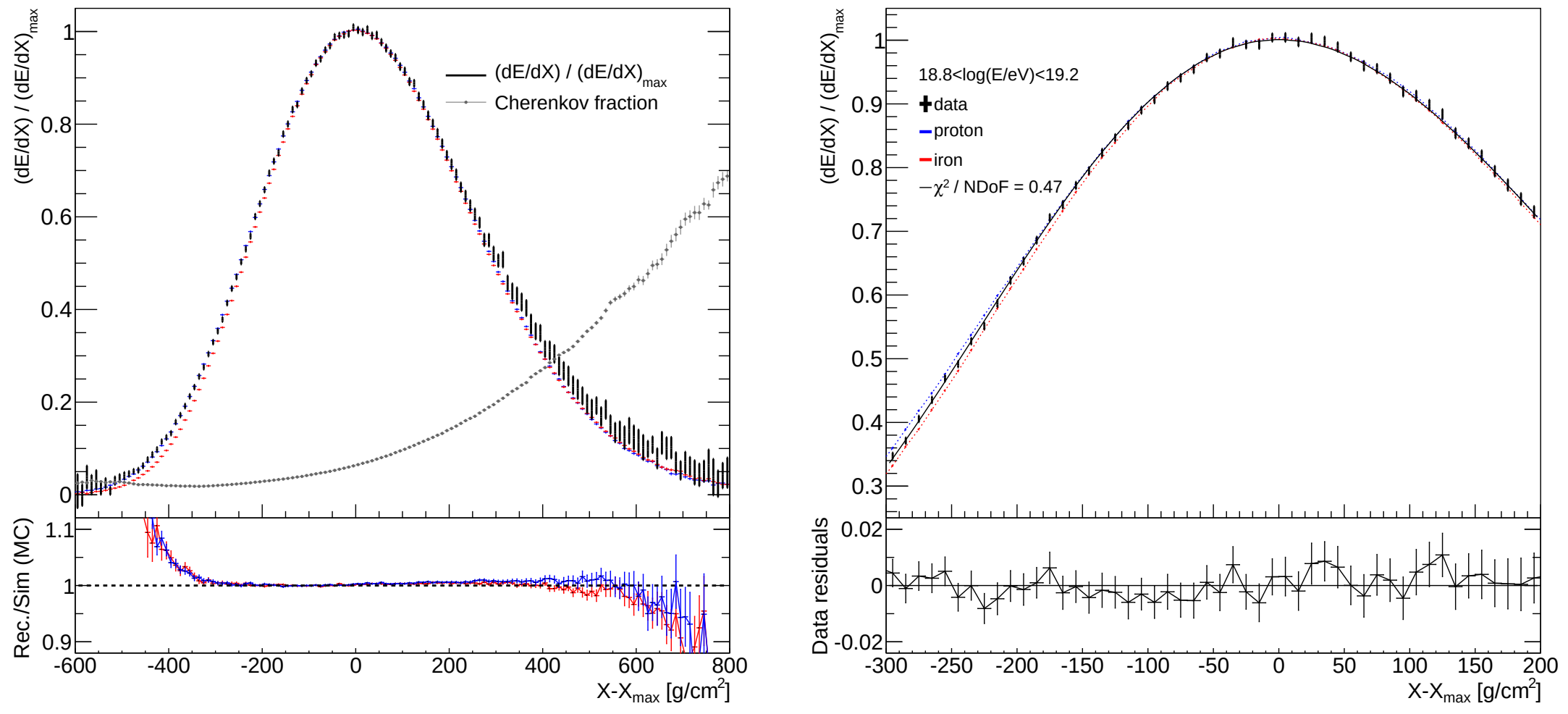

**Figure 2:** Left: Average profiles for energies between $10^{18.8}$ and $10^{19.2}$ eV. Data is shown in black, proton and iron reconstruction in blue and red respectively. The grey points represents the fraction of the total Cherenkov (direct and scattered) contribution in each bin. In the bottom plot is the ratio of reconstructed MC profiles over the generated ones for proton (blue) and iron (red). Right: The same profiles as on the left but only the fitting range is shown and the Gaisser-Hillas fit superimposed. The residuals of the fit to the data profile are shown in the bottom.

## 4. Data selection and Monte Carlo validation

The event selection used here is based on the most recent Auger Collaboration $X_{\max}$ analysis paper ([9]). The cuts used include requiring no clouds and an existing aerosol measurement, as well as a good hybrid geometry reconstruction. On the shower profile very strict cuts are made: at least 300 g/cm$^2$ must be observed, including the $X_{\max}$ depth, for which the expected resolution must be below 40 g/cm$^2$. A fiducial field of view is defined to guarantee the $\langle X_{\max} \rangle$ measurement is unbiased w.r.t. composition. Also, to minimize the amount of Cherenkov light, the minimum angle between a pixel pointing vector and the shower axis has to be larger than 20°. In this work, two additional cuts were used. One of the telescopes was excluded due to alignment problems, detected as significant time residuals found for showers crossing more than one telescope. Events in which more than 25% of the triggered pixels had large time residuals (greater than $3\sigma$) from the geometry fit, were also not used. In the first cut we lose approximately 3% of the events (479), while in the second only 8 events were excluded.

In total 15782 events were selected, and we divide them in 6 energy bins. The shower profiles are constructed in 10 g/cm$^2$ bins in $X'$, in which each energy deposit is accumulated with a weight corresponding to the inverse of its squared error. The profiles for all energies are shown in figure 1. Each average profile is then fitted with function (3.1).

This method was validated with a full detector simulation for energies between $10^{17}$ and $10^{20}$ eV with proton and iron as primary particles. Comparing simulated and reconstructed average showers an excellent agreement was found for $X' > -300$ g/cm$^2$, but reconstruction deviated increasingly below it (see figure 2 (left) for the energy bin around $10^{19}$ eV). Above $X_{\max}$, showers lose most of the primary information (in [6], when trying to maximize proton-iron separation using only simulations, 100 g/cm$^2$ was chosen as upper fit limit) and the fluorescence light fraction falls rapidly (figure 2 (right)). Hence, the fit limits were chosen to be $-300$ to $+200$ g/cm$^2$ since it allows having a statistical error smaller than the proton/iron separation at all energies while keeping the minimum fluorescence fraction around 80%.

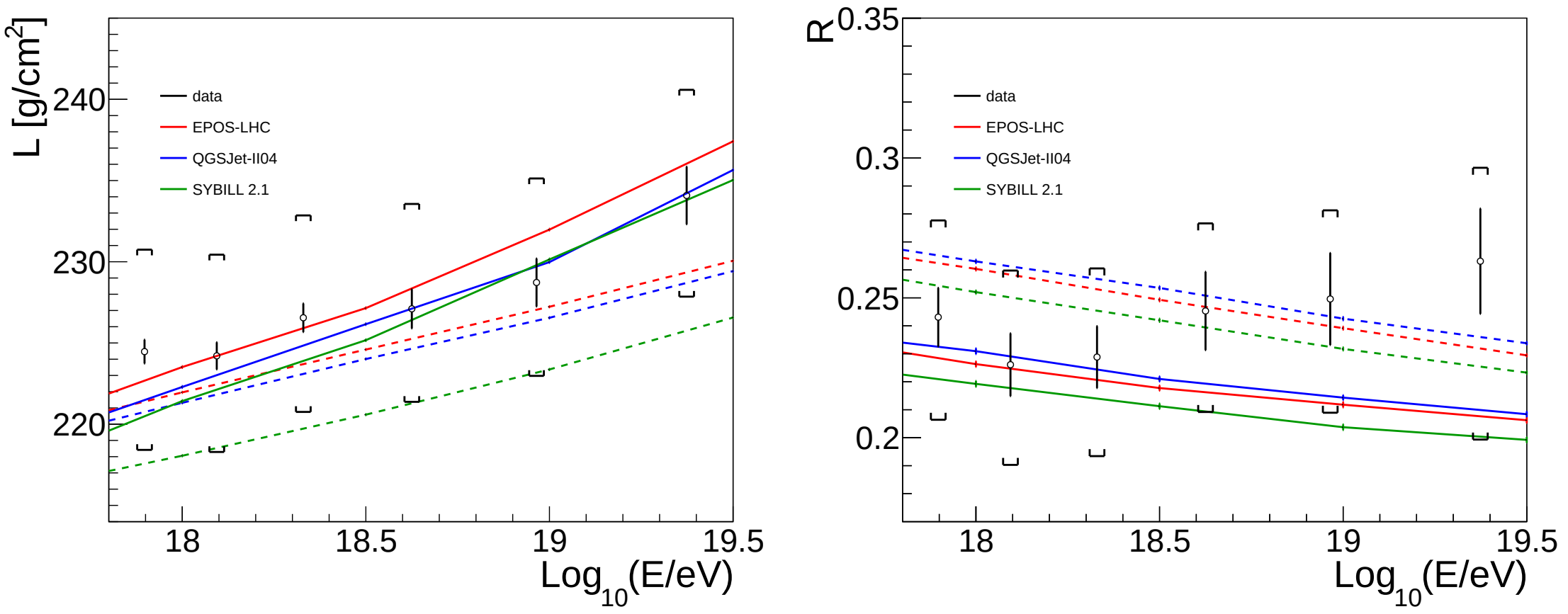

**Figure 3:** *L* (left) and *R* (right) as a function of energy. The data is shown in black, with the vertical line representing the statistical error and the brackets the systematic uncertainty. Hadronic interaction models are shown each with its color (see legend), with full lines being proton predictions and dashed lines iron ones.

The reconstructed and simulated profiles are then fitted with equation (3.1) leaving all parameters unconstrained[2]. The fitted values for the shape parameters agree well between simulation and reconstruction for all energies above $10^{18}$ eV, with a larger difference for the first energy bin ($10^{17.8}$ to $10^{18}$ eV). The average bias is corrected and half the proton-iron difference value is added to the reconstruction systematic uncertainty in Table 1.

## 5. Systematic uncertainties

The atmospheric conditions play a crucial role in the propagation of the light, so several systematic uncertainties related to it were studied: cloud effects, uncertainties of the overall aerosol content as well as its height dependence, the effect of excluding the outermost 1.5° at the border of the camera from the fit and differences found when separating data by the seasons of the year. Since all these effects tend to change mostly either the beginning (clouds) or the end (aerosols) of the profile, they contribute strongly to the systematic uncertainty of the asymmetry (*R*) . We also considered the uncertainties in the determination of the fraction of measured light that corresponds to fluorescence, direct or scattered (Mie or Rayleigh) Cherenkov light and multiple scattering. This includes changing the fluorescence and Cherenkov yield value within its experimental uncertainty in the reconstruction, accounting or not for the multiple scattering corrections, and separating data according to the fluorescence fraction on the event. Among these, the largest effect found was that showers with fluorescence fraction lower than the average (around 90%) are approximately 4 g/cm$^2$ larger in width, $L$. The shape parameters for the individual shower, $X_0$ and $\lambda$ are constrained with a value measured in the data, so we changed these constraints by $1\sigma$ in the reconstruction. We tested for systematic effects of the telescope alignment by studying the telescope-to-telescope differences of the reconstructed shape. Also, a dependence of $L$ and $R$ on the zenith angle or distance from $X_{\max}$ to the telescope was studied, but found to be relatively small in comparison to the previous ones. The uncertainty from the proton-iron difference in the reconstruction bias correction and the uncertainty of the energy scale of 14% [10] are also small.

[2]In addition to *R* and *L*, also the normalization and the maximum are allowed to vary around 1 and 0 respectively, since in data we have smearing and small energy dependent bias. Values in data are always within the MC predictions, i.e., $<0.5\%$ for normalization and less than 1 g/cm$^2$ for maximum

| | $R$ | $L$ [g/cm$^2$] |
|---|---|---|
| Atmosphere | 0.053 | 3.6 |
| Light components & fit | 0.011 | 4.0 |
| Telescope | 0.023 | 3.2 |
| Geometry | 0.018 | 2.0 |
| Bias corr. & Energy | 0.007 | 0.6 |
| **Total** | **0.063** | **6.3** |
| Statistical | 0.019 | 1.8 |

**Table 1:** Breakdown of systematic uncertainties for $R$ and $L$. Uncertatinties are energy dependent and asymmetric so that only the largest value is reported.

## 6. Results

The fit of data profiles to the Gaisser-Hillas function (3.1) is shown in figure 2 (right). The fitted function follows data points through the whole depth range used in this work, $[-300,+200]$ g/cm$^2$, with residuals always within the statistical uncertainty. The overall reduced $\chi^2$ is below 1.5 for all energies, showing that the analysis of the average shape of profiles in terms of the $R$ and $L$ parameters in equation (3.1) is an accurate description. The results of $L$ and $R$ as a function of energy are shown in figure 3. The width, $L$, in data agrees well with the predicted values for all models, and its energy evolution is consistent with a linear increase with $\log_{10}[E/\mathrm{eV}]$. The asymmetry, $R$, is compatible with models. In the data there is an increase with energy not predicted by them, although it is contained within the systematic uncertainty of the measurement.

It is also interesting to see the results in the $(R,L)$ plane for a fixed energy (Figure 4). In these plots we can also represent all possible composition scenarios (as a combination of proton, He, N and Fe) for a given energy and make use of our knowledge about the correlation between $R$ and $L$. In Figure 4 (left), for a low energy bin, we can see the average value in data is in the area occupied by most models for a light composition, while at $10^{19}$ eV (figure 4 (right)) it is within the predictions for heavier primaries. However, they are still fully compatible with all composition scenarios on $2\sigma$ level, so the objective of future work is to decrease the systematic uncertainty and derive further constraints on predictions of hadronic models.

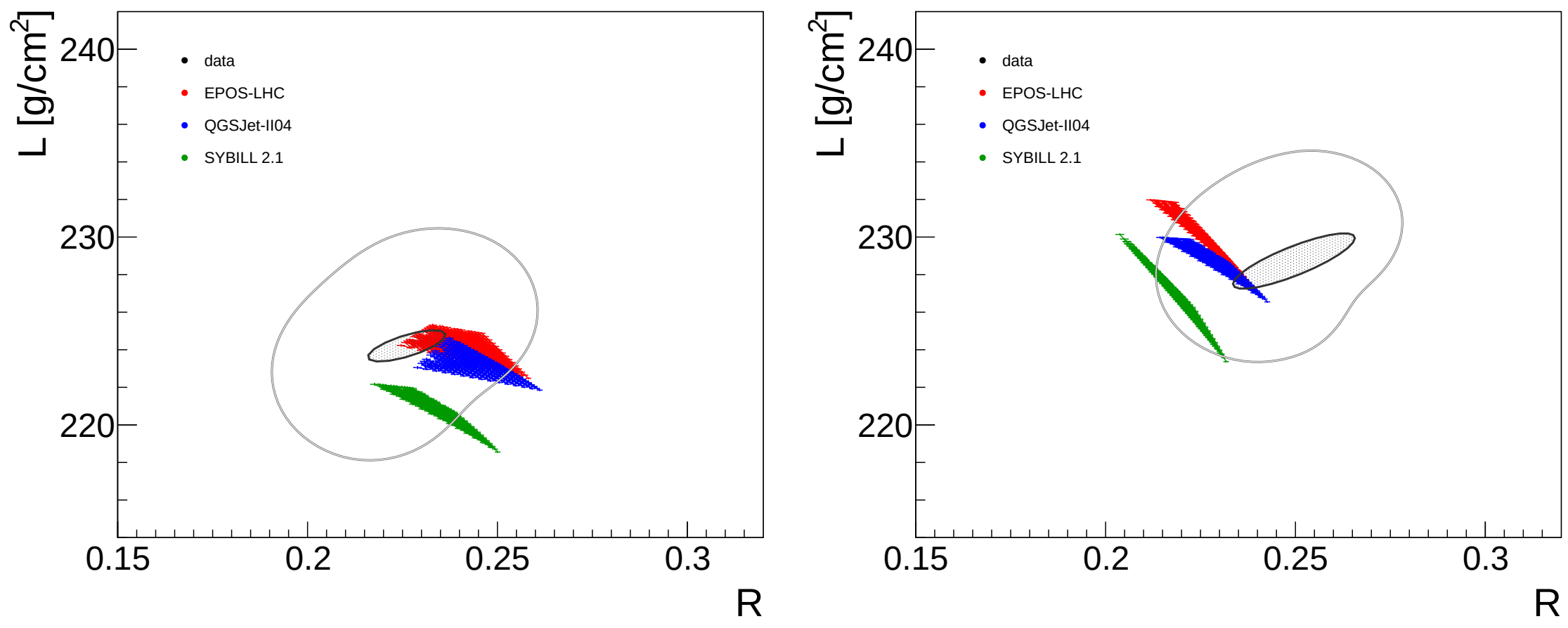

**Figure 4:** $L$ vs $R$ for the energy bin $10^{18}$ to $10^{18.2}$ eV (left) and from $10^{18.8}$ to $10^{19.2}$ eV (right). The inner dark grey ellipse shows the fitted value for data and its statistical error, and the outer light grey area the systematic uncertainty. For each hadronic model all combinations of proton, helium, nitrogen and iron were simulated and are represented by its respective colored area. Pure proton is, for each model, on the upper left side and the transition to iron goes gradually to the lower right one.

## 7. Conclusion and Outlook

In this work, the average shape of the longitudinal profile of the air showers in the Pierre Auger Observatory was measured. We first validated the method in a full detector simulation of proton and iron primaries, which showed that reconstructed and simulated profiles are in very good agreement for all energies above $10^{17.8}$ eV. We have shown that average profiles of the data are well described by a Gaisser-Hillas function through the entire fitting range chosen. We estimated the systematic uncertainties contributing to our measurement, and concluded that the atmospheric description and the Cherenkov contribution are the main factors that affect the asymmetry and the width of the profile, respectively. The two shape parameters, $R$ and $L$, resulting from this fit were compared with model predictions, being fully compatible with them.

# 5

# Cosmology and Geophysics

PIERRE
AUGER
OBSERVATORY

# Combined fit of spectrum and composition data as measured by the Pierre Auger Observatory

**Armando di Matteo**[∗a] **for the Pierre Auger Collaboration**[b]
[a]*INFN and Department of Physical and Chemical Sciences, University of L'Aquila, L'Aquila, Italy*
[b]*Observatorio Pierre Auger, Av. San Martín Norte 304, 5613 Malargüe, Argentina*
*E-mail:* auger_spokespersons@fnal.gov
*Full author list:* http://www.auger.org/archive/authors_2015_06.html

We present a combined fit of both flux and composition of ultra-high energy cosmic rays as measured by the Pierre Auger Observatory. The fit has been performed for energies above $5\times10^{18}$ eV, i.e. the region of the all-particle spectrum above the so-called "ankle" feature. A simple astrophysical model consisting of identical sources has been adopted, where nuclei are injected with a rigidity dependent mechanism and the sources are uniformly distributed in a comoving volume. The fit results suggest a source model characterized by relatively low maximum injection energies and hard spectral indices. The impact of different sources of systematic uncertainties in the above result is discussed.

∗Speaker.

## 1. Introduction and motivation

More than half a century after their first detection, the origin of ultra-high energy cosmic rays (UHECRs), particles (mostly protons and other nuclei) reaching the Earth with energies over $10^{18}$ eV up to $10^{20}$ eV and beyond, is still unknown. Nevertheless, a general consensus has emerged that the most energetic cosmic rays are extragalactic, with the transition between galactic and extragalactic cosmic rays taking place somewhere between $10^{17}$ and a few times $10^{18}$ eV. The flux of cosmic rays above $10^{18}$ eV is of the order of 1 km$^{-2}$ yr$^{-1}$. Therefore, very large arrays of particle detectors are needed to study them; the largest such array is the Pierre Auger Observatory in Argentina [1]. The propagation of such particles across cosmological distances can affect their observed energy spectrum and mass composition in nontrivial ways. For this purpose several Monte Carlo codes have been developed, including CRPropa [2, 3, 4] and *SimProp* [5, 6, 7].

While the energy of single UHECR events can now be estimated with relatively good precision, it is impossible to determine the mass of UHECRs on an event-by-event basis. The distribution of parameters such as $X_{\max}$, the atmospheric depth at which an air shower reaches the maximum particle number, can be used to statistically estimate the mass distribution of UHECRs [8, 9].

In this work, we attempt to simultaneously reproduce both the Auger spectrum [10] and $X_{\max}$ [11] data with a simplified model of UHECR sources, characterized by: identical sources homogeneously distributed in a comoving volume; injection consisting only of $^{1}$H, $^{4}$He, $^{14}$N and $^{56}$Fe nuclei, which are approximately equally spaced in $\ln A$; power-law spectrum with rigidity-dependent broken exponential cutoff,

$$\frac{\mathrm{d}N_{\mathrm{inj},i}}{\mathrm{d}E} = \begin{cases} J_0 p_i \left(E/E_0\right)^{-\gamma}, & E/Z_i < R_{\mathrm{cut}} \\ J_0 p_i \left(E/E_0\right)^{-\gamma} \exp\left(1 - E/Z_i R_{\mathrm{cut}}\right), & E/Z_i > R_{\mathrm{cut}} \end{cases} \tag{1.1}$$

where $J_0$ is a normalization factor, $E_0 = 10^{18}$ eV, $A_i$ and $Z_i$ are the mass number and atomic number of the $i$-th injected nuclide and with normalized element fractions $\sum_i p_i = 1$. Such a simple model cannot reproduce measured data over their entire energy range [12]. For this reason we only fit data at energies above the ankle ($E \geq 10^{18.7}$ eV) and make no hypotheses about the nature of possible extra components accounting for the rest of the sub-ankle UHECR spectrum. For examples of such hypotheses made by different authors see e.g. [13, 14, 15].

The propagated fluxes can be strongly sensitive to poorly known quantities such as the height of the far infrared peak in the extragalactic background light (EBL) spectrum and the cross sections for photodisintegration of nuclei ejecting $\alpha$-particles, as well as on approximations made in simulation codes [16]. In this work we use both *SimProp* and CRPropa each with several different settings to quantify the effect of such differences on our fit (see Table 1). Other EBL models such as Kneiske 2004 [17] or Stecker 2005 [18, 19] give similar results.

| | MC code | $\sigma_{\mathrm{photodisint.}}$ | EBL model |
|---|---|---|---|
| SPG | *SimProp* | PSB | Gilmore 2012 |
| SPD | *SimProp* | PSB | Domínguez 2011 |
| STG | *SimProp* | TALYS | Gilmore 2012 |
| CTG | CRPropa | TALYS | Gilmore 2012 |
| CTD | CRPropa | TALYS | Domínguez 2011 |
| CGD | CRPropa | Geant4 | Domínguez 2011 |

**Table 1:** The various propagation models we used (see Ref. [16] and references therein for details)

## 2. Fit results and physical parameters

The data we attempt to fit consist of 15 measurements of the UHECR energy spectrum in $\log_{10}(E/\text{eV})$ bins of width 0.1 from 18.7 to 20.2 [10], and 110 non-zero measurements of the $X_{\max}$ distribution in $\log_{10}(E/\text{eV})$ bins of width 0.1 from 18.7 to 19.5 plus one of width 0.5 from 19.5 to 20.0 and $X_{\max}$ bins of width 20 g/cm$^2$ [11]. We approximate the probability distribution of each point of the spectrum as a Gaussian with the standard deviation corresponding to the measurement statistical uncertainty, whereas for the $X_{\max}$ distributions we use a multinomial distribution over the $X_{\max}$ bins for each energy bin, where the mass distributions at the Earth from the simulation outputs are converted to the $X_{\max}$ distributions via a Gumbel parameterization [20] based on the EPOS-LHC [21] model of UHECR-air interactions.

The free parameters of the fit are: the injection normalization factor $J_0$, the injection spectral index $\gamma$, the cutoff rigidity $R_{\text{cut}}$, and the element fractions at injection (three free parameters $p_{\text{H}}$, $p_{\text{He}}$, $p_{\text{N}}$; the fourth is bound by $p_{\text{Fe}} = 1 - p_{\text{H}} - p_{\text{He}} - p_{\text{N}}$). In total, we have 125 non-zero data points and 6 free parameters.

Using the propagation model listed as SPG in Table 1, we find the best fit at $\gamma = 0.94^{+0.09}_{-0.10}$, $R_{\text{cut}} = 10^{18.67\pm0.03}$ V, with a deviance[1] (generalized $\chi^2$) per degree of freedom $D_{\min}/n = 178.5/119$.[2] To assess the statistical significance of this, we repeated the fit using $10^4$ mock data sets for the spectrum and $X_{\max}$ distributions generated according to the best-fit model and the same statistic as the real data, and found that $D_{\min}$ exceeds 178.5 for $p = 2.6\%$ of the mock data sets, indicating a slightly statistically significant deviation of the data from the model. The normalization $J_0$ we find corresponds to a total emissivity $\mathscr{L}_0 = 5.15 \times 10^{44}$ erg/Mpc$^3$/yr, and those for the single elements $\mathscr{L}_{\text{He}} = 0.289\mathscr{L}_0$, $\mathscr{L}_{\text{N}} = 0.656\mathscr{L}_0$, and $\mathscr{L}_{\text{Fe}} = 0.055\mathscr{L}_0$.

In Fig. 1 we show the deviance of our fit as a function of $(\gamma, R_{\text{cut}})$, where for each point of the profile likelihood plane the values of the remaining paramenters ($J_0$ and $p_i$) are chosen so as to minimize the deviance (marginalized). The best fit can be seen to be part of a long ‘valley’ extending to lower values of $\gamma$ and $R_{\text{cut}}$ approximately along the shown curve. There also is a second local minimum at $\gamma \approx 2$, but it is much worse than the global minimun ($D_2 - D_1 = 56.5$, corresponding to a $7.5\sigma$ exclusion; $p = 5 \times 10^{-4}$ from mock data sets), mainly due to predicted $X_{\max}$ distributions at most energies that are broader than observed.

The corresponding simulated spectra and the mean and variance of the simulated $X_{\max}$ distributions are shown in the left panels of Fig. 2 for the best fit and in the right panels for the local minimum at $\gamma \approx 2$.

## 3. Systematic uncertainty due to the propagation

Certain assumptions about the processes affecting the propagation of UHECRs can result in substantially different spectra at the Earth for a given spectrum at injection [16]. To study the effect of this on our results, we repeated the fit described in the previous section for each of the various propagation models listed in Table 1. The results are shown in Table 2.

[1] The deviance is defined as $D = -2\ln(L/L_{\text{sat}})$, where $L$ is the likelihood of the model considered and $L_{\text{sat}}$ is that of a hypothetical model perfectly reproducing the data. It coincides with the usual $\chi^2$ statistic for Gaussian likelihoods.

[2] The uncertainties given for $\gamma$ and $R_{\text{cut}}$ correspond to the intervals where $D \leq D_{\min} + 1$ (68% confidence). For the fit with propagation model SPG, the correctness of errors has been checked with *a posteriori* errors obtained from mock data sets, which were also used for the uncertainties on $p_i$ given in Fig. 1.

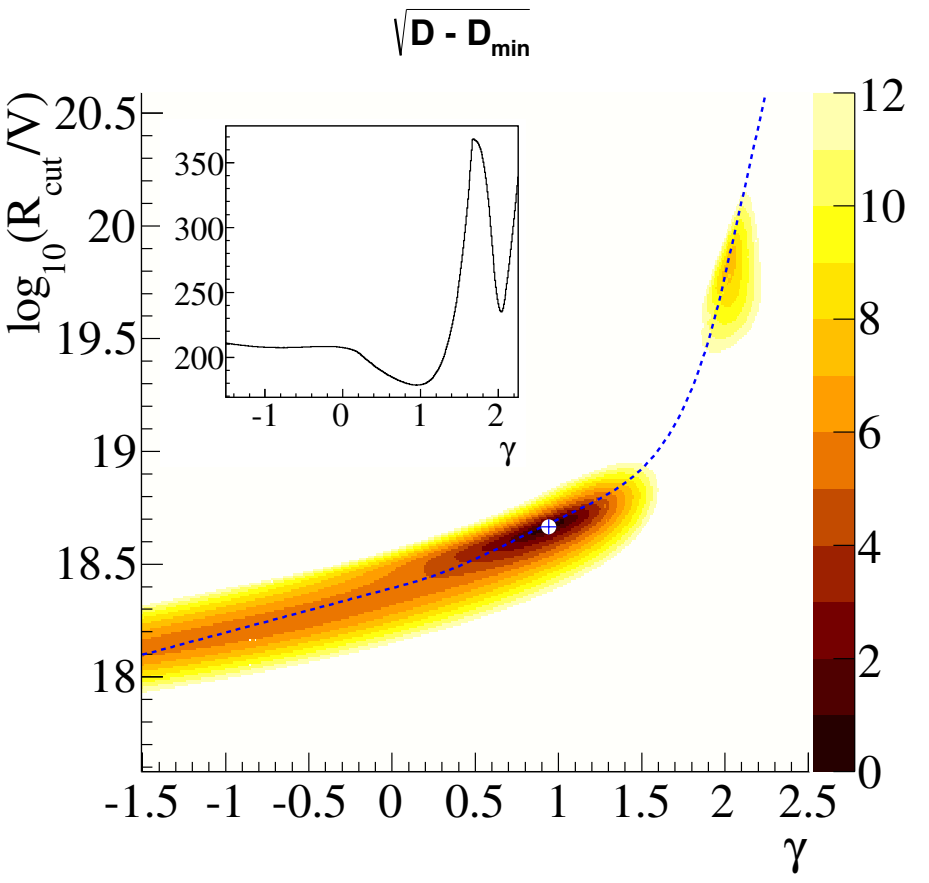

| model SPG | best fit | 2nd local min |
|---|---|---|
| $J_0$ [$\mathrm{eV}^{-1}\ \mathrm{Mpc}^{-3}\ \mathrm{yr}^{-1}$] | $7.17 \times 10^{18}$ | $4.53 \times 10^{19}$ |
| $\gamma$ | $0.94^{+0.09}_{-0.10}$ | $2.03$ |
| $\log_{10}(R_{\mathrm{cut}}/\mathrm{V})$ | $18.67 \pm 0.03$ | $19.84$ |
| $p_{\mathrm{H}}$ | $0.0^{+29.9}\%$ | $0.0\%$ |
| $p_{\mathrm{He}}$ | $62.0^{+3.5}_{-22.2}\%$ | $0.0\%$ |
| $p_{\mathrm{N}}$ | $37.2^{+4.2}_{-12.6}\%$ | $94.2\%$ |
| $p_{\mathrm{Fe}}$ | $0.8^{+0.2}_{-0.3}\%$ | $5.8\%$ |
| $D/n$ | $178.5/119$ | $235.0/119$ |
| $D\ (J)$, $D\ (X_{\mathrm{max}})$ | 18.8, 159.8 | 14.5, 220.5 |
| $p$ | $2.6\%$ | $5 \times 10^{-4}$ |

**Figure 1:** Left: $\sqrt{D - D_{\mathrm{min}}}$ where $D$ is the profile deviance as a function of $(\gamma, R_{\mathrm{cut}})$ and $D_{\mathrm{min}}$ is the best-fit deviance. Each coloured area corresponds to $1\sigma$, $2\sigma$, ... confidence intervals. The inset shows the values of $D$ along the dotted curve. Right: best-fit and second local minimum parameters for model SPG.

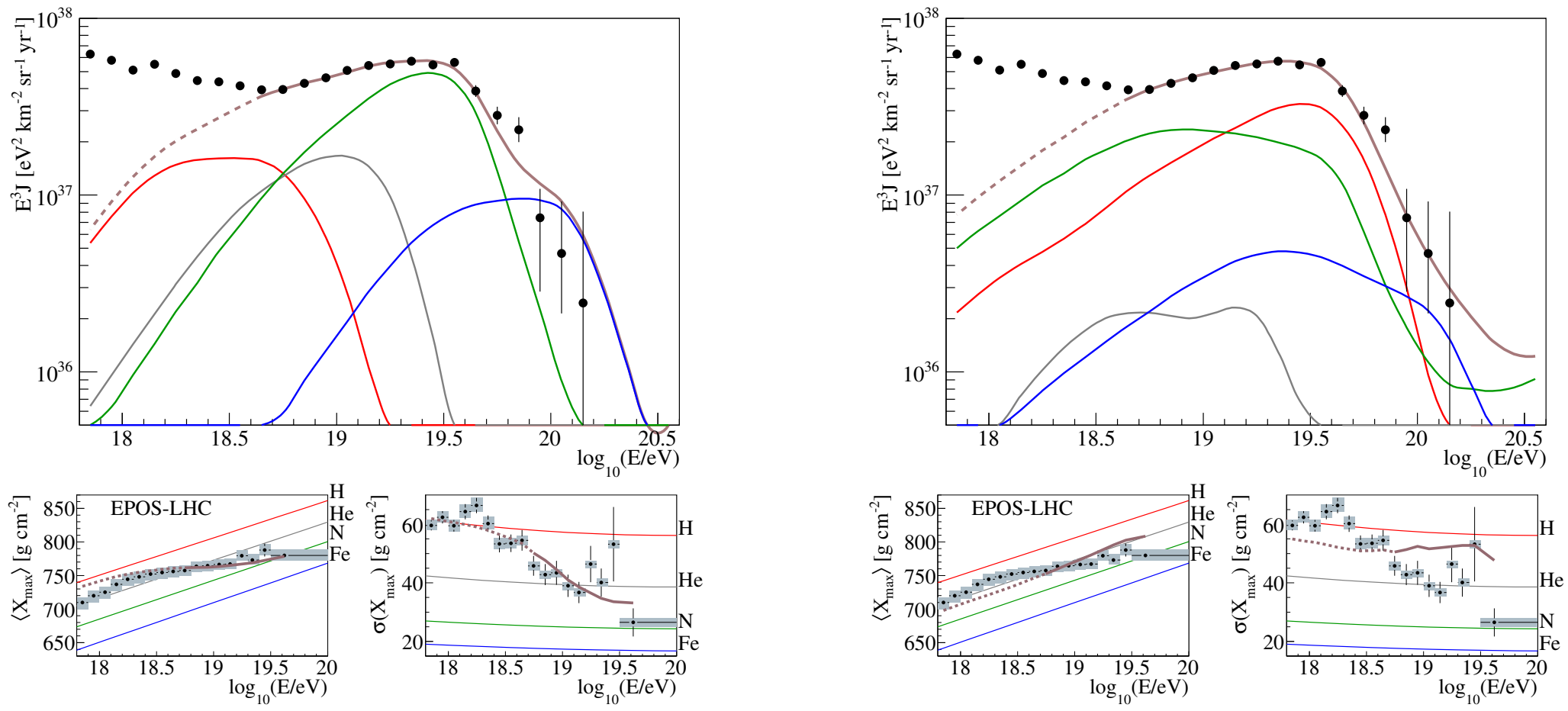

**Figure 2:** Top: simulated energy spectrum of UHECRs (multiplied by $E^3$) at the top of the Earth's atmosphere with the best-fit parameters (left) and the local minimum at $\gamma \approx 2$ (right) for model SPG, along with Auger data points [10]. Partial spectra are grouped according to the mass number as follows: $A = 1$ (red), $2 \leq A \leq 4$ (grey), $5 \leq A \leq 26$ (green), $27 \leq A$ (blue), total (brown). Bottom: average and standard deviation of the $X_{\mathrm{max}}$ distribution as predicted (assuming EPOS-LHC UHECR-air interactions) for the model predictions in the two scenarios (brown), pure $^{1}$H (red), $^{4}$He (grey), $^{14}$N (green) and $^{56}$Fe (blue). Only the energy range where the brown lines are solid is included in the fit.

From Fig. 3, it can be seen that the relationship between $\gamma$ and $R_{\mathrm{cut}}$ and the position of the second local minimum are very similar from one model to another, but the position of the best fit within the 'valley' and the height of the 'ridge' between the two local minima are strongly model-dependent. Furthermore, propagation models with lower photodisintegration rates[3] tend to result in better fits to the Auger data, except at very low values of $\gamma$ and $R_{\mathrm{cut}}$.

[3]The Domínguez EBL model has a stronger far infrared peak than the Gilmore model, and TALYS predicts sizeable

| | $\gamma$ (1st min) | $\log_{10}(R_{\text{cut}}/\text{V})$ | $D^{D(J)}_{D(X_{\max})}$ |
|---|---|---|---|
| SPG | $+0.94^{+0.09}_{-0.10}$ | $18.67 \pm 0.03$ | $178.5^{18.8}_{159.8}$ |
| SPD | $-0.45 \pm 0.41$ | $18.27^{+0.07}_{-0.06}$ | $193.4^{21.1}_{172.3}$ |
| STG | $+0.69^{+0.07}_{-0.06}$ | $18.60 \pm 0.01$ | $176.9^{19.5}_{157.4}$ |
| CTG | $+0.73^{+0.07}_{-0.06}$ | $18.58 \pm 0.01$ | $195.3^{33.6}_{161.7}$ |
| CTD | $-1.06^{+0.29}_{-0.22}$ | $18.19^{+0.04}_{-0.02}$ | $192.3^{21.2}_{171.1}$ |
| CGD | $-1.29^{+0.38}_{*}$ | $18.18^{+0.06}_{-0.04}$ | $192.5^{19.2}_{173.3}$ |

*This interval extends all the way down to $-1.5$, the lowest value of $\gamma$ we considered.

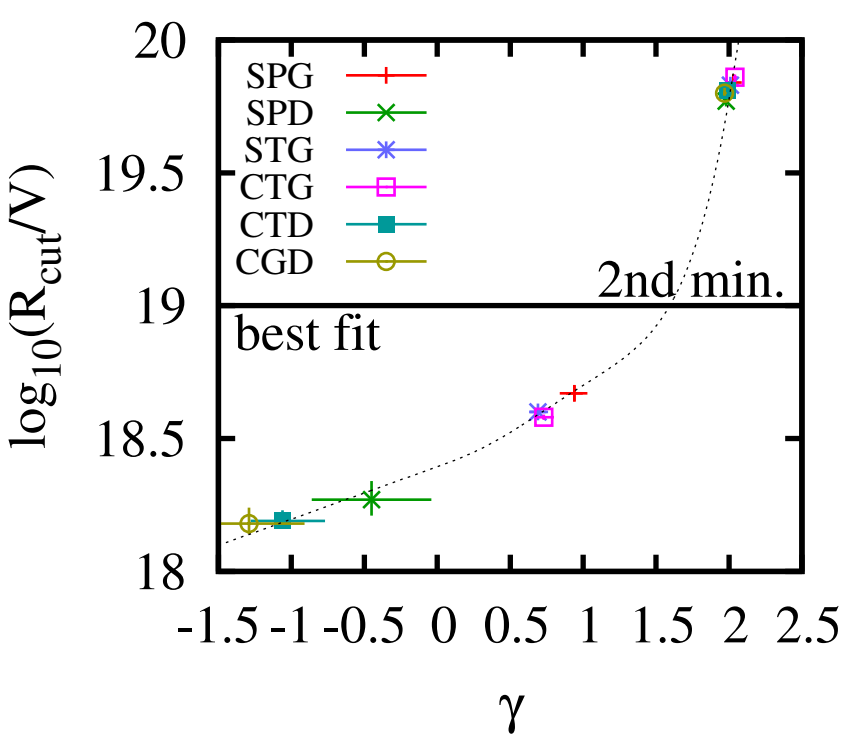

**Table 2:** Best-fit parameters and 68% uncertainties for the various propagation models we used (see Table 1). In the right panel local minima at $\gamma \approx 2$ are also shown. The dotted line is the same as in Fig. 1.

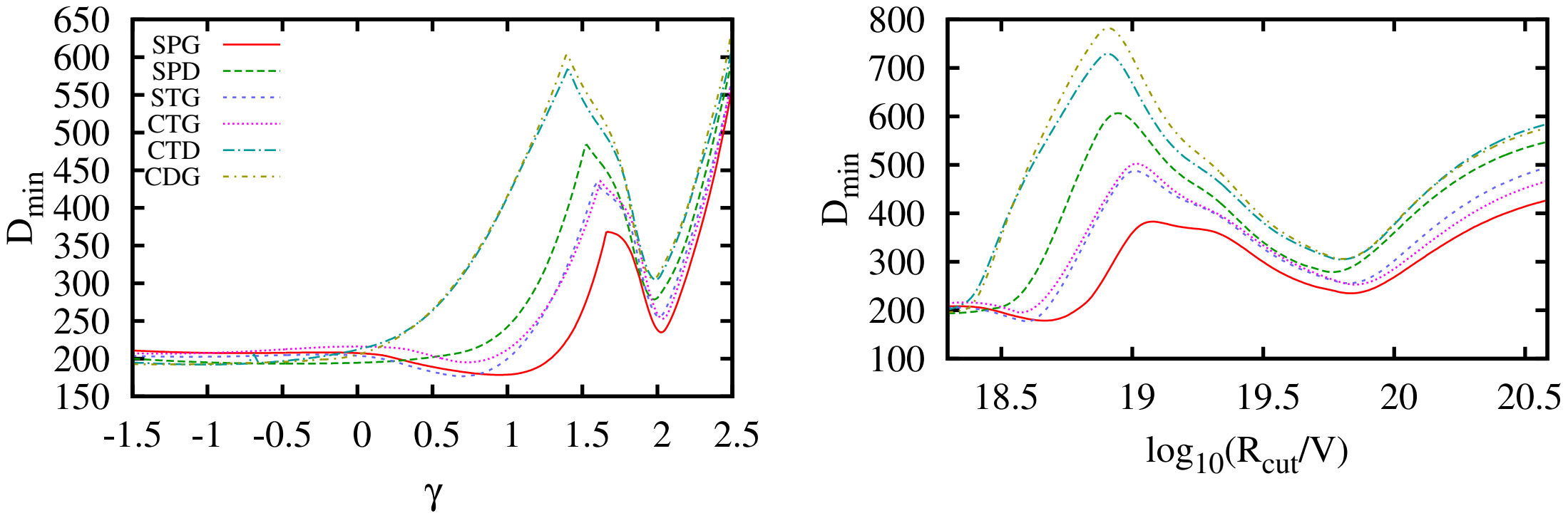

**Figure 3:** $D$ for each value of $\gamma$ (first panel) and $R_{\text{cut}}$ (second panel) for each propagation model, where the other parameters are chosen so as to minimize $D$

## 4. Effect of systematic uncertainties of the measurements

To study the effect of the systematic uncertainty of the energy scale of the Pierre Auger Observatory, we repeated the fit using the propagation model A shifting the energy of all measured data points by $\pm 14\%$, corresponding the systematic uncertainty of the energy scale, and/or shifting the $X_{\max}$ scale by the systematical uncertainties.[4]. The resulting best fit parameters are shown in Table 3 and the relationships between the parameters and the deviance are shown in Fig. 4.

Except at very high values of $R_{\text{cut}}$, the fit is improved by shifting the energies downwards and worsened by shifting them upwards. This effect is largely due to the data points in the measured spectrum around $10^{19.8}$ eV, being shifted towards or away from the descent in the simulated propagated nitrogen spectrum due to photodisintegration interactions. As for the $X_{\max}$ scale, lowering it in higher values of $\gamma$ and $R_{\text{cut}}$ and vice versa. Furthermore, lowering the $X_{\max}$ scale improves the fit and raising it worsens it.

---

cross sections for certain photodisintegration channels (largely in excess of the available measurements) which are neglected altogether in PSB.

[4]The systematical uncertainties on $X_{\max}$ are asymmetric and slightly dependent on energy, ranging from 6.9 to 9.4 g/cm$^2$. See Table IV in Ref. [11] for details.

| $\gamma$<br>$\log_{10}(R_{\text{cut}}/\text{V})$<br>$D^{D(J)}_{D(X_{\max})}$ | $-14\%$ | nominal energy scale | $+14\%$ |
|---|---|---|---|
| $-1\sigma_{\text{syst}}$ | $+1.32^{+0.05}_{-0.07}$<br>$18.68^{+0.05}_{-0.04}$<br>$157.4^{9.0}_{148.4}$ | $+1.35\pm0.05$<br>$18.73\pm0.02$<br>$172.1^{18.4}_{153.7}$ | $+1.39^{+0.05}_{-0.04}$<br>$18.78\pm0.01$<br>$203.5^{46.5}_{157.0}$ |
| nominal $X_{\max}$ scale | $+0.90^{+0.10}_{-0.15}$<br>$18.64^{+0.03}_{-0.04}$<br>$165.5^{9.3}_{156.1}$ | $+0.94^{+0.09}_{-0.10}$<br>$18.67\pm0.03$<br>$178.5^{18.8}_{159.8}$ | $+0.98^{+0.10}_{-0.11}$<br>$18.70\pm0.03$<br>$214.9^{50.4}_{164.5}$ |
| $+1\sigma_{\text{syst}}$ | $\leq -1.50$<br>$\leq 18.22$<br>$\leq 207.0$ | $\leq -1.50$<br>$\leq 18.24$<br>$\leq 217.2$ | $-1.34^{+0.31}_{*}$<br>$18.28^{+0.05}_{-0.08}$<br>$256.0^{55.7}_{200.4}$ |

*This interval extends all the way down to $-1.5$, the lowest value of $\gamma$ we considered.

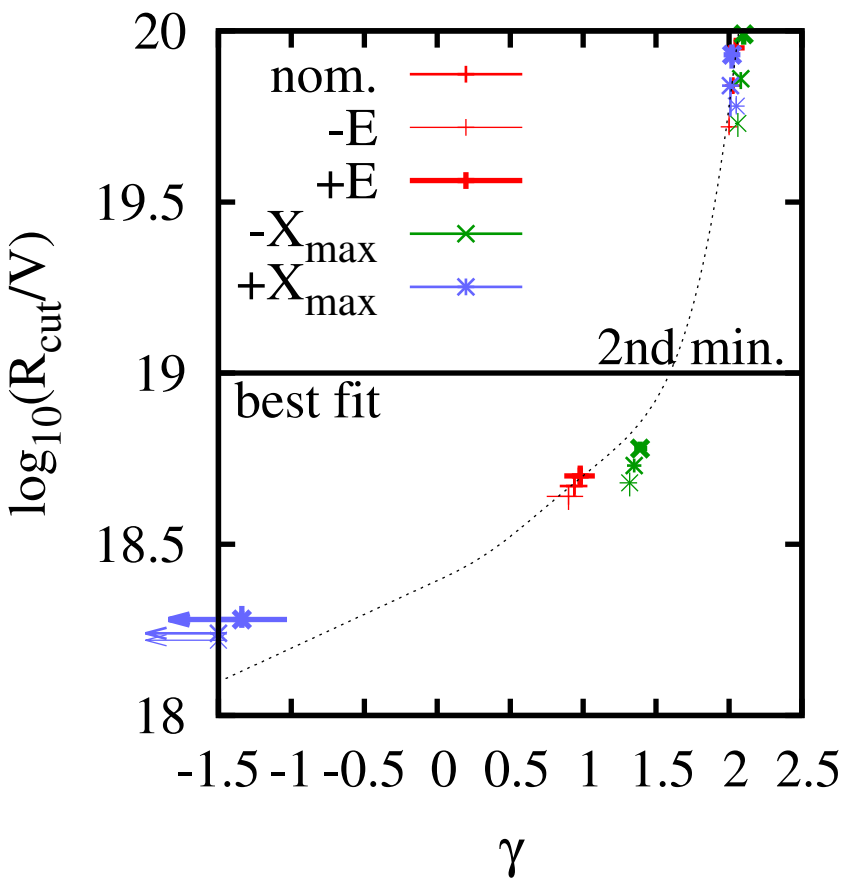

**Table 3:** Same as in Table 2, using propagation model SPG and shifting the Auger energy and/or $X_{\max}$ scales

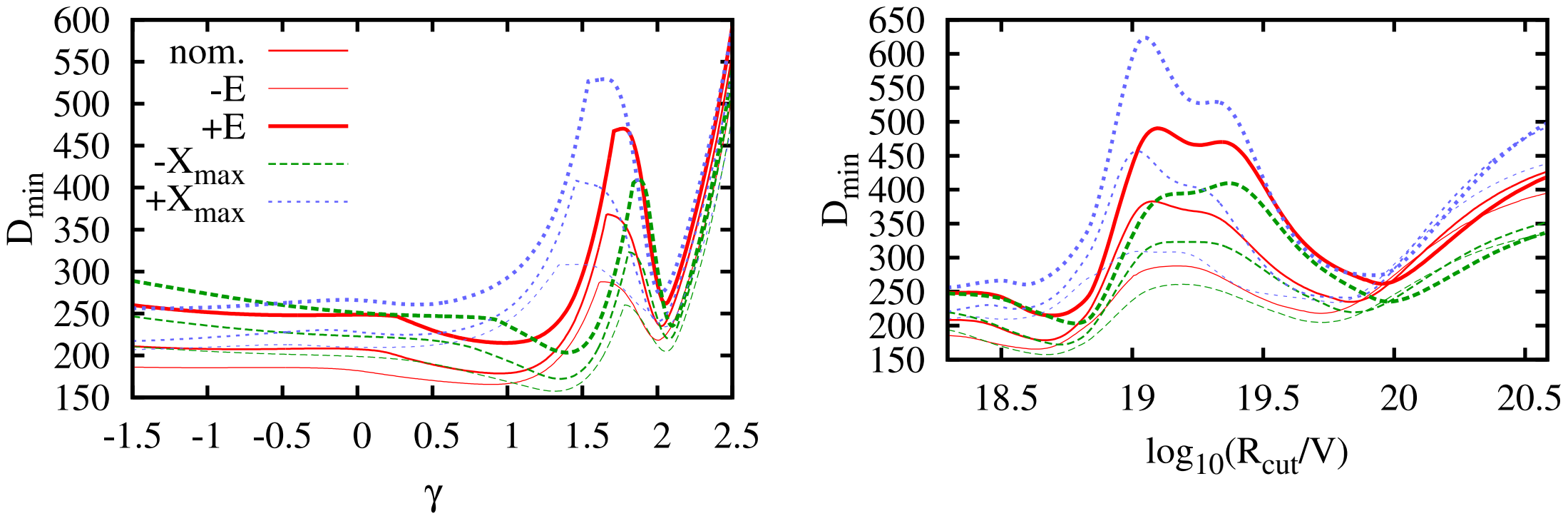

**Figure 4:** Same as in Fig. 3, using propagation model SPG and shifting the Auger energy and/or $X_{\max}$ scales

## 5. Effect of UHECR-air interaction models

In order to study the effect of different models of UHECR-air interactions, we repeated our fit with propagation model SPG using Sibyll 2.1 [22] and QGSJet II-04 [23] instead of EPOS-LHC. The results are shown in Table 4 and in Fig. 5.

|   | $\gamma$ (1st min) | $\log_{10}(R_{\text{cut}}/\text{V})$ | $D^{D(J)}_{D(X_{\max})}$ |
|---|---|---|---|
| E | $+0.94^{+0.09}_{-0.10}$ | $18.67\pm0.03$ | $178.5^{18.8}_{159.8}$ |
| S | $\leq -1.50$ | $\leq 18.27$ | $\leq 256.8$ |
| Q | $\leq -1.50$ | $\leq 18.28$ | $\leq 344.3$* |

*At least in the $\gamma$ range we considered, this minimum is actually worse than that at $\gamma \approx 2$ for this model.

**Table 4:** Same as Table 2, using propagation model SPG and various UHECR-air interaction models (E: EPOS-LHC, S: Sibyll 2.1, Q: QGSJet II-04)

The use of Sibyll 2.1, and to an even larger extent QGSJet II-04, worsens the fit at all considered values of $\gamma$ and $R_{\text{cut}}$; in addition, the low-$\gamma$ minimum is pushed down to values of $\gamma$ outside the range we considered, and with QGSJet II-04 the high-$\gamma$ local minimum is better than at least the visible portion of the low-$\gamma$ one.

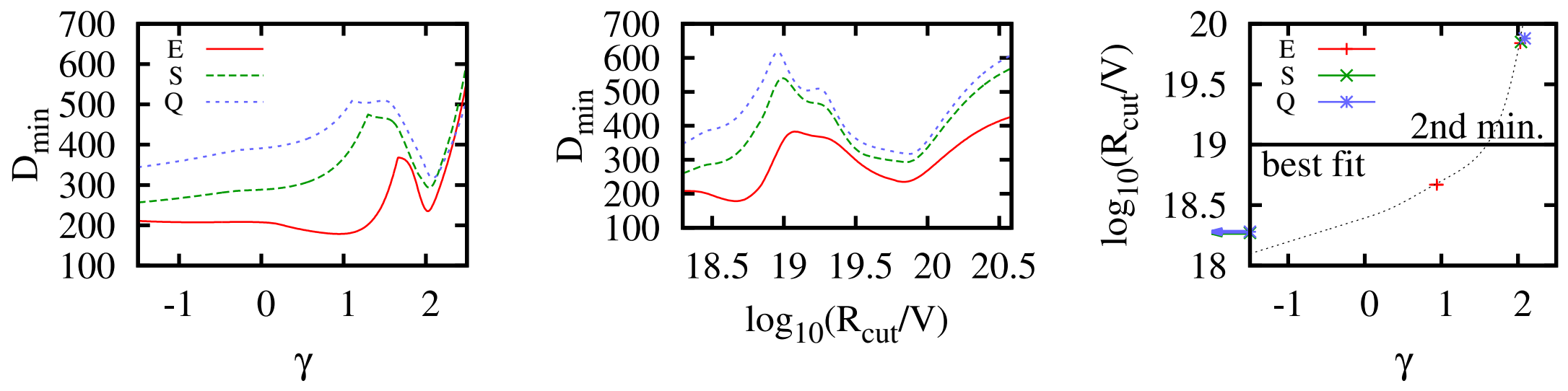

**Figure 5:** Same as in Fig. 3, using propagation model SPG and various UHECR-air interaction models (E: EPOS-LHC, S: Sibyll 2.1, Q: QGSJet II-04)

## 6. Conclusions

When interpreted with a simple model of UHECR injection, the Auger data are best fitted by very hard ($\gamma \lesssim 1$) injection spectra and the flux is mostly limited by the maximum energy at the sources. The local minimum with $\gamma \approx 2$ and large maximum rigidity, which is more in line with standard models of cosmic ray acceleration, predicts wider distributions of UHECR masses at each energy than observed in the data. This conclusion is robust with respect to all the model variations we considered, but the position of the best fit is strongly sensitive to the details of the propagation, though in all cases it is in the same curved strip of the $(\gamma, R_{cut})$ plane. The uncertainty due to our ignorance of details of the propagation are much larger than that due to the statistical uncertainty of measured data.

As for the goodness of fit, it is better for models of UHECR propagation with lower photodisintegration rates (Gilmore 2012 EBL model, PSB cross sections) than with higher rates (Domínguez 2011 EBL model, TALYS cross sections), better for models of UHECR-air interactions predicting deeper showers (EPOS-LHC) than shallower ones (QGSJet II-04), and better for the cases when the Auger energy and $X_{max}$ scale are assumed to be lower than the nominal ones.

Other possible ways of improving the agreement with measured data, such as considering more injection masses, considering non-uniform source distributions, or more complicated injection spectra, are outside the scope of this work.

# Solar Cycle Modulation of Cosmic Rays Observed with the Low Energy Modes of the Pierre Auger Observatory

**J.J. Masías-Meza[a] for the Pierre Auger Collaboration[*,b]**
[a] *Departamento de Física and IFIBA, (FCEN-UBA-CONICET), Buenos Aires, Argentina*
[b] *Observatorio Pierre Auger, Av. San Martín Norte 304, 5613 Malargüe, Argentina*
*E-mail:* auger_spokespersons@fnal.gov
*Full author list:* http://www.auger.org/archive/authors_2015_06.html

The low energy modes of the surface detector array of the Pierre Auger Observatory record variations in the flux of low energy secondary particles with extreme detail. These two modes consist of recording (1) the rate of signals for energies between ∼ 15 MeV and ∼ 100 MeV (the Scaler mode) and (2) the calibration charge histograms of the individual pulses detected by each water-Cherenkov station, covering different energy channels up to ∼ 1 GeV (the Histogram mode). Previous work has studied the flux of galactic cosmic rays on short and intermediate time scales (i.e. from minutes to weeks) using these low energy modes. In this work, after including a long-term correction to the response of the detectors, we present the first long-term analysis of the flux of cosmic rays using scalers and two energy bands of the calibration histograms. We show its sensitivity to the solar cycle variation and its relation to the solar modulation of cosmic rays for an 8-year period.

[*] Speaker.

## 1. Introduction

In the interplanetary medium, the transport of high-energy/non-thermal charged particles with Larmor radius comparable to or smaller than the Heliosphere size, is sensitive to solar magnetic activity, mainly due to its effects (on various space-time scales) on several heliospheric properties.

While many of these effects have been studied since the 1960's, mainly using neutron monitors, during the last decade several new particle detectors (such as muon telescopes or the low energy modes of some high energy observatories) have also begun helping us to better understand the physical processes in the heliosphere that affect the propagation of these energetic particles.

In particular, the solar modulation of galactic cosmic rays (GCRs) on time-scales from minutes to several days, has been studied recently using the flux of secondary particles detected by the surface detector array (SD) of the Pierre Auger Observatory [1, 2, 3]. These studies quantified the sensitivity of GCRs to specific transient interplanetary structures for different energies [4].

The Pierre Auger Observatory [5] is located at Malargüe, Argentina (69.3$^{o}$ W, 35.3$^{o}$ S, 1400 m a.s.l.), and was designed for the study of cosmic rays (CRs) at the highest energies. Its SD array consists of 1660 water-Cherenkov detectors (WCD), arranged in a triangular grid with a spacing of 1500 m, distributed over an area of 3000 km$^2$.

Measurements of low-energy cosmic rays (primary energies from $\sim$ 10 GeV to a few TeV [3]) can be performed at the Pierre Auger Observatory by exploiting two different low energy modes implemented at the Observatory. These produce complementary sets of data [4]: the Scaler and Histogram modes. The Scaler Mode is a particle counter mode. It was implemented in all the detectors of the SD array and these data are recorded for every station every second, reaching typical counting rates of $\sim 1.8 \times 10^8$ counts per minute. From September 2005, it has recorded the number of signals detected above a very low threshold of 3 ADC counts above the station baseline and with an upper cut of 20 ADC counts. This range of ADC counts corresponds to a deposited energy $E_d$ (in the detector volume) between $\approx$ 15 MeV and $\approx$ 100 MeV [2].

The Histogram mode comes from measuring the properties of pulse signals produced by the particles interacting with the water volume. In particular, one-minute histograms are built at each SD station from the peak of each signal (peak histograms), and from the total charge deposited at the photomultiplier tubes (PMTs) of the detectors (charge histograms), which is proportional to the deposited energy within the detector volume [6]. Thus, after a calibration process considering the energy deposited $E_d$ by a central and vertical muon (equivalent to a signal of 1 VEM, for vertical equivalent muon), it is possible to construct histograms of deposited energy up to $\sim$ 1 GeV. The use of histogram data allows us to study the variation of the counting rates in different energy-deposit bins, corresponding to different CR primary energies [4]. There are two particular ranges that will be analyzed in the present paper: one comparable to deposited energies of the scaler mode ($E_d$ between 60 MeV and 120 MeV), and one associated with energies deposited by single muons near the range associated with vertical incidence ($E_d$ between 200 MeV and 280 MeV).

Both Scaler and Histogram modes are complementary in some sense, and so it is also possible to make a cross-check of the two kinds of data for making consistency tests.

By using the data described above, we have studied so far [1, 2, 4, 3] transient phenomena of solar origin such as, for instance, Forbush decreases produced by Interplanetary Coronal Mass Ejections and the daily modulation of the counting rate. In this paper, for the first time, we exploit these data to study CR variations over time-scales close to that of one solar cycle. The data used in the following span an 8-year period from January 2006 to December 2013.

In section 2 we describe the general data treatment and selection criteria applied to Scaler and Histogram data. In section 2.1, we present the corrections that we apply to the data to take into account the long-term variations of the response of the water-Cherenkov detector to particles. This correction is followed by a data rejection, described in section 2.2, and then we describe (2.3) how we account for atmospheric effects (mainly due to pressure) on the rate of observed particles. Then, in Sections 3 and 4, we show the behavior of the corrected counting rate at different energies and over 8 years, and compare them with neutron-monitor data, giving a discussion and our conclusions for this paper.

## 2. Data Treatment

It is well established that the so called Area over Peak (AoP) signals characterize the long-term (years) variations of the response of a WCD to particles (see for example [7, 5] and cited references therein). This AoP is a proxy for the impulse response of the detector to individual muons, and is defined as the ratio of the deposited charge ($\mathrm{VEM_q}$) to the peak ($\mathrm{VEM_p}$) of the pulses associated to the passage of vertical muons trough the water volume:

$$\mathrm{AoP} = \frac{\mathrm{VEM}_q}{\mathrm{VEM}_p}. \tag{2.1}$$

As the signals given by extensive air shower detected in the WCD are calibrated in terms of the instantaneous value of the charge corresponding to the VEM [6], these variations do not affect the standard data reconstruction and analysis. Even so, since the data of the low energy modes are based on the detector response to secondary particles produced during the interaction of CRs with the atmosphere, the AoP variation with time could affect the rates measured in those modes and so has to be corrected when long-term solar modulation effects (such as the eleven year solar cycle) are to be studied.

In this section we explore the effect of the AoP evolution as a function of time on the scaler rate and histograms at the individual station level, and then implement a procedure to correct the long-term behaviour for this effect.

The main interest of this work is to analyze the corrected data from the two modes (Scalers and Histograms) and to find the physical long-term flux of atmospheric secondaries as measured by the entire array. To do this, we introduce a particular data analysis process at the individual detector level.

The high statistics of the array allow us to introduce restrictive filters to guarantee the quality of the data we include in this study. We only include data that satisfy the following conditions: 1) the three PMTs of the detector must be operative (this information is obtained by counting the number of working PMTs in the monitoring data as a function of time), 2) the average number of operative detectors must be: 2.a) for scalers: larger than than 600 within 5 minutes intervals; 2.b) for histograms: larger than 150 within 15 minutes intervals; and 3) for histogram data we include an additional criterion: we only use histograms whose first local maximum (the so called electromagnetic peak) is located at values of deposited energies larger than 20 MeV. Histograms that do not fulfill this condition are very few but have typically noisier rates and are discarded.

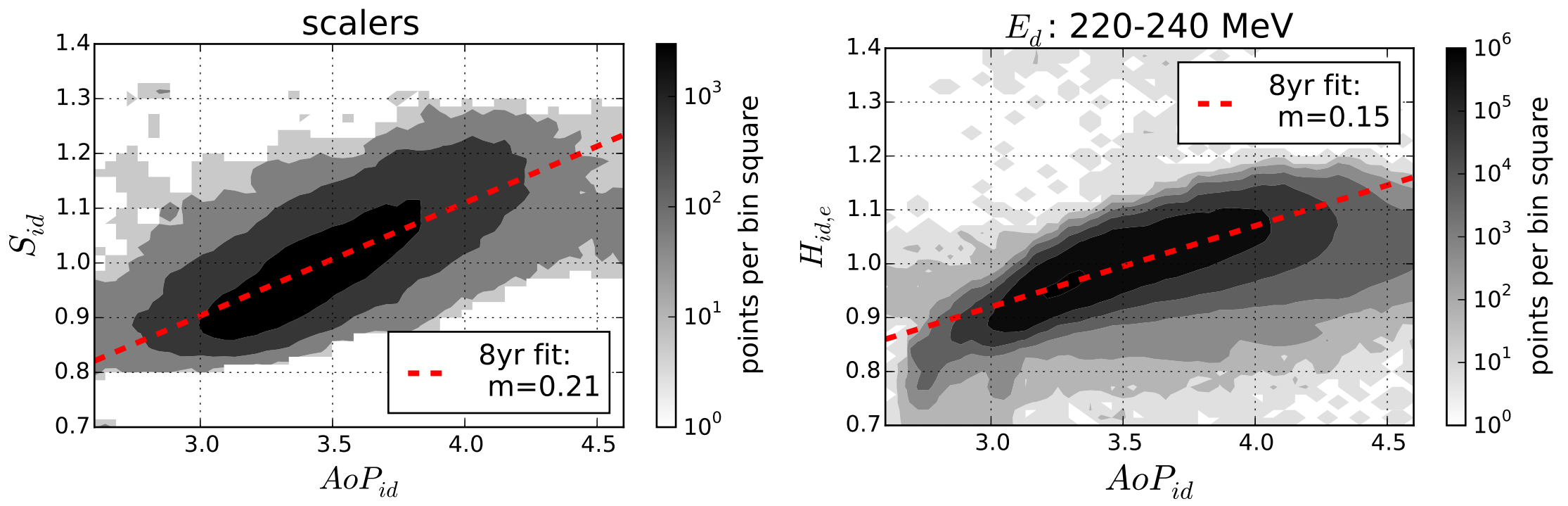

**Figure 1:** Two dimensional histogram showing the correlations between the Scaler rates $S_{id}$ and $AoP_{id}$ at individual WCD station data. Left panel: using scaler data averaged over 5-day time windows. Right panel: we show the rate, for the band where the muon content is dominant, for charge histograms at 15-min resolution level. This band is shown as an example; for the rest of the energy channels, the correlations have different slopes but similar structures, as shown in figure 2. The bin widths in AoP and the normalized rates are respectively 0.04 and 0.014. The dashed red lines represent the linear fits obtained from equation (2.2).

## 2.1 Area over Peak correction

In the following section we present corrections by looking for linear correlations between AoP and count rates (scaler and histograms) at the individual SD level.

A strong correlation is found when the data from the whole array are considered, as can be seen in Figure 1. These data show that detectors with large AoP values also have large scaler rates. This can be understood since large AoP implies a better response of the detector to individual particles. From the Scaler point of view, larger pulses implies that more pulses will reach the counting threshold to be recorded.

In the following notation, it is understood that $S_{id}$ is the scaler rate of the WCD station with identification number *id*, and that it is normalized by the whole array temporal average $\mathcal{S}^{phys} = \langle\langle S_{id}^{phys}(t)\rangle_{\forall \mathrm{id}}\rangle_{\forall \mathrm{t}}$, such that $S_{id} = S_{id}^{phys}/\mathcal{S}^{phys}$, where the superindex *phys* refers to magnitudes in physical units. Similarly, we will assume that the histograms rates $H_{id,e}$ are normalized by the whole array time average at each energy band $E_d$ indexed with $e$, denoted by $\mathcal{H}_e^{phys} = \langle\langle H_{id,e}^{phys}(t)\rangle_{\forall \mathrm{id}}\rangle_{\forall \mathrm{t}}$, such that $H_{id,e} = H_{id,e}^{phys}/\mathcal{H}_e^{phys}$. The deposited energy band rates have been processed in uniform intervals of 20 MeV in the range $E_d = (0-1000)$ MeV; so that energy indices are $e = 0,1,2,\ldots 49$. To account for the AoP effect we fit a linear regression:

$$
\begin{aligned}
S_{\mathrm{id}} &= m^{\mathrm{S,AoP}}\mathrm{AoP}_{id} + b^{\mathrm{S,AoP}} \\
H_{\mathrm{id,e}} &= m_e^{\mathrm{H,AoP}}\mathrm{AoP}_{id} + b^{\mathrm{H,AoP}},
\end{aligned}
\tag{2.2}
$$

obtaining slopes with statistical errors of the order of 1% for scalers and histograms for the 2006–2013 period. The values of $m^{S,AoP}$ and $m_e^{H,AoP}$ are shown in the left panel of Figure 2. The dependence of $m_e^{H,AoP}$ on energy is closely related to the long-term evolution of the charge histograms. An example of the linear fits for the energy range channel between 220 MeV and 240 MeV is shown in the right side of Figure 1.

To perform a first-order correction in the individual scalers ($S_{\mathrm{id}}$) and histograms ($H_{\mathrm{id,e}}$), we apply a linear detrend on the individual $\mathrm{AoP_{id}}$ as follows:

$$\begin{aligned} S_{\mathrm{id}}^{\mathrm{corr/AoP}}(t) &= S_{\mathrm{id}}(t) - m^{\mathrm{S,AoP}}(\mathrm{AoP_{id}}(t) - \overline{AoP}) \\ H_{\mathrm{id,e}}^{\mathrm{corr/AoP}}(t) &= H_{\mathrm{id,e}}(t) - m_e^{\mathrm{H,AoP}}(\mathrm{AoP_{id}}(t) - \overline{AoP}), \end{aligned} \tag{2.3}$$

where $S_{\mathrm{id}}^{\mathrm{corr/AoP}}$ and $H_{id,e}^{corr/AoP}$ are the rates corrected for the AoP modulation. The global average over the considered detectors and the whole time period is given by $\overline{AoP} = \langle\langle AoP_{id}(t)\rangle_{\forall \mathrm{id}}\rangle_{\forall \mathrm{t}}$. The number of detectors included in the average calculation is not necessarily the whole array, as the detectors must pass the filters mentioned earlier in this section. The average term has been added in order to conserve the mean values: $\langle\langle S_{\mathrm{id}}^{corr/AoP}(t)\rangle_{\forall id}\rangle_{\forall t} = 1$, and similarly $\langle\langle H_{\mathrm{id,e}}^{corr/AoP}(t)\rangle_{\forall id}\rangle_{\forall t} = 1$.

### 2.2 Data rejection method

Once we have obtained the mean value for the AoP corrected scalers and bands-of-histogram rates, we need to take into account those outlier values that do not fit within this correction. To do this, we introduce the following data rejection method: we discard all the scalers $S_{\mathrm{id}}^{\mathrm{corr/AoP}}(t)$ outside $2.5\sigma(t)$ from the mean value $\langle S_{\mathrm{id}}(t)\rangle_{\forall \mathrm{id}}$, where $\sigma(t)$ is the standard deviation determined from the set of data of the entire array in a 5 minute window centered at time $t$.

At this point, we note that when averaging over the entire array, the data are still noisy for some time intervals where the number of detectors contributing data is low. Thus, we include additional filtering criteria, to both the Scaler and Histogram rates:

1. for each time bin, using a moving window centered average with a sliding window of 4-months length, we determine the mean and the standard deviation $\sigma_N$ of the number of working detectors as a function of time. Those periods where the instantaneous number of detectors is not within $\pm 1\sigma_N$ respect to that mean value are discarded; and
2. the instantaneous $\langle \mathrm{AoP_{id}}(t)\rangle_{\forall \mathrm{id}}$ value must be inside $\pm 3.3\sigma(t)$ with respect to the mean of the 4-months period.

### 2.3 Pressure correction

To obtain the physical signal we still have to correct for the atmospheric pressure effect in the flux. This effect is evidenced as the usual anticorrelation, and the linear parameters can be obtained from the fit of a linear function for the flux as the function of atmospheric pressure. The corresponding slopes for the scalers rates, $m^{S,press}$, and for the histograms bands, $m_e^{H,press}$, can be seen in the right panel of Figure 2. After applying this correction we finally obtain the AoP and pressure corrected rates $\langle S_{\mathrm{id}}^{\mathrm{corr/AoP}}(t)\rangle_{\forall \mathrm{id}}^{\mathrm{corr/press}}$, $\langle H_{\mathrm{id,e}}^{\mathrm{corr/AoP}}(t)\rangle_{\forall \mathrm{id}}^{\mathrm{corr/press}}$, that will be described in the next section.

For discussions hereafter, we will use the abbreviated notation $\widetilde{H}_e^{phys} = \mathcal{H}_e^{phys} \times \langle H_{\mathrm{id,e}}^{\mathrm{corr/AoP}}(t)\rangle_{\forall \mathrm{id}}^{\mathrm{corr/press}}$, which refers to histogram rates corrected by AoP and barometric effects, with physical units at each energy band $e$.

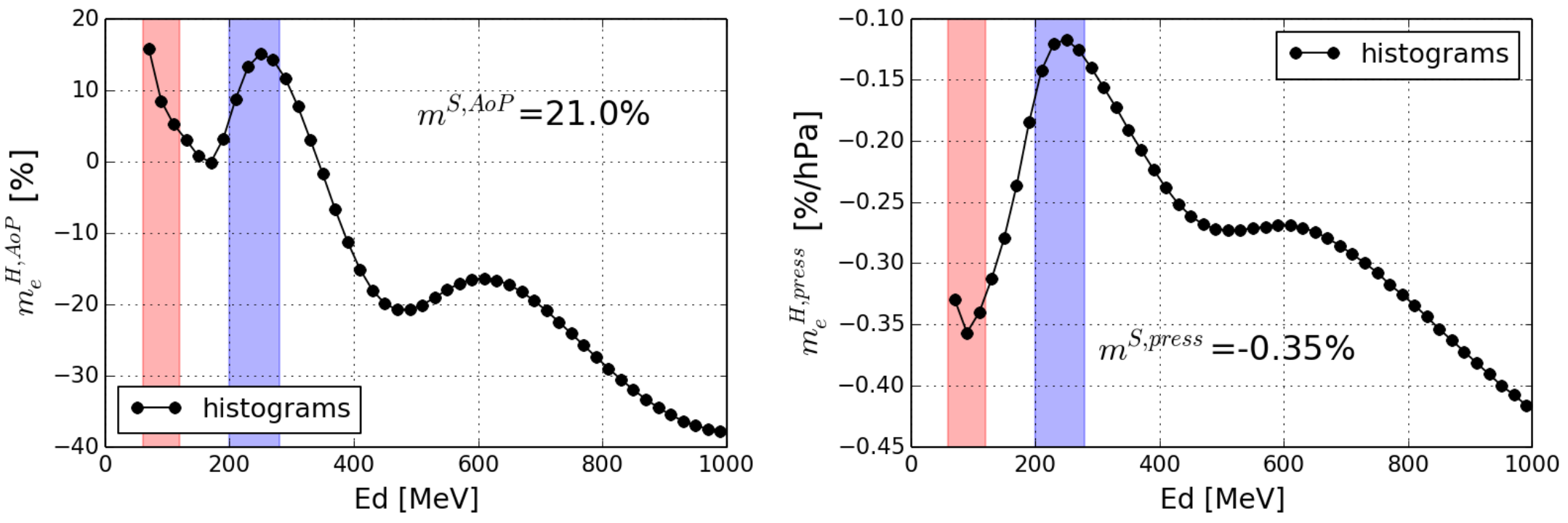

**Figure 2:** Left: slopes $m_e^{\mathrm{H,AoP}}$ (black dotted lines) from fits to the correlations between charge histogram counts and AoP as shown in Figure 1. Units are in percentage per AoP unit, and the errors are of the order of 1%. In the inset the corresponding unique value for scalers $m^{\mathrm{S,AoP}}$ is shown. Right: slopes $m^{S,press}$ and $m_e^{H,press}$ (black dotted lines) from fits to correlations between histogram rates and atmospheric pressure. As in the left pannel, in the inset the corresponding slope for the Scaler rate with pressure is shown. In both cases, red and blue bands correspond to the $H_{sc}$ and $H_\mu$ rates, defined in section 3.

## 3. Results and Discussion

In left panel of Figure 3 we show the long-term pressure-corrected Scaler rates. Grey and black curves show data without and with AoP correction, respectively.

Some differences between the rate of muons and the rate of neutrons (e.g. observed by Neutron Monitors) are known, e.g. due to local atmospheric effects and associated primary energies [8] and because of the different production rates during the shower. However, in the left panel of Figure 3, we present a comparison of observations of the Pierre Auger Scaler mode with data from different neutron monitors (McMurdo, Kiel, and Athens), which have different rigidity cut-offs because of their different geographic locations. Notice that the AoP-corrected Scalers present a global maximum near the middle of 2009, in agreement with the time when neutron monitors observe the maximum flux of GCRs associated with the minimum of solar activity, toward the beginning of the solar cycle 24. As expected, the amplitude of the GCR modulation at different stations decreases monotonically as the geomagnetic rigidity cutoff of each site increases. It should be noted that the rigidity cutoff at Malargüe ($R_c = 9.8\,\mathrm{GV}$) is well below the characteristic energy of GCRs flux observed with the Scaler mode at Pierre Auger Observatory, which has a median value estimated as $\sim 90\,\mathrm{GeV}$ (see [3]).

To explore the effects of the solar modulation in different energy ranges, we consider two different channels of deposited energy: low and high energy.

The low energy channel is obtained from integrating the histograms between 60 MeV and 120 MeV of deposited energy, which is related to the scaler trigger bounds and comes from $H_{sc}(t) = \sum_{e=3}^{5} \widetilde{H}_e^{phys}(t) / \sum_{e=3}^{5} \mathcal{H}_e^{phys}$. For this channel we consider limits at $E_d$ =(60-120) MeV because, for energies $E_d \lesssim 60\,\mathrm{MeV}$, the rate could be affected by the trigger system of the detector.

The high energy channel is related to the second peak of the charge histograms, where vertical muons become the predominant component.It is obtained from $H_\mu(t) = \sum_{e=10}^{13} \widetilde{H}_e^{phys}(t) / \sum_{e=10}^{13} \mathcal{H}_e^{phys}$, corresponding to $E_d$ =(200-280) MeV.

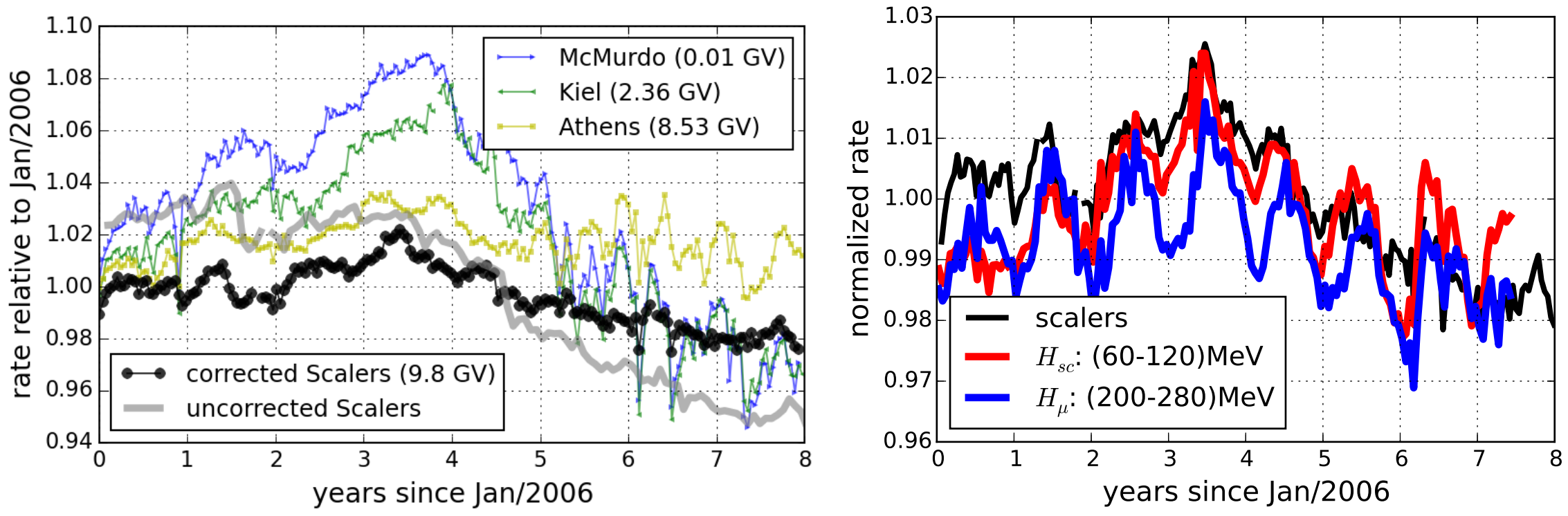

**Figure 3:** Left panel: Long-term of pressure-corrected Scalers with AoP correction (black dots) and without AoP correction (gray line), compared with different NMs (inset shows geomagnetic rigidity cut-off). The lower Auger amplitude of the solar cycle modulation is consistent with the decreasing trend for high rigidity NMs. Right panel: time profiles for $H_{sc}$ (red) and $H_{\mu}$ (blue); both normalized by their time average. In black, the same scaler profile as in left panel. The three profiles show a global peak at the solar minimum period. Note that the global peak at lower energies is stronger than at higher energies, as expected.

The long-term profiles for $H_{sc}$ and $H_{\mu}$ are shown in the right panel of Figure 3. In this figure it is also possible to see a global maximum flux of particles near the middle of 2009, at the same time as in the Scalers.

Furthermore, we find that, as expected, the peak at lower energies ($H_{sc}$) is stronger than at higher energies ($H_{\mu}$). Superposed to the rates of scalers and histograms, there is a pronounced annual modulation, that appears stronger in the Histogram data. This is a negative temperature effect (anticorrelation between the rate and the temperature), an effect that has been previously reported for muon detectors [9, 8]. The increase of the temperature causes a decrease of the air density, and consequently an increase of the volume of the atmosphere; therefore, charged pions are produced higher in the atmosphere and more muons will decay before reaching the detector level.

## 4. Summary and Conclusions

Summarizing, in the right panel of Figure 3 we present for the first time the long-term (8-years) behaviour for Scalers, for $H_{sc}(t)$, and for $H_{\mu}(t)$, measured by the low energy modes of the Pierre Auger Observatory, where the solar cycle (modulation associated with interplanetary physical mechanisms) and the daily modulation (associated with atmospheric physical processes) can be observed for the different energy bands of these modes.

The period observed corresponds to the last, very weak, period of solar activity, with a consequent interest of being associated with the lowest interplanetary effects on GCRs, ever recorded.

Other detectors that can provide such a long recording period, as the one presented here, are neutron monitors. However, the rates that we present correspond to higher energies than the ones observed using neutron monitors (which also are counters that do not distinguish different energies). Then, our observations can help to better understand the effects of different fundamental processes (as drift effects or turbulent diffusion) affecting the GCR transport in the heliosphere, for different primary energies.

Furthermore, these low energy mode data (e.g. Scalers, $H_{sc}$, and $H_{\mu}$) provided by the Pierre Auger Observatory, give the flux of low energy GCRs with the highest statistical significance due to its huge total collecting area from more than 1600 detectors.

Thus, we have presented measurements of scaler and muon fluxes at unprecedented statistical confidence, covering the most interesting period of recent solar cycles. The low energy (scaler and muon) modes of the Pierre Auger Observatory can extend the range of rigidities that are routinely studied using NMs.

# Lightning Detection at the Pierre Auger Observatory

**Julian Rautenberg**[*,a] **for the Pierre Auger Collaboration**[b]
[a]*Bergische Universität Wuppertal, Wuppertal, Germany*
[b]*Observatorio Pierre Auger, Av. San Martín Norte 304, 5613 Malargüe, Argentina*
*E-mail:* auger_spokespersons@fnal.gov
*Full author list:* http://www.auger.org/archive/authors_2015_06.html

The Auger Engineering Radio Array, an extension of the Pierre Auger Observatory with antennas in the MHz range, requires to monitor the atmospheric conditions, which have a large influence on the radio emission of air showers. In particular, amplified signals up to an order of magnitude have been detected as an affect of thunderstorms. For a more detailed investigation and more generally, for detecting thunderstorms, a new lightning detection system has been installed at the Pierre Auger Observatory in Argentina. In addition, an electric-field mill measures the field strength on ground level at the antenna array. With these measurements, data periods affected by thunderstorms can be identified. Additionally, a lightning trigger for the water-Cherenkov detectors was developed to read out individual stations when a lightning was detected nearby. With these data, a possible correlation between the formation of lightning and cosmic rays can be investigated even at low energies of about $10^{15}$ eV. The structure and functionality of the lightning detection are described and first data analyses are shown.

*Speaker.

## 1. Introduction

The Pierre Auger Observatory [1], located in the Province of Mendoza, Argentina, is the world's largest cosmic-ray observatory. The objectives of the observatory are to probe the origin and characteristics of cosmic rays above $10^{17}$ eV and to study the interactions of the most energetic particles observed in nature. The observatory is a hybrid system, a combination of a large surface detector (SD) and a fluorescence detector (FD). The SD is composed of 1660 water-Cherenkov stations placed on a triangular grid with nearest neighbors separated by 1500 m. It is spread over an area of $\sim 3000\ \mathrm{km}^2$ and overlooked by 27 air fluorescence telescopes, as indicated in Fig. 1.

In 2006, the Pierre Auger Collaboration started an R&D program for radio-detection of air showers [2]. With the Auger Engineering Radio Array (AERA) [3] first important results for the understanding of the radio emission process could be provided [4]. A strong enhancement of the detected amplitude in case of thunderstorms has been reported before [5, 6]. Therefore, radio detection requires the monitoring of the environmental electric field (E-field) condition to identify data periods affected by thunderstorms. This can be done using an E-field mill. In case of lightnings, the measured E-field values change within a second by several kV. Algorithms to analyse the time-sequence of the E-field data are used to identify different environmental conditions like thunderstorm and lightning strikes [7]. The E-field mill is limited in time-precision to a one-second sampling. Moreover, it offers no space information except for the position of the E-field mill, and its sensitivity is limited to distances of a few tens of kilometer.

## 2. Lightning detector

The StormTracker from Boltek[1] is a lightning detection device which consists of a PCI card and an external antenna connected via an Ethernet cable. This device is able to detect lightning strikes within a radius of 1000 km. It detects radio signals in the kHz region (ca. 10 to 90 kHz) which are initiated by lightnings.

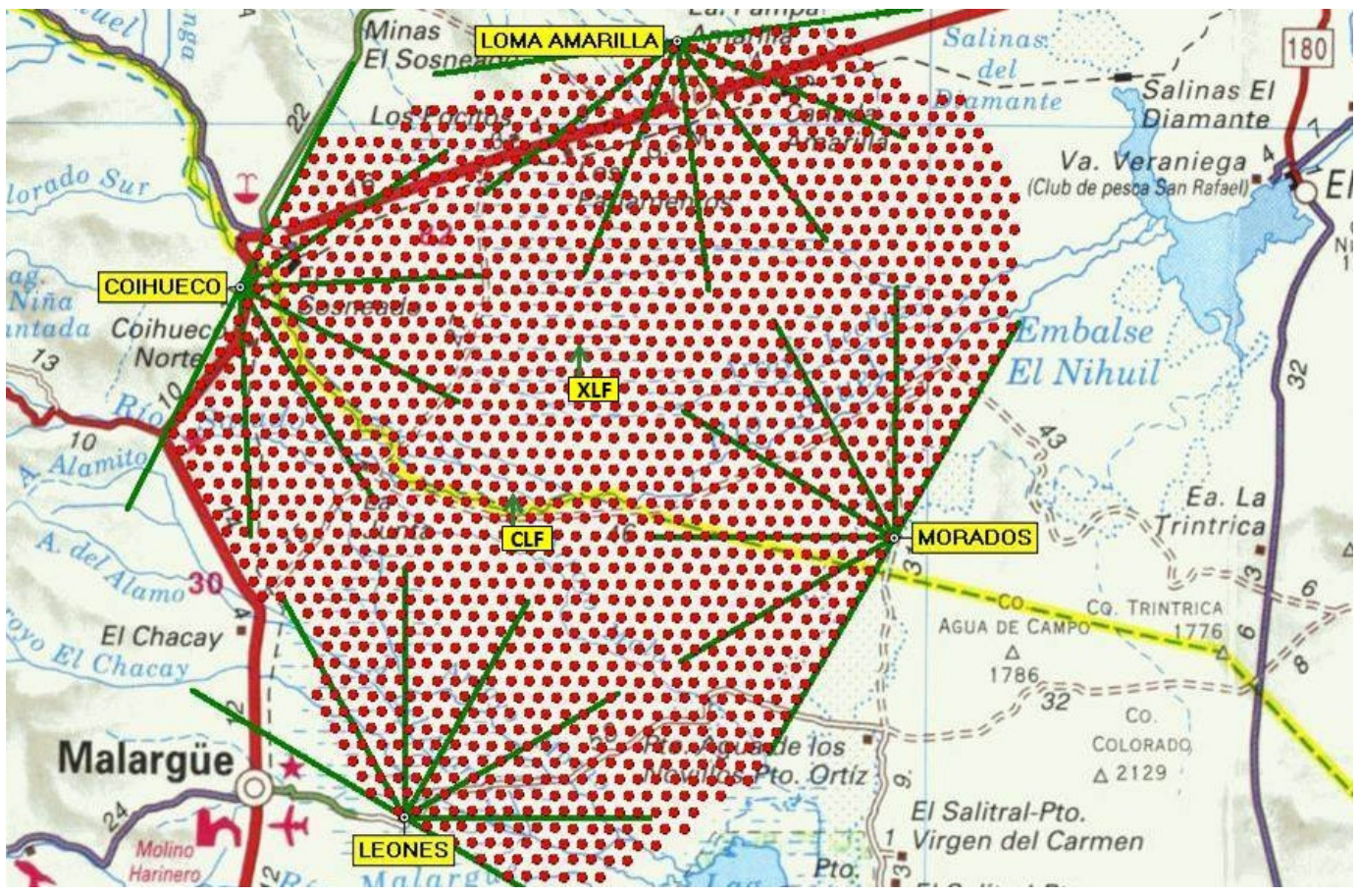

**Figure 1:** The Pierre Auger Observatory. Each dot corresponds to one of the 1660 surface detector stations. The four fluorescence detector enclosures are shown, each with the $30^\circ \times 30^\circ$ field of view of its six telescopes. The five lightning detector stations are located at the four fluorescence buildings and at the campus in Malargüe. From [1].

[1]http://www.boltek.com/product/stormpci-long-range-detection-kit

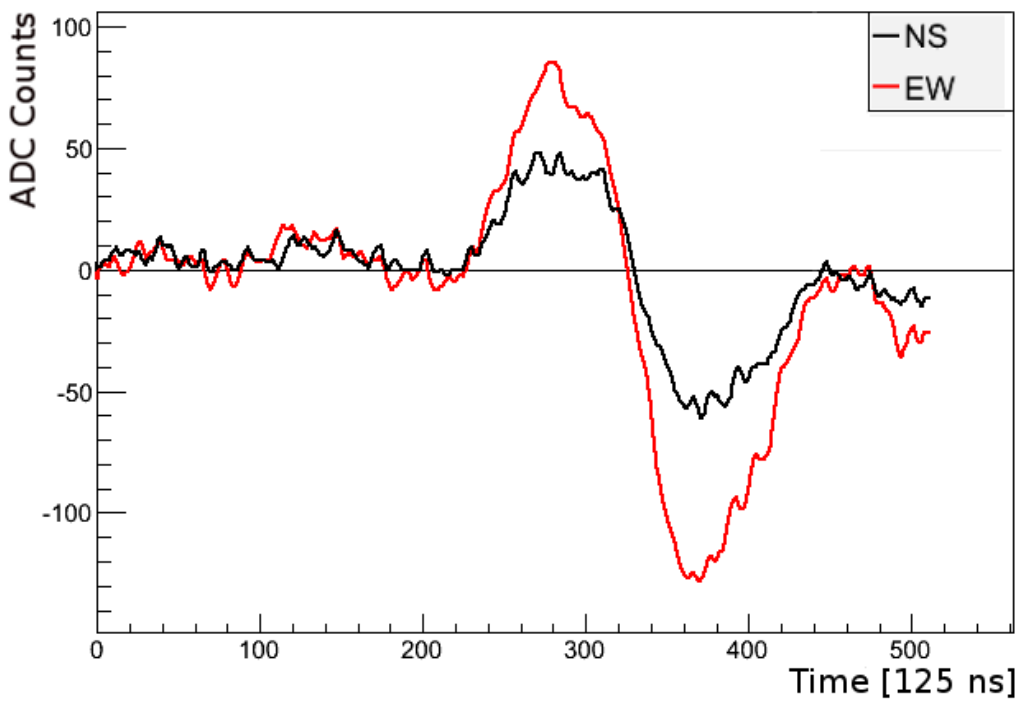

**Figure 2:** Example of the trace of a lightning event as detected with the Boltek StormTracker. The strong deviation from null indicates a lightning. The full length of the trace is 64 $\mu$s.

The PCI card is mounted in a small, low-power consumption PC providing the PCI-port. The data acquisition has been programmed interfacing the Boltek Linux development kit[2]. This allows direct access to the PCI-card and the readout of the full sampling trace measured for the two polarization directions North-South (NS) and East-West (EW). Each input is sampled with 8 MHz. The trace contains 512 samples in a 125 ns binning for a full length of 64 $\mu$s. The card uses a threshold-trigger to generate events. An example of the two traces corresponding to the two antenna polarizations is shown in Fig. 2.

An estimation of the direction of the lightning is done by each single detector via the fraction of the signal amplitude in both polarization directions. The distance itself is estimated from the amplitude, assuming an average lightning intensity and the amplitude to scale as the square of the distance. Traces with a frequency higher than 90 kHz are rejected as noise.

In addition to the simple position reconstruction using the polarization and amplitude of one lightning detection station (LDS) as shown in Fig. 3, the position of the lightning can be determined also from the timing information of several stations measuring the same lightning event. This coincident measurement of several stations significantly reduces the background of transient pulses mainly originating from locations close to one detector station. The necessary absolute time for the measured events is obtained by a GPS module (ublox LEA-6T[3], see Fig. 4). The GPS module receives directly the trigger information from the StormTracker card via a cable. The corresponding exact absolute time can be readout from the LDS via USB.

Five LDS have been deployed at the Pierre Auger Observatory, one at the Campus in Malargüe (MG), and the other ones at each of the Fluorescence Detector buildings indicated in Fig. 1: Los Leones (LL), Los Morados (LM), Coihueco (CO) and Loma Amarilla (LA).

[2]previously available at http://www.boltek.com

[3]www.u-blox.com/de/gps-mouldes/u-blox-6-timing-module/lea-6t.html

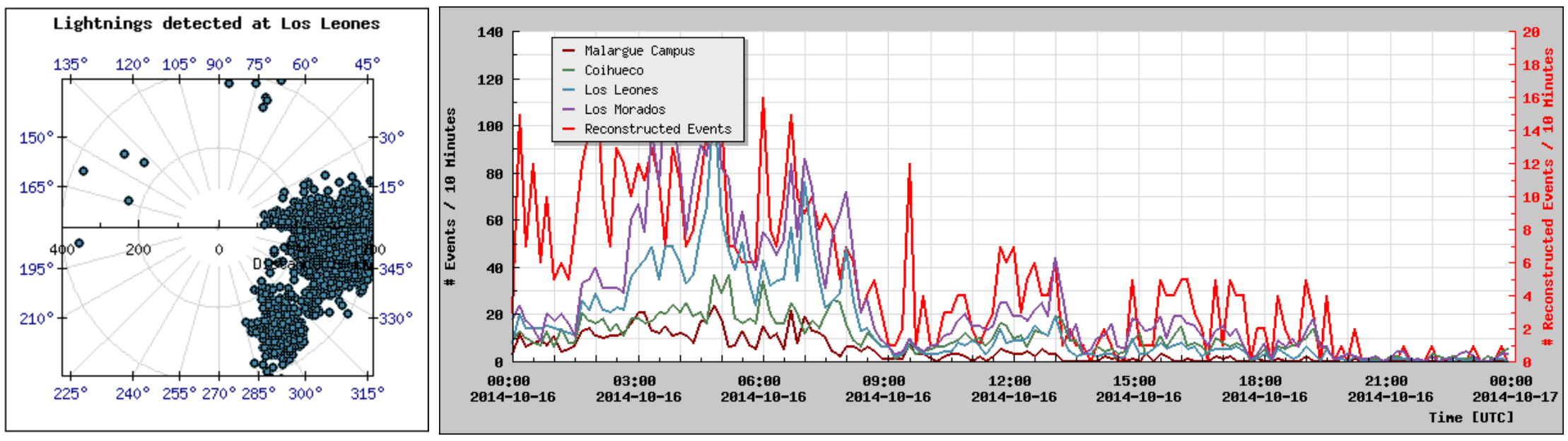

**Figure 3:** Left: Example of the data measured with one LDS (LL) for one day with a thunderstorm mainly east of Los Leones (16th October 2014). Right: Rates of the LDS and the reconstructed events (note the separate scale on the right).

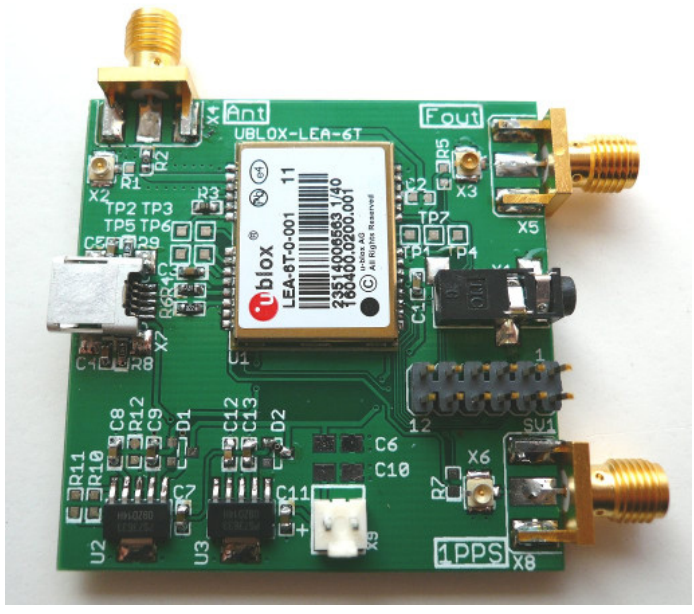

**Figure 4:** Picture of the GPS module integrated in the LDS and triggered by the lightning detection card.

## 3. Reconstruction of lightning events

The position of a lightning strike can be reconstructed using the absolute timing information of several stations. Considering $i$ LDS, the determination of time and place is overconstrained. It is estimated by fitting the time and place assuming the propagation with speed of light to the stations. Actually, the quantity minimized is the deviation w.r.t. one LDS, here LDS 0 is selected arbitrarily, which is compatible with minimizing to the mean time, using:

$$t_i = t + \sqrt{(x_i - x)^2 + (y_i - y)^2}/c. \tag{3.1}$$

The distance from each detector to the lightning position $s_i$ is described as:

$$s_i(x,y) = \sqrt{(x_i - x)^2 + (y_i - y)^2} = s_0(x,y) + (t_i - t_0) \cdot c. \tag{3.2}$$

Rearranging equation 3.2 yields $s_{0,i}$, the distance of the lightning event from detector 0 calculated using data from detector $i$:

$$s_{0,i}(x,y) = \sqrt{(x_i - x)^2 + (y_i - y)^2} - (t_i - t_0) \cdot c \tag{3.3}$$

Without any measurement uncertainties, $s_{0,i}$ is the same for all detectors. Since measurements do have uncertainties, the position of the lightning is estimated by minimization of

$$P(x',y') = \min\left(\sum_{i \neq j} (s_{0,i}(x',y') - s_{0,j}(x',y'))^2\right). \tag{3.4}$$

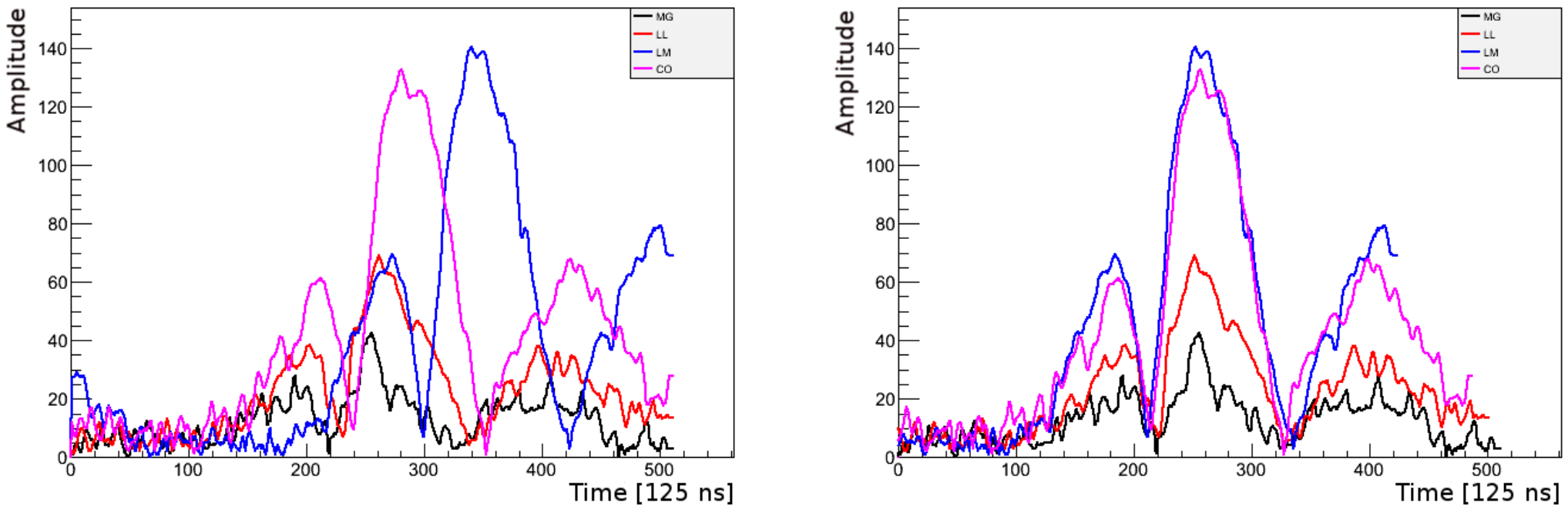

**Figure 5:** Example of a lightning event measured in four LDS (left). Smaller pre-peaks might exceed the threshold at a station resulting in a different reference time of the trace. The same event after shifting to maximize the cross-correlation product (right).

Important for the estimation of the lightning time and position is the exact reference time at the station. Here, a simple time of threshold is not sufficient, since lightnings often show several pulses (see example in Fig. 5), and depending on the distance, a pre-pulse can exceed the threshold while at other LDS a later pulse might be the first superating the threshold, giving different times of references for the lighting estimation. Therefore, the time of reference is estimated by maximizing the cross-correlation product, $CC(\mathrm{offset})$, of the full traces:

$$CC(\mathrm{offset}) = \max\left(\sum_j (S_{0,j} \cdot S_{i,j+\mathrm{offset}})\right) \tag{3.5}$$

$$\Delta t_i = \mathrm{offset} \cdot 125\,\mathrm{ns}. \tag{3.6}$$

The starting-time of each detector is corrected by adding the time offset $\Delta t_i$ to the start-time of the trace for the $i$th LDS. The resulting optimal shifted traces are also shown in Fig. 5. In this way, the optimal time estimation for the different LDS independent of the amplitude of the lightning-signal is achieved.

## 4. Estimation of the reconstruction accuracy

To test the accuracy of the reconstruction method we simulated lightning events. The timing of single stations is smeared by a Gaussian distribution with a width of 30 ns corresponding to the absolute timing uncertainty of the GPS. The events are then reconstructed using the same algorithm. The positions of the LDS are given by the buildings of the fluorescence detectors of the Pierre Auger Observatory and the campus building. These positions are not optimally distributed, so we expect regions and directions with better and worse accuracy. In Fig. 6, the average distance between the simulated and the reconstructed position is shown for the simulated positions. Especially outside of the array, the estimation of the direction and distance are clearly different, making a polar representation more useful, as given in Fig. 7. As can be seen, the accuracy is usually about 2.5% in distance and 0.06° in direction. For lightning strikes over the Pierre Auger Observatory, the position is determined typically better than 100 m.

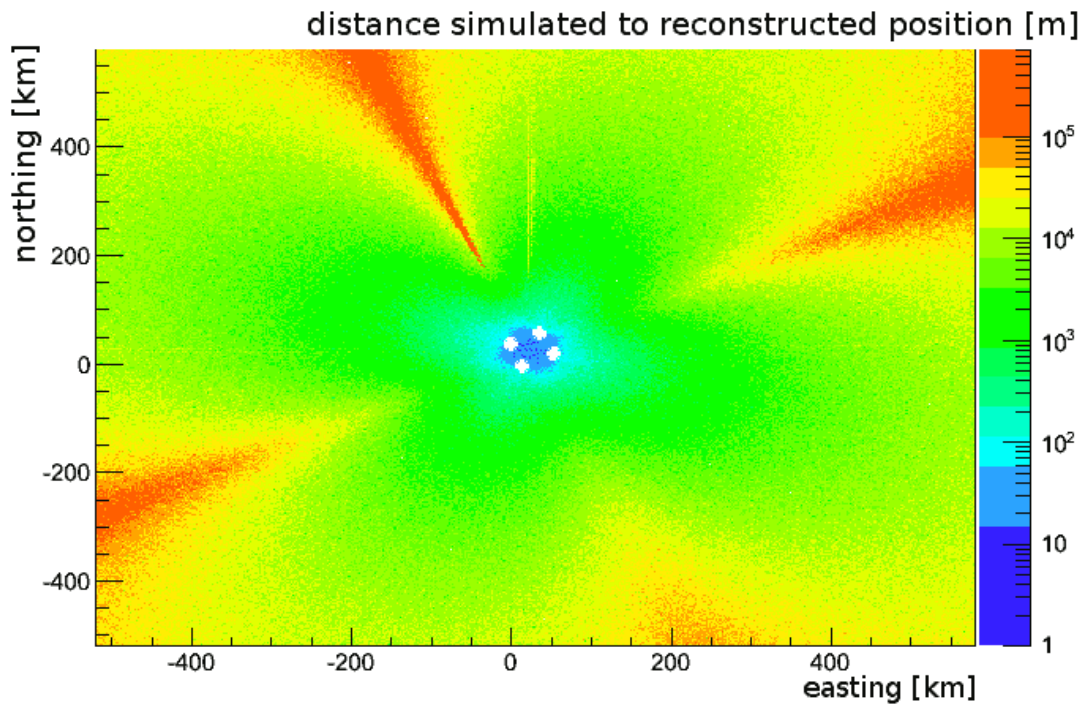

**Figure 6:** Average distance in meter between the simulated and the reconstructed position at the simulated position on the ground for simulated lightning events smeared according to the absolute time accuracy of the GPS modules. The four LDS used in this simulation are indicated by the white circles, which give a scale for the size of the observatory in this plot.

## 5. Comparison with WWLLN data

The *WorldWide Lightning Location network* (WWLLN) provides realtime locations of cloud-to-ground lightning discharges [8]. Currently, more than 50 sensors [9] around the globe detect spherical activity in the VLF band. The accuracy of the network was originally determined to be about 30 km [10], and has been improved to 10 km for regions with higher sensor density. For a test of the lightning detection system at the Pierre Auger Observatory, one month of WWLLN data with lightnings occurring within 1000 km of the array center have been provided by the WWLLN. For the matching candidates, the difference in distance and direction are given in Fig. 8.

## 6. Prospects for cosmic ray measurement

The additional lightning detector system at the Pierre Auger Observatory offers the possibility to study a correlation of lightnings with cosmic-ray showers. This has been an active research field in the last decade. The passage of highly localized distributions of electrons and positrons can create a region with high fractal ion conductivity in the thundercloud, which may enhance the probability to initiate the ground discharge. With the lightning system described in the previous sections, we have an accurate measurement of lightnings that can be combined with the cosmic ray measurement of the Pierre Auger Observatory at the same location.

The triggering of the surface detector of the Pierre Auger Observatory is based on a coincidence of three individual surface detector stations, since this is the minimal required measurement for a reconstruction of an air shower. Alternatively, even a single station is read out in case of a trigger based on the fluorescence detector measurement for a hybrid reconstruction. Due to the spacing of 1.5 km between the surface detectors, the energy threshold for air shower measurements is $3 \times 10^{18}$ eV for zenith angles smaller than $60°$. Lightnings are by far more frequent than cosmic-ray showers with energies above $10^{18.5}$ eV. Therefore, a correlation has to be searched for air shower below energies of $10^{18.5}$ eV, where the shower is too small to hit more than one surface detector. The idea is to use the lightning detection to trigger the read out of individual surface detector stations, in analogy of the hybrid trigger initiated by the fluorescence detector. Yet, the bandwidth to read out additional events is limited. The trigger-condition for the read out of a single station has to be restricted to a maximum window in time and space around the lightning detection that still can be transmitted by the communication system without disturbing the standard operation of the surface detector.

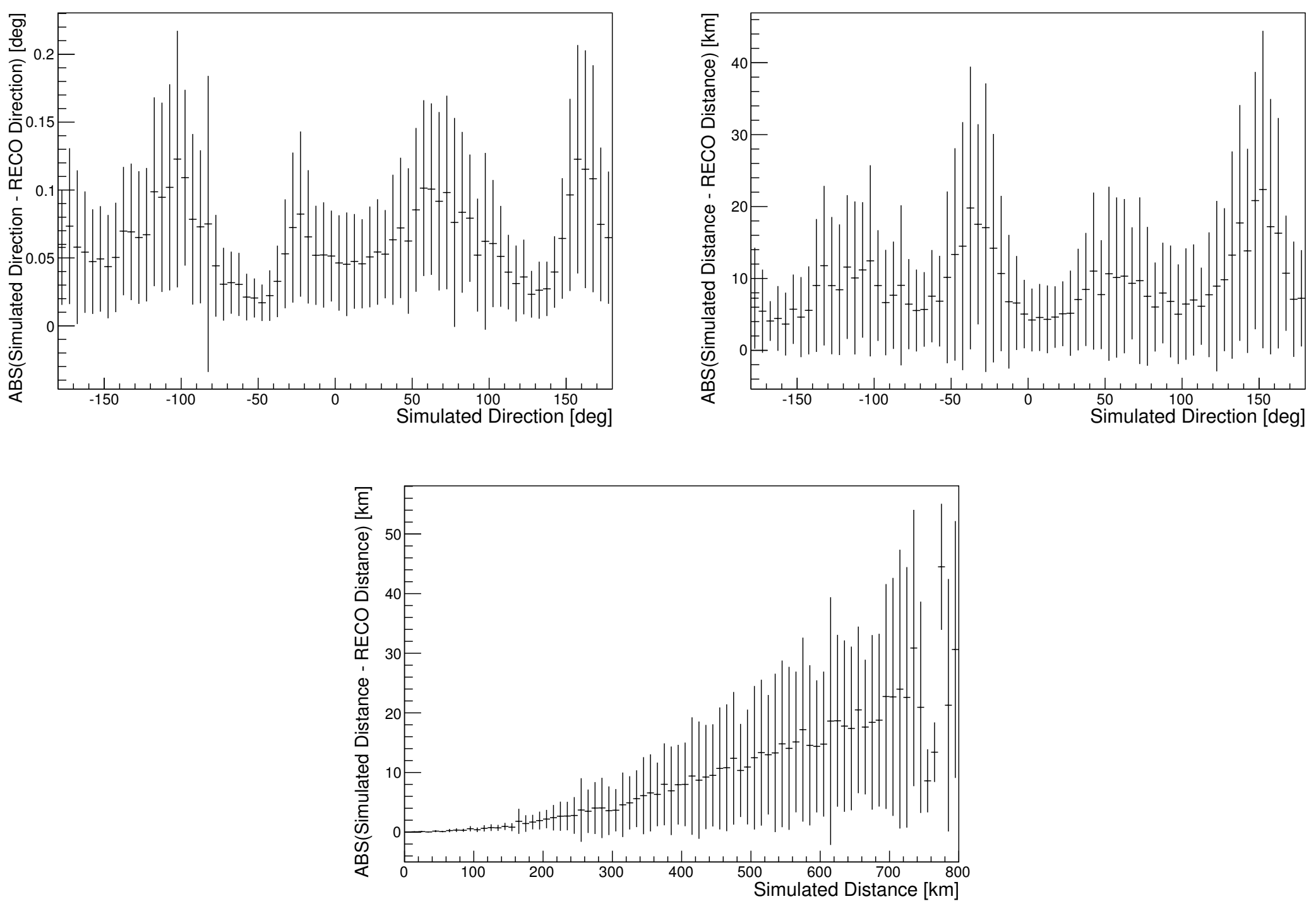

**Figure 7:** Reconstruction resolution for simulated lightning events smeared according to the absolute time accuracy of the GPS modules. Shown is the difference between reconstructed and simulated direction (top-left) and distance (top-right) to the array center vs the azimuth of the direction, as well as the difference in the distance vs the distance (bottom). The error-bars indicate the spread of the underlying distributions.

In addition, AERA itself is capable to measure lightning strikes with high precision due to the higher sampling rate. The operation of AERA to record lightnings is difficult in parallel to the normal cosmic ray detection, though.

## 7. Summary

The radio detection of air showers requires the measurement of the environmental electric field using a E-field mill. This allows to identify data periods affected by thunderstorms. An additional lightning detection system has been installed at the Pierre Auger Observatory to enhance the spatial and time resolution of detected lightnings. Their reconstruction is optimized using the cross-correlation of the measured signals, achieving typically 2.5% resolution in distance and $0.06^\circ$ in direction. This offers possible correlation studies between lightnings and cosmic rays. By providing a lightning trigger information to the data acquisition system of the surface detector, the threshold for a coincident detection of lightning and cosmic ray will be further reduced.

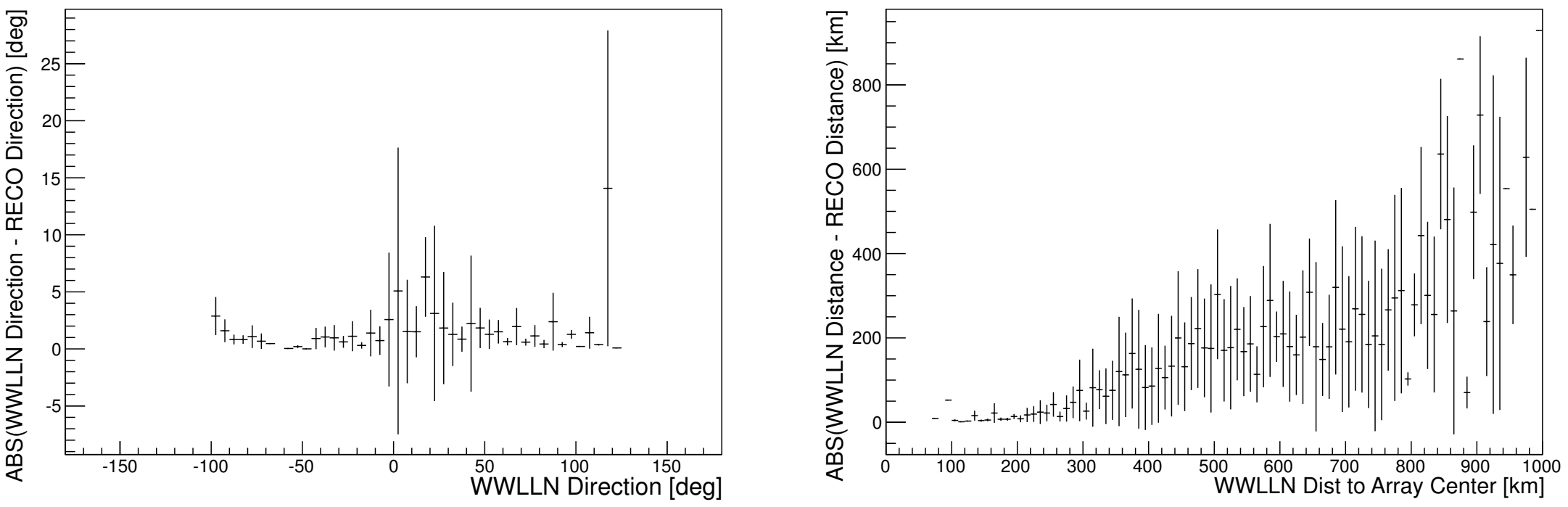

**Figure 8:** Comparison between the position measured with the lightning detection system installed at the Pierre Auger Observatory with the one given by the WWLLN for a sample of lightnings in October 2014. The error-bars indicate the spread of the underlying distributions.

## 8. Acknowledgement

The authors wish to thank the World Wide Lightning Location Network (http://wwlln.net), a collaboration among more than 50 universities and institutions, for providing the lightning location data used in this paper.

# Studies in the atmospheric monitoring at the Pierre Auger Observatory using the upgraded Central Laser Facility

**Carlos Medina-Hernandez*[a] for Pierre Auger Collaboration[b]**
[a]*Department of Physics, Colorado School of Mines, Golden CO, USA.*
[b]*Observatorio Pierre Auger, Av. San Martín Norte 304, 5613 Malargüe, Argentina*
*E-mail:* auger_spokespersons@fnal.gov
*Full author list:* http://www.auger.org/archive/authors_2015_06.html

The Fluorescence Detector (FD) at the Pierre Auger Observatory measures the intensity of the scattered light from laser tracks generated by the Central Laser Facility (CLF) and the eXtreme Laser Facility (XLF) to monitor and estimate the vertical aerosol optical depth ($\tau(z,t)$). This measurement is needed to obtain unbiased and reliable FD measurements of the arrival direction and energy of the primary cosmic ray, and the depth of the maximum shower development. The CLF was upgraded substantially in 2013 with the addition of a solid state laser, new generation GPS, a robotic beam calibration system, better thermal and dust isolation, and improved software. The upgrade also included a back-scatter Raman LIDAR to measure $\tau(z,t)$. The new features and applications of the upgraded instrument are described. These include the laser energy calibration and the atmospheric monitoring measurements. The first $\tau(z,t)$ results and comparisons after the upgrade are presented using different methods. The first method compares the FD hourly response to the scattered light from the CLF (or XLF) against a reference hourly profile measured during a clear night where zero aerosol contents are assumed. The second method simulates FD responses with different atmospheric parameters and selects the parameters that provide the best fit to the actual FD response. A third method uses the new Raman LIDAR receiver in-situ to measure the back-scatter light from the CLF laser. The results show a good data agreement for the first and second methods using FD stations located at the same distance from the facilities. Preliminary results of $\tau(z,t)$ using the Raman LIDAR are presented as well.

*Speaker.

## 1. Introduction

The Pierre Auger Observatory, located in Malargüe, Argentina, explores the nature and origin of Ultra High Energy Cosmic Rays (UHECRs) with energies above $10^{18}$ eV [1]. The observatory (Figure 1) combines a Fluorescence Detector (FD) and a Surface Detector array (SD) in a hybrid technique. The FD consists of 27 telescopes arranged at four sites along the perimeter of the observatory [2]. The SD is a collection of 1660 water-Cherenkov detectors, placed on a triangular grid with a separation of 1.5 km and covering an area of 3000 km$^2$. Two laser facilities are located near the center of the SD array. Each facility is nearly equidistant from three of the four FD stations.

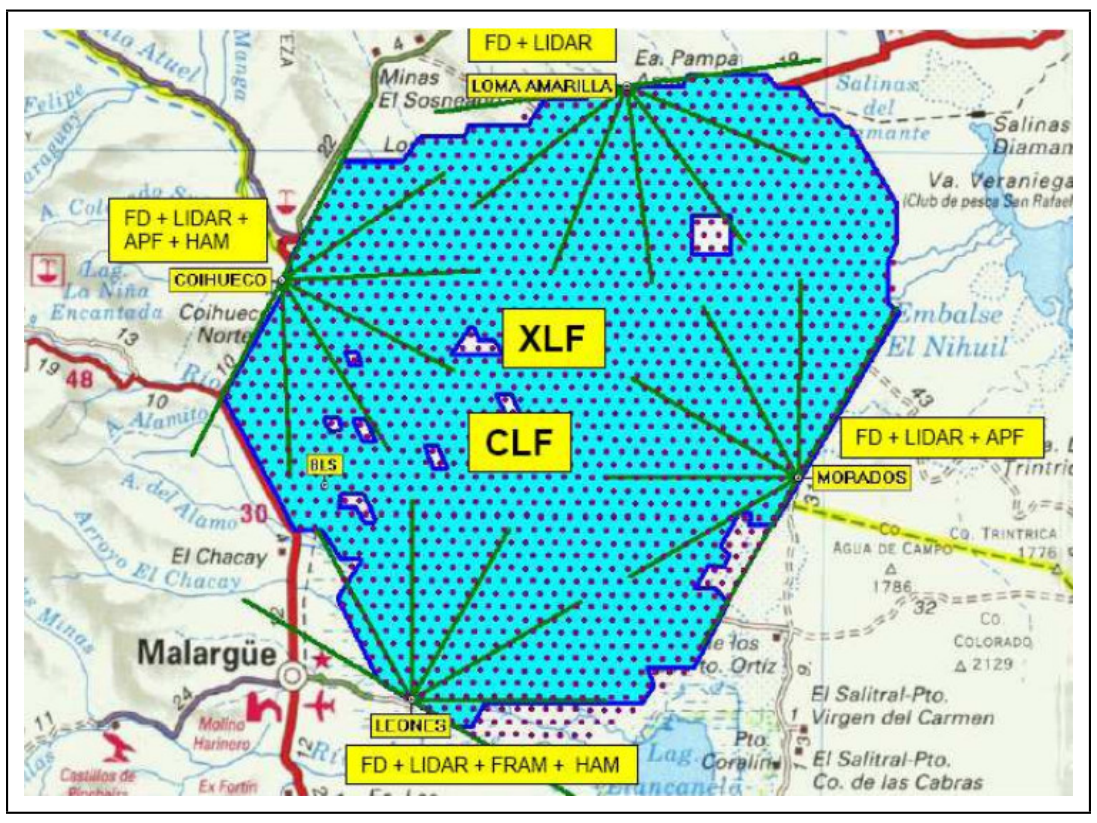

**Figure 1:** The configuration of the Pierre Auger Observatory.

UHECRs entering the atmosphere induce Extensive Air Showers (EASs). The FD uses the atmosphere as a giant calorimeter in which an EAS deposits most of its energy. The reliability of the reconstruction of any EAS requires an accurate description of the atmospheric conditions at the time of detection. The atmospheric description includes the measurement of the vertical aerosol optical depth for different heights and times $\tau(z,t)$ during FD operations. For this purpose, the FD records tracks from the laser pulses generated by the Central Laser Facility (CLF) and the eXtreme Laser Facility (XLF). The CLF and XLF have been operational since 2004 and 2008, respectively. In early 2013, the CLF was upgraded with a Raman LIDAR receiver that measures $\tau(z,t)$ independently.

## 2. The CLF and XLF

The CLF and XLF systems [3] generate 7 ns laser pulses at a nominal energy of 5 mJ and 355 nm wavelength, near the prominent 357 nm band of ultraviolet fluorescence in the air [4]. The laser tracks are distinguished from regular air shower events, via GPS timing identification. For atmospheric monitoring during FD operations, the CLF and XLF fire sets of 50 laser pulses, every 15 minutes, in the vertical direction. Other laser pulses are also fired between the atmospheric laser sets. For example, steered shots are directed toward astronomical objects of interest. The FD response to the scattered light from the fixed vertical laser-pulses at each FD site can be compared either with similar measurements recorded during a night free of aerosols, or with simulations of data under different aerosols conditions. These comparisons are used to calculate $\tau(z,t)$.

Both laser facilities are controlled by Single Board Computers (SBC) that can be operated and programmed remotely. These facilities are powered by solar panels that charge a battery bank. Local data produced during a run are transferred daily to a remote server.

### 2.1 The upgraded CLF

After nearly a decade of service, the CLF [5] was substantially upgraded during the first half of 2013 (Figure 2):

- A back-scatter Raman LIDAR receiver was installed.
- The original flash lamp pumped laser was replaced by a solid state laser.
- A newer GPS clock system improves the timing resolution from 100 to 20 ns.
- The original 20 ft shipping container was replaced by a newer 40 ft unit with tighter doors and better insulation. A 2000 liter thermal reservoir coupled to the optical table was added to reduce thermal variations.
- The better sealed container features a separate room for the laser system to reduce the dust accumulation. This is important because dust accumulation on optical components increases the systematic uncertainty of laser energy delivered into the sky.
- A robotic system for automatic energy and polarization calibrations was added.

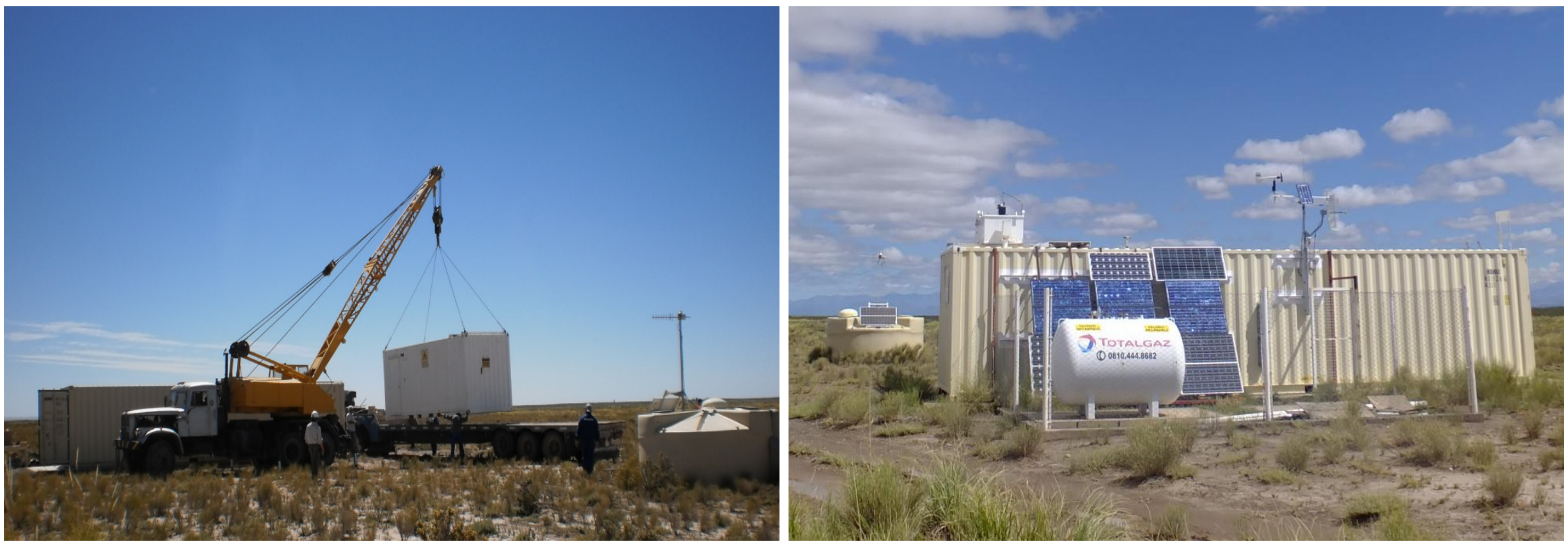

**Figure 2:** (Left) Replacing the old CLF container. (Right) The CLF after the upgrade was completed.

### 2.2 The energy and polarization calibration system

The relative energy of every laser shot is measured by a monitoring energy probe that collects a small percentage of the total laser shot energy ($E_{\text{monitoring}}$). A second probe is temporarily positioned over the beam, for absolute energy calibrations ($E_{\text{calibration}}$). A Calibration Factor ($CF$) is calculated by averaging the ratio of the calibration and monitoring probes for 13 shots:

$$CF = \frac{1}{13} \cdot \sum_{i=1}^{13} \left( \frac{E_{\text{calibration}}}{E_{\text{monitoring}}} \right). \tag{2.1}$$

The calibration system provides *CF* measurements as frequently as needed for vertical and steered modes (Figure 3). The system also includes a second energy probe for cross calibrations. An identical calibration system was installed in the XLF at the time of its construction.

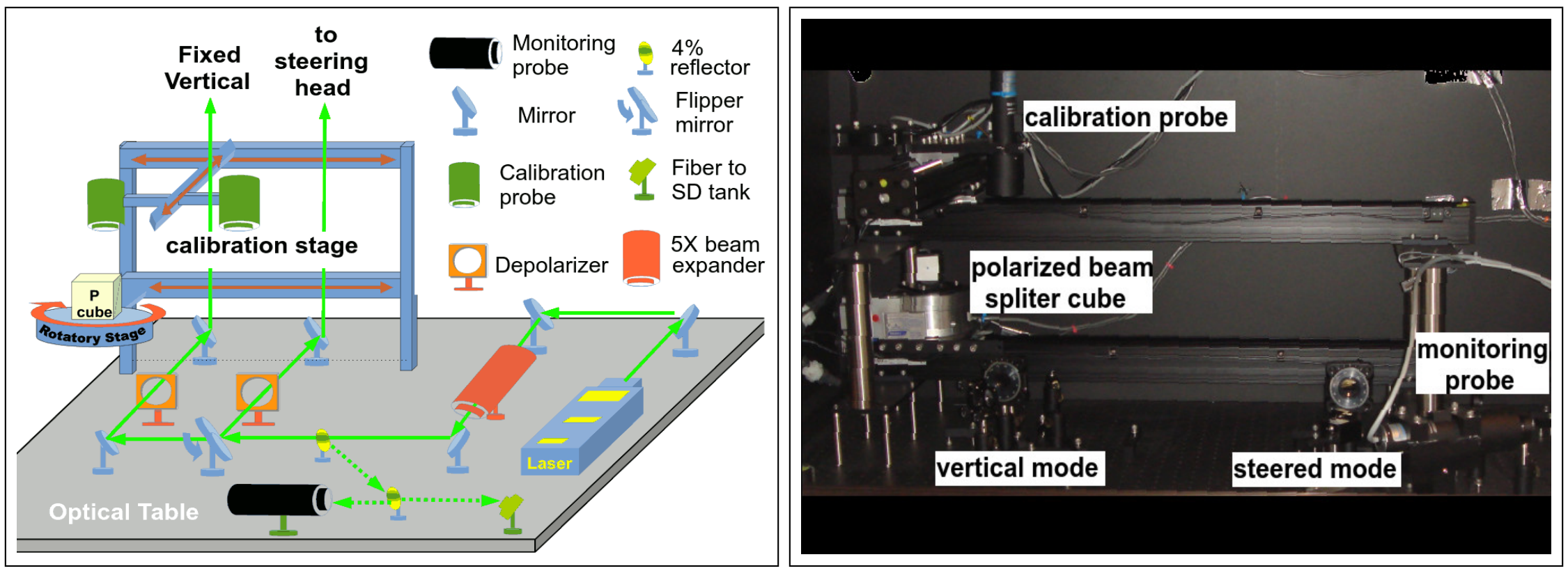

**Figure 3:** The calibration stage. (Left) A schematic diagram of the calibration system components on the optical table. (Right) The calibration system as fabricated.

Before the upgrade, absolute calibrations were performed manually at time intervals that ranged from a couple of months to one year. The available *CF* measurements were linearly fit to provide a calibration function for different epochs. Epochs correspond to dates where the hardware was changed or the optics were cleaned. The calibration function is corrected to account for the amount of light captured by the probe that is not delivered into the sky, because of scattering off dust that had accumulated on the tilted mirror. The *CF* distributions before and after the upgrade (Figure 4) demonstrate the improvement in the stability of the laser energy delivered to the sky and the frequency of the *CF* measurements.

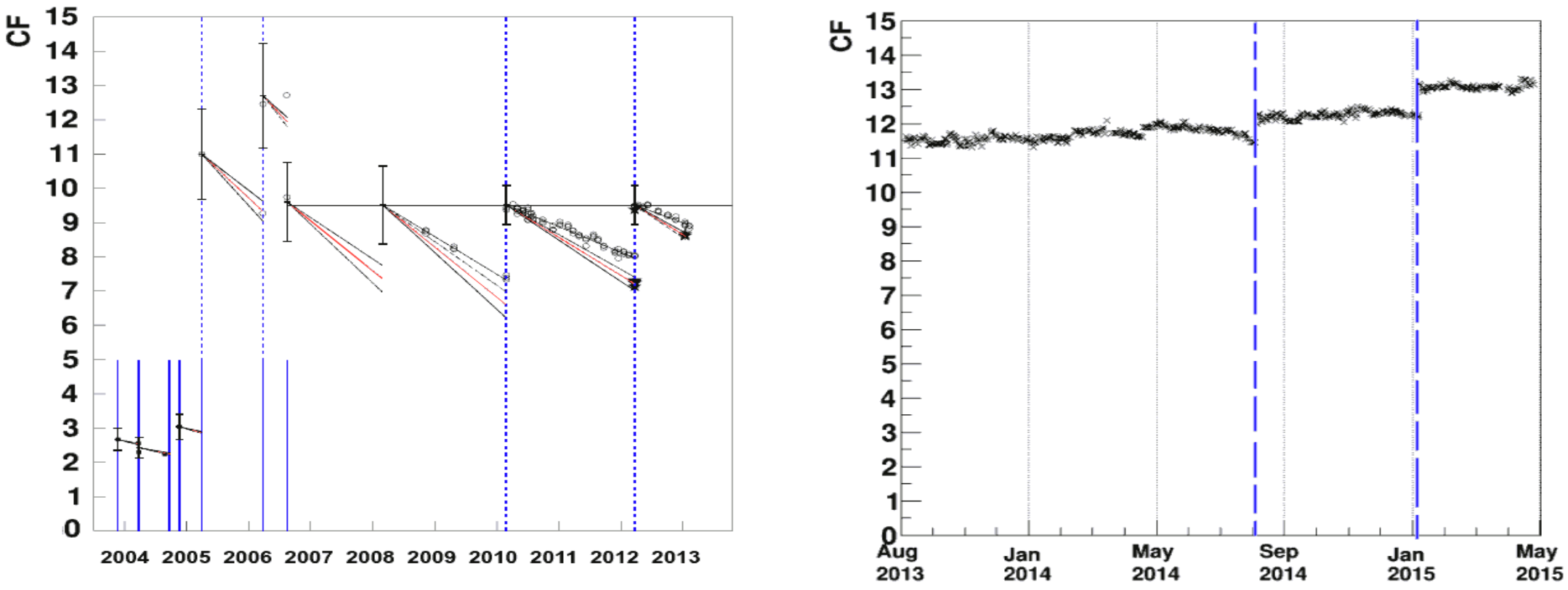

**Figure 4:** (Left, pre-ugrade) *CF* are shown in circles. Black lines represent fit functions including corrections and blue lines represent epochs. (Right, post-upgrade) *CF* are measured every day. Calibration functions and fits are no longer required.

A laser beam without net polarization is preferred because the atmospheric scattering is symmetric to the beam axis. To achieve this, the CLF uses polarization randomizers. The calibration system measures the net polarization, by rotating a polarization beam-splittering cube in front of the calibration probes, and by measuring the $CF$. The data points are fit to the function:

$$P_0 + (P_1 \cdot sin(\theta + P_2)). \tag{2.2}$$

The energy variations due to polarization for 6 months of data are presented in Figure 5 (left). The fitting function is presented in Figure 5 (right) in polar coordinates and overlapping a circle of radius $P_0$ that represents a beam without polarization. The maximum deviation of the fitted line from the circle at angles facing the four FD sites is no larger than 1%.

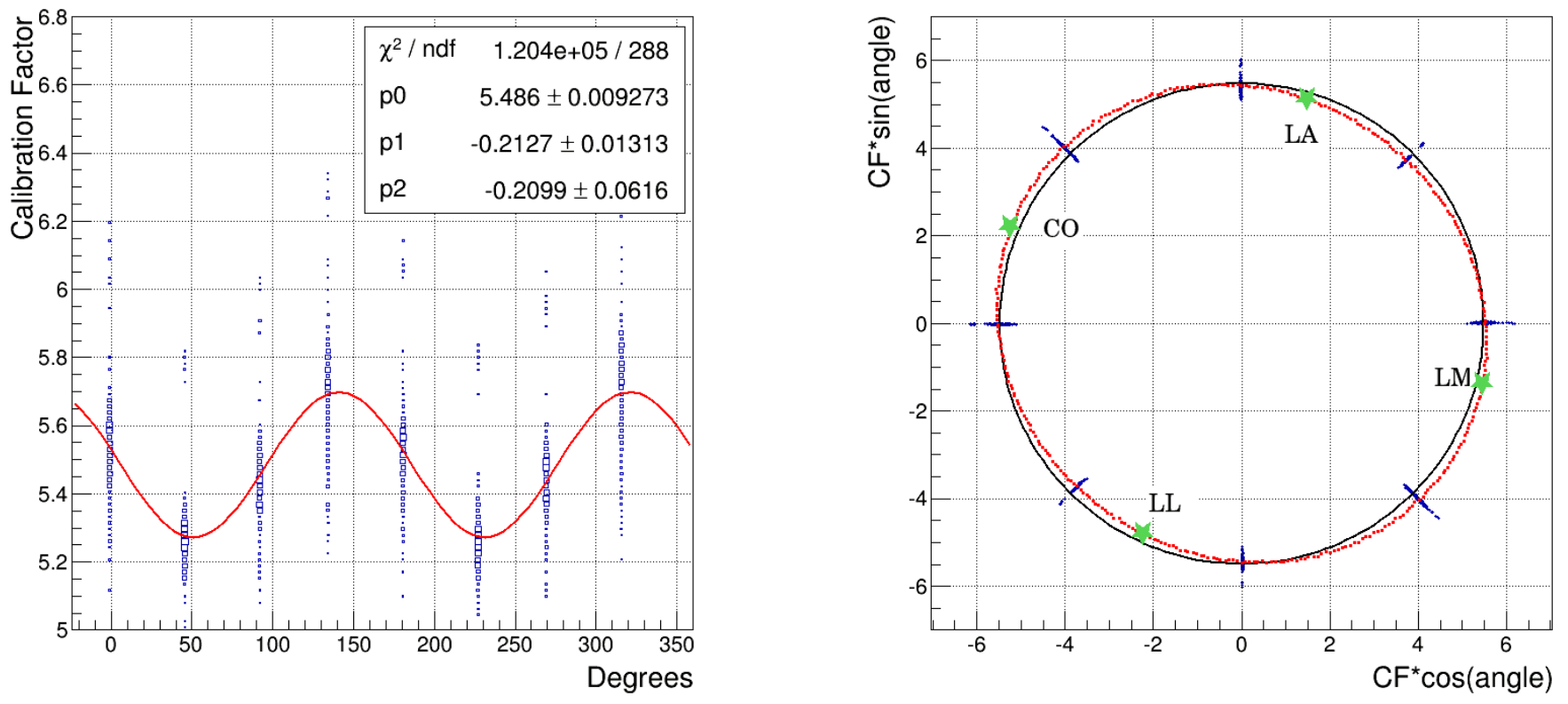

**Figure 5:** (Left) Six months of polarization calibration data. Blue points represent energy measurements at eight different rotation angles of the polarization analyzer. The red line represents the fit function. (Right) Data and fit line are presented in polar coordinates. The black circle of radius $P_0$ represents a beam without net polarization. The four green stars represent the directions toward the four FD sites.

## 2.3 The Raman LIDAR

The Raman LIDAR measures the back-scatter light from the laser to obtain $\tau(z,t)$ independently [6]. The system has three channels: one for elastic scattering, one for Raman scattering off $N_2$ molecules, and one for Raman scattering off water vapor (Figure 6).

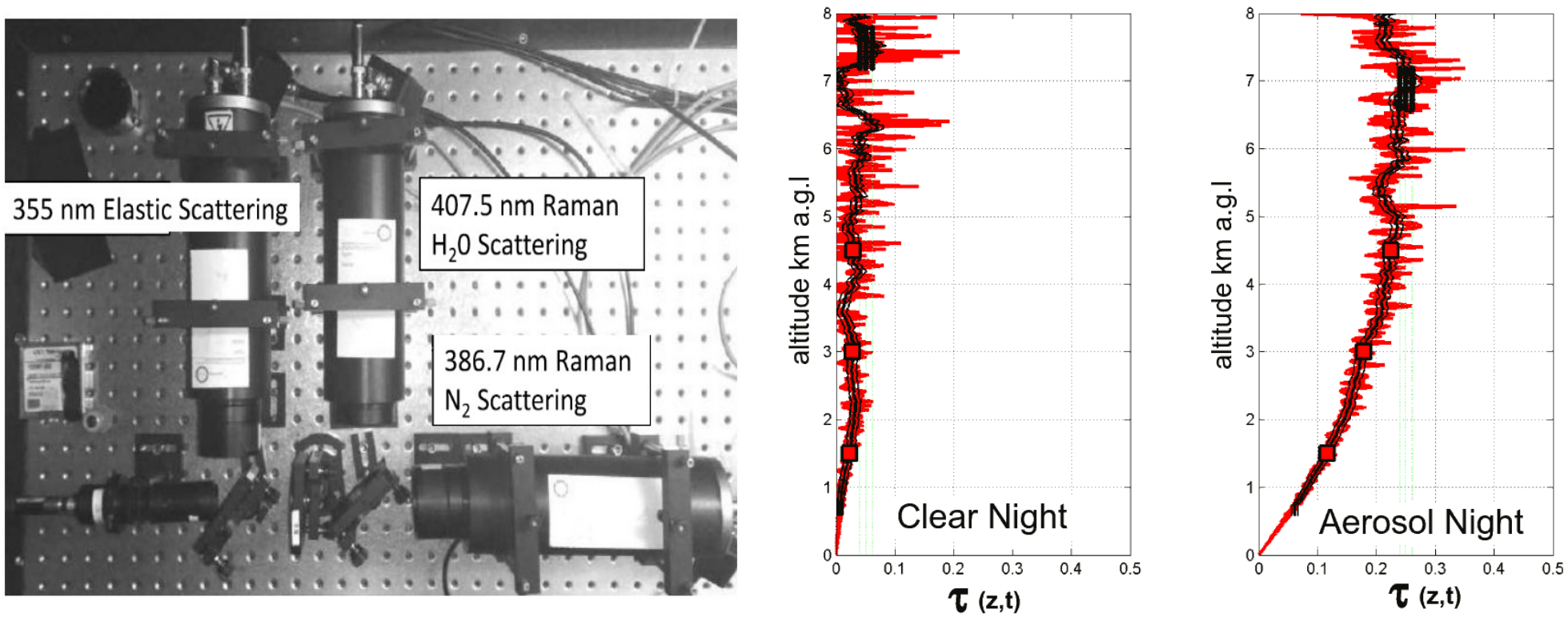

**Figure 6:** (Left) A picture of the Raman LIDAR optical components. (Right) Examples of hourly $\tau(z,t)$ during a clear and a typical night with aerosols.

The low cross section of Raman scattering requires a relatively long laser exposure to reduce statistical uncertainties. The Raman LIDAR runs take about 24 minutes. Shorter and longer runs can also be programmed. The Raman LIDAR has been running nightly before and after every FD shift since November 2013. To increase the number of $\tau(z,t)$ measurements, an additional run was added in the middle of the FD shift beginning November 2014. During this additional Raman run, the FD bays overlooking the CLF are closed to avoid saturating the FD photomultipliers.

## 3. Measurements and comparisons

The $\tau(z,t)$ profiles are calculated using a Data Normalized (DN) method. This method compares the FD response averaged over 200 laser shots (1 hour) with the FD response to 200 laser shots recorded during a clear, aerosol free, reference hour. Each average is normalized to the laser energy on a shot-by-shot basis. The ratio is used to obtain $\tau(z,t)$ with a 400 m and a 1 h resolution. The reference hour is selected by scanning the highest hourly FD response during a one-year period. Most reference hours are found during winter periods. A FD response comparison between the 2014 reference hour (found on July 7th) and a typical aerosol-content-hour is shown in Figure 7 (left). The $\tau(z,t)$ output of this comparison using the DN analysis is shown in Figure 7 (right).

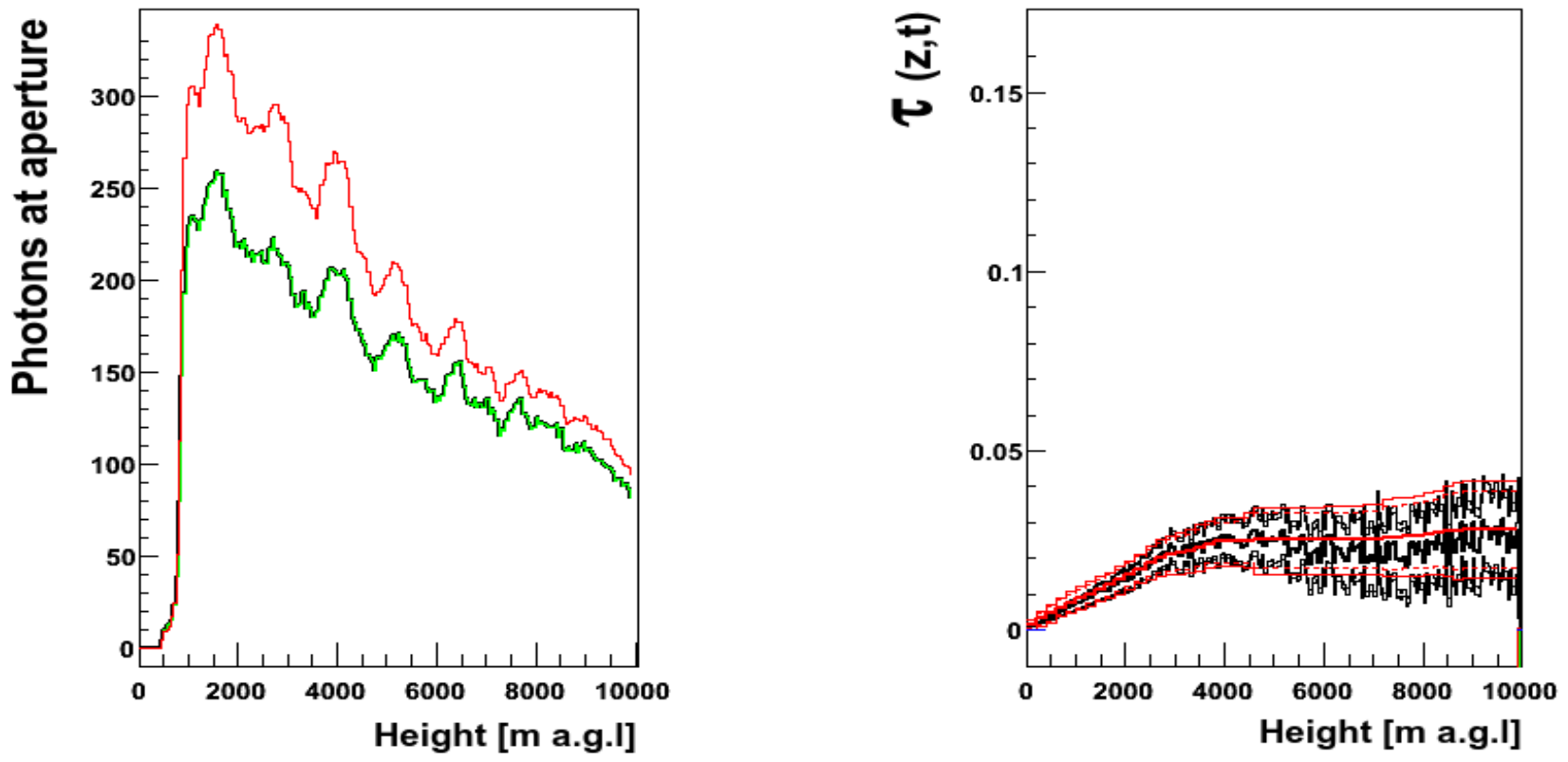

**Figure 7:** (Left) The green line represents the FD response in an hour with typical aerosol content and the red line represents extremely low aerosol content. (Right) The black traces represent $\tau(z,t)$ and its uncertainties for an hour with typical aerosol content. The red traces are the fitted values.

The DN analysis uses the CLF for FD sites Los Leones, Los Morados and Coihueco, and uses the XLF for the FD site Loma Amarilla. This ensures that the scattered photons travel similar distances from the laser source to the FD sites. The following analysis will focus on results obtained with the upgraded CLF.

A comparison of $\tau(z = 3\,\text{km})$ values obtained with the DN analysis at the same hour is presented in Figure 8 (left-center). The DN method presents good correlation for the FD sites Coihueco and Los Morados located at similar distances from the laser source. Similar results are found at different altitudes. Most outlying points are caused by scattering from clouds.

$\tau(z,t)$ can be obtained using a second method called Laser Simulation (LS) analysis. This method compares the observed FD response to a set of simulated responses that are generated with different atmospheric parameters. The best fit to the actual FD response is selected. The LS method is used to fill in values of $\tau(z,t)$ in the database that for various reasons are not obtained with the DN method. The details of the DN and the LS methods are described in [7]. A good correlation between the two methods using the CLF after the upgrade is observed (Figure 8(right)). The level of consistency between the two methods is the same as it was before the upgrade [8]. The $\tau(z,t)$ values from both analyses are merged and registered into the observatory database for use in the reconstruction analysis of cosmic ray events [7].

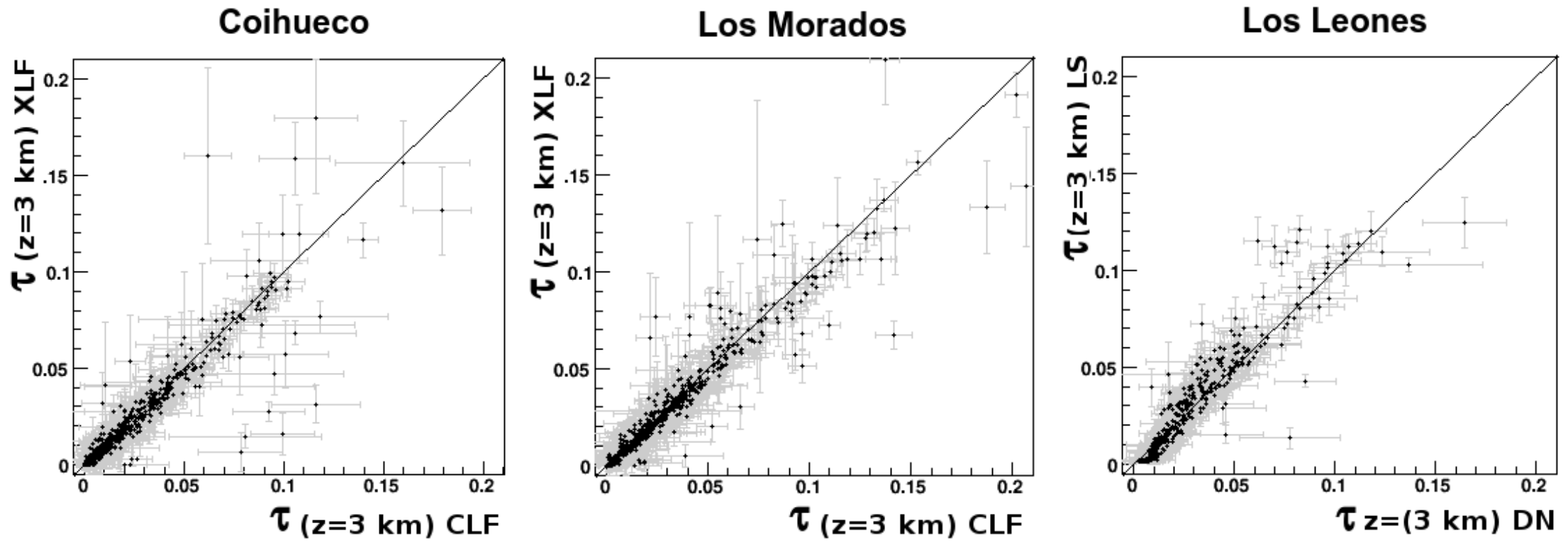

**Figure 8:** (Left and center) $\tau(z=3\,\mathrm{km})$ correlations in 2014 between the CLF and XLF using the DN analysis. (Right) $\tau(z=3\,\mathrm{km})$ correlations in 2014 between the DN and the LS analysis using the CLF.

A preliminary analysis of the Raman LIDAR measurements was used to obtain monthly averages of $\tau(z,t)$ at different heights. An example of the $\tau(z=3\,\mathrm{km})$ monthly average is presented in Figure 9. The data include one measurements of $\tau(z,t)$ every night during 2014. Seasonal variations in the aerosol optical depth are observed.

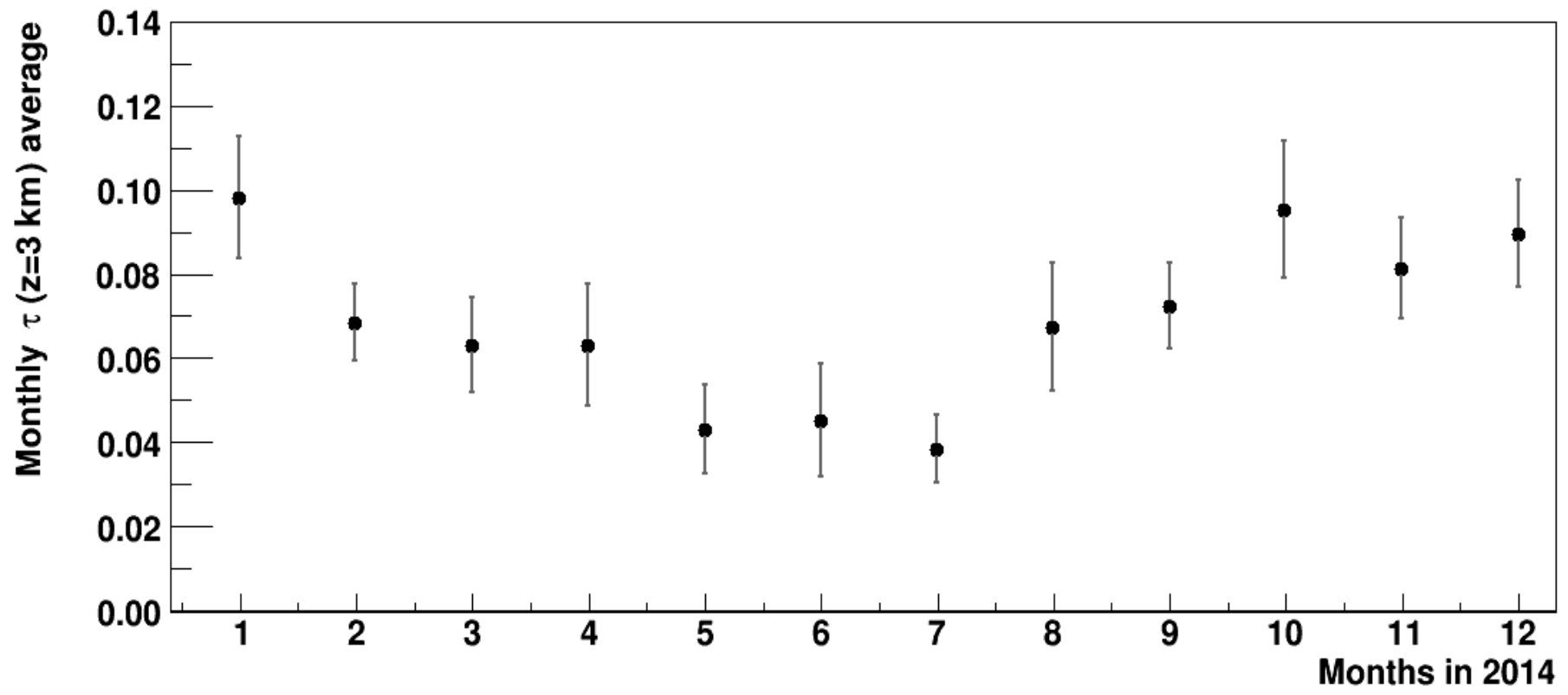

**Figure 9:** Preliminary monthly $\tau(z=3\,\mathrm{km})$ averages measured with the Raman LIDAR. The error bars represent one sigma of statistical uncertainties.

Statistical uncertainties in the Raman LIDAR measurements are due to the small Raman cross section in combination with the limited laser exposure (24 minutes or less). Systematic uncertainties are being evaluated. This includes an examination of multiple scattering effects, an examination of the aerosol content during the reference hour and the use of a longer Raman run.

## 4. Conclusions

The CLF has been upgraded to improve performance and add capabilities. The Raman LIDAR at the CLF is the first of its kind used for atmospheric measurements of $\tau(z,t)$ in an UHECR experiment. The DN analysis provides consistent results for equidistant FD telescopes. The DN analysis results are consistent with the LS analysis. The Raman LIDAR has been used to measure $\tau(z,t)$ every night for a full year. A preliminary analysis of the Raman LIDAR has been used to obtain averaged monthly values of $\tau(z,t)$. Ongoing studies will continue to evaluate the systematic uncertainties and possible physics implications.

# 6

# Detectors

# Upgrade of the Pierre Auger Observatory (AugerPrime)

**Ralph Engel**[*a] **for the Pierre Auger Collaboration**[b]
[a]*Karlsruhe Institute of Technology (KIT), 76021 Karlsruhe, Germany*
[b]*Observatorio Pierre Auger, Av. San Martín Norte 304, 5613 Malargüe, Argentina*
*E-mail:* auger_spokespersons@fnal.gov
*Full author list:* http://www.auger.org/archive/authors_2015_06.html

The data collected with the Pierre Auger Observatory have led to a number of surprising discoveries. While a strong suppression of the particle flux at the highest energies has been established unambiguously, the dominant physics processes related to this suppression could not be identified. Within the energy range covered by fluorescence detector observations with sufficient statistics, an unexpected change of the depth of maximum distribution is found. Using LHC-tuned interaction models these observations can be understood as a correlated change of the fluxes of different mass groups. On the other hand, they could also indicate a change of hadronic interactions above the energy of the ankle. Complementing the water Cherenkov detectors of the surface array with scintillator detectors will, mainly through the determination of the muonic shower component, extend the composition sensitivity of the Auger Observatory into the flux suppression region. The upgrade of the Auger Observatory will allow us to estimate the primary mass of the highest energy cosmic rays on a shower-by-shower basis. In addition to measuring the mass composition the upgrade will open the possibility to search for light primaries at the highest energies, to perform composition-selected anisotropy studies, and to search for new phenomena including unexpected changes of hadronic interactions. After introducing the physics motivation for upgrading the Auger Observatory the planned detector upgrade is presented. In the second part of the contribution the expected performance and improved physics sensitivity of the upgraded Auger Observatory are discussed.

*Speaker.

## 1. Introduction

Measurements of the Auger Observatory [1] have dramatically advanced our understanding of ultra-high energy cosmic rays (UHECRs). Particularly exciting is the observed behavior of the depth of shower maximum with energy, which changes in an unexpected, non-trivial way. Around $3\times10^{18}$ eV it shows a distinct change of slope with energy, and the shower-to-shower variance decreases [2]. Interpreted with the leading LHC-tuned shower models, this implies a gradual shift to a heavier composition [3]. A number of fundamentally different astrophysical model scenarios have been developed to describe this evolution, see, for example, [4–8]. The high degree of isotropy observed in numerous tests of the small-scale angular distribution of UHECRs above $4\times10^{19}$ eV is remarkable [9], challenging original expectations that assumed only a few cosmic ray sources with a light composition at the highest energies. Interestingly, the largest departures from isotropy are observed for cosmic rays with $E > 5.8\times10^{19}$ eV in $\sim20^\circ$ sky windows [9, 10].

Due to a duty cycle of $\sim15\%$ of the fluorescence telescopes, the data on the depth of shower maximum extend only up to the flux suppression region, i.e., $4\times10^{19}$ eV. Obtaining more information on the composition of cosmic rays at higher energies is of central importance for making further progress in understanding UHECRs. Care must be taken, since precision Auger measurements of shower properties, strongly constrained by the hybrid data, have revealed inconsistencies within present shower models, opening the possibility that the unexpected behavior is due to new hadronic interaction physics at energy scales beyond the reach of the LHC.

The aim of the upgrade of the Auger Observatory, collectively dubbed AugerPrime, is to provide additional measurements of composition-sensitive observables to allow us to address the following questions: (i) Elucidate the mass composition and the origin of the flux suppression at the highest energies, i.e., the differentiation between the energy loss effects due to propagation, and the maximum energy of particles injected by astrophysical sources. (ii) Search for a flux contribution of protons up to the highest energies. We aim to reach a sensitivity to a contribution as small as 10% in the flux suppression region. The measurement of the fraction of protons is the decisive ingredient for estimating the physics potential of existing and future cosmic ray, neutrino, and gamma-ray detectors; thus prospects for proton astronomy with future detectors will be clarified. Moreover, the flux of secondary gamma rays and neutrinos due to proton energy loss processes will be predicted. (iii) Study extensive air showers and hadronic multiparticle production. This will include the exploration of fundamental particle physics at energies beyond those accessible at terrestrial accelerators, and the derivation of constraints on new physics phenomena, such as Lorentz invariance violation or extra dimensions.

The needed composition-sensitive information can be obtained by upgrading the Auger Observatory. In Sec. 2 a brief overview of the planned hardware upgrades is given, focusing on new scintillation detectors. The expected physics performance reached with the upgraded Observatory is illustrated in Sec. 3 using detailed simulations of different composition scenarios.

## 2. Planned upgrade

The key element of the upgrade will be the installation of a new detector consisting of a plastic scintillator plane above each of the existing water-Cherenkov detectors (WCD). This scintillation detector will provide a complementary measurement of the shower particles: they will be sampled with two detectors having different responses to muons and electromagnetic particles, allowing the reconstruction of the different components. The design of the surface scintillator detectors (SSD) is simple, reliable and they can be easily deployed over the full $3000\,km^2$ area of the overall Auger Surface Detector (SD). An SSD unit will consist of a box of $3.8\,m \times 1.3\,m$, housing two scintillator modules, each covering an area of $1.9\,m^2$, see Fig. 1 (left). The 1 cm thick scintillators are read out by wavelength-shifting fibers guiding the light of the two modules to a PMT. The deviations from a uniform detector response over the area of the scintillator are smaller than 5%. The SSD is triggered by the larger WCD, resulting in a clean separation of the MIP signal from the background, see Fig. 1 (right). An engineering array of 10 detectors will be installed at the Auger site in 2016.

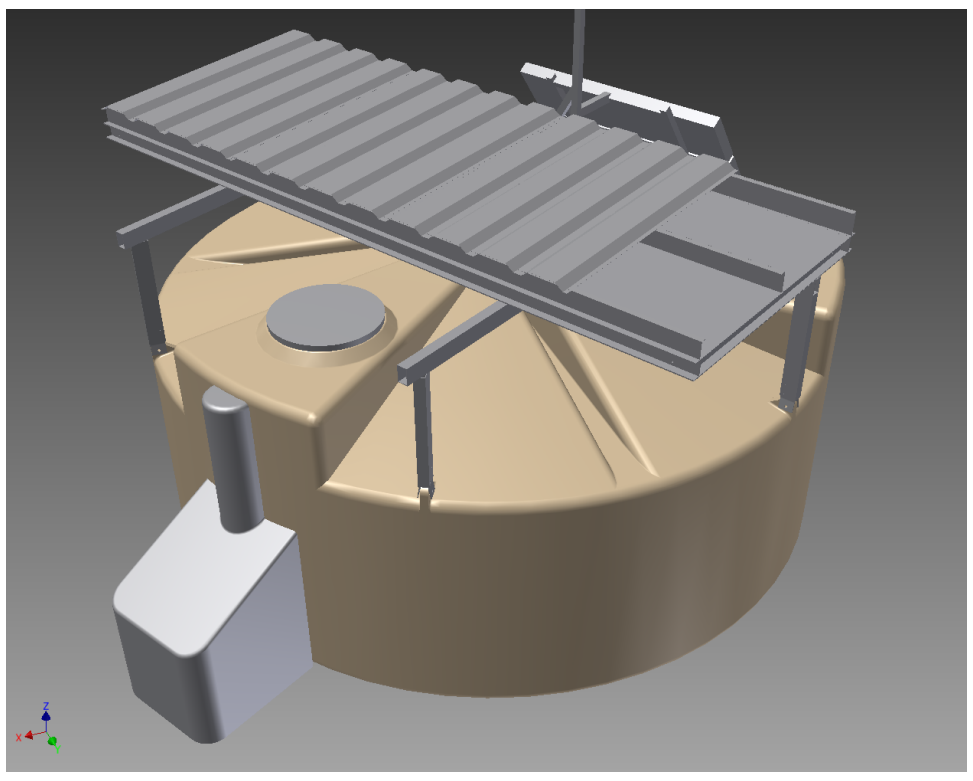

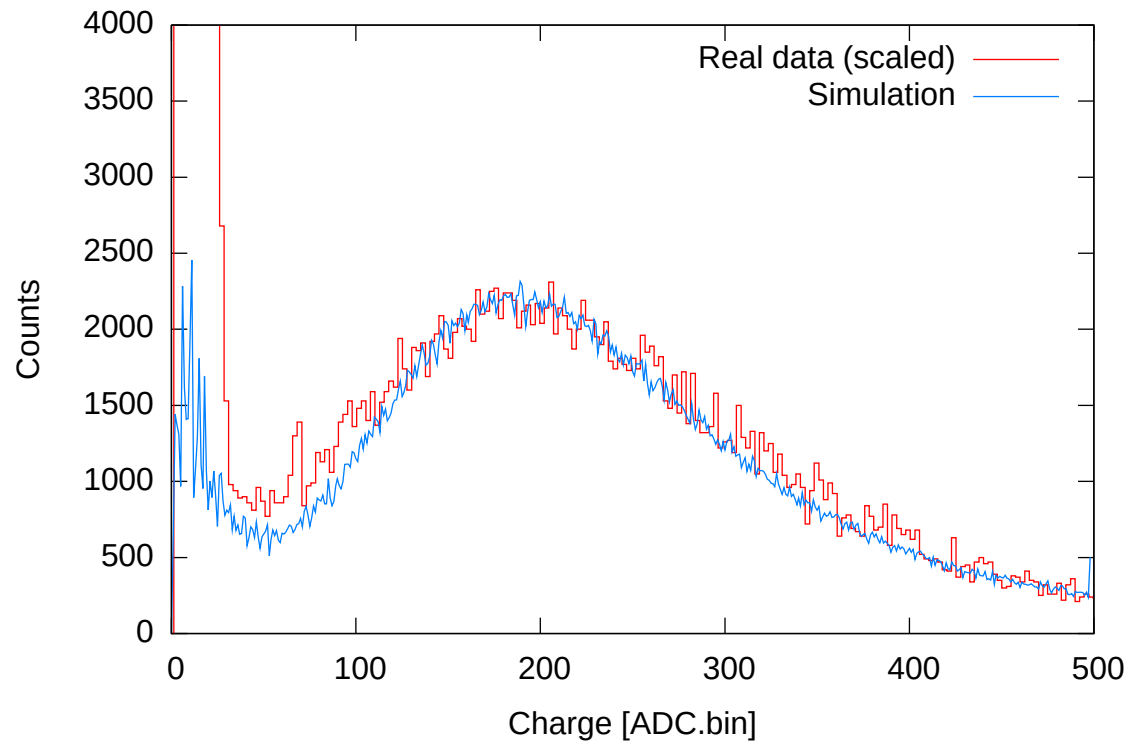

**Figure 1:** Left panel: 3D view of a SSD mounted on a WCD. A double roof, with the upper layer being corrugated aluminum (here shown partially cut away for clarity), is used to reduce the temperature variations. Right panel: MIP histogram taken with a $2\,m^2$ prototype installed in the Auger array. The data correspond to one minute of data taking and are well reproduced by the detector simulation based on Geant 4.

The SD stations will be upgraded with new electronics that will process both WCD and SSD signals. Use of the new electronics also aims to increase the data quality (with faster sampling of ADC traces, better timing accuracy, increased dynamic range), to enhance the local trigger and processing capabilities (with a more powerful local station processor and FPGA) and to improve calibration and monitoring capabilities of the surface detector stations. The signals of the WCDs and SSDs will be sampled synchronously at a rate of 120 MHz (previously 40 MHz) and the GPS timing accuracy will be better than 5 ns. The dynamic range of the WCDs will be enhanced by a factor of 32 due to the new electronics and an additional small 1″ PMT that will be inserted in one of the filling ports. First prototypes of the new electronics have been built and are being tested. Both the SSDs and the electronics upgrade can be easily deployed, and will have only minimal impact on the continuous data taking of the Observatory.

A network of underground muon detectors, part of the AMIGA [1,11] system, is now being deployed in the existing SD infill area of $23.5\,km^2$. This will provide important direct measurements of the shower muon content and its time structure, while serving as verification and fine-tuning of the methods used to extract muon information with the SSD and WCD measurements.

In parallel with the SD upgrade, the operation mode of the Fluorescence Detector (FD) [12] will be changed to extend measurements into periods with higher night sky background. This will allow an increase of about 50% in the current duty cycle of the FD.

## 3. Expected physics performance

In the following we consider different levels of complexity of the information derived from shower data. We first discuss the reconstruction of the muonic shower component, then show the discrimination power for different primary particles, and finally analyze Monte Carlo generated event samples to test the sensitivity to different physics scenarios. As a generic measure of discrimination power for separating primary $i$ and $j$ using the corresponding observables $S_i$ and $S_j$ (with the RMS $\sigma$) we use the merit factor

$$f_{\rm MF} = \frac{|\langle S_i\rangle - \langle S_j\rangle|}{\sqrt{\sigma(S_i)^2 + \sigma(S_j)^2}}. \tag{3.1}$$

### 3.1 Event based observables

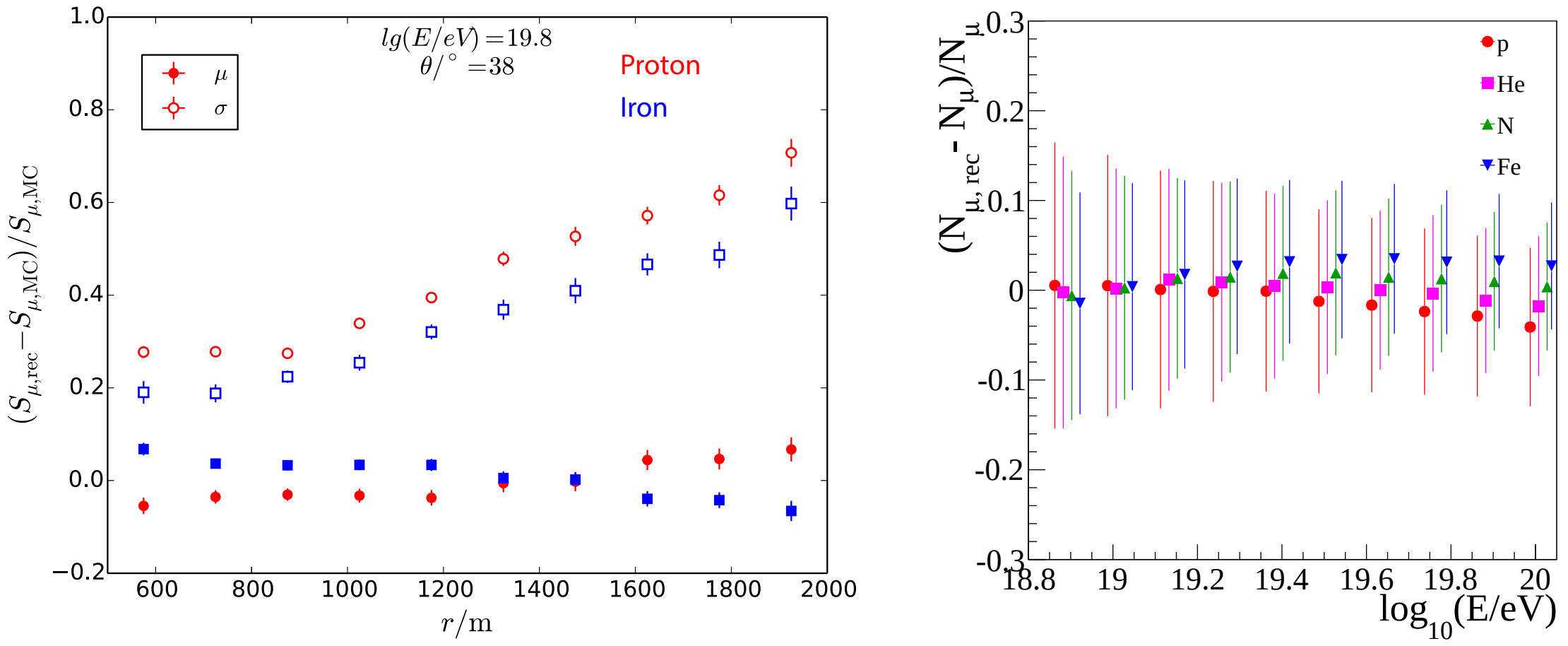

**Figure 2:** Left panel: Reconstruction bias (solid symbols) and resolution (open symbols) of the muonic signal contribution for individual detector stations, as a function of distance $r$ from the shower core. The results for proton and iron showers are shown in red and blue, respectively. Right panel: Number of muons $N_\mu$ reconstructed for individual showers using shower universality, compared with the true $N_\mu$ as a function of energy for different primary species. Error bars represent the RMS of the distributions.

The least model-dependent and most direct composition-sensitive observable that can be obtained from the upgraded detector array is the number of muons (or, equivalent to it, the muonic signal in the WCD) in individual detector stations. Thanks to the signal responses in the SSD ($S_{\rm SSD}$) and the WCD ($S_{\rm WCD}$) to particles of the electromagnetic and muonic shower components, it is possible to derive the muonic signal $S_{\mu,\rm WCD}$ on a station-by-station basis, as described in [13], from

$$S_{\mu,\rm WCD} = a\,S_{\rm WCD} + b\,S_{\rm SSD}, \tag{3.2}$$

where the signals are measured in units of the response to a vertical equivalent muon (VEM) or minimum ionizing particle (MIP), respectively. The factors $a$ and $b$ are derived from detector simulation and have only a very weak dependence on the primary composition and lateral distance from the shower core. Restricted by the limited size of an individual detector station, this method is subject to large fluctuations and is only well suited for deriving mean muon numbers, see Fig. 2 (left).

The performance of the reconstruction of the muonic signal is considerably improved by fitting a lateral distribution function (LDF) to the signals. Eq. 3.2 is then applied to the LDF values for the WCD and SSD to calculate the muonic signal at an 800 m core distance, $S_\mu(800)$. Reconstruction resolutions of the muonic signal of, for example, $\sigma[S_\mu(800)]/\langle S_\mu(800)\rangle \approx 22\%$ for protons and $\sigma[S_\mu(800)]/\langle S_\mu(800)\rangle \approx 14\%$ for iron are reached at $E \approx 10^{19.8}$ eV and $\theta = 38°$. Using $S_\mu(800)$ as a composition estimator, the obtained merit factors for distinguishing between proton and iron primaries are above 1.5 at high shower energies ($E > 10^{19.5}$ eV) and small zenith angles.

An analysis based on shower universality (see [14] and references therein), or a sophisticated multivariate analysis, allows one to correlate the detector signals at different lateral distances and also takes advantage of the arrival time (shower front curvature) and temporal structure of the signal measured in the detectors. We are working on developing a reconstruction using all these observables. Some first results are given in the following, but it should be kept in mind that the corresponding merit factors should be considered as lower limits to what will be reached after having a better understanding of the detectors. The universality-based reconstruction provides, in addition to the shower energy, the depth of shower maximum (mainly from the curvature of the shower front and the steepness of the lateral distribution) and the number of muons relative to the prediction of a reference model. The bias of the $X_{\max}$ reconstruction is below 15 g/cm$^2$ with a resolution improving from 40 g/cm$^2$ at $10^{19}$ eV to 25 g/cm$^2$ at $10^{20}$ eV. The corresponding results for the muon number reconstruction are shown in Fig. 2 (right). Of interest is also the energy resolution for the reconstruction which is about 10% at $10^{19}$ eV down to 7% at $10^{20}$ eV. Examples of merit factors obtained by combining the two observables are 1.54 (proton-iron), 0.41 (proton-helium), and 0.64 (nitrogen-iron), all calculated for $10^{19.6}$ eV, zenith angles $\theta < 60°$ and using QGSJet II.04.

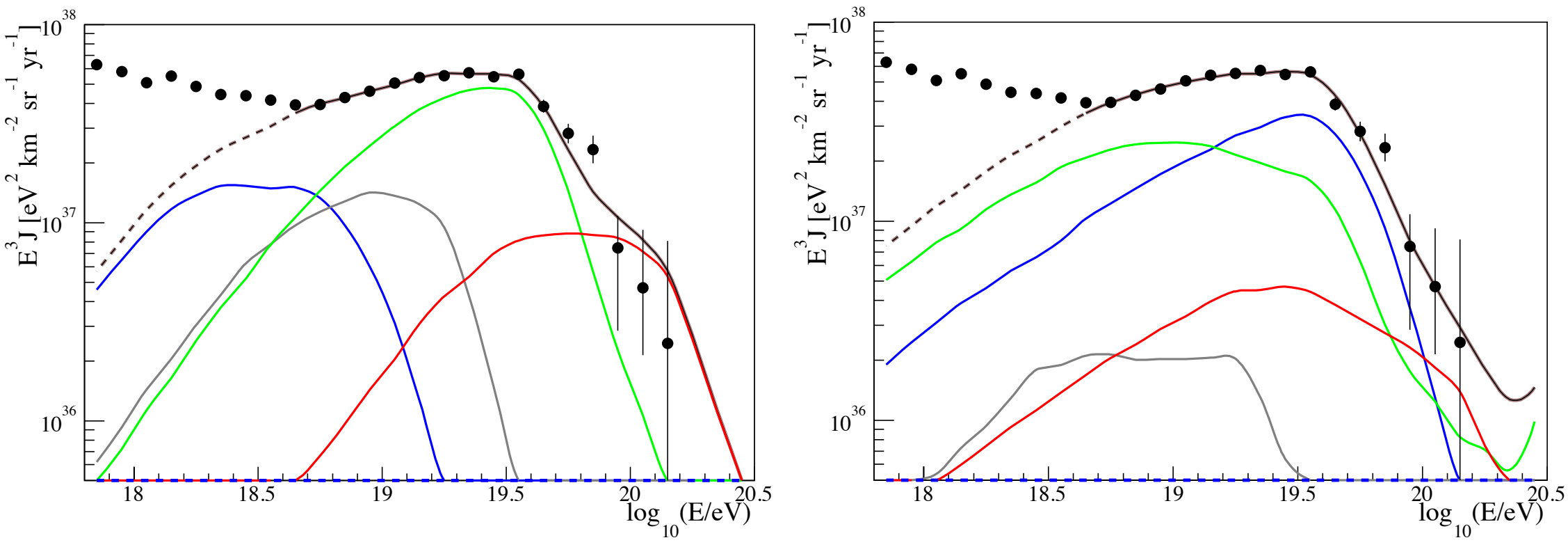

**Figure 3:** Benchmark spectra chosen as representations of a maximum-rigidity scenario [4] (left panel) and one photo-disintegration scenario (right panel). The index of the injection energy spectrum at the sources is about $-1$ ($-2$) for the maximum-rigidity (photo-disintegration) scenario – for details see [15]. The colors for the different mass groups are protons – blue, helium – gray, nitrogen – green, and iron – red.

## 3.2 Application to physics goals

Without knowing what composition to expect in the GZK suppression region it is difficult to demonstrate the potential of AugerPrime. Therefore, we have chosen two physics motivated benchmark models [15] fitted to the Auger flux [16] and composition data [3] for $E > 10^{18.7}$ eV, see Fig. 3, to illustrate the discrimination power of the additional information. Mock data sets were generated for these scenarios with a statistics corresponding to 7 years of data taking with AugerPrime. Only SD data are used in the reconstruction.

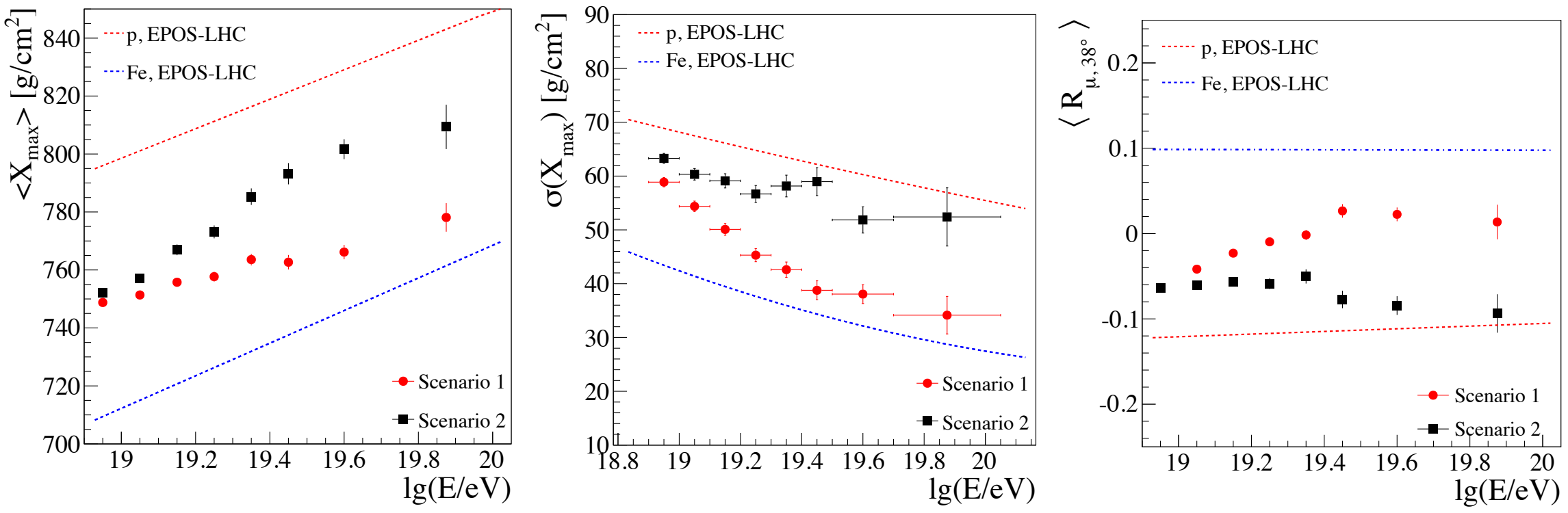

**Figure 4:** Reconstructed mean depth of shower maximum $X_{\max}$ (left) and its fluctuations (center), and the mean number of muons at 38° (right) using only SD data. Shown are the two scenarios: (1) maximum-rigidity model; (2) photo-disintegration model. The $\sigma(X_{\max})$ contains the intrinsic air-shower fluctuations and the detector resolution. The number of muons is given relative to that expected for an equal mix of p-He-CNO-Fe as primary particles.

The mean $X_{\max}$, $\sigma(X_{\max})$, and the relative muon number $R_\mu$ are shown in Fig. 4. The $\sigma(X_{\max})$ contains the intrinsic air-shower fluctuations and the reconstruction resolution. While the mean $X_{\max}$, $\sigma(X_{\max})$ and $R_\mu$ are very similar up to $10^{19.2}$ eV, the energy range that is well covered by data of the fluorescence telescopes [2], the models predict significantly different extrapolations into the GZK suppression region. This difference is well reproduced with the reconstructed $X_{\max}$, $\sigma(X_{\max})$ and $R_\mu$ and the two scenarios can be distinguished with high significance and statistics.

As a next step we want to illustrate the increased sensitivity of AugerPrime with a more specific example. We use the arrival directions of the 454 measured events with $\theta < 60°$ and energy higher than $4\times10^{19}$ eV (see [9]) and randomly assign each event an $X_{\max}$ value according to model 1 (maximum-rigidity scenario). To implement a 10% proton contribution we assigned 10% of the events a proton-like $X_{\max}$. Half of these randomly chosen, proton-like events were given arrival directions that correlate with AGNs with a distance of less than 100 Mpc of the Swift-BAT catalog [17] within 3°. The other half were chosen with larger angular distances. By construction, this artificial data set reproduces many arrival direction features found in the Auger data while at the same time having a model-predicted mass composition.

Analyzing this data set without using any composition information, a correlation with the AGNs of the Swift-BAT catalog is found at a level similar to that reported in [9]. The improvement of the sensitivity in finding the correlation with AGNs in this data set is shown in Fig. 5 for $E > 4\times10^{19}$ eV. The left panel gives the result of the complete data set and the right panel that of a proton-enriched sample. The proton-enriched sample is obtained by selecting events with a reconstructed $X_{\max}$ greater than 770 g/cm$^2$ at $10^{19}$ eV, adjusted to the event energies with an elongation rate of 55 g/cm$^2$ per decade. While the correlation of the arrival directions with that of AGNs in the Swift-BAT catalog is not significant for the complete data set, a correlation well in excess of $3\sigma$ can be found for the proton-enriched samples.

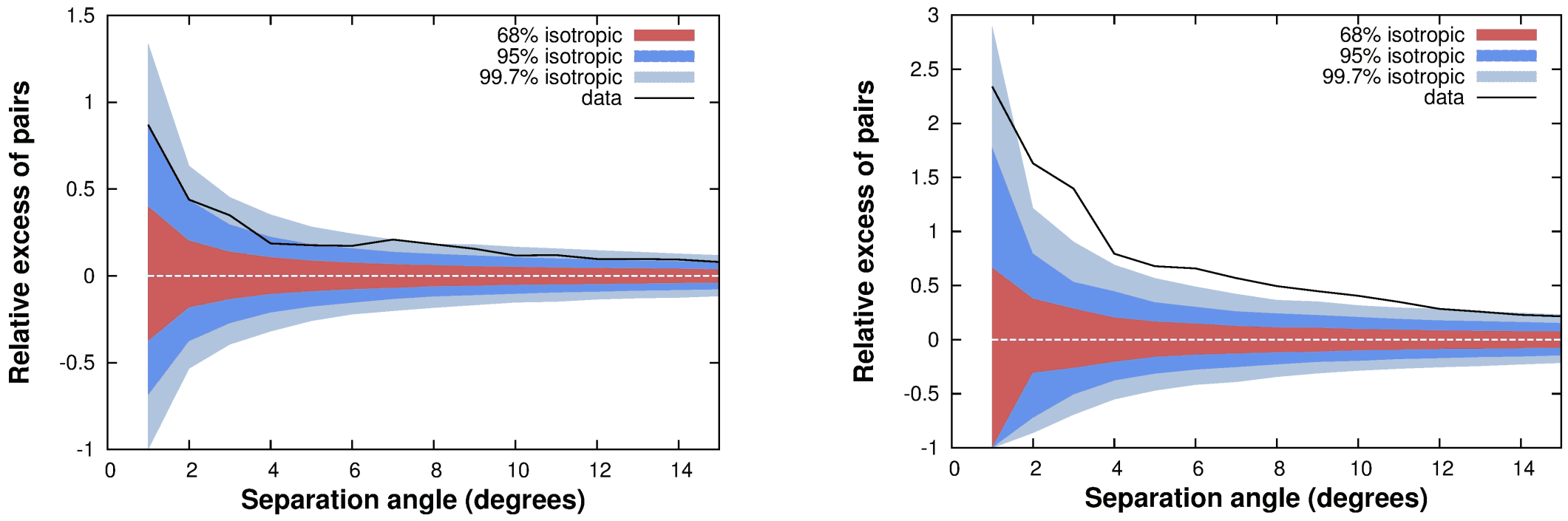

**Figure 5:** Angular correlation of cosmic rays of the modified Auger data set with AGNs of the Swift-BAT catalog [17]. Shown are distributions of relative excesses of pairs of events as a function of angular separation between them, for the complete data set (454 events, left) and the proton-enriched selection (128 events, right).

In addition to these studies, the availability of muon information on an event-by-event basis allows for many ways of studying features of hadronic interactions. For example, the correlation between the number of muons and the depth of shower maximum can be used to study general features of muon production, including the search for exotic interaction scenarios at very high energy [18].

## 4. Conclusions

The Auger upgrade promises high-quality future data, and real scope for new physics uses of existing events. With operation planned from 2018 until 2024, event statistics will more than double compared with the existing Auger data set, with the critical added advantage that every event will now have mass information. This will allow us to address some of the most pressing questions in UHECR physics, including that of the origin of the flux suppression, the prospects of light particle astronomy and secondary particle fluxes, and the possibility of new particle physics at extreme energies.

Obtaining additional composition-sensitive information will not only help to better reconstruct the properties of the primary particles at the highest energies, but also improve the measurements in the important energy range just above the ankle. Furthermore, measurements with the new detectors will help to reduce systematic uncertainties related to modeling hadronic showers and to limitations of reconstruction algorithms. This improved knowledge of air-shower physics will likely then also allow a re-analysis of existing data – for improved energy assignments, for mass composition studies, and for photon and neutrino searches.

Finally it should be mentioned that the addition of scintillator detectors across the entire Observatory will also make possible direct comparisons of Auger measurements with those of the surface detectors of the Telescope Array experiment. This will strengthen the already productive cooperation between the two Collaborations, which has an aim of understanding the highest energy cosmic ray flux across the entire sky.

# The AMIGA Muon Counters of the Pierre Auger Observatory: Performance and Studies of the Lateral Distribution Function

**Brian Wundheiler**[*,a] **for the Pierre Auger Collaboration**[b]
[a]*Instituto de Tecnologías en Detección y Astropartículas (CNEA, CONICET, UNSAM), Av. Gral. Paz 1499 (1650) Buenos Aires, Argentina.*
[b]*Observatorio Pierre Auger, Av. San Martín Norte 304, 5613 Malargüe, Argentina.*
*E-mail:* auger_spokespersons@fnal.gov
*Full author list:* http://www.auger.org/archive/authors_2015_06.html

The AMIGA enhancement (**A**uger **M**uons and **I**nfill for the **G**round **A**rray) of the Pierre Auger Observatory consists of a 23.5 km$^2$ infill area where air shower particles are sampled by water-Cherenkov detectors at the surface and by 30 m$^2$ scintillation counters buried 2.3 m underground. The Engineering Array of AMIGA, completed since February 2015, includes 37 scintillator modules (290 m$^2$) in a hexagonal layout. In this work, the muon counting performance of the scintillation detectors is analysed over the first 20 months of operation. Parametrisations of the detector counting resolution and the lateral trigger probability are presented. Finally, preliminary results on the observed muon lateral distribution function (LDF) are discussed.

*Speaker.

## 1. Introduction

The Pierre Auger Observatory [1], located in the province of Mendoza in Argentina, is a hybrid detector covering 3000 km$^2$ with 1660 surface stations (the surface detector, SD) and 27 fluorescence telescopes (the Fluorescence Detector, FD). The SD stations are separated by 1.5 km, while the telescopes are split in four buildings at the edge of the surface array, and point towards the atmosphere and the centre of the array. Currently, the Auger Observatory is being upgraded and AMIGA is one of the enhancement projects [2].

AMIGA consists of 61 detector pairs, each composed of a surface water-Cherenkov detector (SD infill) and a buried 30 m$^2$ Muon Counter (MC). The AMIGA MCs are arranged on a 750 m triangular grid to directly measure the muon content of showers with primary energies $\geq 3 \times 10^{17}$ eV. The complete AMIGA array will cover an area of 23.5 km$^2$ providing sufficient statistics given the higher rate of the sub-EeV showers. Important results on cosmic ray physics by means of muon detection techniques have been obtained previously by the Haverah Park [3], Akeno [4], Yakutsk [5] and AGASA experiments [6] and more recently by the KASCADE [7] and KASCADE-Grande [8] experiments.

The Engineering Array of AMIGA, called the Unitary Cell (UC), is a hexagon in the Observatory infill area (750 m spacing) with 2.3 m-deep buried MCs at each hexagon vertex and at its centre (see Fig. 1). It has been completed since February 2015. Each of these 7 MCs is composed of 4 scintillation modules (SM), of either $2 \times 5$ m$^2$ or $2 \times 10$ m$^2$ area, comprising 64 plastic scintillation bars sealed in a PVC casing, containing wavelength-shifting optical fibres, a 64 multianode photomultiplier tube, and acquisition electronics. The light produced in these bars is collected and propagated along the fibres which couple to the multi-pixel PMTs. Segmentation was selected since it permits the AMIGA counter electronics to just count pulses above a given threshold (see Sec. 3), without a detailed study of signal structure or peak amplitude/charge. In two UC positions, twin detectors are deployed consisting of four extra SMs at the KT position identified in Fig. 1 and three 10 m$^2$ extra ones at the PC position. At the latter, there are also two extra 10 m$^2$ SMs deployed at 1.3 m depth. There is a SM with 4 SiPMs installed so that in total there are 37 deployed and working SMs.

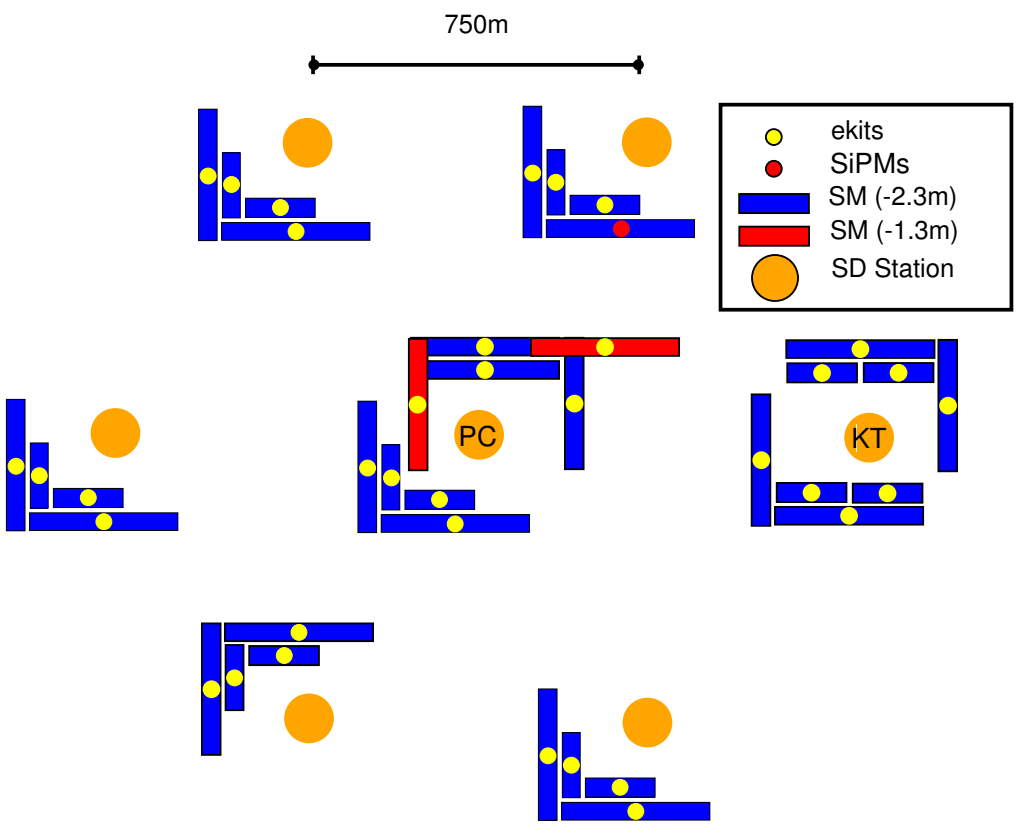

**Figure 1:** AMIGA Unitary Cell as of February 2015. The 5 m$^2$ detectors at the KT and PC positions have been used for this analysis.

In this work we analyse the data set from the first two years of UC operation. We evaluate the response of the twin position at KT to assess the muon counting resolution of the MCs, and get a parametrisation that allows us to establish the counting uncertainty of single events. We find the scaling between the muon number at a reference distance from the shower core and the energy of the primary cosmic ray, and we study the lateral distribution function (LDF) of the observed muons and obtain an effective value for its slope. This value can be used for single event fitting, leaving only the relevant composition-sensitive observable of the LDF, i.e. the number of muons at an optimal distance from the shower axis, to be determined from the fit.

## 2. Design Considerations

Having described the scintillator modules it is worth briefly describing the path followed by the light travelling through the optical fibre until the signal is processed by the Central Data Acquisition System (CDAS) at the Observatory campus. The 64 optical fibres are matched to a 64-pixel PMT through a custom-tailored optical alignment device. The PMT chosen is the Hamamatsu ultra bi-alkaline H8804-200MOD, a H7546 type PMT but with a different casing and an increased quantum efficiency peaking around 350 nm.

The electronics of the MCs is split into two components, the underground electronics installed in each buried module and a reduced electronics at the surface. Both are powered by solar panels. The underground electronics includes the PMT, the front-end, digital, slow control, and power distribution boards and a data transmission unit. The surface electronics comprises an interface with the SD electronics (to get the trigger from the SD and to transfer muon data), the wireless communication to CDAS, the network switch, and the power regulator. The analogue front-end holds the pre-amplifiers and discriminators which are remotely set to an adjustable fraction of the average Single Photo Electron (SPE) amplitude of each PMT pixel. Thus, PMT pulses are converted into a train of digital 0's and 1's corresponding to the presence or absence of a signal above the aforementioned threshold. One bit per channel is saved in the front-end memory forming a 64-bit string. This conversion is performed in 3.125 ns time bins by a field-programmable gate array (FPGA). The memory consists of two circular buffers that store 2048 bins of 64 bits. These bit trains are stored following an SD trigger, recovered and transmitted upon a request from the CDAS. The MC event acquisition is synchronised at the lowest (hardware) level to the surface stations through a dedicated triggering line. The MC electronics maintains synchronisation, through a time-tagging scheme, mostly implemented in the FPGA. An event data trigger request, received by the surface radio, is sent from the surface to the underground microcontroller through an Ethernet line. The FPGA searches for the requested event and retrieves data. A counting strategy searches offline for the muon traces by inspecting the individual SPE signatures. As the vast majority of contaminating events produce only a sole SPE, by requiring at least a 1X1 string (with X either a 1 or a 0) for a muon footprint, accidentals are removed. Consequently, most of the accidental data, such as crosstalk or thermal photoelectrons are discarded. Muons are counted in time windows of 25 ns, the duration corresponding to the detector dead time given by the width of the muon pulse due to the scintillator and fibre decay times (for details on counting techniques and performance see [9]).

## 3. Calibration Routine and Data Set

Due to the one-bit electronics technique, the discriminator levels, set relative to the mean pixel SPE amplitude, play a major role. Setting them too high will cause a loss of counting efficiency since the digital output of the discriminator might be too short for the 3.125 ns sampling period. Too low thresholds might produce two or three adjacent positive digital samples (i.e. 1's) for a SPE. The aim of the one-bit background-radiation calibration is to identify a feature (e.g. a plateau) in the behaviour of the rates as a function of the SPE threshold. Setting each individual threshold on its plateau will render a fully efficient detection system. The calibration induces the MCs to self-trigger which is not performed in the usual air-shower detection mode when the counters make use of their accompanying SD trigger.

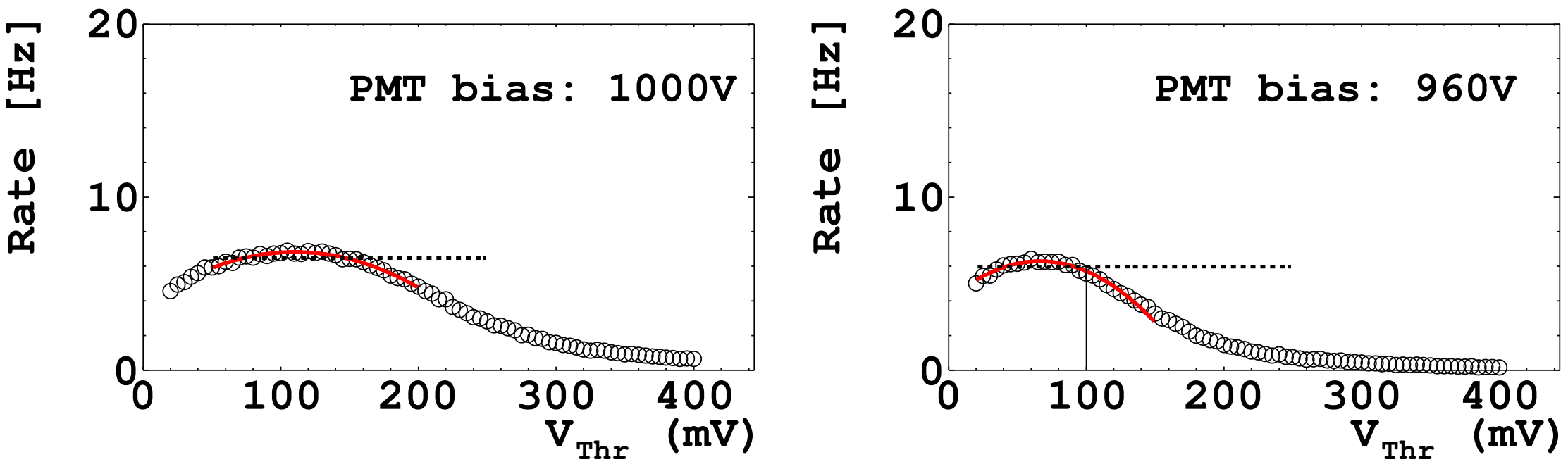

**Figure 2:** Background-radiation rates as a function of threshold, measured at KT for a 5 m$^2$ module. Also shown are fits to the plateau regions in red (dotted line at 5% below the maxima). (Left) PMTs were biased at 1000 V as of March 2015. (Right) PMTs were biased at 960 V in the previous period, the full line showing the 100 mV threshold fixed for all channels in the data set analysed in this work.

In Fig. 2 (left panel) we show a typical distribution of the background-radiation rate from a single channel of a 5 m$^2$ module measured with its PMT biased at 1000 V. Each PMT was characterised prior to deployment with a mounted opto-electronics device. A plateau-like structure is clearly apparent within the range from 80-190 mV (i.e. ∼25% to 45% of the SPE). In stable operational mode, the pixel calibration is performed either to any given plateau maximum rate or to a SPE fraction level measured in the field in real time.

During the period that preceded the implementation of this calibration procedure, the UC was deployed and operated to validate the muon detector design. The voltage at which the deployed SMs were operated was 960 V and with a uniform discrimination level of 100 mV set for every channel of every module. We performed a calibration run at 960 V to evaluate how these fixed parameters (high voltage and threshold level) are reflected in the quality of data. There is a counting efficiency loss as thresholds move away from the plateaus and this loss occurs more rapidly in the 10 m$^2$ modules. The loss of efficiency arises from the fact that as we inspect SPEs with higher thresholds some are rejected since either they fall short of the threshold of the digital discriminator or the digital signal is too short for the 3.125 ns sampling period. So both types of modules are in principle affected by this efficiency loss, but the loss is more pronounced for the 10 m$^2$ module due to the reduced number of SPEs coming from the far end [10]. Also the plateau regions become narrower with lower PMT bias voltages, more evidently so for the 10 m$^2$ modules. A uniform 100 mV threshold was found to be essentially working for the 5 m$^2$ modules in the data acquisition period mentioned above, shown in Fig. 2 (right panel). Therefore we will restrict the analysis presented in this work only to these modules. The complete data set considered in this work comprises all the cosmic ray events recorded during the first 20 months of the UC operation from March 2013, during the period previous to the implementation of the calibration routines. In this period six 5 $m^2$ modules were operational: four in KT and two in PC. The UC has modules of two different sizes to study the counting efficiency and the saturation range with respect to the shower core position. AMIGA is designed to work with 30 $m^2$ MCs divided in three 10 $m^2$ modules. We combined the information of 2×5 $m^2$ modules for each event, to analyse the muon counts for 10 $m^2$ of detection area. We added muon counts from the 5 $m^2$ SMs at the KT South twin position, at the KT North twin, and at the PC South twin. In this way we performed our analyses over three independent 10 $m^2$ detection areas. The infill stations provide the geometry and energy reconstruction of the cosmic showers, 1235 events with zenith angle up to 45°, with energies higher than $10^{17}$ eV. Stations at least at 200 m away from the shower core were considered. The reconstruction algorithm for the events triggering the infill array is based on the code for the regular surface detector array. After selecting the signals which are generated by air showers, the direction and the energy of the primary cosmic ray are deduced from the timing information and from the total recorded signal in the stations [1].

## 4. Detector Resolution

As done to study the accuracy of the SD stations [11], the twin detectors can be used to study the fluctuations of the measured signals and preliminary results for the MC counting accuracy have been reported [10]. Having two detectors measuring basically the same spot on the shower allows us to estimate the signal fluctuation by analysing the difference of their signals for a given event. Considering that the separation between twin detectors is ∼20 m, only AMIGA signals at least 200 m away from the shower axis are considered. In this way we can assume that both detectors are measuring samples coming from the same muon density.

The linear correlation between KT-North and KT-South is shown in Fig. 3 (left panel), where only events with at least 1 muon over a 10 $m^2$ area are considered. From the slope of the linear fit of $1.05 \pm 0.02$, we can consider that the detectors respond in the same way for each event, a hypothesis needed to determine the resolution based on these data.

For each event we construct the resolution estimator Δ based on the sample variance and mean, define as follows: $\Delta^2 = \left(\sigma/\overline{N}\right)^2$, where $\sigma^2$ and $\overline{N}$ are the variance and mean estimators respectively calculated from the number of measured muons by each twin.

We extract the mean value of $\Delta^2$ within bins of average number of muons. In Fig. 3 (right panel), the values obtained from each $\Delta^2$ bin are displayed as a function of the muon count. For an ideal Poisson counter, the resolution should be $\Delta^2_{\mathrm{poisson}} = N_\mu^{-1}$. The blue line is the fit of our model to the data. The muon detector resembles the ideal behaviour in this range of signals. From this analysis we can establish the counting uncertainty as a function of the number of counted muons as $\Delta^2\left(N_\mu\right) = (0.8 \pm 0.2)/N_\mu^{(0.9\pm0.1)}$.

## 5. Muon Lateral Distribution

For studying the lateral distribution of muons we choose a KASCADE-Grande-like LDF [8], since we are also counting muons over 10 $m^2$ areas and the energy range of both experiments

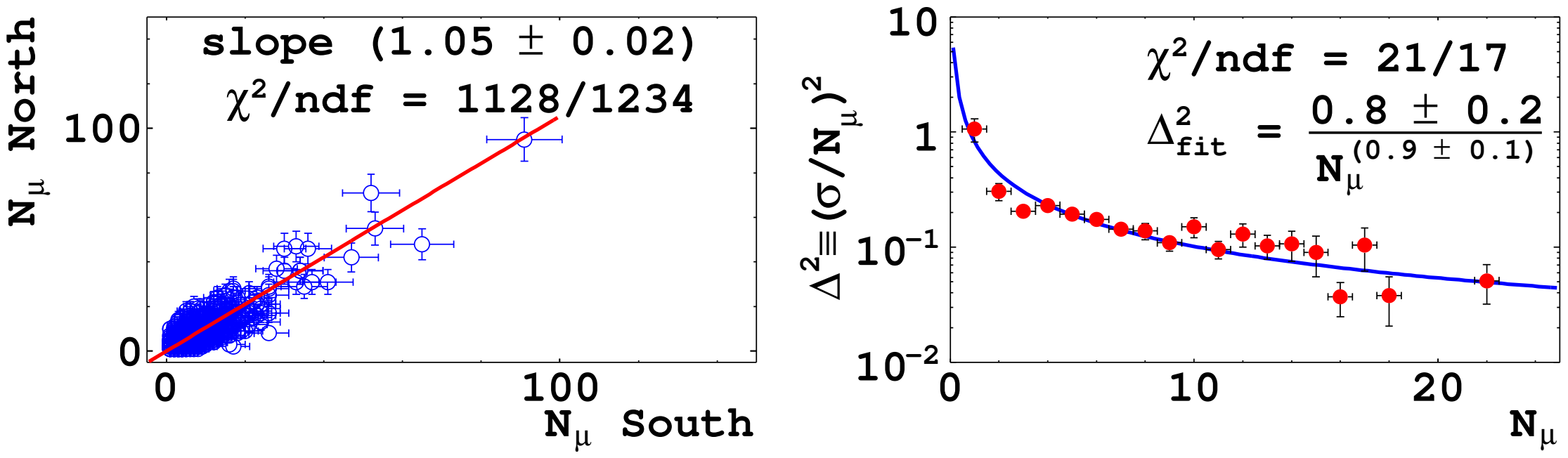

**Figure 3:** (Left) Muon counting comparison between twin counters at KT for events with $\theta \leq 45°$, $\log(E/\mathrm{eV}) \geq 17.0$, at least 200 m from the shower core, and at least 1 muon over the $10\,\mathrm{m}^2$ area considered. (Right) Parametrisation of the muon counting resolution. The fitted model is displayed by the blue line.

overlaps. This is given by

$$N_\mu(r,E) = N_0(E) \cdot f_\mu(r) / f_\mu(r_0)$$
$$f_\mu(x) = \left(\frac{x}{r^\star}\right)^{-\alpha} \left(1+\frac{x}{r^\star}\right)^{-\beta} \left(1+\left(\frac{x}{10 \cdot r^\star}\right)^2\right)^{-\gamma} \quad (5.1)$$

where $\alpha = 1$, $\gamma = 1.85$ and $r^\star = 150\,\mathrm{m}$ are fixed parameters from Monte Carlo simulations. $N_0(E) = N_\mu(r_0 = 450\,\mathrm{m}, E)$ is the number of muons at the optimal distance from the shower axis. At $r_0$, the fluctuations of the LDF fit are minimised, this distance being mostly determined by the spacing between the detectors [12]. We analyse the data set for getting an effective value for the slope $\beta$ instead of using a simulation-driven result, leaving only the composition-sensitive parameter $N_0$ to be fitted for each event.

The method of finding an effective $\beta$ has two iterations. Firstly, we determine $N_0(E)$, i.e. we study the relation between the number of muons at the optimal distance and the energy. Secondly, we use this relation to normalise each event and, in this way, we are able to fit $f_\mu(r)$ of eq. 5.1 to all the events together and get an effective value of $\beta$ as a result.

The relation between $N_0$ and $E$ is derived from a subset of events for which the highest SD signal is found within the UC. We defined a Lateral Trigger Probability (LTP) for the muon detector, and demand LTP $\geq 90\%$ at 450 m from the shower core, which implies an energy cut of $E > 10^{17.3}$ eV. Above this energy threshold the trigger efficiency of the infill is $\geq 90\%$. We define the LTP at different core distances in the range of 200-1000 m. Trigger probabilities are also obtained within energy bins. For a fixed energy range, we calculate in distance range the ratio of the number of muon counters with at least 1 count to the total number of muon counters that received an SD trigger. Uncertainties in the trigger probability are given by the binomial error. As already mentioned, we are only considering three independent $10\,\mathrm{m}^2$ muon-detection areas as muon counters in this work (see Sec. 3). In Fig. 4 (left panel) the muon trigger probability is shown for several energy ranges as a function of the distance to the shower axis.

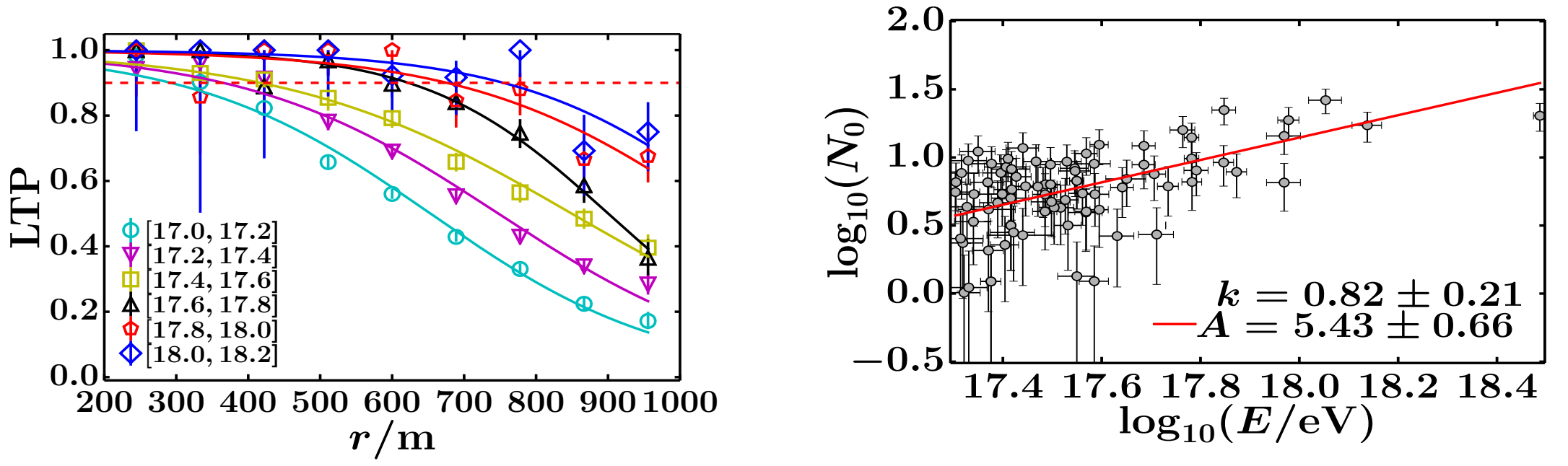

**Figure 4:** (Left) Muon counter lateral trigger probability as a function of distance from the shower core, for different energy ranges in $\log(E/\mathrm{eV})$. (Right) Determination of $N_0(E)$ with 100 selected events from muon counters within a distance $(450 \pm 25)$ m from the shower core, and trigger probability higher than 90%.

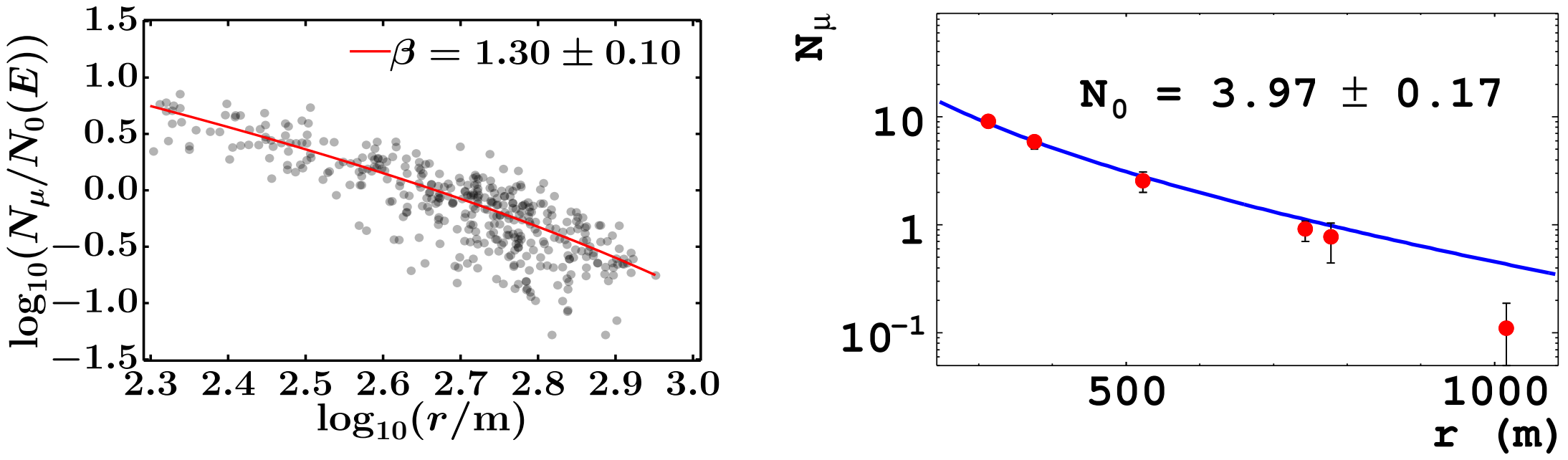

**Figure 5:** (Left) Muon lateral distribution for normalised data; an effective value for the slope $\beta$ is found and can be use for individual event fits. (Right) As a preliminary result, an LDF fit fixing $\beta$ is shown for a $10^{18.3}$ eV event, where only $N_0$ remains as a free parameter.

The events shown in Fig. 4 (right panel) are fitted with the function,

$$N_0(E) = A \cdot \left(\frac{E}{10^{17.5}\,\mathrm{eV}}\right)^k \tag{5.2}$$

where $A = (5.43 \pm 0.66)\,\mathrm{m}^{-2}$ and $k = 0.82 \pm 0.21$. A general maximum-likelihood approach developed in [13] is maximised to obtain the parameters. It is encouraging to note that the value of the scaling exponent $k$ is in agreement within statistical uncertainties with the one found by Akeno for the conversion between the number of muons and the energy [4].

To determine an effective value for $\beta$ we take a subset of events above the infill full efficiency, i.e. $10^{17.5}$ eV, and those muon detectors within a core distance for which their LTP $\geq 90\%$. Each of the 389 inputs of this subset is normalised to $N_0(E)$ of eq. 5.2 thus eliminating the energy dependence. In Fig. 5 (left panel) the normalised number of muons is shown as a function to the shower axis. The subset is fitted with $f_\mu(r)$ of eq. 5.1 leading to an effective value of $\beta = 1.3 \pm 0.1$ to be used for individual events. In Fig. 5 (right panel) we show an example of the application of this value of $\beta$ on a single event fitting. The event has $E = 10^{18.3}$ eV, and only $N_0$ remains as a free parameter.

## 6. Summary

The Unitary Cell of AMIGA has been completed since February 2015. It includes 37 scintillator modules in a hexagonal layout. In this work, we analysed the muon data over the first 20 months of operation.

We showed a method for finding a parametrisation of the muon counting resolution, to be used on an event by event basis, i.e. $\Delta^2\left(N_\mu\right) = (0.8 \pm 0.2) / N_\mu^{(0.9\pm0.1)}$. The scintillation module response resembles that of a Poisson counter.

Estimating the muon lateral trigger probability, we selected a fair sample for the parametrisation of the number of muons over $10\,\mathrm{m}^2$ at the optimal distance for the shower axis, as a function of the energy of the primary cosmic ray particle. This relation allows us to normalise the muon data in order to obtain an LDF fit performed on an energy-independent data set. This method led us to a value for the slope $\beta = 1.3 \pm 0.1$ to be used for single event fitting.

Future work is needed that includes a more detailed comparison with results from previous experiments cited in the introduction. Such analysis will have to deal with the different conditions in which these results were found: different muon energy threshold, different lateral distribution and optimal distances, and different shielding.

# The Energy Content of Extensive Air Showers in the Radio Frequency Range of 30-80 MHz

**Christian Glaser**[*a] **for the Pierre Auger Collaboration**[b]
[a] *RWTH Aachen University, Aachen, Germany*
[b] *Observatorio Pierre Auger, Av. San Martín Norte 304, 5613 Malargüe, Argentina*
*E-mail:* auger_spokespersons@fnal.gov
*Full author list:* http://www.auger.org/archive/authors_2015_06.html

At the Auger Engineering Radio Array (AERA) of the Pierre Auger Observatory, we have developed a new method to measure the total amount of energy that is transferred from the primary cosmic ray into radio emission. We find that this radiation energy is an estimator of the cosmic ray energy. It scales quadratically with the cosmic ray energy, as expected for coherent emission. We measure 15.8 MeV of radiation energy for a 1 EeV air shower arriving perpendicular to the geomagnetic field at the Auger site, in the frequency band of the detector from 30 to 80 MHz. These observations are compared to the data of the surface detector of the Observatory, which provide well-calibrated energies and arrival directions of the cosmic rays. We find energy resolutions of the radio reconstruction of 22% for the complete data set, and 17% for a high-quality subset containing only events with at least five stations with signal.

*Speaker.

## 1. Introduction

The detection of ultra-high energy cosmic rays via short radio pulses that are emitted during the air shower development has become an active field of research in recent years [1, 2, 3, 4, 5, 6, 7]. One advantage of this technique is a detection of cosmic rays that is sensitive to the particle type at close to a 100% duty cycle. The potential for a large scale radio detector is currently being explored in several experiments.

Much progress has been achieved in the understanding of the radio emission process. The dominant process is the geomagnetic emission that is due to the deflection of charged shower particles in the Earth's magnetic field which is polarized in the direction of the Lorentz force that acts on the shower particles [1, 3, 8]. The second component is polarized radially with respect to the axis of the air shower and results from the negative charge excess in the shower front [9, 10, 11, 12]. Its relative strength is on average 14% at the Auger site for an air shower arriving perpendicular to the geomagnetic field [13]. The constructive and destructive interference of the two emission processes lead to a radial asymmetry of the lateral signal distribution function (LDF) [14, 15, 16] that can be modeled with a two-dimensional LDF [17, 18].

The Auger Engineering Radio Array (AERA) [19] at the Pierre Auger Observatory [20] is the world's largest cosmic ray radio detector. In its latest stage of expansion, it consists of 153 radio detector stations that cover an area of approximately 17 $\text{km}^2$. AERA can take advantage of the fluorescence and surface detectors (SD) of the Pierre Auger Observatory that provide well-calibrated energies and arrival directions of the cosmic rays. The radio detector (RD) stations of AERA are located in an area of denser detector spacing of the SD array. This region, with SD station spacing of 0.75 km, allows the detection of cosmic ray energies down to about 0.1 EeV.

## 2. Data selection and event reconstruction

In this work we are using RD and SD data recorded between April 2011 and March 2013 when AERA was operating in its first phase. In this phase, AERA consisted of 24 antenna stations with a spacing of 144 m. The stations are equipped with logarithmic-periodic dipole antennas and measure over a frequency range of 30 to 80 MHz [21]. We use only events that have been measured by the radio and surface detectors in coincidence.

The raw data are reconstructed using the software framework $\overline{\text{Off}}\underline{\text{line}}$ of the Pierre Auger Collaboration [22, 23]. We correct for the influence of the analog signal chain using the absolute calibration of the AERA station and reconstruct a three-dimensional electric field by using the direction of the shower and applying the simulated antenna response [21]. An example of a reconstructed electric field trace $\vec{E}(t)$ is shown in Fig. 1.

We determine the energy density $u$ of the incoming electromagnetic radio pulse at each radio station by calculating the time integral over the absolute value of the Poynting vector. We add up the square of the magnitude of the electric field trace in a time-window of 200 ns ($[t_1, t_2]$) around the pulse maximum. The pulse maximum has been determined from the Hilbert envelope of the trace. We subtract the contribution of background noise (determined in the noise window $[t_3, t_4]$) under the assumption that the main contribution is white noise. The energy density $u$, in units of

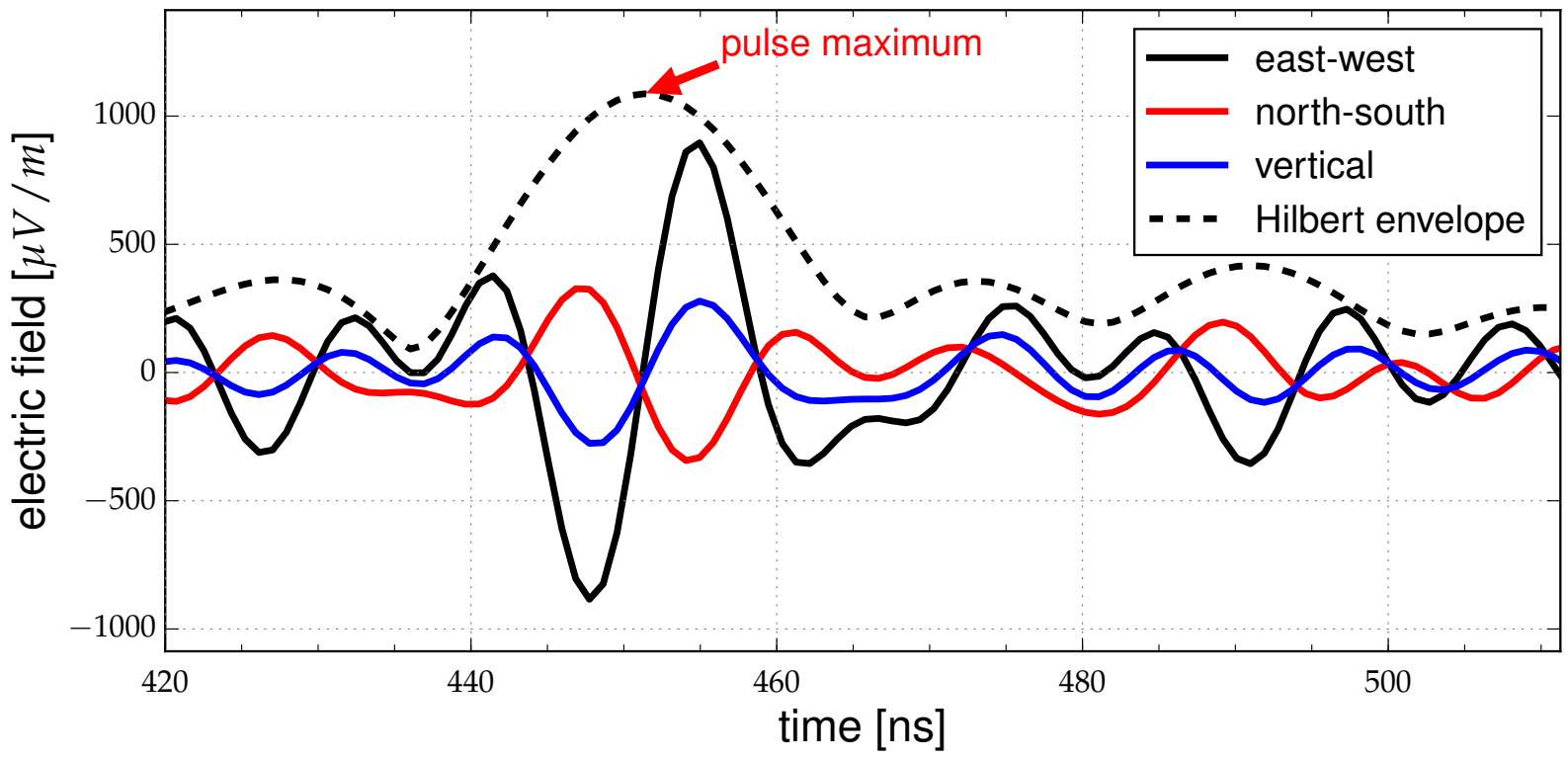

**Figure 1:** Reconstructed electric field trace of one of the measured cosmic ray radio events. An upsampling by a factor of five was applied. The shown Hilbert envelope (dashed line) is the square root of the quadratic sum of the Hilbert envelopes of the three polarization components.

eV/m$^2$, is given by

$$u = \varepsilon_0 c \left( \Delta t \sum_{t_1}^{t_2} |\vec{E}(t_i)|^2 - \Delta t \frac{t_2 - t_1}{t_4 - t_3} \sum_{t_3}^{t_4} |\vec{E}(t_i)|^2 \right), \tag{2.1}$$

where $\varepsilon_0$ is the vacuum permittivity, $c$ is the speed of light in vacuum, $\Delta t$ is the size of one time bin. To approximate the uncertainty the noise level is used. In addition, the amplification of the signal chain adds 5% uncertainty in $u$ that is added in quadrature.

We apply quality cuts on the data of the surface detector [24]. The most important cuts are that the core position be surrounded by a hexagon of active stations and that the zenith angle of the incoming direction be less than 55°. Furthermore, we require that the reconstructed incoming direction from the radio and the surface detectors agree within 20°, that the event not be recorded during thunderstorms and we impose a cut on the measured polarization to efficiently reject noise pulses. 134 events remain after all cuts.

## 3. Radiation energy and energy estimator

To reconstruct the radiation energy, i.e., the energy that has been transferred from the cosmic ray into radiation in the 30 to 80 MHz regime, the energy density – measured at the individual radio stations – is interpolated spatially and integrated. We use a two-dimensional lateral signal distribution function that describes all main features seen in simulated and measured cosmic ray radio events [17]. We parametrize the function with four free parameters, namely the amplitude $A$, the slope parameter $\sigma$ and the particle core position $\vec{r}_{\mathrm{core}}$. The remaining constants $C_0 - C_4$ are estimated from Monte Carlo simulations for the AERA site (see Tab. 1). Thus we find

$$u(\vec{r}) = A \left[ \exp\left( \frac{-(\vec{r} + C_1 \vec{e}_{\vec{v}\times\vec{B}} - \vec{r}_{\mathrm{core}})^2}{\sigma^2} \right) - C_0 \exp\left( \frac{-(\vec{r} + C_2 \vec{e}_{\vec{v}\times\vec{B}} - \vec{r}_{\mathrm{core}})^2}{(C_3 e^{C_4 \sigma})^2} \right) \right]. \tag{3.1}$$

All coordinates are in the shower plane and $\vec{r}$ denotes the station position. $\vec{e}_{\vec{v}\times\vec{B}}$ is the unit vector pointing into the polarization direction of the geomagnetic emission. The LDF is fitted to the data using a chi-square minimization. In case of low station multiplicity, the particle core position is taken from the SD reconstruction, which enables us to also use events with signals in only three or four radio stations. After a quality cut on the LDF fit, 126 events remain. An example of one air shower of our data set is shown in Fig. 2.

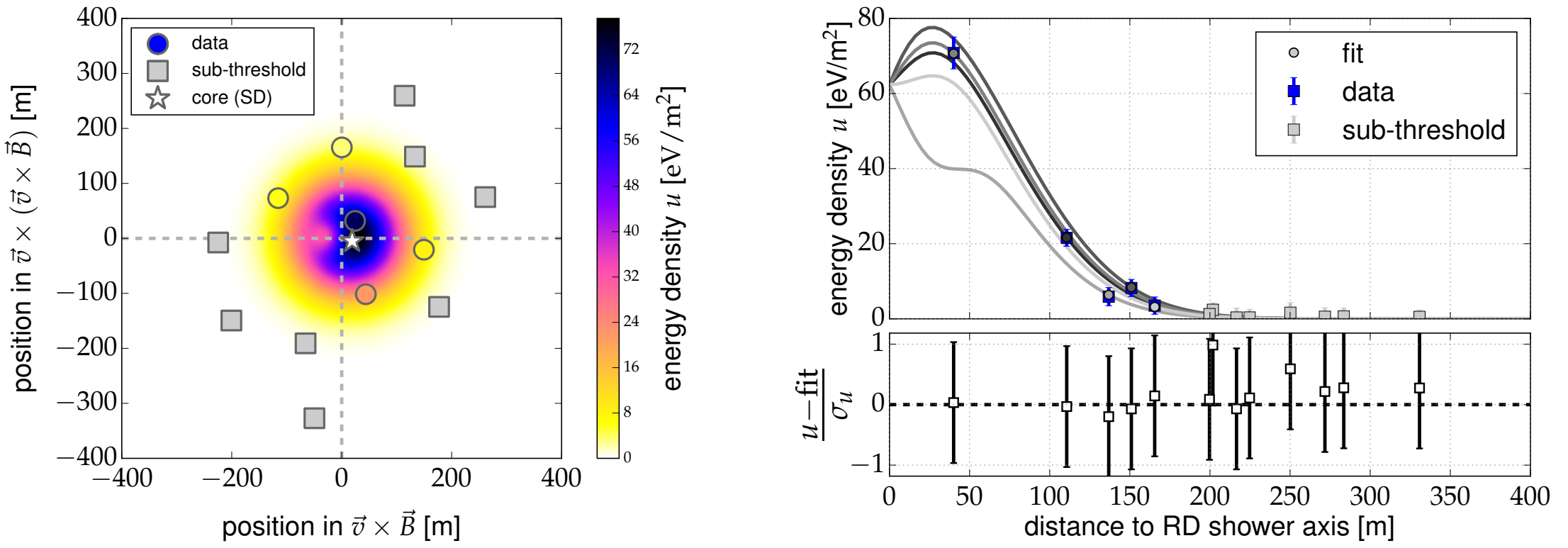

**Figure 2:** Lateral signal distribution of an air shower with an energy of 0.65 EeV and arriving at a zenith angle of 21° and from 31° east of north. Left: Two dimensional distribution of the RD energy density in the shower plane. The coordinate origin is the RD shower core. The star indicates the SD core position. Circles and squares mark the positions of the radio stations. The color shows the measured energy density. Stations with signal below threshold are marked with gray squares. The background map shows the LDF parametrization. Right: One-dimensional projection of the LDF. Blue squares show the measured energy density and gray squares denote the sub-threshold stations. Gray circles denote the value of the LDF parametrization at the position of the measurement. The radial fall-off of the LDF along a line connecting the radio core position with each station position is shown as gray curves. Also shown are the residuals in units of the uncertainty of the measurement.

The radiation energy is the spatial integral of the LDF and will be given in units of eV. To obtain a cosmic ray energy estimator we correct for the different emission strengths at different angles $\alpha$ between the shower axis and the magnetic field by dividing the radiation energy by $\sin^2 \alpha$:

$$S_{\text{radio}} = \frac{1}{\sin^2 \alpha} \int_{\mathbb{R}^2} u(\vec{r})\, \mathrm{d}^2\vec{r} = \frac{A\pi}{\sin^2 \alpha} \left( \sigma^2 - C_0 C_3^2\, e^{2C_4\sigma} \right) , \tag{3.2}$$

where $\mathbb{R}^2$ denotes the shower plane. The formula is valid for values of $\alpha > 10°$. For smaller values of $\alpha$ the geomagnetic emission is not the dominant contribution and the $\sin^2 \alpha$ correction becomes invalid. However, due to the reduced emission strength the number of detections for arrival directions within 10° of the geomagnetic axis is suppressed. In our data set all events have arrival directions further away from the magnetic field axis.

The dominant event by event uncertainties of the energy estimator are the uncertainty of the LDF fit, the antenna response pattern (10%) and the temperature dependence of the amplifiers (8%) that is not yet corrected for. The average fit uncertainty of $S_{\text{radio}}$ is 46% and reduces to 24% in case of events with five or more stations with signal. The absolute scale uncertainty of the energy estimator is dominated by the absolute scale uncertainty of the antenna response pattern (25%) [21] and the analog signal chain (12%) and amounts to 28%. All uncertainties are added in quadrature. Please note that as the energy estimator scales quadratically with the cosmic ray energy, the resulting uncertainty of the cosmic ray energy will be only half of the uncertainty of the energy estimator.

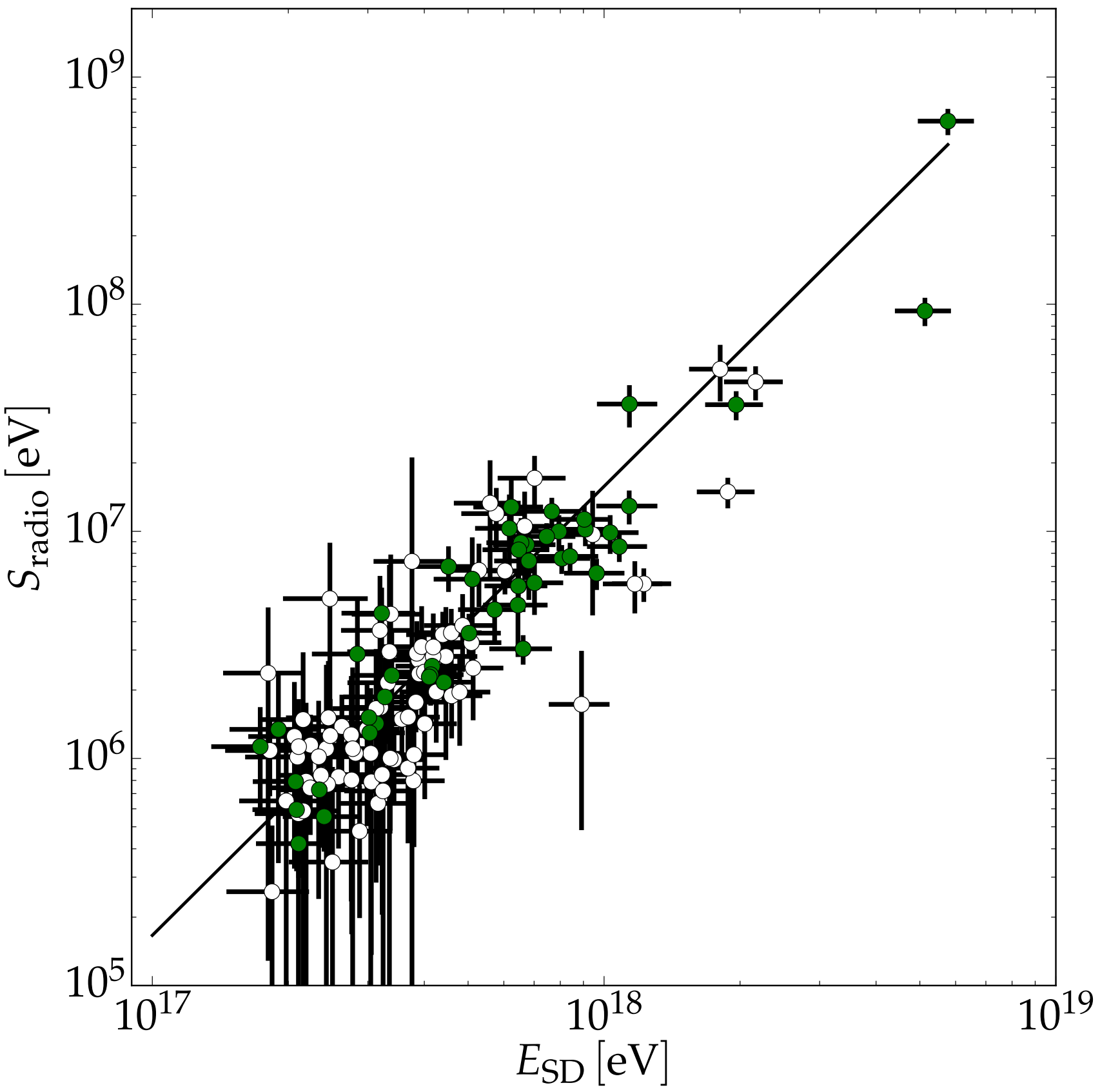

**Figure 3:** Correlation between the radio energy estimator $S_{\mathrm{radio}}$ and the cosmic ray energy measured with the surface detector. Green filled circles denote air showers where the core position has been determined in the radio LDF fit, i.e., all air showers with at least five stations with signal. Open circles denote events with less than five stations with signal which use the SD core position. The black line is the best fit of the calibration function.

The energy estimator $S_{\mathrm{radio}}$ is calibrated using air showers measured with AERA and the surface detector in coincidence. $S_{\mathrm{radio}}$ shows a clear correlation with the cosmic ray energy measured by the surface detector (cf. Fig. 3). The calibration function $S_{\mathrm{radio}} = A \times 10^7\,\mathrm{eV}\,(E_{\mathrm{SD}}/10^{18}\,\mathrm{eV})^B$ is obtained by maximizing a likelihood function that takes all measurement uncertainties, detector efficiencies and the steeply falling energy spectrum into account. This method was previously used in [25, 26] and is documented in [27]. The result of the calibration fit is $A = 1.58 \pm 0.07$ and $B = 1.98 \pm 0.04$. For a high quality subset of events with at least five radio stations with signal (47 events) the fit gives a compatible result of $A = 1.60 \pm 0.08$ and $B = 1.99 \pm 0.05$.

To estimate the energy resolution of AERA we subtract the known SD energy resolution from the scatter around the calibration curve, assuming that the energy estimators of the SD and RD are uncorrelated for a fixed cosmic ray energy. We find a resolution of 22% for the full data set and 17% for the high quality set. We also compare the observed and the expected scatter that can be calculated from the likelihood function and the uncertainties of the SD energy estimator and $S_{\mathrm{radio}}$ discussed above and we find that the observed scatter is compatible with our expectation. As the radio energy estimator is calibrated with SD data, we inherit the systematic uncertainty of the SD energy scale. It is 14% at energies $\geq 10^{18}$ eV [28] and increases to 16% at $10^{17.5}$ eV.

## 4. The energy content of extensive air showers in the radio frequency range of 30 to 80 MHz

We can generalize our results to other locations by normalizing the calibration function to the local magnetic field. Then, we can predict the radiation energy in the frequency range of 30 to 80 MHz by

$$E_{30-80\,\mathrm{MHz}} = (15.8 \pm 0.7(\mathrm{stat}) \pm 6.7(\mathrm{sys}))\,\mathrm{MeV} \left( \sin\alpha \, \frac{E}{10^{18}\,\mathrm{eV}} \, \frac{B_{\mathrm{Earth}}}{0.24\,\mathrm{G}} \right)^2 , \tag{4.1}$$

where $E$ is the cosmic ray energy, $B_{\mathrm{Earth}}$ denotes the local magnetic field strength and 0.24 G is the magnetic field strength at the AERA site. We found that it is sufficient to correct only for the dominant geomagnetic emission process. However, this formula will become invalid for radio detectors at high altitudes because the amount of radiation energy decreases as – depending on the zenith angle – a significant part of the air shower is clipped away at the ground.

The direct measurement of the radiation energy holds further potential for a precise determination of the cosmic ray energy scale. Radio emission originates only from the electromagnetic part of an air shower. Therefore, the radiation energy can be predicted by Monte Carlo simulations from first principles [16, 29], as the emission arises from the acceleration/deceleration of charges which is described by classical electrodynamics [30]. In addition, no relevant propagation uncertainties arise as the atmosphere is transparent for radio emission. Then, in turn, the energy scale of the radio detector can be propagated to the other detection techniques.

## 5. Conclusions

We have presented a direct measurement of the radiation energy of an extensive air shower in the 30 to 80 MHz regime. The AERA radio stations are well calibrated which enables the reconstruction of the energy density of the radio pulse at each detector position. The radiation energy is then the integral of the two-dimensional lateral signal distribution function that is fitted to the data. We measure 15.8 MeV of radiation energy for a 1 EeV cosmic ray arriving perpendicular to the local magnetic field.

As a cosmic ray energy estimator we use this radiation energy corrected for different emission strengths at different angles between the shower axis and the geomagnetic field. We calibrate the energy estimator using the reconstructed energy from the surface detector of the Pierre Auger Observatory and find that the radio energy estimator scales quadratically with the cosmic ray energy. We analyze the scatter around the calibration and find an energy resolution of the radio detector of 22%. The resolution improves to 17% for a high quality subset of the data where only events with at least five radio stations with signal are used.

We generalize our result by normalizing the calibration function to the local magnetic field strength at the Auger site. This leads to a universal prediction of the radiation energy that can be used at any location.

**Table 1:** Parameters $C_0$ - $C_4$ of Eq. (3.1). $C_3 = 16.25\,\text{m}$ and $C_4 = 0.0079\,\text{m}^{-1}$. The zenith angle dependent values used to predict the emission pattern are given for zenith angle bins up to 60°.

| zenith angle | $C_0$ | $C_1$ [m] | $C_2$ [m] |
|---|---|---|---|
| $0° - 10°$ | 0.41 | $8.0 \pm 0.3$ | $-21.2 \pm 0.4$ |
| $10° - 20°$ | 0.41 | $10.0 \pm 0.4$ | $-23.1 \pm 0.4$ |
| $20° - 30°$ | 0.41 | $12.0 \pm 0.3$ | $-25.5 \pm 0.3$ |
| $30° - 40°$ | 0.41 | $20.0 \pm 0.4$ | $-32.0 \pm 0.6$ |
| $40° - 50°$ | 0.46 | $25.1 \pm 0.9$ | $-34.5 \pm 0.7$ |
| $50° - 60°$ | 0.71 | $27.3 \pm 1.0$ | $-9.8 \pm 1.5$ |

# Measurement of the water-Cherenkov detector response to inclined muons using an RPC hodoscope

**Pedro Assis**[*a†] **for the Pierre Auger Collaboration**[b]
[a] *LIP / IST, Lisboa, Portugal*
[b] *Observatorio Pierre Auger, Av. San Martín Norte 304, 5613 Malargüe, Argentina*
*E-mail:* auger_spokespersons@fnal.gov
*Full author list:* http://www.auger.org/archive/authors_2015_06.html

The Pierre Auger Observatory operates a hybrid detector composed of a Fluorescence Detector and a Surface Detector array. Water-Cherenkov detectors (WCD) are the building blocks of the array and as such play a key role in the detection of secondary particles at the ground. A good knowledge of the detector response is of paramount importance to lower systematic uncertainties and thus to increase the capability of the experiment in determining the muon content of the extensive air showers with a higher precision. In this work we report on a detailed study of the detector response to single muons as a function of their trajectories in the WCD. A dedicated Resistive Plate Chambers (RPC) hodoscope was built and installed around one of the detectors. The hodoscope is formed by two stand-alone low gas flux segmented RPC detectors with the test water-Cherenkov detector placed in between. The segmentation of the RPC detectors is of the order of 10 cm. The hodoscope is used to trigger and select single muon events in different geometries. The signal recorded in the water-Cherenkov detector and performance estimators were studied as a function of the trajectories of the muons and compared with a dedicated simulation. An agreement at the percent level was found, showing that the simulation correctly describes the tank response.

[*] Speaker.
[†] P. Assis gratefully acknowledges the financial support by FCT/COMPETE/QREN

## 1. Introduction

The water-Cherenkov detector (WCD) is the main building block of the Surface Detector (SD) of the Pierre Auger Observatory. The WCD samples the shower charged particles that arrive at ground [1]. Through the number of particles that reach each SD station and their arrival times it is possible to infer the nature and arrival direction of the primary cosmic ray which initiated the extensive air shower (EAS). Moreover, using sophisticated analyses that explore the WCD response to different shower components, it is also possible to assess the EAS muon content, an important quantity to evaluate our current understanding of the shower description. Naturally, these analyses require a deep understanding of the detectors.

In this work, which extends the measurements performed in Orsay with a test WCD equipped with scintillators [2], we aim to improve the understanding of the parameters governing the light propagation inside the WCD, namely the light reflection and absorption. For that, a hodoscope was mounted, enabling the selection of particular particle trajectories inside the WCD.

The detailed information from these measurements can be used to fine-tune the simulation of the detector and contribute to a reduction of the present systematic uncertainties deriving from the optical properties of the WCD.

## 2. Description of the experimental apparatus

The SD [1] is composed of an array of WCD which are molded plastic containers (tanks) with a sealed liner filled with purified water. The inner side of the liner is surfaced with Tyvek® to reflect the light generated in the water by the Cherenkov effect of the crossing relativistic particles. At its top, three PMTs are installed by means of transparent plastic windows. A detailed description of the WCD can be found in [3]. The Gianni Navarra detector, the focus of this work, is a test WCD located in the Malargüe central campus and thus disconnected from the actual SD.

Resistive Plate Chambers (RPCs) are gaseous detectors capable of achieving very high detection efficiency and high timing accuracy. The RPCs are made of a sensitive gas volume, where the primary ionization and avalanche multiplication occurs, and by a signal pickup module. The pickup is segmented into 64 pads $\left(\sim 15 \times 19\,\mathrm{cm}^2 \text{ each}\right)$, allowing the estimation of the active region. A review of RPCs and their principle of operation can be found in [4, 5]. The detectors used in the hodoscope are a result of a specific development of RPCs for the Pierre Auger Observatory [6, 7] that was focused on the operation of autonomous stations equipped with RPCs. As such it was necessary to develop RPCs with low gas consumption and simple electronics to meet stringent requirements on price and power consumption.

The hodoscope is formed by two stand alone low gas flux RPC detectors with the testing WCD placed in between. A photograph of the setup is shown in fig. 1. The readout of the RPCs is segmented into small pads, which allows for an accurate reconstruction of the individual muon trajectories. The hodoscope is used to trigger and select single trajectory events in different geometries. Both RPCs are installed in moveable structures allowing for different configurations and hence probing different regions of angles of incidence.

## 3. Data Acquisition system

The data acquisition (DAQ) system of the setup has two main components: the WCD DAQ and the hodoscope DAQ. The WCD DAQ uses the standard Surface Detector Electronics (SDE) [8]. The Gianni Navarra station is not included in the standard array and is, as such, not connected to the Central Data Acquisition System. Instead, the control and readout of the electronics is performed through a direct link to the console. The largest part of the event data consists of the traces of the three photomultipliers. Occasionally the station records calibration histograms that are computed internally using a dedicated self-trigger. These calibration data consist in the set of the distributions of the baseline and of the charge amplitude of single particles.

The hodoscope electronics is based on a prototype discrete electronics system (PREC) developed for the readout of the RPCs. The system uses an architecture with front-end boards and one motherboard. In the Front End (FE), the signal from each pad is amplified and then a simple threshold is applied to perform a 1-bit digitization. These signals are then fed into a purely digital motherboard comprising 14 FPGAs (Field-Programmable Gate Array) organized in a mother-daughters configuration. The communication between the mother FPGA and a PC is accomplished with an USB connection. For each acquired pad this scheme allows us to record whether there was a signal above the imposed threshold in a window of 1 μs before the WCD trigger.

Two types of trigger systems are used: the WCD-based trigger and the hodoscope-based trigger. The WCD-based trigger uses the SDE triggering system lowering the threshold to 0.2 VEM to catch signals as low as possible. A VEM is defined as the most probable value for the signal from a center going vertical muon. This SDE trigger was then fed to the PREC system to record the data from the hodoscope. Data synchronization was achieved by means of a global identifier generated by the WCD electronics and communicated to the PREC using auxiliary digital data lines. Upon a trigger, the PREC electronics saves the data for all the pads.

To increase the rate of events with particles crossing both the hodoscope and the WCD, a new triggering scheme was set up, based on the hodoscope signals and implemented in the motherboard. In short, one FPGA from the motherboard looks for a pattern where at least one pad on the top RPC and at least one pad on the bottom RPC were activated in a time window of 40 ns. The generated trigger is then sent internally into all FPGAs of the motherboard and also to the WCD, causing it to acquire the trace of the PMTs and to generate the event identifier.

## 4. Dedicated simulation

To assess the results obtained with the hodoscope, a dedicated simulation was developed. This simulation tries to reproduce the conditions at which the measurements were performed, accounting for geometrical and detector efficiencies.

In order to use a realistic set of events, CORSIKA simulations of the atmospheric particles were run for the Malargüe site. The expected primary cosmic ray fluxes in the range energy $\log(E/\mathrm{eV}) \in [11, 15]$ and an isotropic arrival direction were considered in the simulation.

The resulting energy distribution is shown in fig. 1. The all-particle distribution is displayed by the black curve while the different components of the distribution are shown by the filled his-tograms. It can be seen that the low energy region is dominated by photons while above a few GeV, only muons contribute.

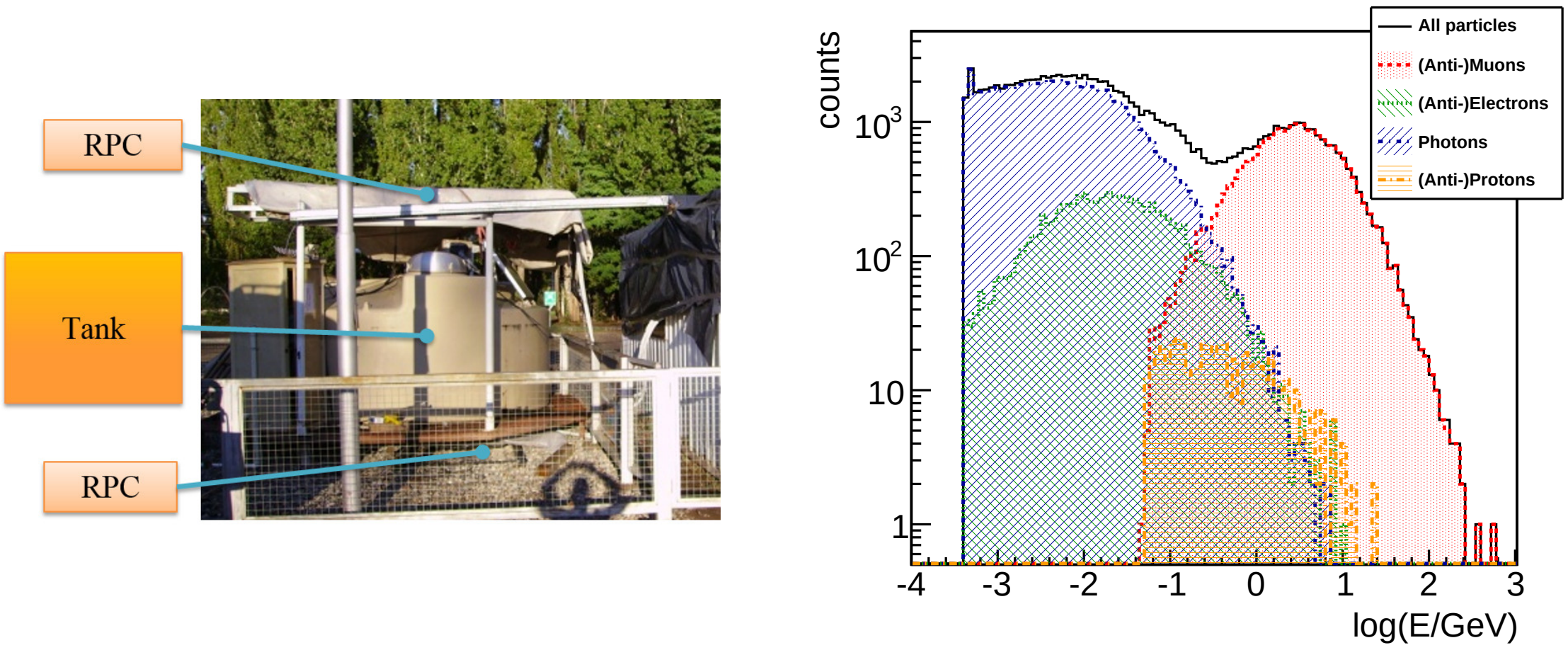

**Figure 1:** Left: photograph of the muon hodoscope setup at the Gianni Navarra WCD. Right: atmospheric particle energy spectrum used to produce the dedicated simulation.

The simulated atmospheric particles are then injected into a simulation of the WCD+RPC setup using GEANT4. A detailed description of the RPC structure was implemented by taking advantage of the GEANT4 capabilities to describe complex geometries and the physical properties of the materials (fig. 2).

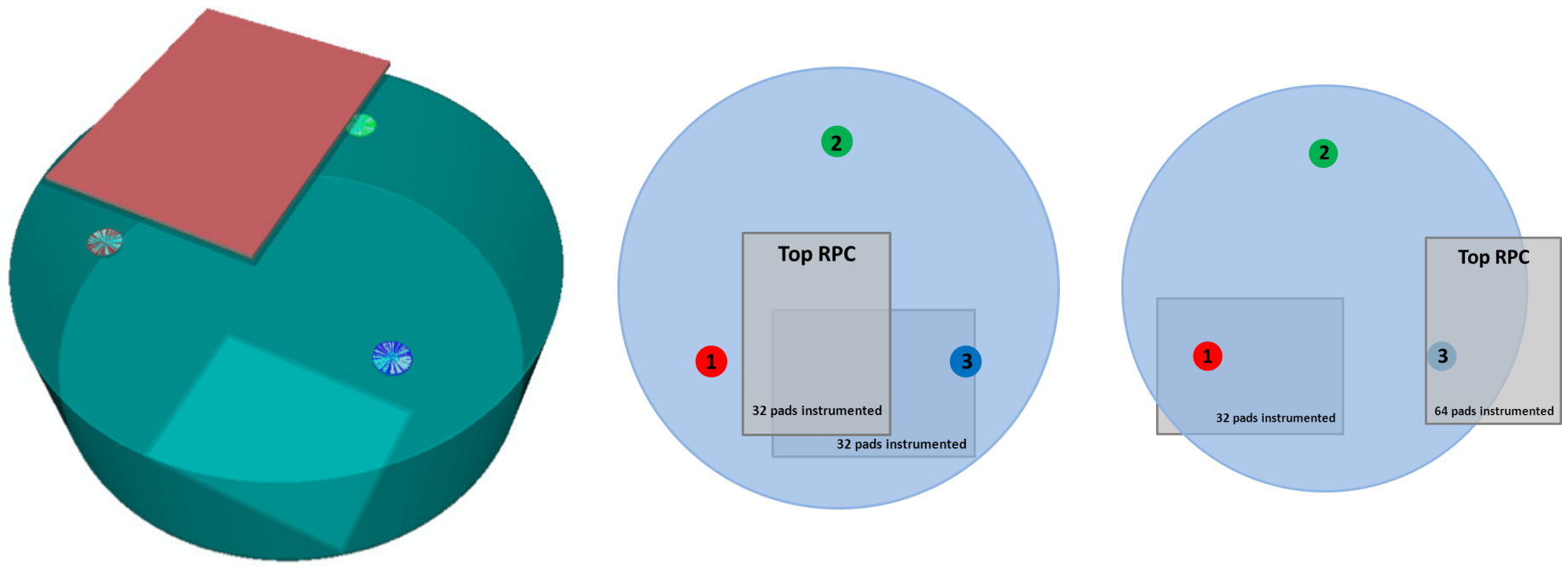

**Figure 2:** Left: view of one of the RPC configurations as implemented in GEANT4. Middle and right: top view of configuration 1 and 2 respectively (see text for details).

The particle tracking in the WCD, the Cherenkov light generation and propagation is performed exactly in the same way as in the standard simulation of the WCD in the Auger Offline software [9]. The tracking in the RPCs enables the recording of the incoming particle position, direction and the time, as well as the ionizing energy deposited in the gas, which is then used to generate charge pulses. The electronic signals are then effectively simulated by a parameterisation.

## 5. Acquired Data and Analysis

The data presented in this work correspond to three experimental campaigns: in campaign 1 the WCD trigger was used and the RPC were almost on top of one another allowing to study zenith angles in $\sim[0^\circ, 40^\circ]$ ; in campaign 2 the RPCs were moved to cover zenith angles in $\sim[20^\circ, 55^\circ]$; in campaign 3 the later RPC configuration was kept and the hodoscope-based trigger was used. The number of collected events in each campaign was, respectively, $1.2\times10^3$, $2.3\times10^3$ and $3.5\times10^5$.

The RPC data were analyzed with the goal of determining the trajectory of single atmospheric muons passing through the hodoscope by reconstructing the muon zenith and azimuth angles, as well as the muon tracklength $(L)$ inside the WCD water. A three-step analysis was developed for this purpose.

Firstly, noisy and dead RPC pads were identified and removed from the analysis. In fig. 3 the selected pads for the data analysis, for the third data collection campaign, are shown.

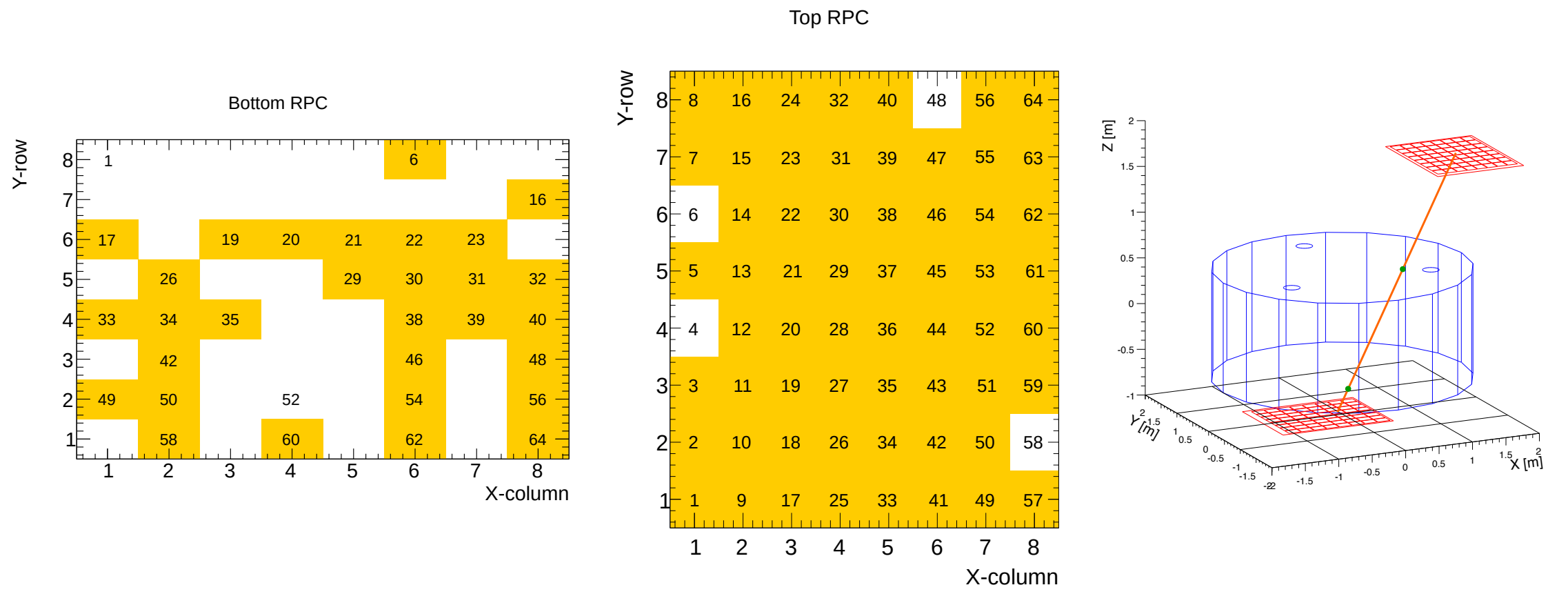

**Figure 3:** Left and middle: scheme of the bottom and top RPC pads in experimental campaign 3. The pads with readout instrumentation are numbered and those that passed the selection cuts are shown in orange. Right: schematic view of a muon track going through the hodoscope mounted on the Gianni Navarra WCD.

The second step of the data analysis is the event selection. Since the goal of this work is to study the WCD response as a function of the muon trajectory, the interesting events are those in which a single muon triggers the WCD and both RPCs. The event selection criteria are therefore to require *one and only one* hit in the top RPC and *one and only one* hit in the bottom RPC. This defines the "single-hit" selection. The contamination of the sample by random coincidences was studied and found to be only of a few percent, thus having a negligible impact on the results.

In the third step the muon trajectory, shown schematically in fig. 3 (right), is reconstructed from the positions of the hit pads.

The direction of the traversing particles is calculated from the center of the two activated pads. Since each pad has a finite area $\left(\sim 15\times 19\,\mathrm{cm}^2\right)$ an uncertainty is introduced in the estimation of the geometry which is the solid angle subtended by the two pads. The resulting hodoscope resolution in terms of tracklength, zenith angle and azimuth angle is of the order of a centimeter, $1^\circ$ and $2^\circ$, respectively for near vertical trajectories.

Finally, the baseline, peak time-bin, current and charge for high gain traces of individual PMTs and for the total trace were estimated using standard analysis methods [10] of the Pierre Auger Collaboration. These quantities were studied as a function of the muon trajectory.

The WCD calibration data were used to make the conversion from charge (in integrated ADC counts) to signal in VEM units. The muon peak was fitted by adjusting Gaussian functions to the calibration data to determine the position of the maximum and therefore to obtain the calibration constants, relating the charge in integrated ADC counts with the maximum of the muon peak that corresponds to 1.04 VEM [10]. While in campaigns 1 and 2 the calibration histograms were available and the above mentioned calibration procedure was followed, in campaign 3 they were not. In this case, the data were rescaled to the simulation value for $L \in [1.25, 1.3]$ m.

## 6. Preliminary Results

The measured total signal distribution (charge distribution) upon applying the VEM calibration procedure is shown in fig. 4 (left). The red histogram is the signal distribution for all recorded events while the black one is the distribution after the single-hit selection cut.

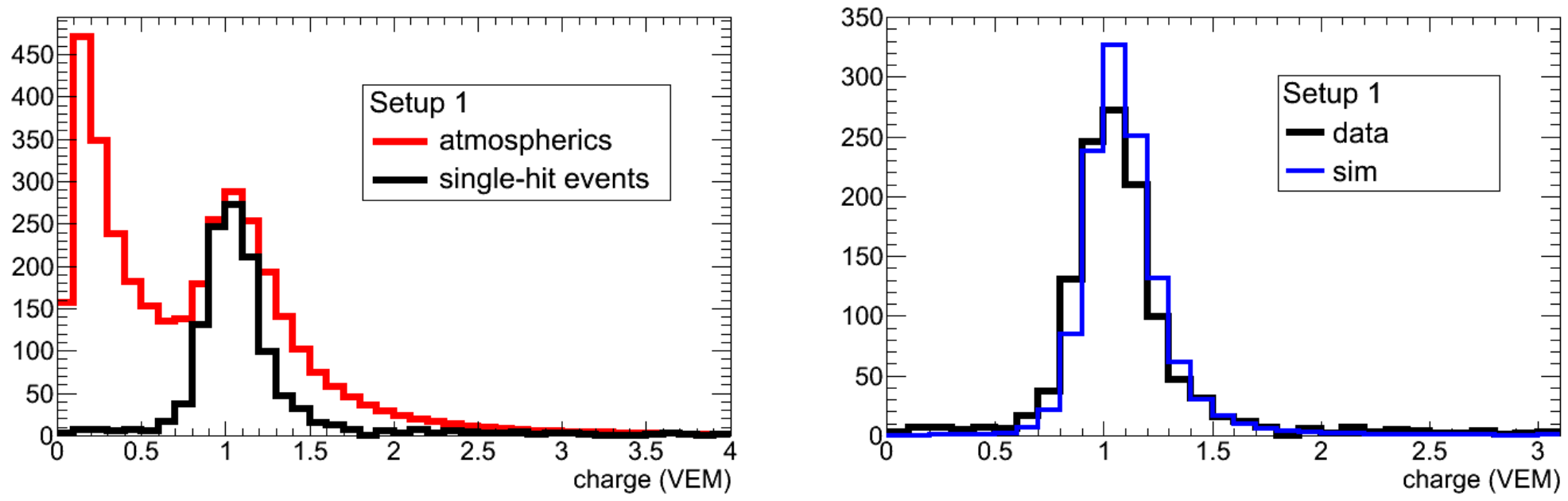

**Figure 4:** Left: charge distribution in data, before and after selecting single hit events; Right: charge distributions for single hit events in data and simulation. The number of simulated events was normalized to the number of acquired events.

As a consequence of this cut, the electromagnetic component of the atmospheric particle flux is severely reduced. This leads to the disappearance of the first, low energy, peak shown in this plot, confirming the selection of muons. The events with high charge are characterized by having large tracklengths or/and high multiplicity. As the RPCs configuration limits the maximum tracklength, and the required multiplicity of the cut is 1, the tail at the right of this plot is also reduced.

In fig. 4 (right), a comparison between data and simulation is shown for the total signal distribution requiring single hit events. It can be seen that the histograms have a good qualitative agreement with both peaks close to 1 VEM.

Further insight on the details of the WCD response can be gained by studying the dependency of the signal as a function of the muon tracklength ($L$) in the WCD. Fig. 5 shows the average trace for different ranges of $L$, showing as expected an increase of the signal amplitude with tracklength.

In fig. 6 the dependency of the signal with $L$ is shown. Data are plotted in black and simulation in blue. As expected, the signal increases as a function of the tracklength in the WCD. The experimental data are in agreement with the simulation to the few percent level. Moreover, the geometry of the selected tracks introduces some structures that are well described by the simulation.

## 7. Summary and Prospects

In this work, we reported on the hodoscope system installed at the Gianni Navarra WCD for the detailed study of the detector response to individual muons traversing the WCD. We focus on the setup, the simulation and the results from the first collected data. The system is comprised of a hodoscope realized with RPCs that is able to tag single muon events and to estimate the trajectory, namely the tracklength in the WCD, with an accuracy of a few cm. A dedicated simulation was developed to take into account not only the geometry of the setup but also a realistic flux of atmospheric particles. The signal evolution with the tracklength behaves as expected when compared to the simulation. A maximum deviation of $\sim 2\%$ was found for the acquired data, showing that the simulation correctly describes the tank response.

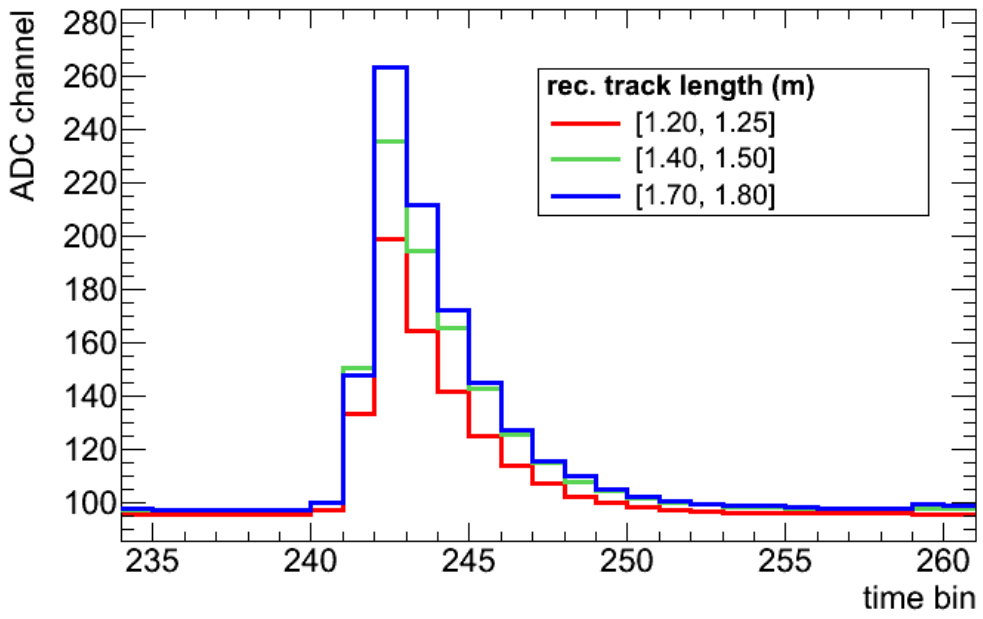

**Figure 5:** Average trace (PMTs 2 and 3) for a few selected tracklength intervals.

In the future we foresee making several upgrades to the setup in order to stabilize and enhance its acquisition capabilities. The most important upgrade will be a change in the mechanical setup that will allow us to rotate the RPCs and increase the limit in the zenith angle up to almost 90°.

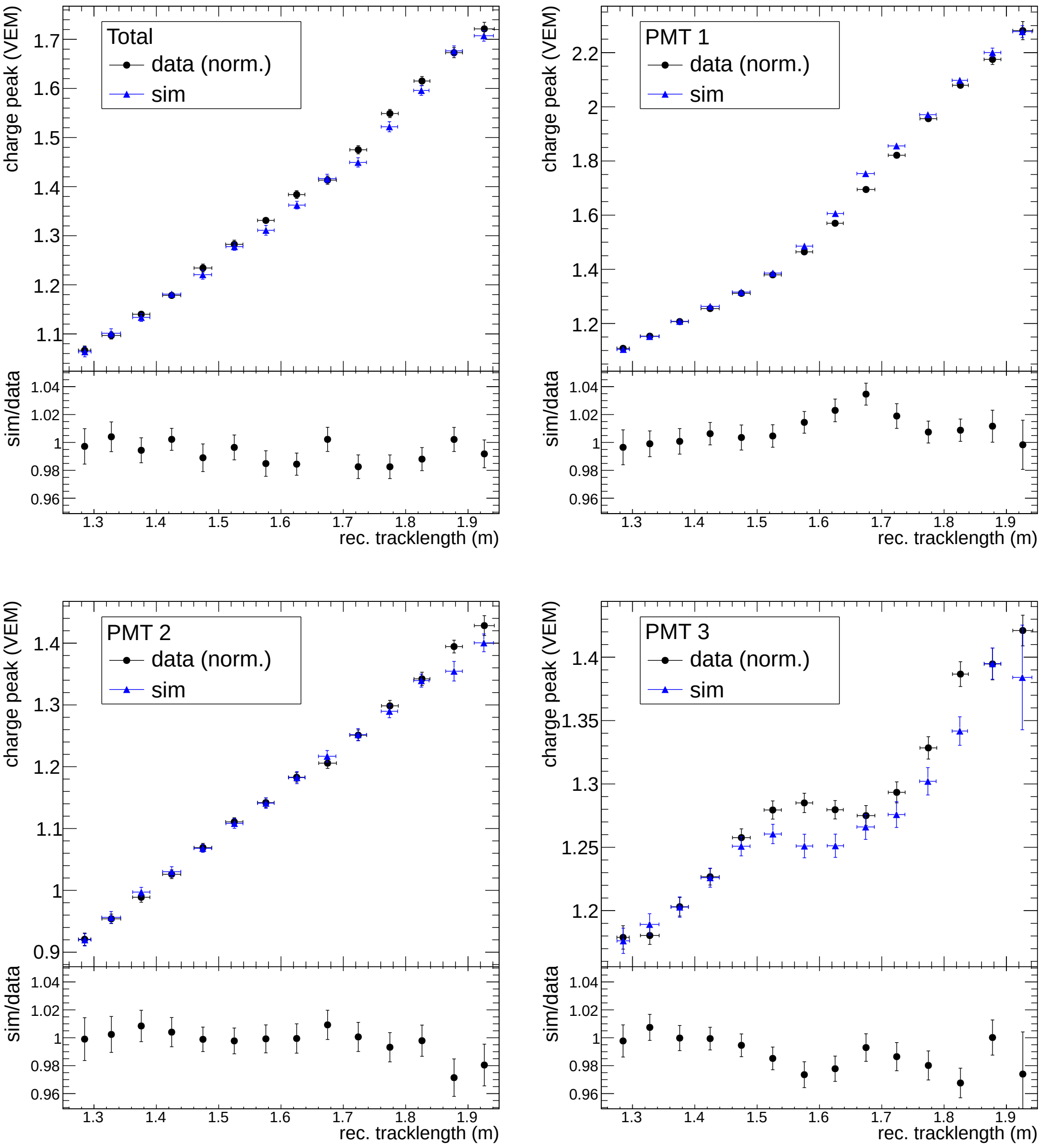

**Figure 6:** Peak of the charge distribution as a function of the muon tracklength in the WCD water, for the total PMT signal (top left) and for individual PMTs in campaign 3. Data were rescaled to the simulation value for $L \in [1.25, 1.3]$ m. The ratio between simulation and data is also shown at the bottom.

# Automated procedures for the Fluorescence Detector calibration at the Pierre Auger Observatory

**Gaetano Salina[*] for the Pierre Auger Collaboration[a]**
[*]*Istituto Nazionale di Fisica Nucleare Sezione di Roma Tor Vergata*
*Via Della Ricerca Scientifica 1, Rome, Italy*
[a]*Observatorio Pierre Auger*
*Av. San Martín Norte 304, 5613 Malargüe, Argentina*
*E-mail:* auger_spokespersons@fnal.gov
*Full author list:* http://www.auger.org/archive/authors_2015_06.html

The quality of the physics results, derived from the analysis of the data collected at the Pierre Auger Observatory depends heavily on the calibration and monitoring of the components of the detectors. It is crucial to maintain a database containing complete information on the absolute calibration of all photomultipliers and their time evolution. The low rate of the physics events implies that the analysis will have to be made over a long period of operation. This requirement imposes a very organized and reliable data storage and data management strategy, in order to guarantee correct data preservation and high data quality. The Fluorescence Detector (FD) consists of 27 telescopes with about 12,000 phototubes which have to be calibrated periodically. A special absolute calibration system is used. It is based on a calibrated light source with a diffusive screen, uniformly illuminating photomultipliers of the camera. This absolute calibration is performed every few years, as its use is not compatible with the operation of the detector. To monitor the stability and the time behavior, another light source system operates every night of data taking. This relative calibration procedure yields more than $2\times10^4$ raw files each year, about 1 TByte/year. In this paper we describe a new web-interfaced database architecture to manage, store, produce and analyse FD calibration data. It contains the configuration and operating parameters of the detectors at each instant and other relevant functional parameters that are needed for the analysis or to monitor possible instabilities, used for the early discovery of malfunctioning components. Based on over 10 years of operation, we present results on the long term performance of FD and its dependence on environmental variables. We also report on a check of the absolute calibration values by analysing the signals left by stars traversing the FD field of view.

* Speaker.

## 1. Introduction

The quality of the physics results, derived from the analysis of the data collected at the Pierre Auger Observatory depends heavily on the calibration and monitoring of the components of the detectors. The low rate of the physics events implies that the analysis will have to be made over a long period of operation. This requirement imposes a very organized and reliable data storage and data management strategy, in order to guarantee correct data preservation and high data quality. The Fluorescence Detector (FD) consists of 27 telescopes with about 12,000 phototubes which have to be calibrated periodically [1]. The 27 telescopes are located in five different sites, i.e. Los Leones, Los Morados, Lama Amarilla, Coihueco and HEAT. The absolute calibration (Drum constants) is performed every few years, as its use is not compatible with the operation of the detector [2]. To monitor the stability and the time behaviour, another light source system operates (Led Control Unit, LCU) every night of data taking [1,2]. In this paper we present the structure and the performance of the Fluorescence Detector Database (FD-DB).

## 2. The Fluorescence Detector Database Architecture

### 2.1 The Data Base Structure

The structure of the FD-DB prototype is shown in Fig.1. It consists of nine different databases, each one containing structurally homogeneous data. We have chosen this structure in order to optimize the management (updating, checks and display) of the data and reduce the load for the SQL Manager and improve the efficiency of the analysis applications, which operate on data that are essentially homogenous. Special care was taken to optimize the queries involving data that are resident in different databases. The data stored in each database are:

- **Global:** Security data: user, password and access level. DB Manager Data: type and time of operation, user executing operation, path of log file and status of the operations.
- **PMTS:** Production and acceptance test on all PMTs. Configuration of the 440 PMTs of each telescope, i.e. the position identified by its absolute geographical address, the serial number and the HV class. The history of the configuration of the 440 PMTs of each telescope. Information on maintenance actions performed on the telescopes: type of action (i.e. cleaning of the mirrors and filters, breakdown of some components and subsequent replacement etc.), and GPS time.
- **DB_Mirrors:** Positions and reflectivity of the mirrors at different wavelengths for each camera.
- **RelCal:** The time response of each PMT to the relative calibration operation (A, B and C type), the absolute geographical address, its value and error and the GPS time. Information on the time stability of the light sources (LED and Xe lamp), i.e. value, error and GPS time.
- **AbsCal:** The nightly time response of each PMT to the absolute calibration, i.e. the absolute geographical address of the PMT, its response value and error and the time in GPS. Values calculated from RelCal values, LCU correction and Drum constants.
- **FDCalib:** The selected nightly time response of each PMT to the absolute calibration, i.e. the absolute geographical address of the PMT, its response value and error and the time in GPS. This is the database used in the reconstruction of the air shower detected by FD.
- **Aux:** Drum constants, telescope reference correction, reference night identification [2].
- **Catalogue:** The absolute path of the relative calibration raw files.

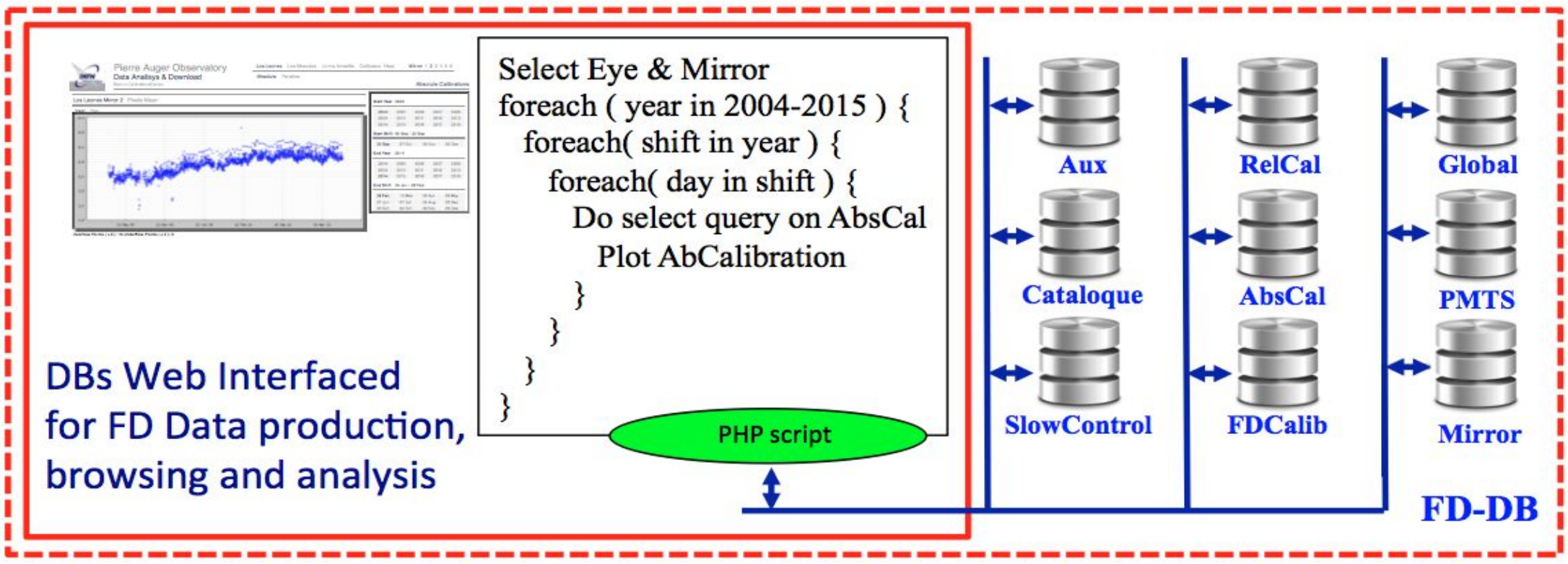

Fig. 1: The structure of FD-DB.

- **SlowControl:** The time response of each PMT to exposure to the background sky light during data taking, i.e. the absolute geographical address of the PMT, its value and error and the GPS time. The time response of an external sensor to the background sky light during data taking, i.e. its value with error and GPS time. The time evolution of the status of the components of each detector: shutters open/close and GPS time, curtains open/close and GPS time, high voltage on/off and GPS time. The high voltage value with error and GPS time. The GPS time of sunset, sunrise, moonset, moonrise, twilight start and end time for each day.

### 2.2 The Data structure

In order to monitor the status and the correct operation of the Fluorescence Detector, we perform several systematic measurements which are stored in FD-DB. Most of these data are time dependent so it is easy to find a common structure to collect, store and manage them. This introduces some complexity in the management and in the data presentation. It is possible to divide these data into two classes:

**Constants**: Production and acceptance test of the PMT's, characteristics of the mirrors, HV classes and all the other construction parameters. For this kind of data, a static architecture has been created as none of these is supposed to change or just minor maintenance variations are foreseen (i.e. the rare event of replacement of a PMT which is malfunctioning).

**Time-dependent**: absolute, relative calibration and LCU values, building/calibration-condition monitoring quantities (i.e. internal temperature, sky light, etc.). The large amount of these data, diversity of sources and the possibility to add new datasets has forced us to adopt a more flexible structure. We have fixed all the common values, like location, bay, calibration run number or GPS time, and used them as unique keys, to have easier access and readability for the graphic interfaces, that we use to show or find relations between different observables.

The total number of records produced every year turns out to be about 50 million. By far the largest amount of data comes from critical, time-dependent quantities.

### 2.3 Data Updating and Consistency

We perform different procedures to put data into the DB, depending on sources, data type and datasets. Cross checks have been made on all raw data since the beginning to ensure data reliability and consistency and to find possible failures in the data taking process.

**Relative Calibration Data:** We analyse the relative calibration raw files data to generate Relative Calibration constants. On the raw files some consistency filters are applied (calibration type, missing records, file size, record values). The files are then parsed and the relative calibration constants are stored into FD-DB tables with different flags, depending on the outcome of the analysis. This procedure is automated and it is performed via the web interface, Fig. 2.

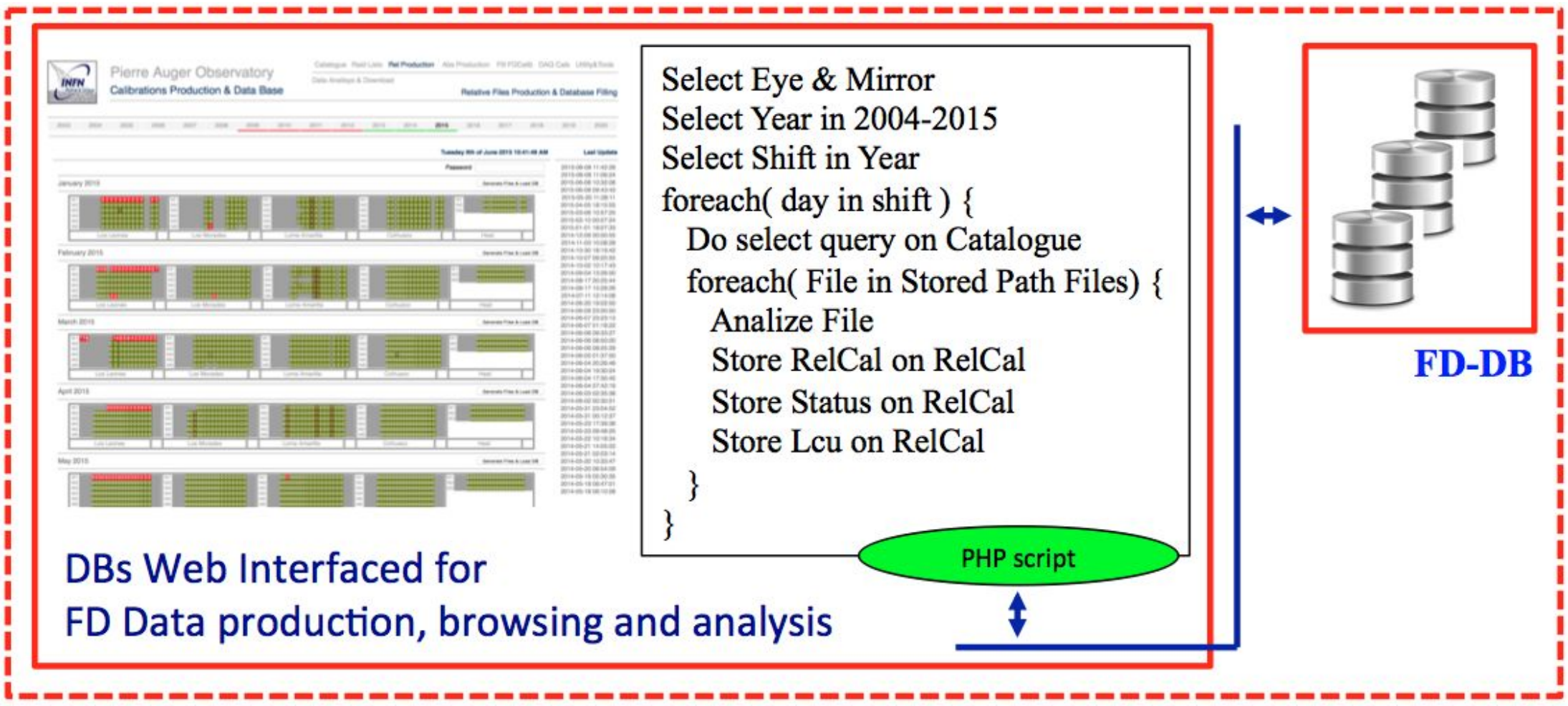

Fig. 2: The automated procedure for the production of Relative Calibration.

**Nightly Absolute Calibration Data:** We generate the Nightly Absolute Calibrations Constants. Some consistency filters are applied (missing records LCU values, record values). The values are stored into FD-DB tables with different flags, depending on the outcome of the analysis. This procedure is automated and it is performed via the web interface. The structure of this procedure is similar to those described above.

**Other Data:** The first step is to download data from web pages or archives managed by the Auger collaboration groups involved in the different activities of the experiment (i.e. Slow Control, hardware status, etc.). In some case a data reduction procedure is applied. An example is the background sky light. The measurement of this parameter is collected every five minutes, as it is important for monitoring the data acquisition and alerting against system malfunctioning; but such a frequent sampling rate is not useful (and would be a waste of disk space) for the off-line analysis. At present we have pre-processed background files, extracting a mean background sky light over an adjustable and reasonable time period and storing only this value.

Except for the production procedure, the other datasets follow basically the same logical flow. Because of the different origins of data, they are both in text file or SQL format. The main reason of this is that the difficulty of building an effective distributed database has an incentive at selecting the best-suited ad-hoc solution, without struggling for harmonization and standardization. For the purpose of the present activity, we have used different tools to perform reliability cross checks. These can be summarized as follows: missing data; out of range data; *orphan* records (i.e. records without or with wrong connection keys); data without physical meaning. It is not always possible to recover these problems by a simple remedy action. In this case, the "wrong" data are either removed during the upload process, discarding them, or marked by associating to them error codes, that allow their filtering at "query-level" by means of static or statistic filters (i.e. fixed range, standard deviation, etc.). We have decided to pre-process data in this preliminary phase to avoid filling database with too much information that is useful for on-line and real time system monitoring, but is completely useless and heavy for time behaviour analysis or long time displaying.

## 3. Web Graphical Interface

Associated to the FD-DB a Web graphical interface has been developed for a fast and easy access to the stored information. The structure of this interface is shown in Fig. 3.

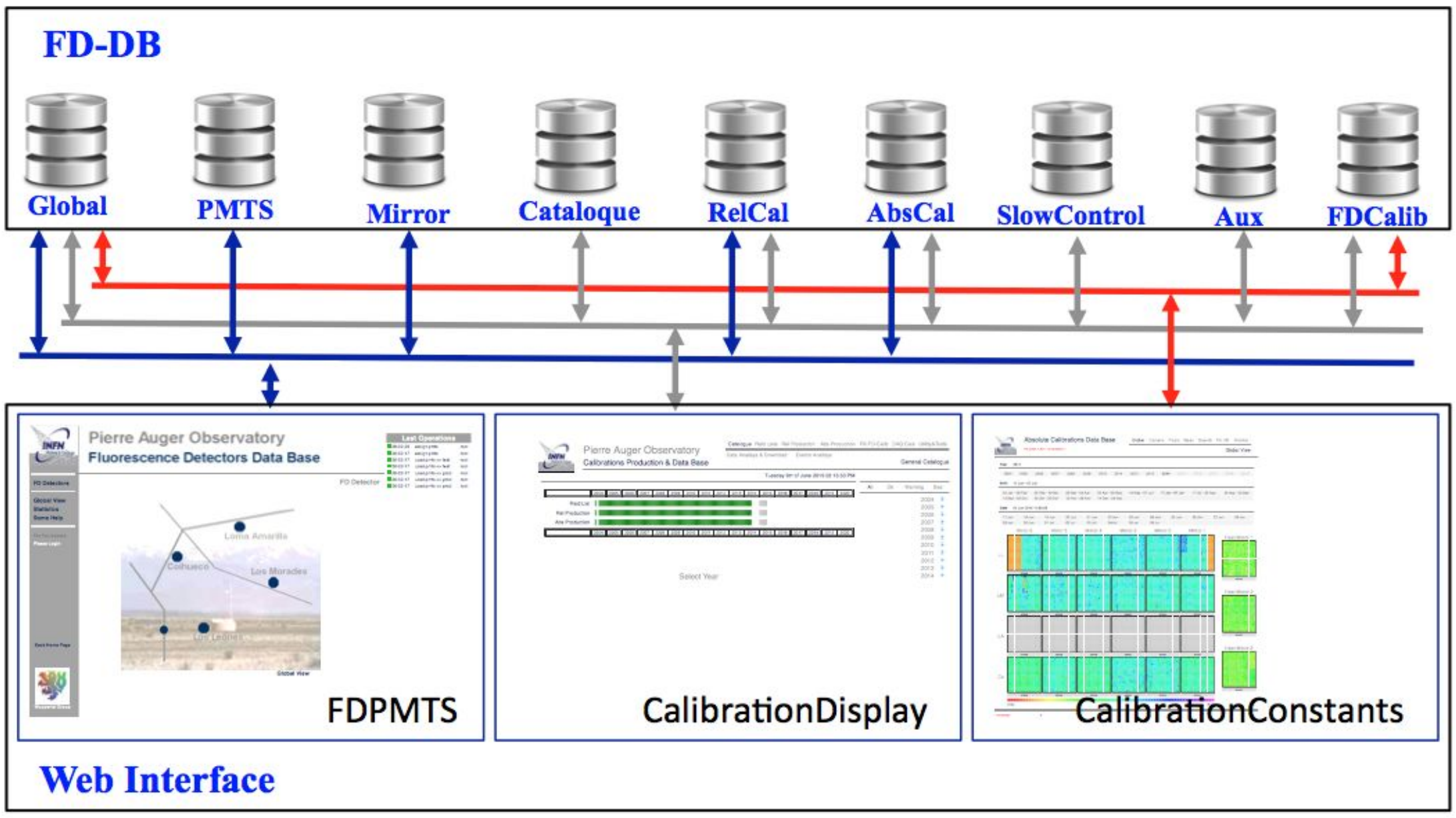

Fig. 3: The structure of the Web Graphical Interface.

**FDPMTS:** This section provides an intuitive interface to get information about FD telescope parameters. It is divided in two sections, the Mirror section containing information on the position and reflectivity of each single element of the mirrors and the Telescope section containing information on the PMTs and associated electronics. In the Telescope section there are two global views that display the installation status and high-voltage classes. It is foreseen to choose mirror and single PMT to have a detailed report on the PMTs position, production parameters, operation time and history and two graphs displaying the time evolution of Absolute and Relative Calibrations.

**CalibrationConstants:** This section provides some graphical representations of the Selected Absolute Calibration data. It is possible to have a global view where all the telescopes are displayed in the same page. It is also possible to display a single telescope page where, at a given time, the absolute calibration data are presented on the telescope picture with colour code and on an values histogram. For each pixel, a graph of the time dependence is also available. This section contains also a useful analysis tool where it is possible to create images of each camera applying filters on pixel position or values. This option gives the possibility of identifying malfunctioning sectors or single pixels. This is very helpful to debug the raw data. Fig. 4a shows the page containing information on the data of the Absolute Calibration done on 26 July 2014 for Los Leones Telescope 4. The space distribution over the camera and the histogram showing the distribution of the measured values are presented.

**CalibrationDisplay:** This is the most powerful and complex database interface. It contains the display, the production tools and the analysis tools for the Relative and Absolute Calibration data. Most of these features are public but some, like data manipulation and DB management are password protected. This interface contains a summary page that shows all the calibrations performed on each telescope with statistics on production progress and failures. It is possible to have details on a single calibration, an analysis section that provides time plot of calibration values, data fitting, data manipulation, a Bad night page that contains analysis tools for "warning" and "bad" configurations, a telescope functioning monitor where several parameters are displayed as camera temperature, shutters status, sky light background and astronomical data and DB management section where some loading, checking and data handling utilities are stored. Fig. 4b shows the page containing information on the time behaviour (2004-2015) of the mean value (averaged over all pixels) of the Absolute Calibration data and the internal temperature values for Coihueco telescope 5.

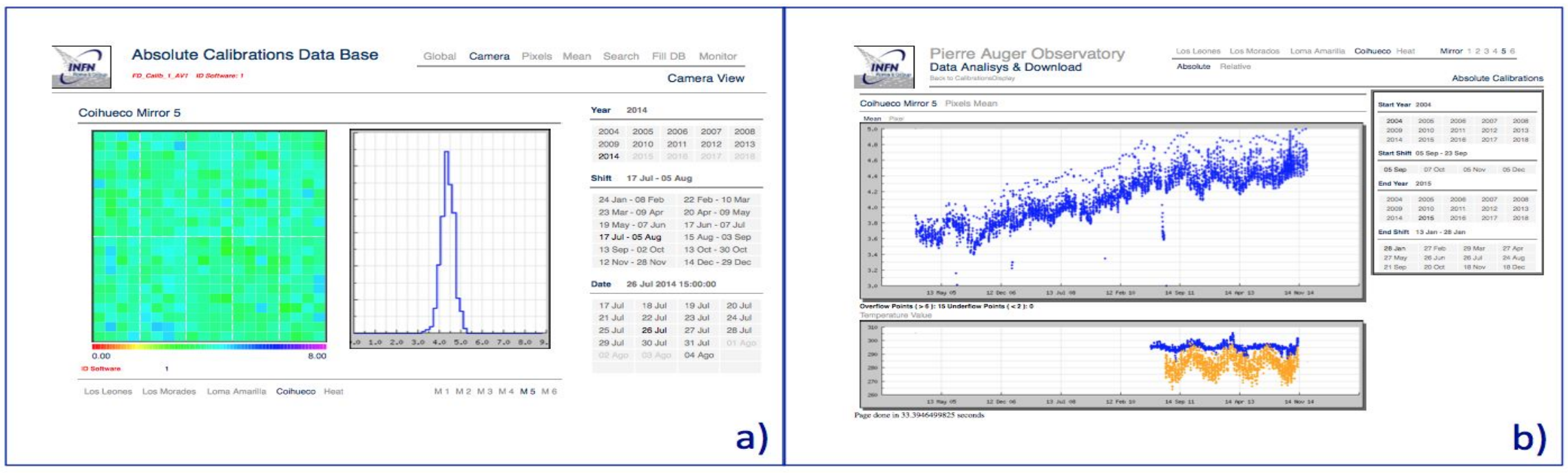

Fig. 4: a) Some information on the 26 July 2014 Absolute Calibration for telescope 4 of Los Leones and b) The time evolution of the Relative Calibration and Temperature for telescope 5 of Coihueco.

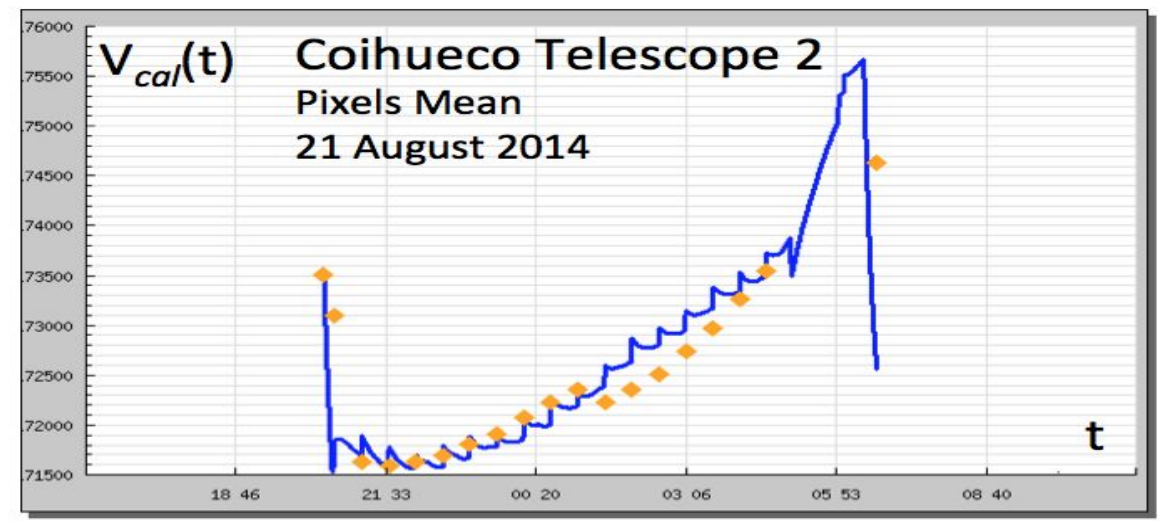

$$\tau \frac{dV_{cal}(t)}{dt} = -V_{cal}(t) + V_{in}(t)$$

where

$$V_{in}(t) = V_0 + \beta\left(T_{Mir}(t) - T_{Ref}\right) + \alpha V_{BG}(t)$$

Fig. 5: The Relative Calibrations performed on 21 August 2014 on Coihueco Telescope 2 (in orange) and the prediction of RC model (in blue). $T_{\mathrm{mir}}(t)$ is the internal temperature and $V_{\mathrm{BG}}(t)$ is the sky background.

## 4. Data Base Functionality and Performance

We show the results of some studies performed on the data stored in FD-DB. The aim is to show the functionality of the database in realistic situations of use. The analysis requires the display, handling and elaboration of a large amount of data. We have set as a condition that they should be performed online and in real time. These are strong requirements, which indicates a good functionality of FD-DB.

### 4.1 Short Time Response of the Fluorescence Detector

This study could be done only after having developed a powerful DB system coupled with a graphical high-level real-time interface. The graphical interface allows visualization and handling of all data describing the physical observables which are relevant for the calibrations. A module that allows numerical solutions of differential equations is associated to the database. It turns out that the time behaviour of the response of the telescopes as measured by the Relative Calibration is very complex. To understand the time behaviour we have constructed a model for short time behaviour assuming that the cameras behave as a RC circuit. In Fig. 5 the qualitative agreement between Relative Calibration Data and the model is shown. This is a strong evidence for a temperature and sky exposure dependence of the short time response of the FD.

### 4.2 Long Time Response of the Fluorescence Detector

Based on over 10 years of operation in Fig. 6 we show the long term time behaviour of the average absolute calibration, done over the 440 pixels, for Los Leones telescope 4. This analysis is systematically performed on all the telescopes with the scope of monitoring their stability. The behaviour of the telescope is characterized by an initial drift (2004-2007) and by seasonal oscillation, due to dependence of the response on temperature and light exposure, see Fig. 7. This long term analysis has been used to optimise the procedure of data-taking improving the stability of the detector and its expected time-life.

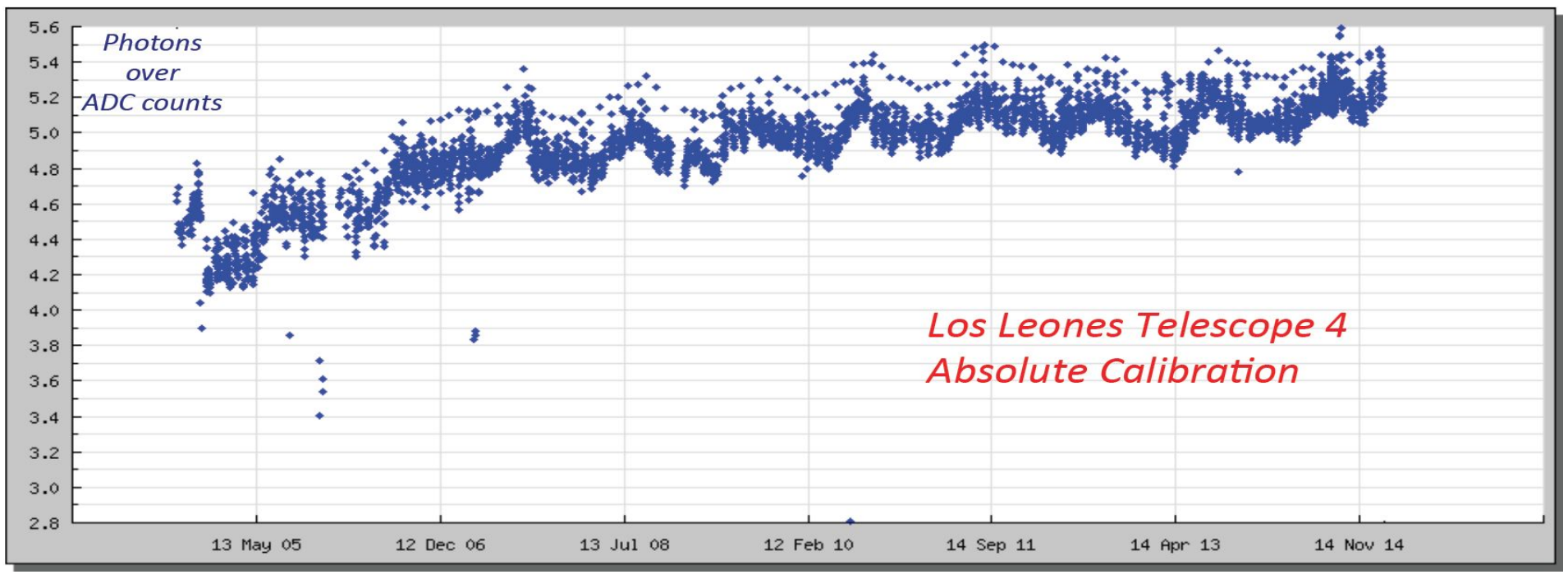

Fig. 6: The long term behaviour of the average, over the 440 pixels, of the absolute calibration for Los Leones telescope 4 based over 10 years of operation.

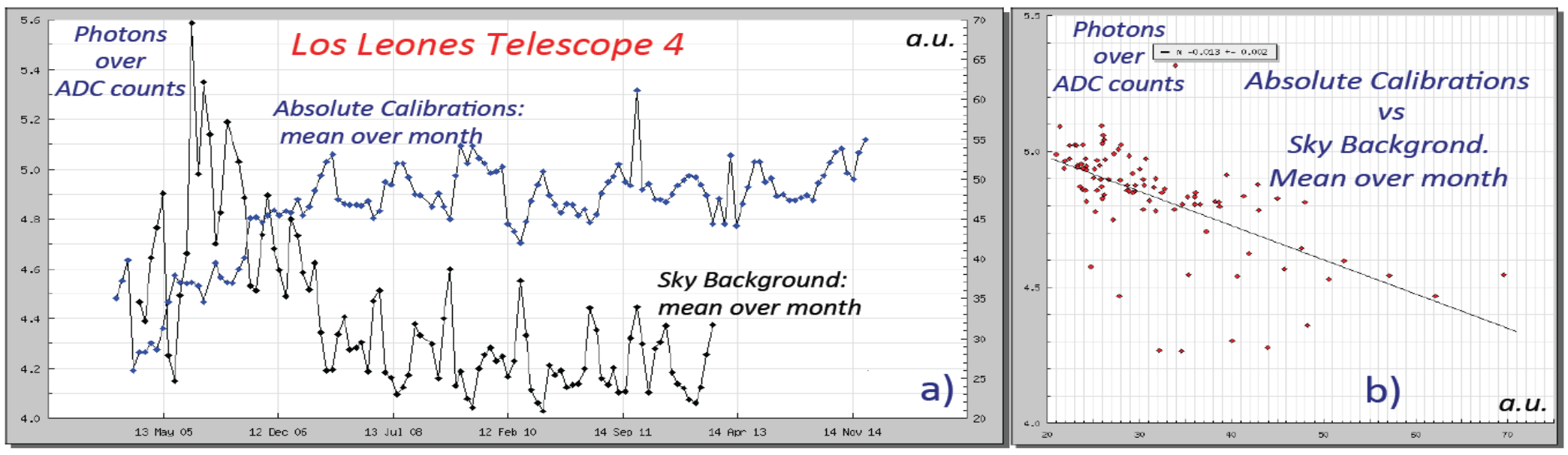

Fig. 7: a) The mean over month of average absolute calibration (in blue) and the mean over month of sky background (in black) in the 2004-2015 years. b) The absolute calibration – sky background correlation.

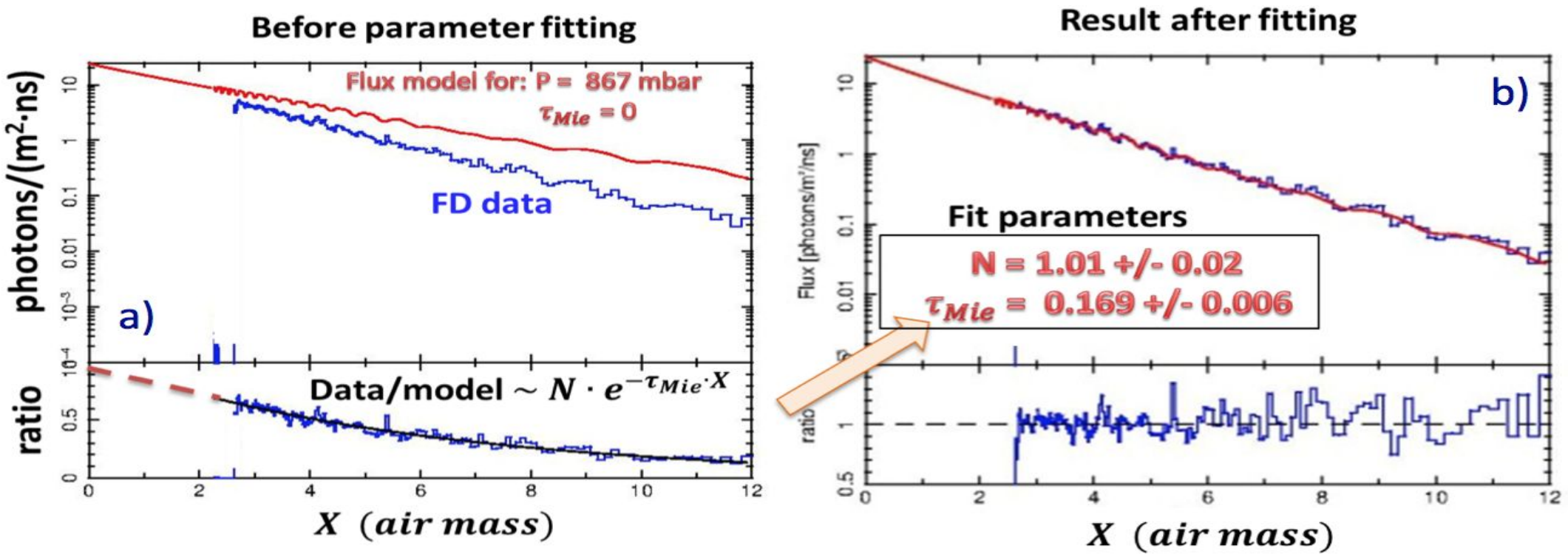

Fig. 8: a) The flux measured and the one simulated without inclusion of the aerosol scattering in the atmosphere model; b) The result after the fitting procedure in which only the aerosol optical depth and a normalization factor are free to vary. Best fit result is shown and residuals after fitting.

More strict prescriptions were systematically adopted at the beginning of 2009. Due to this the response of the telescopes is stable in the 2008-2015 period.

## 5. Check of the absolute calibration using stars traversing the FD field of view

In order to provide an independent verification of the FD absolute calibration status, a procedure has been developed based on the signals generated by reference stars crossing the field of view of the FD. Without requiring any dedicated hardware device, and without any interference with normal telescope operations, the method provides a simple way to verify the long-term stability of FD [3]. The absolute calibration of a telescope can be verified by observations of non-variable stars whose absolutely calibrated spectral distribution is known. The primary reference star for absolute astronomical photometry is Vega (α Lyr), which has repeatedly been calibrated against terrestrial standard sources [4], however, in the southern hemisphere Sirius (α CMa) is preferred to Vega since it can be observed at higher elevation angles and then with lower atmospheric attenuation. Both of these sources have recently been observed by the Space Telescope Imaging Spectrograph on board the Hubble Space Telescope with a spectral resolution of 0.25 nm and a stated absolute uncertainty of 1%.

The FD telescopes have a field of view of $30^\circ$x$30^\circ$ and point to the sky at an elevation angle of $\approx 15^\circ$ degrees and covering, with adjacent field of view, a total of $180^\circ$ in azimuthal range. The three HEAT (High Elevation Auger Telescopes) telescopes point to the sky at an elevation angle of $\approx 45^\circ$ and are located at the Coihueco site. Given the position and orientation of the FD telescopes, any star is usually detected during the same night by two or more FD telescopes in overlapping or adjacent field of view; as example, Sirius crosses the field of view of three different FD telescopes, one of Loma Amarilla, and the other two of Coihueco site but pointing to the sky at different elevation angles. As a first step, we compare on a Langley plot (Fig. 8a), the star flux measured from the FD data to the star flux simulated without including the aerosol scattering in the atmosphere model as can be seen a part of the undulation in the curves corresponding to the star transit between adjacent pixels, the slope of the measured flux is much higher than the slope of the simulated curve, thus indicating the necessity to add a significant amount of aerosol absorption in the atmosphere model. Assuming the aerosol optical depth is independent from wavelength, it can be shown that the ratio between measured and simulated data is expected to be an exponential function exp $(-\tau_{aer} X)$ as effectively observed. By fitting the ratio between measured and simulated flux we obtain the value of aerosol optical depth to be added to the atmosphere model and a multiplicative scaling factor to correct the normalization a new simulated curve is then computed and the procedure is eventually repeated, adjusting the two parameters until a good match between the simulated measured values is obtained. The final result of the iterative fitting process, showing the perfect match between measured and simulated curves, Fig. 8b, where also the best fit parameters are indicated. In this case, notwithstanding the quite high amount of aerosol scattering, the scaling factor is consistent, within a few % of statistical uncertainty, with a unitary value, thus indicating the excellent calibration status of this telescope.

## 6. Conclusion

We have described the architecture of FD-DB and all relevant data stored. The FD-DB architecture is at present still evolving and will be evolving in the future. The modular structure allows implementation of new and at present unforeseen features. The graphical interface is a very useful tool for the presentation of the data and for the study of the time dependence of the different parameters and their correlations. We have demonstrated that star signal extracted from the FD background data can be used to verify the FD absolute calibration. The method described is based on the capability of the FD to be sensitive to signals left by stars traversing the telescope field of view. Using the known absolutely calibrated spectra of reference stars, we compared expected value of photon flux to what the detector actually measures, correcting for the influence of the atmospheric transmission. We confirmed that the absolute calibration used for the FD telescopes does not have large systematic errors and is correct within its estimated error.

# Status and Prospects of the Auger Engineering Radio Array

**Johannes Schulz**[*a] **for the Pierre Auger Collaboration**[b]
[a]*Department of Astrophysics/IMAPP, Radboud University, Nijmegen, The Netherlands*
[b]*Observatorio Pierre Auger, Av. San Martín Norte 304, 5613 Malargüe, Argentina*
*E-mail:* auger_spokespersons@fnal.gov
*Full author list:* http://www.auger.org/archive/authors_2015_06.html

The Auger Engineering Radio Array (AERA) is an extension of the Pierre Auger Observatory. It is used to detect radio emission from extensive air showers in the 30 - 80 MHz frequency band. A focus of interest is the dependence of the radio emission on shower parameters such as the energy and the atmospheric depth of the shower maximum. After three phases of deployment, AERA now consists of 153 autonomous radio stations with different spacings, covering an area of about 17 km$^2$. The size, station spacings, and geographic location at the same site or near other Auger extensions, are all targeted at cosmic ray energies above $10^{17}$ eV. The array allows us to explore different technical schemes to measure the radio emission as well as to cross calibrate our measurements with the established baseline detectors of the Auger Observatory. We present the most recent technological developments and selected experimental results obtained with AERA.

*Speaker.

## 1. Introduction

The subject of radio detection of extensive air showers has developed rapidly over the last years [1]. The performance, in terms of energy and composition measurements, is quickly approaching a competitive level compared to the air-fluorescence technique [2, 3]. In terms of duty cycle, however, the radio technique is advantageous as the operation is only limited in case of strong atmospheric electric fields (e.g., during thunderstorms) which significantly affect the radio emission processes [4]. Therefore a duty cycle close to 100% can be obtained which is approximately seven times higher than that for air-fluorescence measurements that can only take place during clear moonless nights.

By acquiring detailed information about the polarization of the radio pulses [5], an understanding of the two fundamental emission mechanisms, geomagnetic and charge excess emission, has been established. Together with the convergence between predictions of Monte Carlo simulation codes and the radio measurements [3], significant improvements were achieved in the parametrization of the lateral energy density distribution function [6, 7]. The shape of the radio wave front has been found to be hyperbolic [8, 9] and the shape of the measured pulses is also the subject of intense research [10]. With these developments, parameters which are sensitive to the air shower development (often characterized by the atmospheric depth of maximum shower development: $X_{max}$) and to the properties of the initial cosmic ray have been identified. Their cross calibrations with established detection techniques can only be performed with hybrid detector systems.

The Auger Engineering Radio Array is dedicated to multi-hybrid detections of cosmic rays at the Pierre Auger Observatory [11]. It is conveniently co-located within the low-energy enhancement AMIGA and the high elevation fluorescence telescopes (HEAT). AMIGA is an infill region of the surface detector array (SD) with a spacing of 750 m and six associated muon counters. HEAT is an extension of the fluorescence detector (FD) which has an elevated field of view. All together, four different detection techniques are used to detect the same showers which allows us to perform multi detector cross calibrations and analyses. In addition, the systems can trigger each other. This constellation provides unique and optimal conditions for cross calibrations, further technical developments and pathfinder studies for future large scale experiments.

## 2. The Auger Engineering Radio Array

The AERA project has been carried out in three phases. AERA24, an array of 24 radio detection stations (RDS) was deployed in September 2010 to prove the feasibility of radio+particle hybrid air shower measurements, to develop the techniques to reconstruct air shower parameters from the radio data and to investigate the radio emission mechanisms [2, 5]. In a second stage, leading to AERA124, 100 RDS were deployed with a modified design in May 2013. With different grid sizes (144 m, 250 m, 375 m), multiple improved hardware and trigger concepts and an instrumented area of 6 km$^2$, AERA124 detected several thousand events per year. In the third stage, AERA153 covers about 17 km$^2$ since April 2015, with 375 m and 750 m spacing between the additional RDS. The targets of this extension are mainly horizontal showers ($> 60^\circ$ zenith angle) with large scale radio footprints. The layout of AERA including all three stages is shown in figure 1 together with the other co-located detector systems.

The AERA24 stations are equipped with two logarithmic-periodic dipole antennas (LPDA) [12], oriented in magnetic North-South and East-West directions. The signals are amplified and filtered before they are introduced into a filter-amplifier and digitization chain. There, they are band pass filtered between 30-80 MHz, digitized and further processed by a field programmable gate array (FPGA) and a central processing unit (CPU). This chain exists in various forms to test and optimize the technical realization of different detection strategies. For AERA24, two versions were deployed focusing on different triggering methods. One has a sampling rate of 180 MHz and features a 4 GB ring buffer which allows for external triggering by other detectors of the Pierre Auger Observatory. The other has a sampling rate of 200 MHz and is designed for self-triggering. In case of self-triggering, a first trigger decision can be made in the FPGA. This trigger information is then forwarded to the CPU where further processing and signal verification are done. Remaining triggers are GPS time stamped and sent to the central radio data acquisition system where all time stamps of triggered stations are processed. In case of coincidences between RDS, the data are read out, formed into an event and saved to disk. Data from externally triggered stations are directly saved to disk for offline analyses. The communication and data transfer of AERA24 is handled via optical fiber connections. The antenna type used for the new stations of AERA124 and AERA153 is an active bowtie antenna also called "butterfly" [13, 12]. This antenna is highly sensitive towards the ground which enhances the antenna gain. The general layout of the digitizing chain is similar to the one used in AERA24. In total, 89 butterfly RDS are equipped with deeply buffering hardware and 40 butterfly RDS are employing internal triggering only. This internal trigger is based on radio self-triggering and on small scintillation counters in the electronics compartment of the radio station itself. The 40 internally triggered stations (see fig. 1) are currently not included in the data set presented here. Due to the increased size of the array, wireless links were introduced for the new stations. All 153 stations operate autonomously employing solar power systems.
The AERA data are merged with data from the other detector systems and subsequently analyzed with the software package Offline including the dedicated extension for radio [14]. During the event reconstruction, the signals are corrected for the radio detector response and the time-dependent electric-field strength vector is calculated for further analyses. Shower parameters reconstructed with data from the other detector systems can be used as input during reconstruction.
Great efforts were made to calibrate the different AERA stations to determine absolute signal response and relative timing. All components of the analog signal processing chain were calibrated individually before deployment. For the antennas, detailed electromagnetic simulations were performed and various airborne calibration campaigns were carried out on site to calibrate the direction and frequency-dependent responses. The relative timing of the AERA stations is continuously monitored via their phase offset by measuring sine waves from a reference beacon [15]. The beacon is installed at the closest FD site and emits four precisely defined sine wave signals. Furthermore, signals from commercial aircrafts detected by the AERA stations are used to determine the timing. In combination, a precise timing calibration of about 2 ns is possible, whereas the station built-in GPS clocks exhibit drifts of tens of nanoseconds. Without precise timing information, analyses to determine the direction of the shower axis, interferometric approaches to enhance the detector sensitivity and analyses based on the radio wavefront would not be feasible.

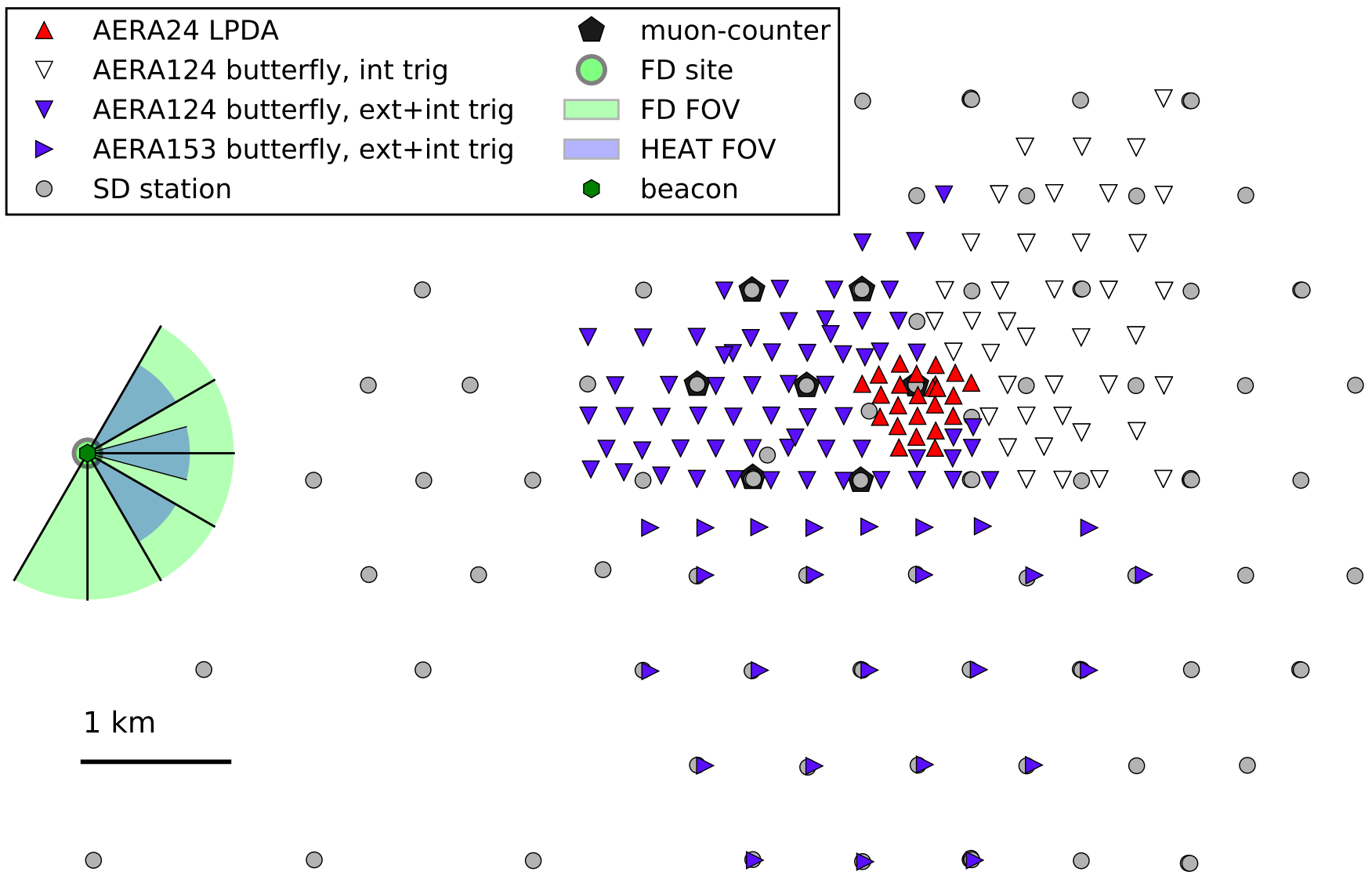

**Figure 1:** Schematic overview of the Auger Engineering Radio Array and part of the surface detector array and AMIGA. One site of the fluorescence detector and the HEAT extension are also indicated with their field of view. The reference beacon for timing calibration is located at the FD site.

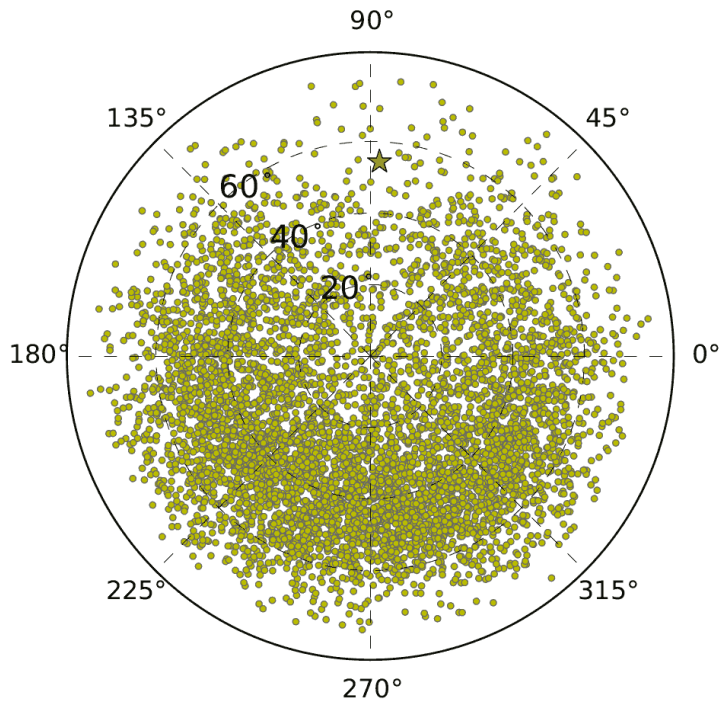

**Figure 2:** Skyplot of AERA+SD hybrid data. The star indicates the direction of the geomagnetic field.

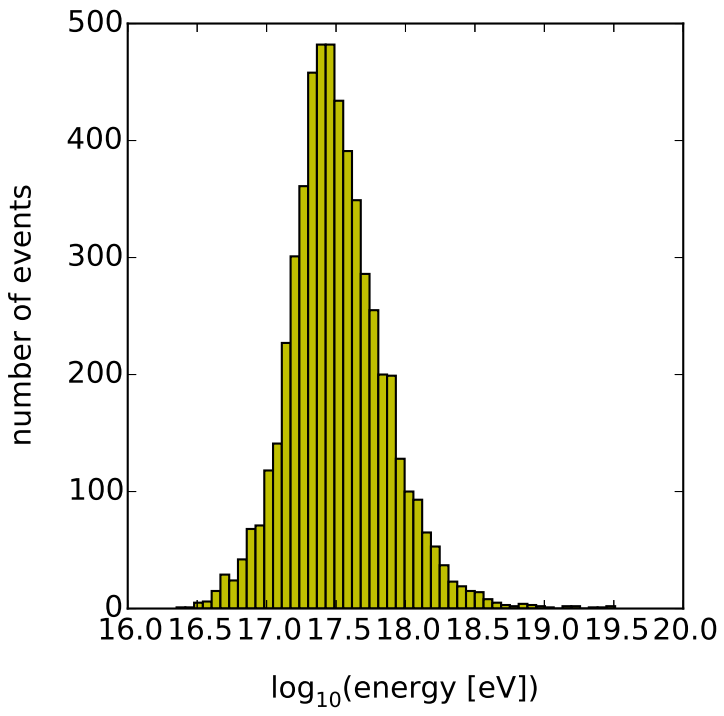

**Figure 3:** Energy distribution of AERA+SD hybrid data measured with the surface detector array.

## 3. Data Set

With AERA we have recorded a large radio data set of extensive air showers. Until the second of March 2015, more than 5500 air showers were detected by three or more antenna stations in coincidence with the surface detector array (AERA+SD hybrid events). The reconstructed arrival directions and the energy distribution as measured by the SD are shown in figures 2 and 3. During reconstruction a low level cut is made on a successfully reconstructed shower axis in agreement with the SD reconstruction within 20°. Most of the air showers are detected with 3 radio stations, but there are also air showers which are detected by more than 40 signal stations. During the stable operation phase of AERA24, about 50 air showers were recorded with three or more RDS per month which has increased to about 250 air showers per month for AERA124 after a short commissioning phase. In addition, 324 air showers were simultaneously detected with the fluorescence detector (AERA+SD+FD hybrid events). For these events, $X_{max}$ is within the FD field of view and the measured profile allowed for an FD-based energy reconstruction. We also detected air showers in coincidence with the AMIGA muon counters. The data set contains 509 AERA+SD+AMIGA hybrid events and 51 AERA+SD+FD+AMIGA hybrid events. An event display of a quadruple detection is shown in figure 4. For about 5% of the data set, increased atmospheric electric fields have been measured. These data will be treated separately in upcoming analyses.

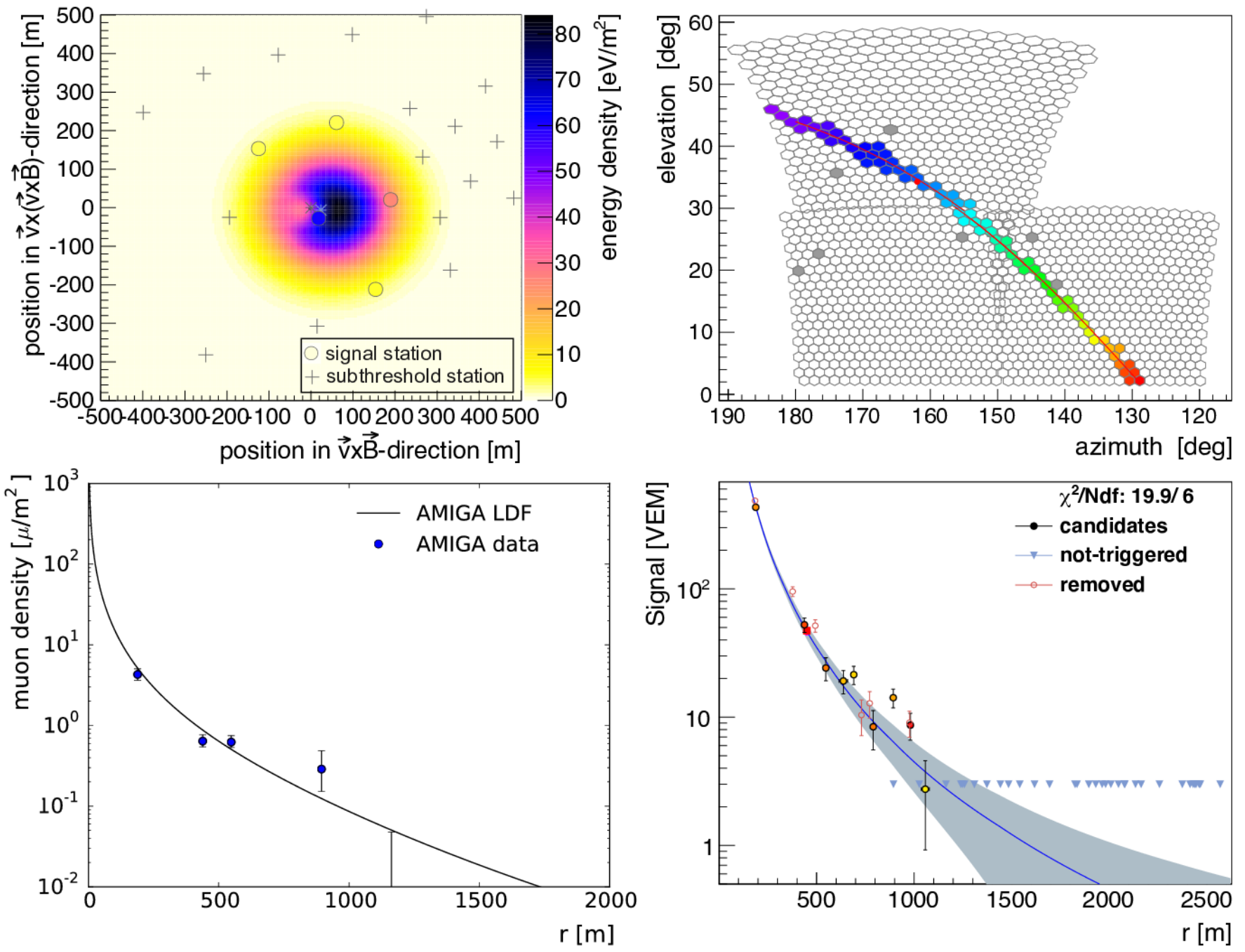

**Figure 4:** An extensive air shower detected with four different detector systems simultaneously. Top left: Radio footprint projected into the shower plane represented by the energy density measured with AERA124 and the fitted parametrization of the lateral distribution function (LDF) (background color). Top right: Longitudinal shower development measured with the fluorescence detector. Bottom left: Footprint of the air shower as measured with the muon counter of AMIGA. Bottom right: Particle signal in the surface detector stations including the classical one dimensional LDF fit.

## 4. Status and Prospects on Multi-Hybrid Analyses

The three main properties of a cosmic ray are arrival direction, energy and particle type. AERA shows a sensitivity to all of these parameters which are discussed in the following.
The direction of the cosmic ray is taken from the axis of the air shower. This axis can be reconstructed by minimizing the difference between measured time at the RDS positions and the predicted time assuming a wavefront model and varying the shower geometry. A good first order approximation of the wavefront is a plane wave. Studies on the angular resolution of AERA are ongoing and only a comparison to the reconstructed axis from the SD can be given. The mean deviation between the two is 1.5° and the distribution can be fitted by a Rayleigh function with $\sigma$=0.45°.
The energy of the cosmic ray can be reconstructed from the radio signals by reconstructing the energy emitted in the measured radiation. Therefore, the electric field traces are converted into the energy density of the electromagnetic wave via the Poynting vector. The measured energy density at all positions of the signal stations is fitted with a two dimensional lateral distribution function [6]. The spatial integral over this LDF gives the energy in 30-80 MHz radiation. The radiated energy is calibrated against the energy reconstructed by the SD. With this method, a resolution of 22% is obtained for a low station multiplicity (<5 signal RDS) data set and 17% for a high station multiplicity set (>5 signal RDS), both based on AERA24 measurements [2]. The radiation energy is subject of ongoing work as it can be predicted by Monte Carlo simulations without being strongly influenced by the uncertainties of the hadronic interaction models. Therefore the energy scale of cosmic rays can be studied based on classical electrodynamics. Four main studies targeting the reconstruction of $X_{max}$ are being carried out on the AERA+SD+FD hybrid data set. They are aiming at radio parameters and their calibration against the FD measurements or using those as comparison.

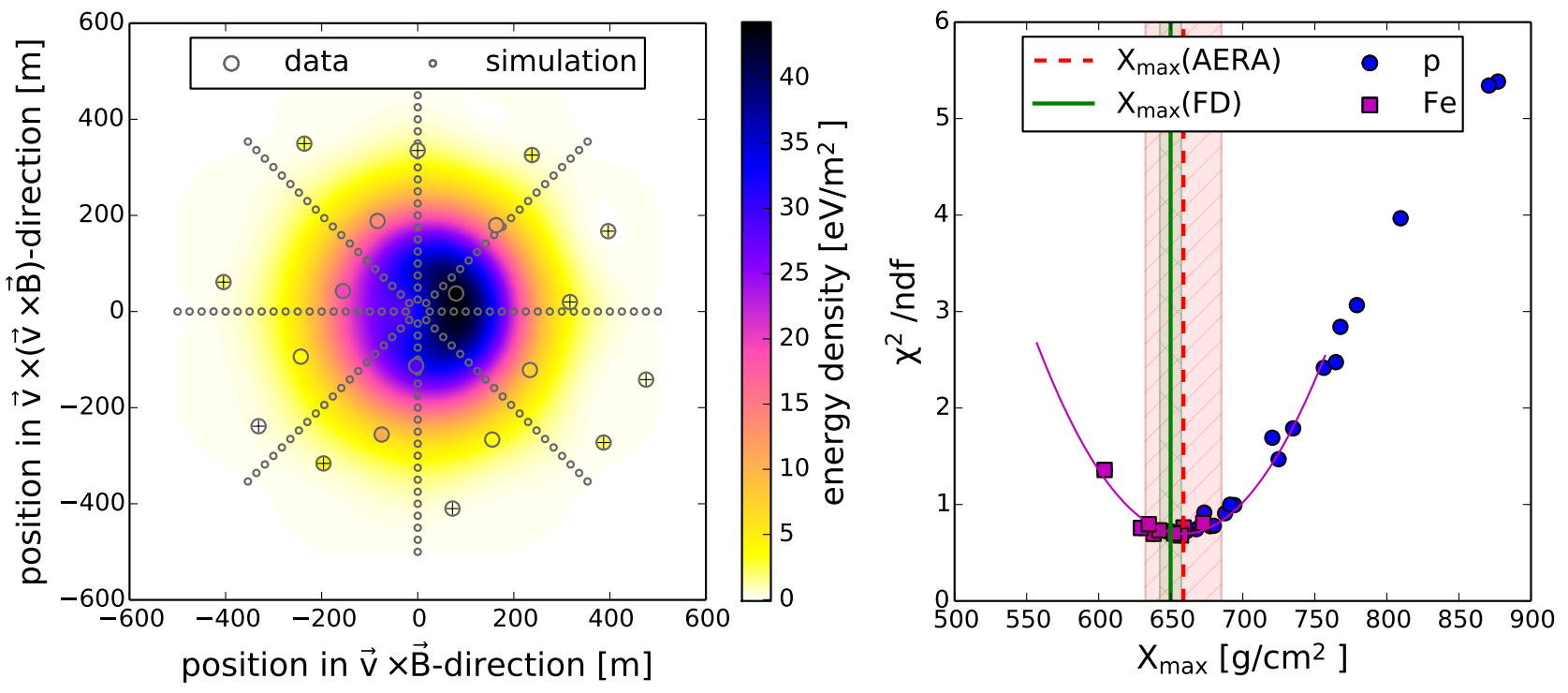

**Figure 5:** Reconstructed $X_{max}$ value based on the fit quality of a set of CoREAS [16] simulations. Left: The two dimensional radio lateral distribution function. It displays the best fitting simulation (small circles) interpolated in the shower plane (background color map) as fitted to the energy densities measured by the AERA stations (colored circle), sub-threshold stations are additionally marked by a "+". Right: Distribution of reduced $\chi^2$ values for the $X_{max}$ value of the individual simulations. The purple line is a parabolic fit to the distribution and the minimum is taken as the value of $X_{max}$. The vertical lines indicate the reconstructed $X_{max}$ values from radio and FD together with the statistical uncertainties.

1. The width of the two dimensional LDF [7]. The further the emission region is away from the observer position, the larger the projection of the emission pattern gets.

2. The shape of the radio wavefront [9, 17]. The precise appearance of the hyperbolic front is determined by the distance of the observer to the endpoint of the emission region. The closer this endpoint is to the observer, the more conical the wavefront gets.

3. The signal pulse shape [10]. Depending on the path length differences for radiation from early and late parts of the air shower, the pulses broaden if the emission region is closer to the observer.

4. The fourth study is targeting a combination of the parameters mentioned above by investigating the agreement between measurement and individual simulations of a set with various $X_{max}$ values customized for the measured direction and energy. Therefore, the method used in [3] is modified to be applicable to the AERA+SD hybrid data. Figure 5 displays an air shower radio footprint and the best fitting simulation. The simulations are performed with CoREAS [16] using antenna positions oriented on the displayed star pattern. The output is interpolated and fitted to the data by varying the core position and the overall scaling. Simulations with different $X_{max}$ values result in different $\chi^2$ values for the fit. The distribution of $\chi^2$ values as a function of the Monte Carlo $X_{max}$ values is fitted with a parabola and the minimum is taken as the value of the reconstructed $X_{max}$.

Another interesting target for hybrid analyses are horizontal air showers as complementary infor-mation can be obtained from radio and particle detectors. As horizontal air showers experience a larger column density of atmosphere before reaching the observer, the electromagnetic part of the shower is already absorbed and only the muonic component is detected by particle detectors. The radiation produced in the air shower is not absorbed however and can be detected. Therefore, the muonic and the electromagnetic parts of the shower can be measured independently by radio and particle detectors. Furthermore, horizontal air showers are interesting as they have large radio footprints due to the large distance to the radio source and projection effects on ground. Therefore, inclined air showers can still be detected with a rather sparse radio detector array allowing for large scale observatories and high statistics, even at the highest energies. The extension performed from AERA124 to AERA153 is dedicated to measuring horizontal showers. An event display of an air shower detected with AERA153 is shown in figure 6. 44 RDS have measured a significant signal for this air shower with an energy of $1.3\times10^{18}$ eV and a zenith angle of 78°.

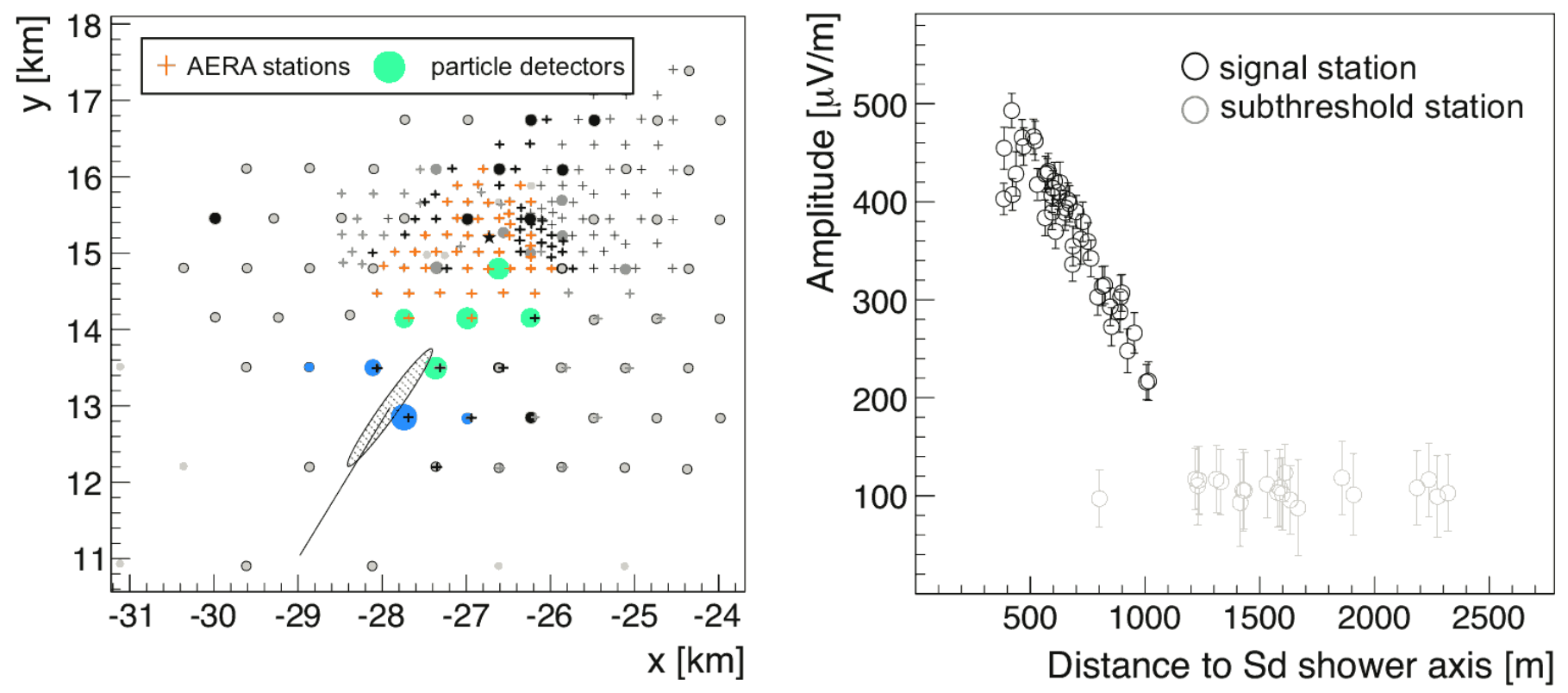

**Figure 6:** Event display for a horizontal shower measured with AERA153. Left: Array plot, where the crosses indicate radio antenna stations and the circles indicate surface detector stations. The size of the circle is scaled with the measured particle signal. The color indicates the signal timing. The reconstructed direction based on the surface detector data is indicated by the line together with the core position (ellipse). Right: LDF in terms of maximum amplitude as function of the distance to the air shower axis.

## 5. Conclusions

At the Pierre Auger Observatory radio emission from air showers is routinely measured simultaneously with other shower observables. With the data of AERA124 already recorded, high statistics radio data are now available in coincidence with the surface detector array. More than 300 air showers are also measured with the fluorescence detector. Combined, these data form the base for analyses targeting arrival direction, energy, and composition of the cosmic rays. The direction reconstruction is in good agreement with the surface detector reconstruction. We also developed a method for the energy reconstruction based on the measured energy density with a resolution of 17%. Based on this method, the determination of the cosmic ray energy scale from first principles comes within reach. The mass determination is a target of intense research and various composition sensitive parameters have been identified. The experimental calibration of these parameters is carried out against fluorescence detector data. Horizontal air showers are targeted with the third stage of AERA and events with more than 50 signal stations have already been detected.

# AugerNext: R&D studies at the Pierre Auger Observatory for a next generation ground-based ultra-high energy cosmic-ray experiment

**Andreas Haungs**[∗a] **for the Pierre Auger Collaboration**[b]
[a]*Karlsruhe Institute of Technology - KIT, Institut für Kernphysik, 76021 Karlsruhe, Germany*
[b]*Observatorio Pierre Auger, Av. San Martín Norte 304, 5613 Malargüe, Argentina.*
*E-mail:* auger_spokespersons@fnal.gov
*Full author list:* http://www.auger.org/archive/authors_2015_06.html

The findings so far of the Pierre Auger Observatory and those of the Telescope Array define some requirements for a possible next generation global cosmic ray observatory: it needs to be considerably increased in size, it needs enhanced sensitivity to composition, and it has to cover the full sky. At the Pierre Auger Observatory, AugerNext aims to conduct some innovative initial research studies on a design of a sophisticated hybrid detector fulfilling these demands. Within a European supported ASPERA/APPEC (Astroparticle Physics European Consortium) project for the years 2011-2014, such R&D studies primarily focused on the following areas: i) consolidation of the detection of cosmic rays using MHz radio antennas; ii) proof-of-principle of cosmic ray microwave detection; iii) test of the large-scale application of new generation photo sensors; iv) generalization of data communication techniques; and v) development of new schemes for muon detection with surface arrays. The AugerNext Consortium consists of 14 principal investigators from 9 countries.[†] This contribution summarizes some achievements of the R&D studies within the AugerNext project.

[∗]Speaker.

[†]Andreas Haungs, KIT, Germany; Johannes Blümer, KIT, Germany; Martin Erdmann, RWTH Aachen, Germany; Karl-Heinz Kampert, University of Wuppertal, Germany; Ad M. van den Berg, KVI Groningen, The Netherlands; Zbigniew Szadkowski, University of Lodz, Poland; Henryk Wilczynski, Institute of Nuclear Physics PAN, Cracow, Poland; Antoine Letessier-Selvon, IN2P3/CNRS, France; Mario Pimenta, LIP - Laboratory of Instrumentation and Experimental Particles Physics, Portugal; Enrique Zas, University of Santiago de Compostela - USC, Spain; Valerio Verzi, INFN Roma Tor Vergata, Italy; Iliana Brancus, IFIN-HH Bucharest, Romania; Masahiro Teshima, MPI für Physik, München, Germany; Martina Bohacova, Institute of Physics - FZU, Prague, Czech Republic.

## 1. Introduction

The Pierre Auger Observatory [1] in Argentina is the largest cosmic-ray experiment in the world and in operation since 2004. Results from this observatory have shown that the spectrum of cosmic rays exhibits a suppression at about 50 EeV [2, 3]; that events with an energy larger than 50 EeV show largest departure from anisotropy [4, 5]; and that data are consistent with a gradual increase of the average mass of cosmic rays for energies above 59 EeV, at least [6]. Presently, the Observatory is initiating a program of upgrades to achieve a better mass sensitivity at the highest energies [7]. At the same time, the Telescope Array Collaboration, with an experiment located in the Northern Hemisphere, has also seen the suppression, but the composition inferred is interpreted as being in agreement with light mass primaries, and a large hotspot in the distribution of the arrival direction of the highest energy cosmic rays has been observed [8, 9, 10]. In coming years the Telescope Array will be enlarged in area by a factor of four to study this hotspot with better statistics.

The findings of these two experiments define already some requirements for a next generation experiment: (i) To acquire sufficient statistics for the highest energies, a sensitive area of several tens of thousands of square kilometers is necessary. (ii) If it is true that a mixed composition around the suppression energy exists, an improved sensitivity to the elemental composition is required. (iii) Due to substantial structural differences with direction in the mass distribution of the nearby Universe, the full sky needs to be investigated with equally high quality. A comprehensive R&D study is needed to fulfill these requirements for the design of such a next generation cosmic-ray experiment.

First ideas are evaluated within the global cosmic-ray community towards such an observatory, i.e., GCOS, the "Global COSmic-ray observatory", is in discussion, where approximately $90,000\text{km}^2$ shall be instrumented. GCOS will be organized in several arrays in a few countries of the Southern and Northern Hemisphere, respectively. These arrays need autonomous detectors with sophisticated features and monitoring modes to operate for ca. 30 years in a maintenance-free way. GCOS will have synergies with the high-energy neutrino detectors IceCube, KM3NeT and the gamma observatory CTA for complementary measurements needed to perform multi-messenger particle astronomy. A realization of GCOS at some time within the next decade needs efforts to commence immediately, including the development of new detection technologies. The AugerNext project as well as the intermediate step of upgrades of the Pierre Auger Observatory and of the Telescope Array experiment are preparatory work in this direction. AugerNext contributes to this process with the aim of performing innovative research studies, primarily at the Pierre Auger Observatory, in order to prepare a proposal fulfilling the aforementioned demands.

AugerNext is a project which was supported by 9 European funding agencies for three years from the end of 2011 to the end of 2014. The AStroParticle ERAnet (ASPERA) was a network of national European government agencies responsible for Astroparticle Physics [11] funded by the European Community. One element of ASPERA was to organize targeted R&D and design studies in view of the realization of future astroparticle infrastructures identified in the ASPERA/APPEC[1] Roadmap (available at [11]), where the proposals should demonstrate a clear added value to the applicants' partnership over and above what could be achieved individually. Within the Pierre Auger

[1] APPEC is the interest group of the Astroparticle Physics European Consortium [12].

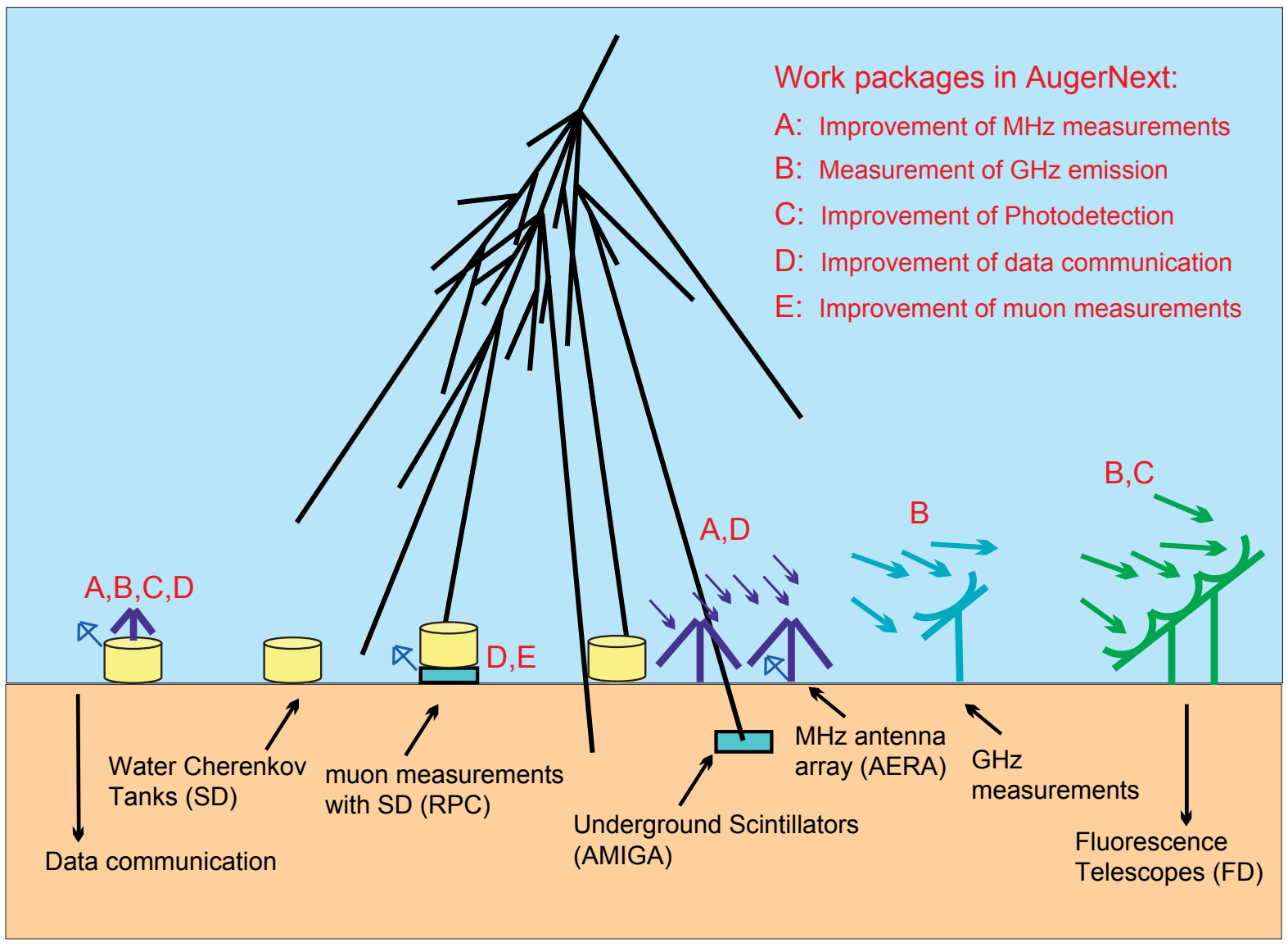

**Figure 1:** Schematic view of the detection of extensive air showers with a Pierre Auger Observatory-like experiment and the connection of the work packages proposed in the AugerNext project.

Collaboration a consortium was formed to propose studies for a next-generation cosmic-ray experiment and the utilization of new innovative detection methods. AugerNext aims to significantly strengthen R&D efforts in specific areas. Fig.1 sketches the topics explored in the context of the existing Pierre Auger Observatory located in Argentina. These topics were chosen by available expertise of the consortium partners and are defined as work packages where specific five-year goals have been formulated, all of a highly innovative character. The work will be continued by support on an institutional level as best as possible.

## 2. The work packages and results so far

**(A) Investigation of MHz radio emission in air showers** The general goal for the 5 year project was to establish the detection method as a hybrid or even standalone technique, and in particular to estimate the resolution and sensitivities of the technique to energy, mass, and arrival direction of the cosmic ray. The hope is that radio can replace the fluorescence detection technique with similar reconstruction quality on primary parameters, but with significantly higher duty cycle. Within the AugerNext project specific topics have been worked on:

1. A hybrid demonstrator was set up (i.e., surface detector including integrated radio antenna) for common trigger and combined analysis of the complementary information at individual surface detectors. This approach named EASIER [13] could reveal lateral distributions of the radio signal with a very large distance between the antennas and could give the proof-of-concept for the particle-radio hybrid technique.
2. The Auger Engineering Radio Array (AERA) at the Pierre Auger Observatory has been set up in order to investigate the radio signal from air showers in detail [16]. The radio emission gives information complementary to particle detectors as the pure electromagnetic component is selected. Self-trigger has been proven, but is difficult due to radio frequency interference (RFI) faking the radio shower signal. The hardware used in AERA (Fig. 2) was and is still being optimized for large-scale applications, where further cost reduction is an important issue. Radio detectors can be built in a cost effective way, but this depends on two main aspects, the cost of an individual radio detector station and the needed density of these sensors. A caveat of the radio technique lies in the relatively small footprint of the detectable signal for vertically arriving air showers. However, the footprint increases drastically when more inclined events are measured.

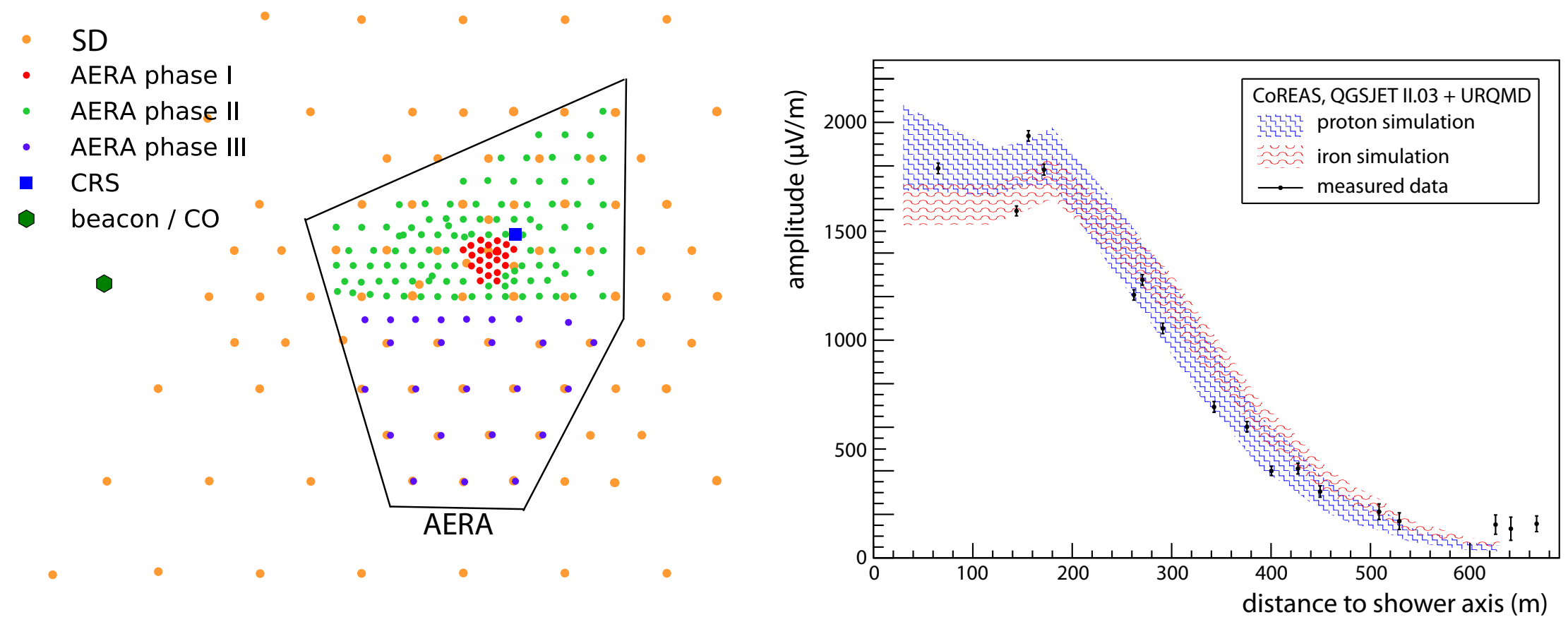

**Figure 2:** Map of the Auger Engineering Radio Arrays. AERA consists of 153 antenna stations and covers an area of 17 $\mathrm{km}^2$ (left panel). Right: lateral distribution of the radio signal of one of the high-energy events measured by AERA in comparisons with detailed simulations for proton or iron primaries (CoREAS [14] in this case, but ZHAires [15] leads to similar results). This shows that even complicated emission patterns are understood and reproduced by the sophisticated simulations.

3. The AERA array was extended to be sensitive also to horizontally arriving air showers. With radio the electromagnetic component can still be measured for those showers, where particle detectors see the muon component only.

The data of AERA and other radio experiments in combination with sophisticated simulation tools have led to a deep understanding of the radio emission in air showers [17]. Recent data analyses [18] have shown that, with the radio signal, a very good angular resolution as well as an energy resolution of better than 20% can be reached. In addition, the lateral distribution and the form of the wavefront of the radio signal are sensitive to the longitudinal shower development. This allows the determination of the position of the shower maximum and, therefore, a reconstruction of the elemental composition. AERA, covering the largest area and highest energies in radio hybrid experiments can investigate the emission mechanisms via detailed polarization measurements. Two radio emission mechanisms, the geomagnetic and the Askaryan effects, contribute mainly to the radio signal from air showers. For individual showers AERA can determine the contribution of the Askaryan mechanism to the total signal to 14% [19] on average for the measured events.

**(B) Detection of the microwave emission in air showers** P. W. Gorham et al. [20] proposed in 2008 the possible observation of extensive air showers in the microwave band. This idea was motivated by the observation of a signal in the 1.5-6 GHz band during the passage of an electron beam in an anechoic chamber. This recorded signal was interpreted as molecular Bremsstrahlung radiation (MBR), i.e., an emission produced by low-energy electrons scattering on the atmospheric molecules. Radio emission produced by this mechanism presents interesting features: it is expected to be isotropic and unpolarized, with a very low natural background. If true, its main advantage is the possibility of instrumenting a large area with 100% duty cycle with a negligibly small atmospheric attenuation, using relatively cheap equipment.

One defined five-year goal in AugerNext concerned the question of whether this technique provides the possibility of measuring the primary parameters of high-energy cosmic rays.

1. A hybrid demonstrator called EASIER was set up (i.e., Auger tanks including integrated horn antennas). A few events have been detected [21], and a detailed analysis has been performed of a possible contribution to the measured signals by the molecular bremsstrahlung effect.
2. A non-imaging detection system, called CROME, was externally triggered by the KASCADE-Grande detector. The CROME setup was well-suited for the detection of pulses of a few nanoseconds duration as expected for cosmic-ray showers. For more than 30 air-shower events a signal in the Gigahertz range could be identified in coincidence with KASCADE-Grande. The cores of these showers are distributed in a ring around the detectors, hinting to an emission which is boosted in the forward direction (Fig. 3). Applying simulations developed for the MHz frequency range was able to confirm the pattern and the polarization of the measured radiation. An isotropic, unpolarized component of the signal (which would hint to the predicted molecular Bremsstrahlung radiation) is still not excluded, but definitely smaller than expected by Gorham et al. in 2008. Therefore, the measurement of the Gigahertz emission in air showers seems not to give a reasonable alternative to the fluorescence measurement for large-scale cosmic-ray experiments [22]. The setup measuring simultaneously in the L-band could not detect a shower signal, mainly due to the increased noise background in this frequency band [23].
3. The feasibility of detecting extensive air showers (EAS) by a radar technique has been investigated. It is based on the observation of the radio waves scattered off the short-lived plasma produced in the atmosphere by the high-energy particles of the shower. Simulations show that the signal received by the detector is strongly dependent on the geometry of the detection system. Moreover, the strength of the received signal depends on the frequency of the emitted radar wave. The crucial point is to choose the optimum combination of the radar system parameters. Based on the first results, however, no high sensitivity will be reached by any reasonable equipment [24].
4. The aim of FDWave is to develop a microwave telescope equipped with a matrix of radio receivers looking to a different part of the sky. The idea is to install GHz sensors in two fluorescence telescopes (FD) of the Auger Observatory. The optimal antenna that could be placed in the FD camera is a Low Noise Block working in the Ku band (11 GHz). A detailed simulation of the detector equipped with this radio sensors shows that a good effective aperture of the telescope can be attained. Of course the feasibility of this new detection technique depends on the emission intensity and the degree of coherence of the underlying process. The latter has been studied by the AMY experiment using the 510 MeV electron beam at the Beam Test Facility of Frascati INFN National Laboratories. The MBR has been studied in a wide frequency range, between 1 and 20 GHz, within a huge (2 m x 2 m x 4 m) anechoic Faraday chamber. Contrary to what has been reported in [20], the signal has been found very prompt in time like the Cherenkov radiation and it has been demonstrated that the observed coherence is caused by the particular time structure of the LINAC beam [25]. A preliminary analysis of the AMY data shows that the intensity of the observed signal is significantly less than the one reported in [20].

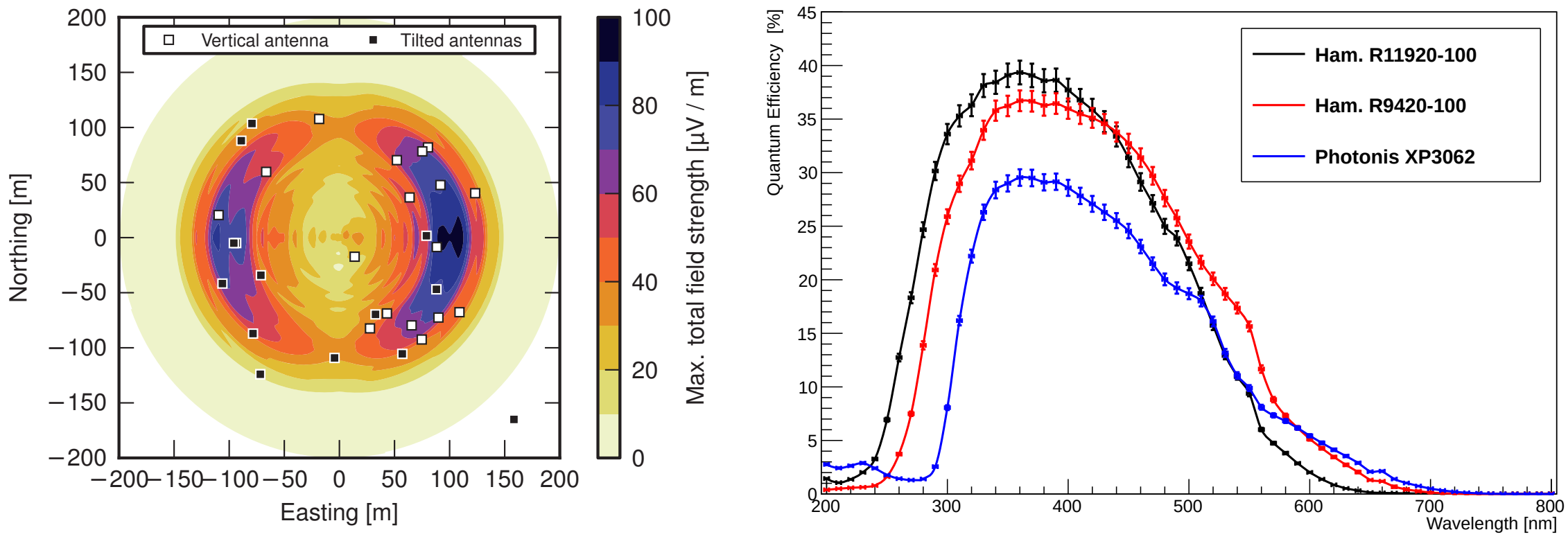

**Figure 3:** Left: Distribution of the cores of the detected events by CROME relative to antenna positions. The underlying color distribution is the footprint of the C-band (3.4-4.0 GHz) of an iron-induced cosmic-ray shower predicted by a CoREAS simulation [22]. Right: Quantum efficiencies measured in the laboratory as a function of wavelength for different types of photomultiplier tubes [26].

**(C) Improvement of photo sensors** There are structured activities worldwide to improve photo detectors. These concern vacuum photomultiplier tubes as well as the development of silicon photomultipliers (SiPM). One objective in this overall package is the development of a new generation of focal planes for cosmic-ray air-fluorescence and Cherenkov imaging, achieving better light detection efficiency and a better spatial and angular resolution. Therefore, within AugerNext we want to test the latest developments, in order to study their capability for a next-generation experiment.

1. After many years of not very significant advances in quality parameters of vacuum photomultiplier tubes, important progress has been made very recently with photo cathodes becoming available with improved quantum efficiencies. Super-bialkali (SBA) photo cathodes reach quantum efficiencies (QE) of about 35%, while ultra-bialkali (UBA) ones reach QEs of even up to 43% (Fig. 3, right panel). The aim is to gain experience with devices in real environments and to cooperate with manufacturers for improving their overall characteristics. High quantum efficiency PMTs are tested at the Pierre Auger Observatory in both fluorescence telescopes and surface detectors [26].
2. A SiPM focal-plane element for the FD (FAMOUS, see Fig. 4) was designed, and a first prototype built and tested, as a prototype fluorescence telescope with a special light collecting optical system of Winston cones to increase the sensitive area [27]. The experience in photodetection with SiPMs is also the basis of tests aiming at the potential implementation of those devices as a new readout system for the upgraded detectors.

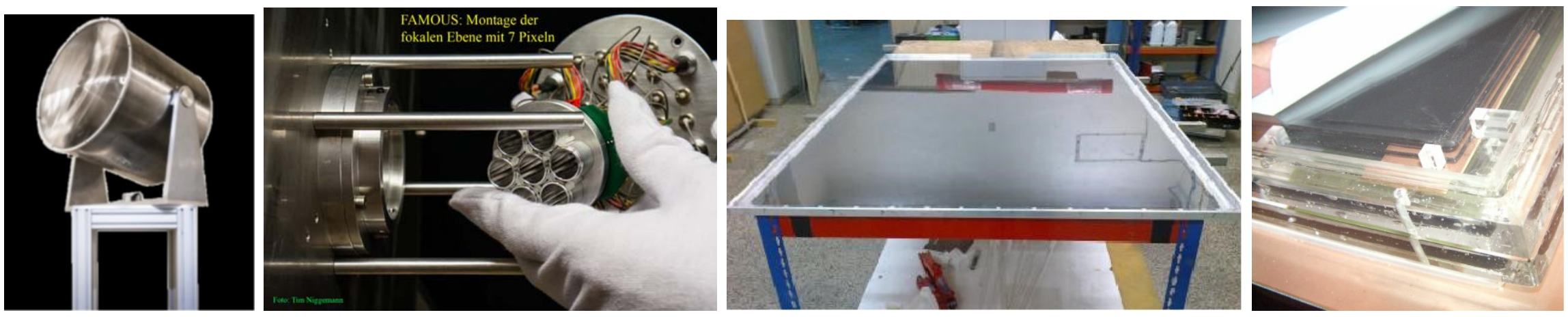

**Figure 4:** From left to right: photo of the compact FAMOUS prototype camera with the Fresnel lense and of the installation of a 7-pixel camera to FAMOUS, the overall photo of the RPC engineering prototype and a detail of the RPC structure.

**(D) Generalization of the data communication system** A crucial aspect in large arrays is the data transfer. An advanced technique has to be developed or adapted from commercial applications (e.g., mobile phone technique). Autonomous detector stations in large-scale air-shower arrays will be part of a distributed sensor network connecting all local stations to a central data-storage unit. Within the project we have developed new strategies for data communication systems with flexible application possibilities for large-scale remote-detector arrays. We compared different approaches in both custom-made in-house or commercially-provided systems to reach a general, worldwide applicable remote-controlled communication system. In particular:

1. One investigation focused on the adoption of a commercial, self-healing communication systems implemented in AERA. We procured a commercial system of 160 radio-communication antennas. They have been installed and tested for their performance with respect to data rates and systems reliability. A high-bandwidth 802.11n commercial wireless system in the 5.8 GHz frequency band from Ubiquiti is used [28]. The system has shown within AERA that it provides a stable operation and with some additional investment can easily be scaled to larger applications.
2. Detailed simulation studies are performed to develop suitable network topologies. Due to the structure of the communications network which interconnects the detectors of a large observatory, a mesh topology has been considered and a gossip-type communication protocol has been analyzed. The performances are analyzed and compared with the ones obtained from the measurements.
3. A new, custom-made communication system in the 2 GHz range was developed which can transfer data over distances of 3 to 6 km with low power consumption. A long-term test was performed to investigate the stability, availability, failure rates, external pollution signals, etc, as well as to optimize the required signal strength at the sender and the receiver. The system was working, but was ultimately abandoned due to the success of the cheaper commercial system.

**(E) Studies for a hybrid muon detector** The objective in this work package is to show that muon detectors based on the Resistive Plate Chamber (RPC) technique (see Fig. 4) can operate under field conditions, i.e., demanding a low energy budget, low cost per unit area, and mechanical toughness. In particular, we want to explore the capabilities to measure new shower observables, or to improve the resolution of actual observables, namely sensitivity to the details of the hadronic cascade through muons. RPCs are fast and have a very good time resolution which could nicely be used to reconstruct single particle tracks and the longitudinal evolution of muons in a shower. In a hybrid operation with SD tanks this can considerably improve the composition sensitivity of a future detector. Within AugerNext we produced and tested with a first prototype the capability and robustness of RPC-based detectors for large-scale applications. We showed that RPCs can serve as a large, cheap, and robust timing detector. A first prototype (including readout electronics) is deployed in Argentina, where the data is presently being analyzed [29].

**Acknowledgement** Besides the Pierre Auger Collaboration, providing the scientific and technical environment for the AugerNext project, and besides the engaged students working on the various problems, the AugerNext consortium acknowledges the funding agencies supporting the project. Namely, Bundesministerium für Bildung und Forschung (BMBF 05A2011), Germany; Nikhef, Netherlands; The National Centre for Research and Development (NCBiR), Poland; Centre de Calcul IN2P3/CNRS, France; Fundação para a Ciência e a Tecnologia, Portugal; Ministerio de Ciencia e Innovación and Consolider-Ingenio 2010 (CPAN), Spain; Istituto Nazionale di Fisica Nucleare (INFN), Italy; and Romanian Authority for Scientific Reseach, UEFICDI, Ctr.Nr.1/ASPERA2 ERA-NET, Romania.

# 7

# Outreach

# Education, Outreach and Public Relations of the Pierre Auger Observatory

**Charles Timmermans*[a] for the Pierre Auger Collaboration[b]**
[a] *Nikhef and Radboud University, Nijmegen, the Netherlands*
[b] *Observatorio Pierre Auger, Av. San Martín Norte 304, 5613 Malargüe, Argentina*
*E-mail:* auger_spokespersons@fnal.gov
*Full author list:* http://www.auger.org/archive/authors_2015_06.html

The scale and scope of the physics studied at the Pierre Auger Observatory continue to offer significant opportunities for original outreach work. Education, outreach and public relations of the Auger Collaboration are coordinated in a dedicated task whose goals are to encourage and support a wide range of efforts that link schools and the public with the Auger scientists and the science of cosmic rays, particle physics, and associated technologies. We focus on the impact of the Collaboration in Mendoza Province, Argentina and beyond. The Auger Visitor Center in Malargüe has hosted over 95,000 visitors since 2001, and a fifth Collaboration-sponsored science fair was held on the Observatory campus in November 2014. The Rural Schools Program, which is run by Observatory staff and which brings cosmic-ray science and infrastructure improvements to remote schools, continues to broaden its reach. Numerous online resources, video documentaries, and animations of extensive air showers have been created for wide public release. Increasingly, collaborators draw on these resources to develop Auger related displays and outreach events at their institutions and in public settings to disseminate the science and successes of the Observatory worldwide. We also highlight education and outreach activities associated with the planned upgrade of the Observatory's detector systems and future physics goals.

*Speaker.

## 1. Introduction

Education and public outreach (EPO) have been an integral part of the Pierre Auger Observatory since its inception. The collaboration's EPO activities are organized in a separate Education and Outreach Task that was established in 1997. With the Observatory headquarters located in the remote city of Malargüe, population 28,000, early outreach activities, which included public talks, visits to schools, and courses for science teachers and students, were aimed at familiarizing the local population with the science of the Observatory and the presence of the large Collaboration of international scientists in the isolated communities and countryside of Mendoza Province. As an example of the Observatory's integration into local traditions, the Collaboration has participated in the annual Malargüe Day parade since 2001 with collaborators marching behind a large Auger banner (see Fig. 1). Close contact with the community fosters a sense of ownership and being a part of our scientific mission. The Observatory's EPO efforts have been documented in previous ICRC contributions [1, 2]. We report here highlights of recent activities, as well as plans for future activities related to the upgrade of the Observatory.

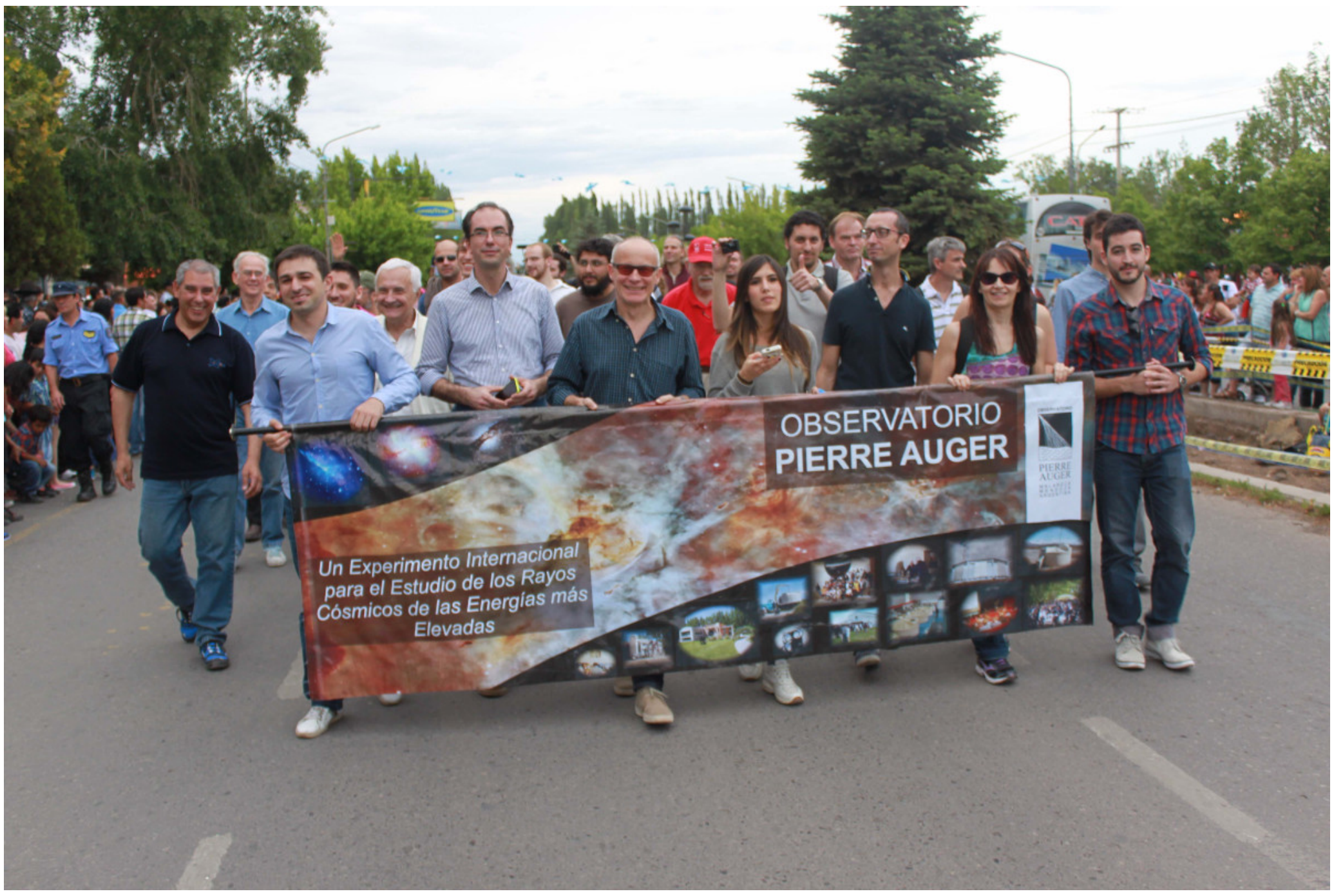

**Figure 1:** Auger collaborators participating in the November 2014 Malargüe Day parade.

## 2. The Auger Visitor Center in Malargüe

The Auger Visitor Center (VC), located in the central office complex in Malargüe, continues to be a popular attraction. Through February 22, 2015, the VC has hosted 93,475 visitors. Fig. 2 shows the integrated number of visitors since Nov. 2001. The noticeable increase of visitors since 2008 occurred after the opening of a nearby planetarium [3] in August of that year. The VC is managed by a small staff led by an Observatory employee; they share the task of giving presentations and tours to visitors and school groups. The upgrade of the Observatory provides an opportunity to modernize the local visitor center, and thereby re-inviting people who have seen the Observatory in the past to relive this experience. It shows a genuine interest of the Collaboration to communicate with the people from Malargüe about the science and tools of the Observatory. The modernization of the center will enable visitors to tour the exhibit on their own. Audio and video displays explain the Observatory. Several objects and 3D models will help to understand the different topics. At the same time, the setup of the VC will be flexible enough to allow for lectures to complete school classes. A possible new layout is shown in Fig. 3.

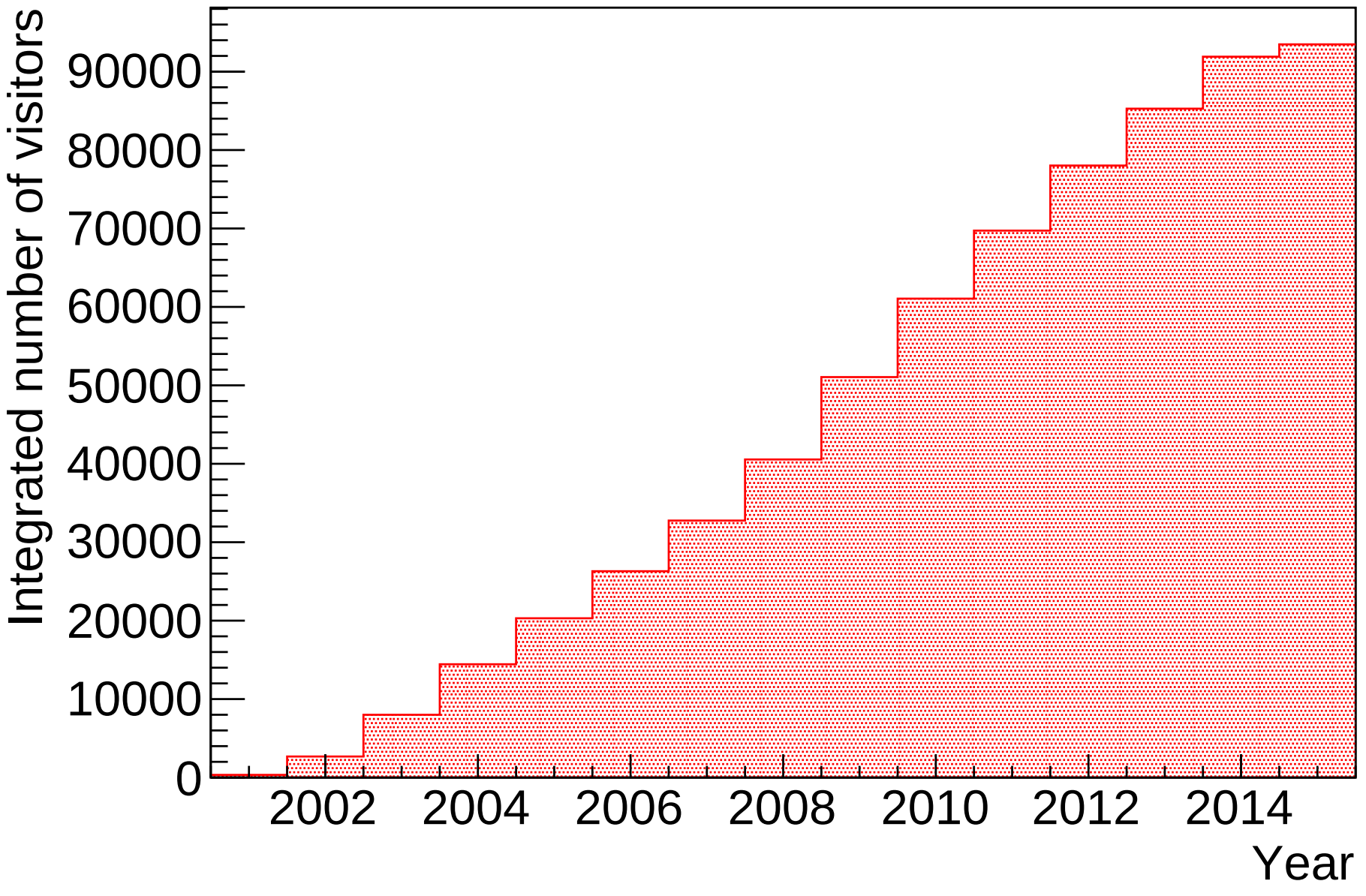

**Figure 2:** The integrated number of visitors logged at the Auger Visitor Center since 2001.

## 3. The Rural Schools Program and Education Fund

The Rural Schools Program, initiated by the Observatory staff who volunteer their time, continues to bring information about the Observatory and needed infrastructure improvements directly to remote schools that otherwise have difficulty exposing their students to the Observatory. The Rural Schools Program is supported by an Education Fund managed by the Observatory staff who collect voluntary financial contributions from collaborating institutions and individuals.

## 4. The 2014 Auger Science Fair

The Observatory hosted its fifth biannual Science Fair in the Assembly Building on November 19-21, 2014, as shown in Fig. 4. Thirty-three student teams from all over Mendoza Province, with ages ranging from primary school through high school, presented research projects in the areas of natural science, exact science, and technology. More than 30 Auger collaborators, from different nationalities, and a few invitees served as judges for the student projects. Prizes were awarded to the top teams in several categories in the closing ceremony on November 21. The November 2014 Science Fair owes its success to the Observatory staff, the collaborators who served as judges, the Municipality of Malargüe, the participating teachers and students, and special mention goes to the lead local organizers: Miguel Herrera, Fabian Amaya, and Alicia Piastrellini.

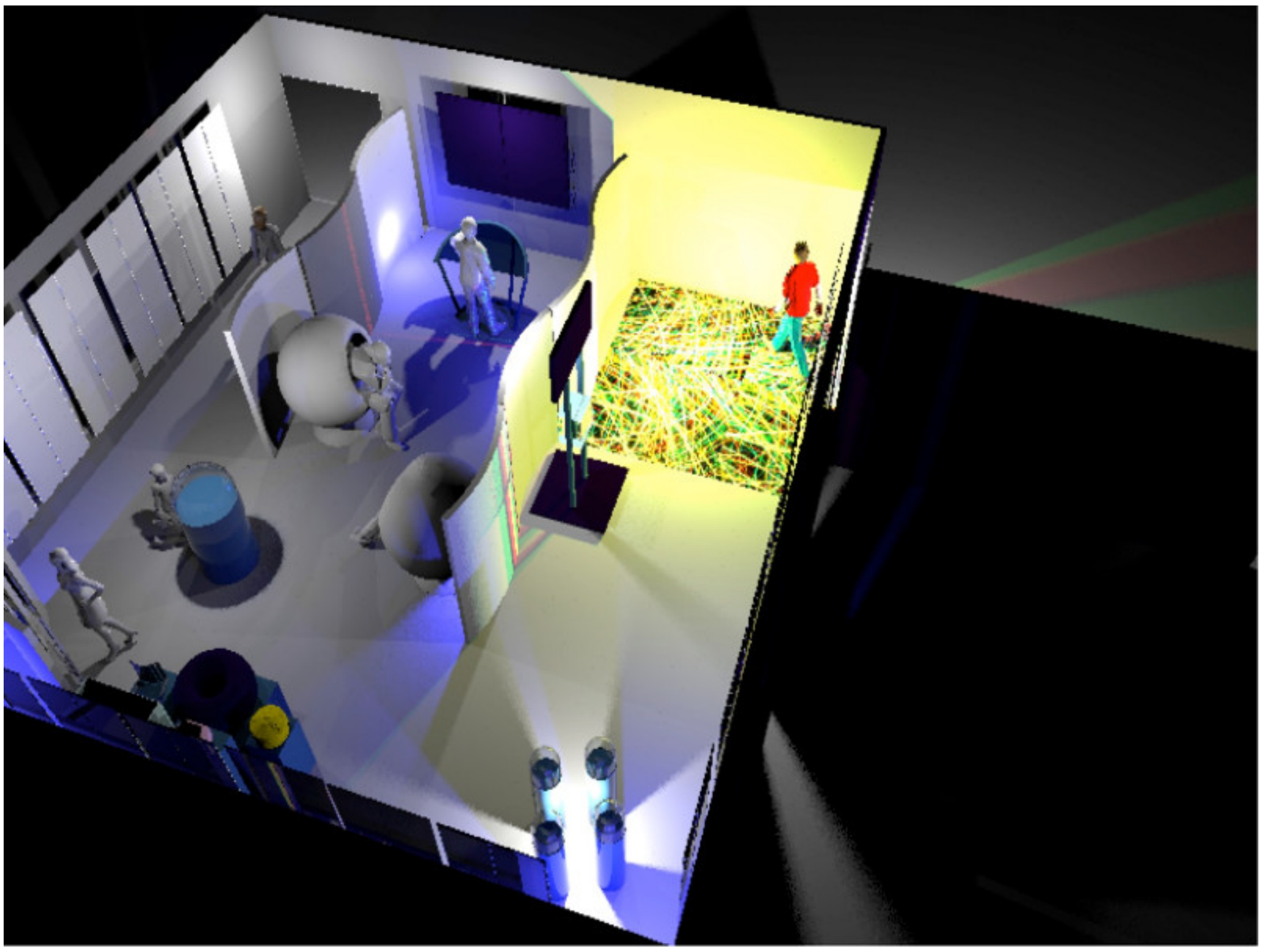

**Figure 3:** Impression of a possible upgraded Auger Visitor Center in Malargüe.

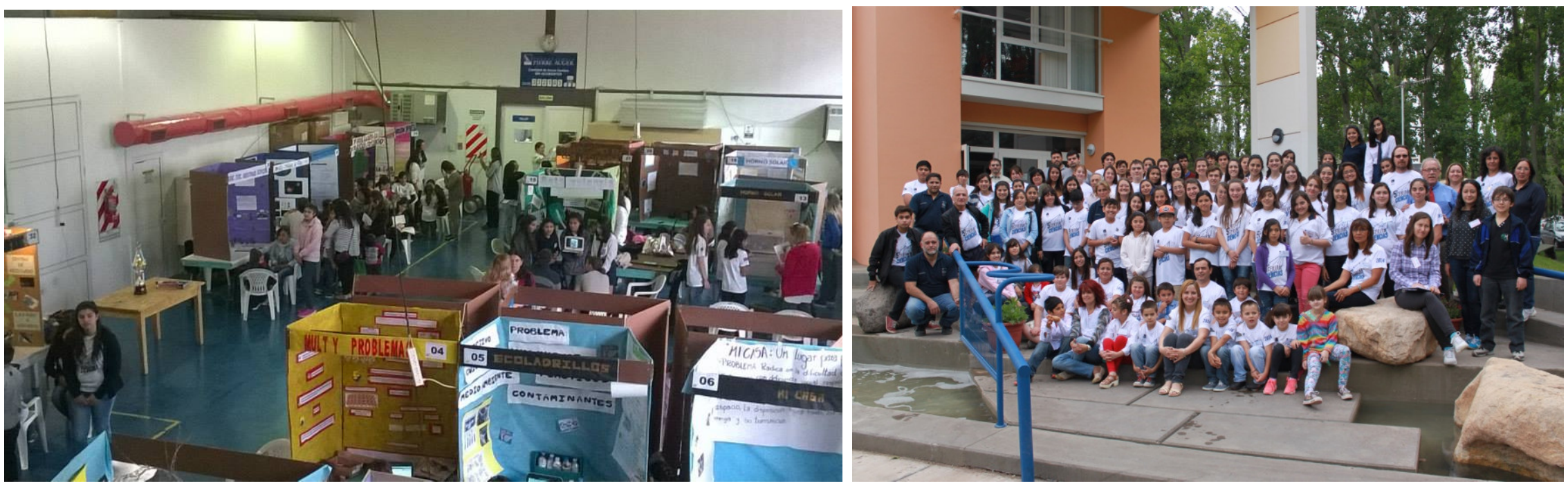

**Figure 4:** Left: A photo of the 2014 Science Fair. Right: The participants of the Fair.

## 5. Public Event Display and Data for Outreach Purposes

The Auger public event display allows the general public to see what information is recorded from an incoming air shower, and grasp the steps involved into obtaining information on the incoming cosmic ray creating this air shower. Until now, 1 % of the data has been available for outreach purposes [4]. However, the Collaboration has committed to increasing this fraction to 10 % of the data without an upper limit on the event energy. This widens the target group for outreach purposes to include even university students for whom statistical tools and analysis packages such as VISPA [5, 6, 7, 8] are available to handle substantial amounts of data. At the same time, high school students are able to make online selections on the events they would like to see or use for their own purposes thereby reducing the data volume while increasing the number of interesting events substantially with respect to the current situation.

## 6. Selected Outreach Activities Outside Malargüe

### 6.1 VISPA

With the VISPA internet platform [5, 6, 7, 8], physics analysis can be performed in a web browser without the need of any software installation. On the start page of the VISPA platform, an overview of different physic examples is presented as shown in the screenshot of Fig. 5. The examples can directly be executed and the analysis code can be modified or extended by the user.

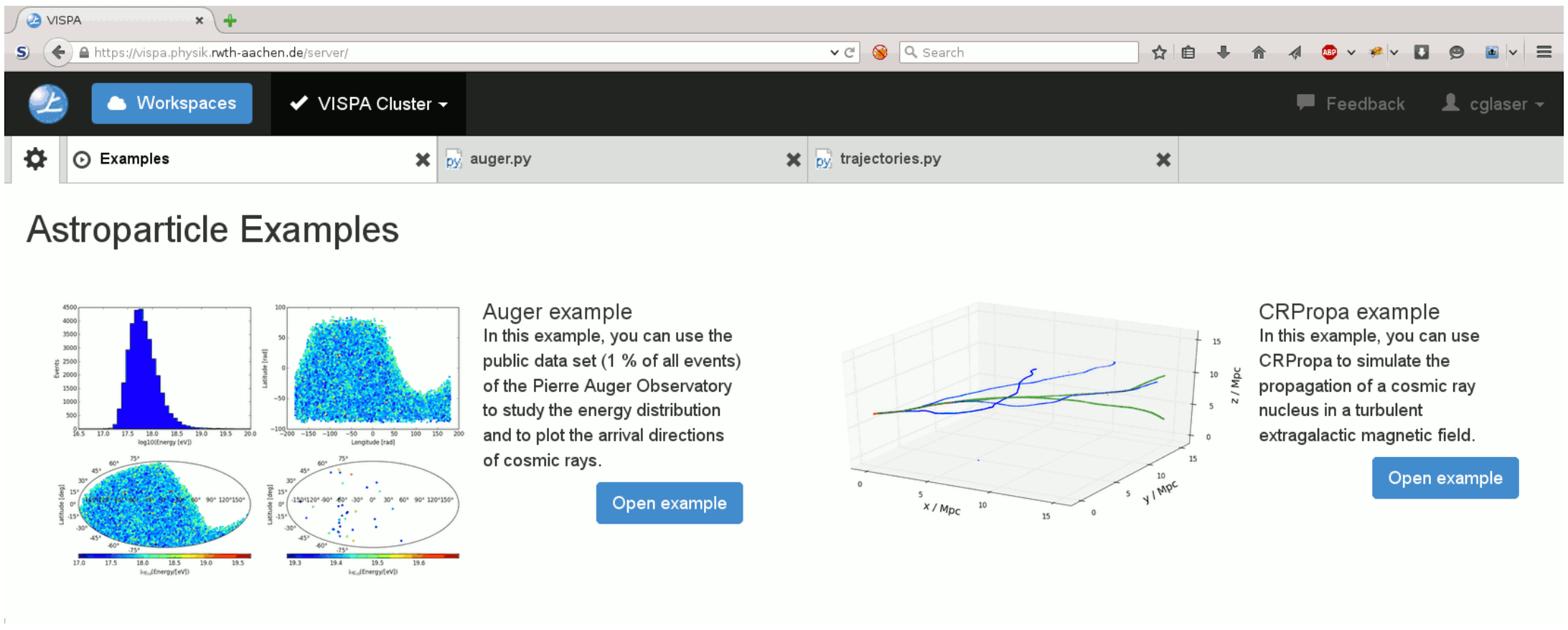

**Figure 5:** Left: Examples using the Auger public data set. Right: Simulations using the CRPropa package.

In the analysis of the Auger public data set, the energy distribution and the arrival directions of cosmic rays at the highest energies are visualized. Furthermore, the propagation of cosmic rays through the universe can be simulated using the CRPropa software package [9]. It reveals that the propagation of UHECRs is essential to understand the data distributions and to find the sources of cosmic rays.

### 6.2 Outreach Activities in Argentina

The Pierre Auger Observatory has played a prominent role in temporary exhibits in Mendoza (La Brújula, "The Compass" [10]) and Technopolis in Buenos Aires [11]. These exhibits attract in total several million visitors, and are important events in Argentina that promote science for the general public. In addition to these temporary exhibits, the Observatory is permanently present in the planetarium in Buenos Aires [12], as can be seen in Fig. 6.

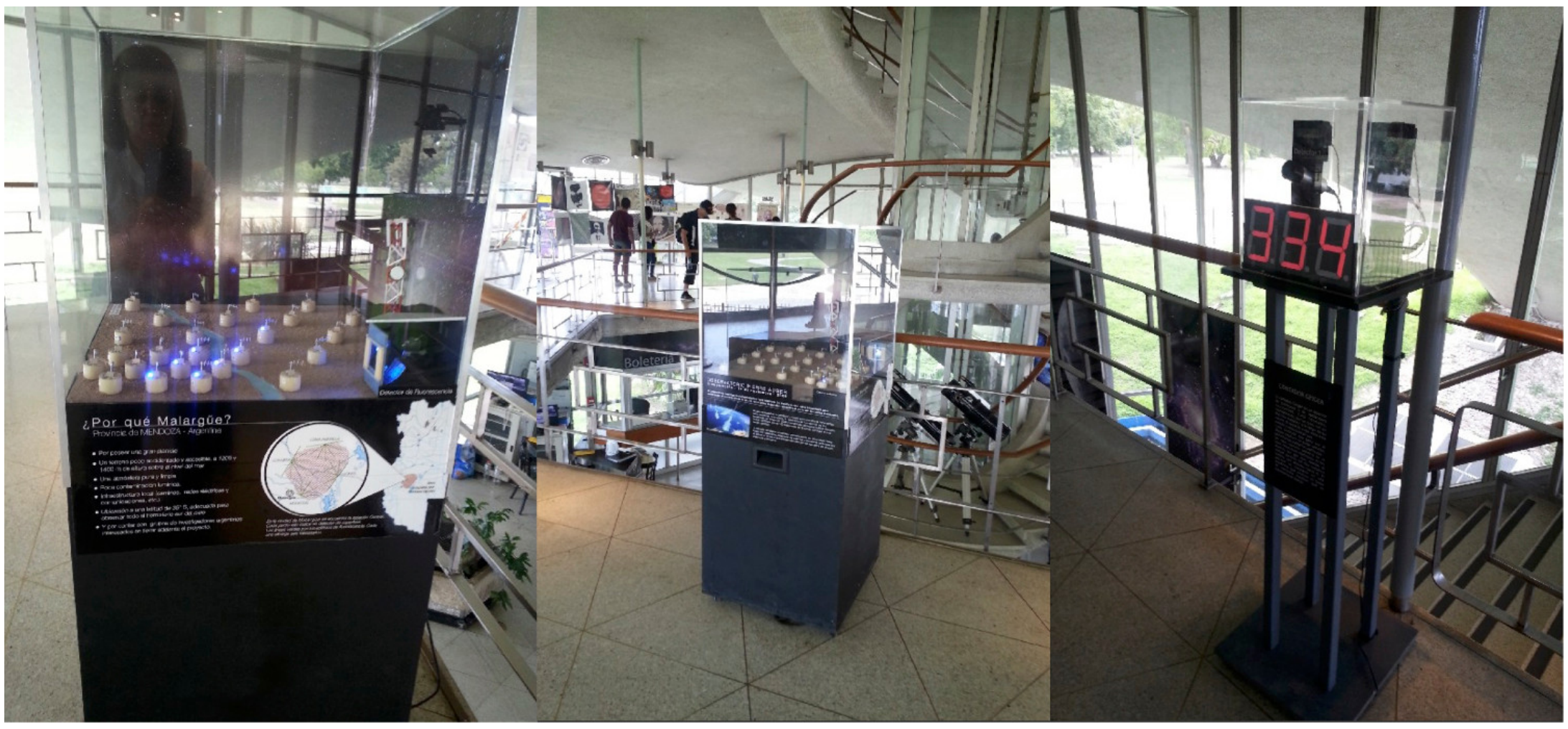

**Figure 6:** Permanent setup of the Pierre Auger Observatory at the planetarium in Buenos Aires.

### 6.3 Outreach Activities in Romania

The popular European Space Expo, an initiative of the European Commission, made a stop in Craiova, Romania, April 19-27, 2014 [13], and attracted over 50,000 visitors. The Expo highlights numerous European space programs using interactive touch-screen displays and attractive signage, providing information on space science to the general public in an accessible way. Auger Collaborators from the Institute of Space Science (ISS) in Bucharest have given several presentations about the Pierre Auger Observatory during the Expo's visit, as shown in Fig. 7. Furthermore, this group provided a featured speaker who discussed the Pierre Auger Observatory during the European Researchers Night [14] which takes place each year simultaneously in many European cities. The September 24, 2014, event in Bucharest was held in a public park and attracted hundreds of people.

During the ISS open house event of 2015 [15], which targets young students and teachers who want to learn about the Institute's scientific and education programs, the Pierre Auger Observatory was prominently featured by our collaborators.

These activities have led to several interviews on radio and television about the Observatory and the upgrade plans for the Observatory [16, 17], reaching wide audiences throughout the country. The ISS group recently released an informative YouTube video [18] about the Central Raman Lidar Facility at the Observatory.

## 7. Conclusions

The Pierre Auger Observatory continues to provide unique education and outreach opportunities which expose people of all ages to the excitement of astroparticle physics. Its Visitor Center,

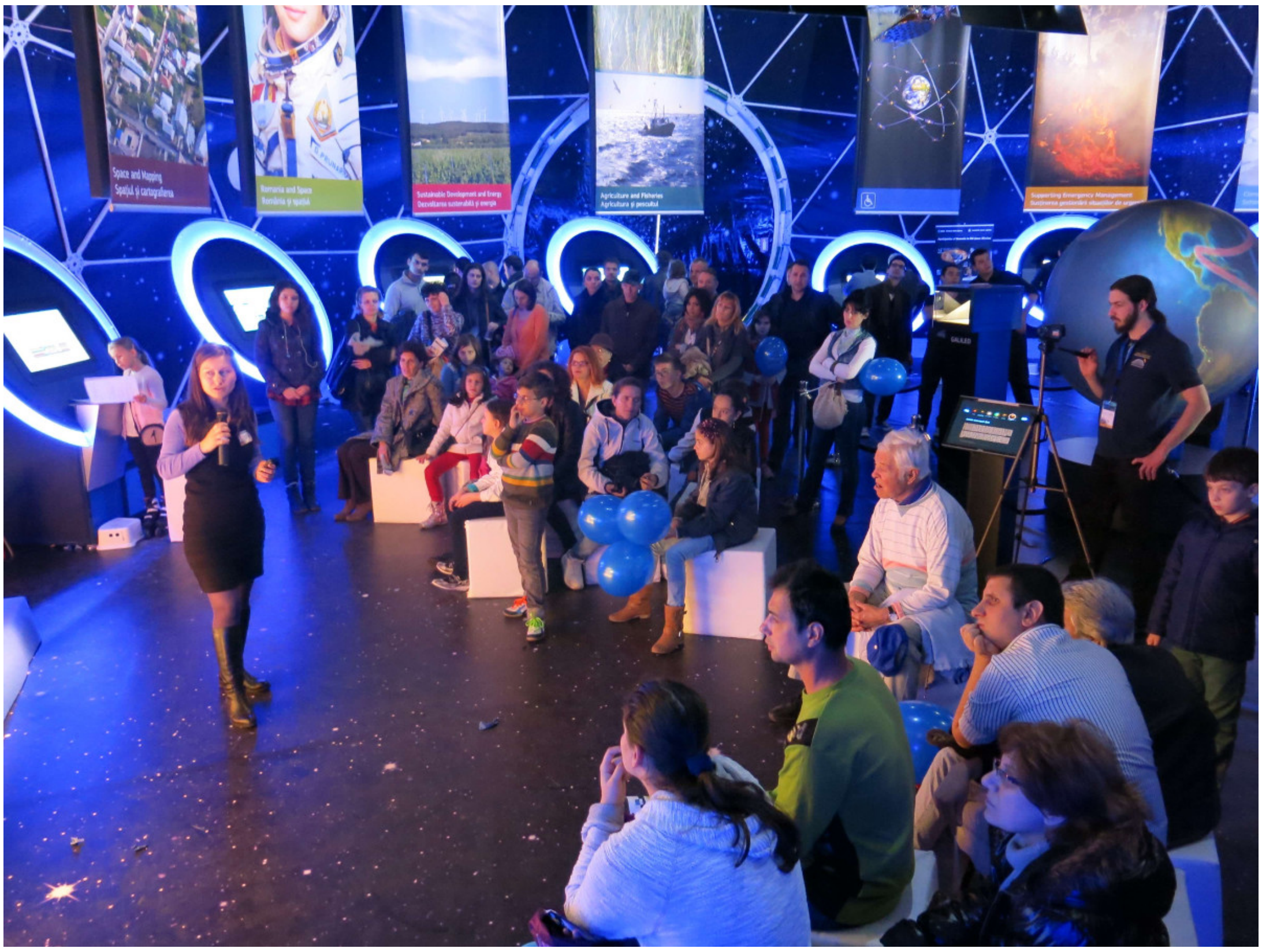

**Figure 7:** Romanian collaborator P. Gina Isar speaking about the Pierre Auger Observatory at the European Space Expo in Craiova in April 2014.

Rural Schools Program, and Science Fairs have great local impact near Malargüe, while collaborators around the world ensure that the Observatory's science and successes have international reach. The planned upgrade of the Observatory provides an excellent opportunity to modernize the Visitor Center and discuss the physics of the upgrade on the internet and in outreach events around the globe.